\documentclass[10pt]{article}
\usepackage{amsmath,amssymb,graphicx}
\usepackage{amsthm}
\usepackage{euscript}
\usepackage{eufrak}

\newtheorem{defn}{Definition}
\newtheorem{lem}{Lemma}
\newtheorem{thm}{Theorem}
\newtheorem{prop}{Proposition}

\newcommand{\pr}{\noindent{\bf Proof}. }
\newcommand{\rem}{\noindent{\bf Remark}. }
\newcommand{\rems}{\noindent{\bf Remarks}. }

\newcommand{\da}{\dagger}
\newcommand{\pa}{\partial}
\newcommand{\one}{\cO(1)}
\newcommand{\bpsi}{\bar \psi}
\newcommand{\bPsi}{\bar \Psi}

\newcommand{\hs}{ \hspace{1cm}}
\newcommand{\tr}{\textrm{ tr }}

\newcommand{\loc}{ \textrm{loc}}
\newcommand{\Vol}{\textrm{Vol}}
\newcommand{\nat}{\natural}
\newcommand{\tk}{\bbT^{-k}_{N -k}}
\newcommand{\tz}{\bbT^0_{N -k}}

\newcommand{\im}{\textrm{Im }  }
\newcommand{\re}{\textrm{Re } } 

\newcommand{\B}{\Big}
\newcommand{\blan}{\Big  \langle} 
\newcommand{\bran}{\Big  \rangle}

\newcommand{\sq}{\square}
\newcommand{\diam}{ \textrm{ diam }}
\newcommand{\crit}{\textrm{crit}}

\newcommand{\be}{\begin{equation}}
\newcommand{\ee}{\end{equation}}
\newcommand{\bs}{\begin{split}}
\newcommand{\es}{\end{split}}

\newcommand{\bom}{\mathbf{\Omega}}
\newcommand{\bla}{\mathbf{\Lambda}}
\newcommand{\bpi}{\mathbf{\Pi}}

\newcommand{\bb}{\mathbf{b}}
\newcommand{\bh}{\mathbf{h}}

\newcommand{\ud}{\underline}

\newcommand{\sZ}{\mathsf{Z}}

\newcommand{\sx}{\mathsf{x}}
\newcommand{\sy}{\mathsf{y}}

\newcommand{\al}{\alpha}
\newcommand{\De}{\Delta}
\newcommand{\de}{\delta}
\newcommand{\ga}{\gamma}
\newcommand{\Ga}{\Gamma}
\newcommand{\ka}{\kappa}
\newcommand{\La}{\Lambda}

\newcommand{\Om}{\Omega}
\newcommand{\om}{\omega}

\newcommand{\ep}{\epsilon}
\newcommand{\si}{\sigma}

\newcommand{\Up}{\Upsilon}
\newcommand{\vep}{\varepsilon}

\newcommand{\st}{\star}

\newcommand{\cA}{ \EuScript{A} }
\newcommand{\fD}{\EuFrak{D}} 
 
\newcommand{\fS}{\EuFrak{S}} 
\newcommand{\fH}{\EuFrak{H}} 

\newcommand{\fG}{\EuFrak{G}}

\newcommand{\cB}{{\cal B}}
\newcommand{\cC}{{\cal C}}
\newcommand{\cD}{{\cal D}}

\newcommand{\cO}{{\cal O}}
\newcommand{\cI}{{\cal I}}
\newcommand{\cH}{{\cal H}}
\newcommand{\cS}{{\cal S}}
\newcommand{\cE}{{\cal E}}
\newcommand{\cR}{{\cal R}}

\newcommand{\cF}{{\cal F}}

\newcommand{\cG}{{\cal G}}
\newcommand{\cM}{{\cal M}}

\newcommand{\cW}{{\cal W}}
\newcommand{\cQ}{{\cal Q}}
\newcommand{\cL}{{\cal L}}

\newcommand{\cJ}{{\cal J}}
\newcommand{\cZ}{{\cal Z}}

\newcommand{\bbZ}{{\mathbb{Z}}}

\newcommand{\bbT}{{\mathbb{T}}}

\newcommand{\bbI}{{\mathbb{I}}}

\begin{document} 
\title{Ultraviolet   stability   for     QED in d=3}
\author{ 
J. Dimock
\thanks{dimock@buffalo.edu}\\
Dept. of Mathematics \\
SUNY at Buffalo \\
Buffalo, NY 14260 }
\maketitle

\begin{abstract}
We   continue    the   study of the ultraviolet problem  for  QED in d=3    using Balaban's formulation of the renormalization group.
The model is defined on a fine   toroidal  lattice and  we  seek  control as the lattice spacing goes to zero.
Drawing on earlier papers in the series the  renormalization group flow is completely controlled   for weak coupling.   
  The  main result is   an ultraviolet stability bound in a fixed finite volume.
   \end{abstract}

 \section{Introduction }
 
 This is the third  paper in a series in which 
 we study the ultraviolet problem for  quantum electrodynamics (QED)  on   a finite  Euclidean space-time of dimension  $d=3$.
 The method is a renormalization group analysis  due to Balaban and collaborators
   featuring block averaging  \cite{Bal82a} - \cite{BOS91}.     
 In the first   paper \cite{Dim15b}  the  renormalization group flow  was  controlled with a bounded field approximation. 
In the second paper \cite{Dim20} 
a number of technical results were developed in preparation for control of the large field corrections. 
  In this  this paper  we gain    control over  the large field corrections and the overall renormalization group flow. 
 This leads to a proof of    an ultraviolet stability bound in a fixed finite volume.

 This paper should be read in conjunction with the papers   \cite{Dim15b}, \cite{Dim20}, and    we freely use the notation, definitions, and results
 therein; see also a guide to the notation in Appendix \ref{notation}.   Nevertheless here is  a brief orientation. 
 The model is initially defined  as a Euclidean functional integral  on  toroidal lattices    $  \bbT^{-N} _0  = ( L^{-N} \bbZ / \bbZ )^3$ with spacing $L^{-N}$
 and  unit volume, and 
with bare coupling constant $e$ and fermion mass $\bar m$.
 We seek control as $N \to \infty$.  However we immediately scale up to a  lattice   $  \bbT^0 _N  = ( \bbZ / L^N\bbZ )^3$ with dimension $L^N$
 and unit spacing.
 On this lattice the initial density  for a gauge field $A_0$ and a  Grassmann fermi field $\Psi_0, \bPsi_0$
 is
\be 
  \rho_0  (A_0, \Psi_0)  =  
  \exp  \B( - \frac12  \|  d A_0 \|^2    - \blan   \bPsi_0, ( \fD_{A_0}  + \bar m_0 )   \Psi_0 \bran  -   m_0  \blan  \bar \Psi_0,   \Psi_0   \bran +   \vep_0  \Vol (\bbT_N^0 )    \B) 
    \ee
  Here $dA_0$ is the field strength, $\fD_{A_0}$ is the covariant lattice Dirac operator with tiny  coupling constant $e_0 = L^{- \frac12 N} e$. The  $\bar m_0= L^{-N}\bar m$ is   the tiny  scaled bare mass, and $m_0$, $\vep_0$ are counterterms which will be chosen to  depend on $N$.   The ultraviolet problem has become an infrared problem with scaled 
  parameters. 
  The partition function is  (a gauge fixed version of)
  \be Z(N,e)  = \int   \rho_0  (A_0, \Psi_0) DA_0 D\Psi_0
  \ee
  The goal is to obtain bounds  independent of  $N$ above and below for the relative partition function of the form
  \be
  K_{-} \leq  \frac {Z(N,e)}{Z(N,0)} \leq K_+
  \ee

Starting with $ \rho_0  (A_0, \Psi_0) $   we    generate a sequence  of densities   $\rho_k( A_k, \Psi_k)$ with fields defined on the smaller lattices
 $\bbT^0_{N-k}=( \bbZ / L^{N-k}\bbZ )^3$ and which yield the same partition function.  Given  $\rho_k( A_k, \Psi_k)$ we generate  $\rho_{k+1}( A_{k+1}, \Psi_{k+1})$  in two steps.  First  
we apply a   block averaging transformation  and define for fields $ A_{k+1}, \Psi_{k+1}$ on the $L$-lattice $  \bbT^1 _{N-k}  = (L \bbZ / L^{N-k}\bbZ )^3$
 \be    \label{basic1} 
 \begin{split}
 &   \tilde \rho_{k+1} ( A_{k+1}, \Psi_{k+1}   )  \\
   & =      
 \int   D\Psi_k  DA_k  \ 
\de_G( \Psi_{k+1} - Q( \tilde \cA  ) \Psi_k )      \de(  A_{k+1} - \cQ A_k  )  \de (\tau A_k ) 
  \rho_{k}(A_k, \Psi_k)  \\
 \end{split}  
   \ee
  Here $Q( \tilde  \cA), \cQ$ average over cubes with side $L$,  $\tilde  \cA$ is a background field to be specified,    $\de_G$ is a Gaussian approximation to a delta function (see (\ref{deltafunction})), 
  and  $\de (\tau A_k )$ enforces axial gauge fixing in each $L$ block.  In the second step we scale down to fields $A_{k+1}, \Psi_{k+1}$  on    $  \bbT^0 _{N-k-1}  = ( \bbZ / L^{N-k-1}\bbZ )^3$
  by   defining $A_{k+1,L}(b) = L^{-\frac12}  A_{k+1}(L^{-1}b)$ and $  \Psi_{k+1,L}  (x) = L^{-1} \Psi_{k+1}(L^{-1}x)$ and 
    \be     \label{scaleit} 
   \rho_{k+1} ( A_{k+1}, \Psi_{k+1}   )   =   L^{-8 (s_N - s_{N-k-1} )  } \   L^{\frac12 (b_N- b_{N-k-1})   -\frac12 (s_N- s_{N-k-1}) }  \tilde \rho_{k+1} ( A_{k+1,L}, \Psi_{k+1,L}   )  
\ee
Here    $s_N$ is the number of sites  and $b_N$ is
the number of bonds in a lattice with $L^N$ sites on a side.

 The   effect is to simultaneously  reduce the number of degrees of freedom and impose a global gauge fixing.   Our overall goal is to show that
with a suitable choice of counterterms the flow of these densities can be controlled.   This is the content of theorem \ref{maintheorem}. 
Once the number of degrees of freedom have been reduced one integrates  and gets the partition function with good estimates.
This yields the ultraviolet stability  in theorem \ref{theorem2}.

\section{Statement of results} 

\subsection{general structure} 

To study  the densities $\rho_k$ a basic idea is that at each step we split the
expression  into large and small field regions and sum over all possible ways of doing this.   In the small field regions
the density can be written as the exponential of an action which can be renormalized and treated by perturbative means.   The large field
regions make a tiny contribution  which does not require renormalization and which are sufficient to control the sum over regions.

Our   first  main result says  that    after  $k$  steps  the 
density can be represented  on the lattice $\tz$ in the form
\begin{equation}     \label{startrep1}
\begin{split}
&  \rho_{k}(A_k,   \Psi_k)  =    \sum_{\bpi}   \sZ'_{k, \bom}(0)    \sZ'_{k, \bom} \  \int Dm_{k, \bom} (A )Dm_{k, \bpi} (Z )  Dm_{k, \bom}( \Psi)   Dm_{k, \bpi}( W)   \   \\
&     \cC_{k, \bpi} \     \chi_k( \La_k)        \exp \B( -\frac12 \| d \cA_{k, \bom}\|^2    -         \fS^+_{k, \bom}\B(\La_k, \cA, \Psi_{k,\bom},  \psi_{k, \bom} (\cA_{k, \bom})   \B )   
  +E_k(\La_k ) +  B_{k,\bpi}   \B)   \\
\end{split}
\end{equation}
 which we now proceed to explain in detail.    Actually  the statement is that at each step  $ \rho_{k}(A_k,   \Psi_k) $ can be modified to a density of this form. 
The modifications are required to preserve integrals over the fermi fields.  This ensures that we are still generating an expression for  the partition function when all integrations
are completed.  

\begin{itemize}

\item Many of the terms in this expression are characterized by reference to a running coupling constant $e_k$ given by 
\be
e_k = L^{\frac12 k}e_0 = L^{-\frac12(N-k) } e
\ee
The fact that this natural running coupling constant is  growing means that $e=0$ is a repelling fixed point which is what one wants for ultraviolet
renormalization  problems.   The fact that the growth is exponential is  characteristic of super-renormalizable models.  

\item  In this expression the sum is over  a sequence of regions $\bpi=(\Pi_1, \dots, \Pi_k ) $  where
\be    
\Pi_j   =  ( \Om_j,  \La_j; P_j, Q_j, R_j, U_j )
\ee
and  each entry  is  a union of  $L^{-(k-j)}M r_j$ cubes     in $\tk$   where
$ r_j \approx    (-\log  e_j)^r$  for a positive integer $r$;    more  precisely it  is the smallest power of $L$ greater  than or equal to  $  (-\log  e_j)^r$. 

\item
 The   $\Om_j$ are the basic  small field regions  defined  by imposing   that certain gauge fields that appear in the action are bounded by  $p_j = (- \log e_j )^p$ for some positive constant $p$  greater than $r$ .    The $\La_j$ have further restrictions    and    are the  regions where the fluctuation integrals are actually carried out.  
  They satisfy 
  \be
  \Om_1 \supset \La_1 \supset \Om_2 \supset \La_2 \supset    \cdots  \supset   \Om_k \supset  \La_k
  \ee
   The domains have the  separation  condition, stronger than that in \cite{Dim20}:    
    \be
    d(\La^c_{j-1},  \Om_j)    \geq  5  L^{-(k-j)}M  r_j \hs    d(\Om^c_j,  \La_j)    \geq  5 L^{-(k-j)}M r_j
    \ee
    However the cases  $\La_{j-1} =  \Om_j$ and $\Om_j =  \La_j$ are not excluded.
  Special subsequences of  $\bpi$ are 
\be
\bom  =   (\Om_1,  \Om_2,  \dots,   \Om_k)   \hs    \bla =   (\La_1,  \La_2,  \dots,   \La_k) 
 \ee
     
  \item 
  The   $ ( P_j, Q_j, R_j, U_j ) $ are large field regions  in which  the $p_j$ bounds are   violated.   They  determine the $\Om_j, \La_j$  (so the  terms in our sum are labeled redundantly).   Contributions from these regions can be estimated by tiny factors $e^{- p_j} $ which are smaller than any power of $e_j$.  
  Given $\La_k $  it is shrunk and then  subsets $P_k,Q_k$ are deleted to give $\Om_{k+1} $, see (\ref{not}) - (\ref{wordy}) for the exact definition. 
  Then  $\Om_{k+1}$  is shrunk and    subsets  $R_k, U_k$ are  deleted to give $\La_{k+1} $, see (\ref{sister1}) - (\ref{wordy2}) and (\ref{sister2}) - (\ref{1781}) for the precise definitions.

  \item  The factors $\sZ'_{k, \bom}(0) ,   \sZ'_{k, \bom}$ are normalization factors for fermions and bosons respectively.  They have the form  
  \be \label{nuisance} 
  \begin{split}
  \sZ'_{k,\bom} (0) = &  \prod_{j=0}^{k-1}    L^{-8 (s_N - s_{N-j-1} )  }  N_{j+1,L^{k-j}  \Om_{j+1}}\de \sZ_{j,L^{k-j} \bom}(0) \\
 \sZ'_{k,\bom}  = &  \prod_{j=0}^{k-1}    L^{\frac12 (b_N- b_{N-j-1})   -\frac12 (s_N- s_{N-j-1}) }  \de \sZ_{j,L^{k-j} \bom} \\
   \end{split}
   \ee   
Here an expression like $\de \sZ_{j, \bom}$ only depends on the first $j+1$ entries $(\Om_1, \dots, \Om_{j+1})$ of $\bom$. 

\item 
There are gauge fields  which after $k$ steps  have  the form 
 $
 A= (A_{0,\Om^c_1}, A_{1, \Om^c_2}, \dots, A_{k-1,  \Om^c_k})
 $
where $A_{j,  \Om^c_{j+1}}$ is a function on  $   \Om^{(j),c}_{j+1}  \subset \bbT^{-(k-j)}_{N-k}$.  Here  $ \Om^{(j)}_{j+1}   \subset \bbT^{-(k-j)}_{N-k}$ are the centers of cubes in $\Om_{j+1}$ with  $L^j$ sites on a side.   The integrals over the large field regions are defined by  measures $Dm_{k, \bom}(A)$ on such fields defined  recursively
  with $\bom^+ = ( \bom, \Om_{k+1}) $  by 
 \be \label{post1} 
 \begin{split}
 Dm_{k+1, \bom^+} (A) =   &   (Dm^0_{k+1, L\bom^+} )_{L^{-1}}  (A^+) \\
  Dm^0_{k+1, L\bom^+} (A) = & Dm_{k, L\bom}(A)\  \de_{L\Om^c_{k+1} } (A_{k+1} - \cQ A_k) \de_{L\Om^c_{k+1}}(\tau A_k ) DA_{k, L\Om^c_{k+1} }
 \end{split}
 \ee
 starting with $Dm_0(A) = DA_{0, \Om_1^c} $.  Here fields scale by $A_L (b) = L^{-\frac 12} A(b/L)$, functions of fields scale by $f_{L^{-1}} (\cA )= f(\cA_L )$,
 and   measures scale by $ \int f_{L^{-1}} (\cA )  d \mu_{L^{-1}} (\cA ) =  \int f (\cA )  d \mu (\cA ) $.  ( The scaling factor $L^{-\frac12} $ is chosen to preserve the free action,  but here in this inactive portion of the integral it is arbitrary.) 
 
\item  There are gauge  fluctuation fields which after $k$ steps have the form 
$Z=   Z_{k, \bpi}$
where
\be 
 Z_{k, \bpi} =  ( \tilde Z_{0, \Om_1- \La_1},\tilde Z_{1, \ \Om_2-\La_2},  \cdots, \tilde Z_{k-1,  \Om_k - \La_k} )
 \ee   with $\tilde Z_{j, \Om_{j+1} - \La_{j+1}}$
 defined on (subsets of)  $\Om^{(j)}_{j+1} - \La^{(j)}_{j+1} \subset \bbT^{-(k-j)}_{N-k}$. Integrals over these variables  are defined by  a measure $Dm_{k, \bom} (Z )  $ defined recursively by 
  \be \label{post2} 
 \begin{split}
 Dm_{k+1, \bpi^+} (Z) =   &   (Dm^0_{k+1, L\bpi^+} )_{L^{-1}}  (Z) \\
  Dm^0_{k+1,L\bpi^+} (Z) = &Dm_{k, L\bom}(Z)\ d \mu_{I,L(\Om_{k+1} - \La_{k+1} )} ( \tilde Z_k )  
 \end{split}
 \ee

 \item 
There are fermi  fields  which after $k$ steps have the form 
 $\Psi= (\Psi_{0,\Om^c_1}, \Psi_{1, \Om^c_2}, \dots, \Psi_{k-1,  \Om^c_k})$
where $\Psi_{j,  \Om^c_{j+1}}$ is a function on $    \Om^{(j),c}_{j+1} $.
 The integrals over the large field regions are defined by  "measures" $Dm_{k, \bom}(\Psi)$ on such fields defined  recursively
by 
 \be \label{post3} 
 \begin{split} 
 Dm_{k+1, \bom^+} (\Psi) =   &   (Dm^0_{k+1, L\bom^+} )_{L^{-1}}  (\Psi) \\
  Dm^0_{k+1,L \bom^+} (\Psi) = & Dm_{k, L\bom}(\Psi)\ \de_{G, L\Om^c_{k+1} } (\Psi_{k+1} - Q(0) \Psi_k)  D\Psi_{k, L\Om^c_{k+1} } 
 \end{split}
 \ee
 starting with $Dm_0(\Psi) = D\Psi_{0, \Om_1^c} $. Here $\de_G $ is a Gaussian approximation to a delta function. 
 Fermi fields scale by $\Psi_L ( x) = L^{-1} \Psi (x/L) $.

\item 
There are fermi  fluctuation fields which after $k$ steps have the form
 $W=   W_{k, \bpi}$
 where
 \be
 W_{k, \bpi} =  ( W_{0, \Om_1- \La_1},W_{1, \ \Om_2-\La_2},  \cdots, W_{k-1,  \Om_k - \La_k} )
 \ee
 with $W_{j, \Om_{j+1} - \La_{j+1}} $
 defined on $\Om^{(j)}_{j+1} - \La^{(j)}_{j+1} $.
 Integrals over these variables  are defined by  "measures" $Dm_{k, \bom} (W )  $ defined recursively  by 
  \be \label{post4} 
 \begin{split}
 Dm_{k+1, \bpi^+} (W) =   &   (Dm^0_{k+1, L\bpi^+} )_{L^{-1}}  (W) \\
  Dm^0_{k+1, L\bpi^+} (W) = &  Dm_{k, L\bpi}(W)\  d \mu_{I, L(\Om_{k+1} - \La_{k+1} )} ( W_k )
 \end{split}
 \ee

\item  The free gauge field action is $\frac12\| d \cA_{k,\bom}  \|^2  $ where   $\cA_{k, \bom}=  \cH_{k, \bom}A_{k,\bom}$ on $\tk$   
is a linear function of the fundamental fields. 
\be
 A_{k, \bom}=  ( A_{0, \Om_1^c},A_{1, \de \Om_1},  \cdots, A_{k-1, \de \Om_{k-1}}, A_{k, \Om_k} ) 
 \ee
Here $\de \Om_j = \Om_j - \Om_{j+1}$.   It is the minimizer of the  relevant part of the effective action in the previous step as constrained by 
the block averaging, see \cite{Dim20}  for the precise definition.  The $\cA_{k,  \bom} $ can be taken   either in the 
axial gauge or  the Landau gauge.  The axial gauge is preferred for stability questions, while the Landau gauge is preferred for good ultraviolet
properties,  i.e. smoothness.

 \item 
The free fermi action 
has the form  
\be \label{picquant}
 \fS^+ _{k, \bom} ( \La_k, \cA,  \Psi_{k, \bom},   \psi    ) =\fS_{k, \bom} ( \cA,   \Psi_{k, \bom},   \psi    ) + m_k   < \bar \psi,  \psi  >_{\La_k} +( \vep_k + \vep_k^0)  \Vol (\La_k) 
\ee
Here  $\fS_{k, \bom} ( \cA,  \Psi_k,   \psi    ) $ is the global free fermion action 
\be      
\fS_{k, \bom} ( \cA,  \Psi_{k, \bom},   \psi    )   = b_k \blan  \bPsi_{k, \bom}- Q_k (- \cA )    \bpsi   ,     \Psi_{k, \bom}- Q_k (\cA )   \psi    \bran_{\Om_1}
+   \blan     \bpsi   ,   \B(\fD_{ \cA }+ \bar   m_k \B)        \psi  \bran    
\ee
defined    with      $e_k= L^{\frac{k}{2} } e_0$ and  $\bar m_k  = L^k \bar m_0$. 
The terms  with $m_k, \vep_k$ are counterterms  
for the mass and vacuum energy respectively, and are localized in the final small field region $\La_k$.   The  $\vep_k^0 $ is a tiny correction. 

The action is evaluated at $\psi = \psi_{k, \bom} (\cA) =  \psi_{k, \bom} (\cA, \Psi_{k, \bom})  $ on $\tk$  which is a linear function of the fundamental fields
 \be
  \Psi_{k, \bom}=  ( \Psi_{0, \Om_1^c},\Psi_{1, \de \Om_1},  \cdots, \Psi_{k-1, \de \Om_{k-1}},  \Psi_{k,  \Om_{k}} ) 
\ee
It is a critical point of the relevant part of the effective action in the previous step  as modified by block averaging; see $\cite{Dim20}$ for the precise
definition.

  \item   $E_k(\La_k ) =   E_k (\La_k, \cA_{k, \bom}, \psi^{\#}_{k, \bom} (\cA_{k, \bom})$ is  localized in  the current small field region $\La_k$ and
  depends on dressed fermi fields 
  \be 
   \psi^\#_{k, \bom}(\cA) = ( \psi_{k, \bom}(\cA), \de_{\al, \cA} \psi_{k, \bom}(\cA))
  \ee
  where $\de_ {\al, \cA}$ is a covariant H\"{o}lder derivative defined in (277) in \cite{Dim15b}. 
 It contains  corrections to the bare action and plays a role in  renormalization. 
 $
   B_{k,\bpi} = B_{k, \bpi}( \cA_{k, \bom} ,  Z_{k, \bpi},  \psi^\#_{k, \bom}(\cA_{k, \bom}), W_{k, \bpi})$ is a boundary term which
  plays no role in  renormalization.
  Both   $E_k(\La_k )$ and  $ B_{k,\bpi} $   will be further specified in great detail. 
  
 \item The  characteristic function    $\chi_k(\La_k)$  is defined  as a product over $Mr_k$ cubes in $\La_k$  by    
\be   \label{cinnamon}
  \chi_k ( \La_k)  =  \prod_{  \square    \in  \La _k}  \chi_k (\square) \hs 
 \chi_k (\square )   =       \chi  \B( \sup_{p \in  \tilde \square}    \B| (d \cA_{k, \bom(\square)} )(p)  \B|  \leq  p_k   \B)  
\ee
where $\tilde \sq$ is the enlargement by a layer of $Mr_k$ cubes.   The field   $\cA_{k, \bom(\square)} $ is a local version of $\cA_{k, \bom}$ defined as follows.
  The sequence is the decreasing sequence 
   \be
  \bom(\square) = (\sq_1 , \sq_2 ,  \dots, \sq_k ) \hs    \sq_k  \supset \tilde  \sq
  \ee
  with  $d(\sq_k^c, \tilde \sq) = Mr_k$ and  $d(\sq_j^c, \sq_{j+1}) = L^{-(k-j)}Mr_k$.  (This is similar to some constructions in \cite{Dim20}, but now $M$ scale is enlarged to $Mr_k $
  scale.)     Then with a surface averaging operator $\cQ_k^s$
  \be \label{pool}
  \cA_{k, \bom(\square)}  = \cH_{k,\bom(\sq)} \B( \cQ_{k, \bom(\sq) } \cQ_k^{s,T} A_k \B)
  \ee
On $ \tilde \sq$ (or even $\Om_1(\sq)= \sq_1$)   the $\cA_{k, \bom(\square)} $  only depends on $\cQ_{k, \bom(\sq) } \cQ_k^{s,T} A_k$ and hence on  $A_k$  on  $\Om_1(\sq) \subset \sq^{\sim 3}$.
  (See the discussion in section III.F.3 in \cite{Dim20}  ). 

Note that on $\tilde \sq  $ we have   $dA_k =d\cQ_{k, \bom(\sq) } \cQ_k^{s,T} A_k = d \cQ_k  \cA_{k, \bom(\square)} =   \cQ^{(2)}_{k } d\cA_{k, \bom(\square)} $.  Hence  
 the characteristic function  (\ref{cinnamon})  requires that  
$| dA_{k}| \leq p_k$    on each $\tilde \sq \subset \La_k $ and so on  $\La_k$.

\item $ \cC_{k, \bpi} $ is a collection of characteristic functions introduced earlier in the expansion and limiting the strength of the gauge field.  
 The precise  definition is  given  in the course of the proof. 
It does have the property   that it enforces   for $j=1, \dots k$
\be \label{simplesimon}
 | dA_j| \leq L^{\frac32(k-j)} p_j \hs \textrm{ on } \hs  (\Om^{\sim5}_j )^{(j)} - \Om_{j+1}^{(j)}
\ee
 Furthermore $\cC_{k, \bpi} $  does not depend on $A_k$ in $\La_k^{3 \nat}$.

\end{itemize}

\subsection{the flow}   \label{theflow}

The function $E(\La_k)$ has a polymer expansion 
 $
  E_k(\La_k) = \sum_{X \in \cD_k, X \subset \La_k} E_k(X)
  $
 where  $\cD_k$  is connected unions of $M$  cubes (polymers), and where $E_k(X)$ 
depends on the fields only in the polymer $X$ and is independent of the history $\bpi$ and invariant under lattice symmetries. 
We will see that as long as everything stays sufficiently small,  the  polymer functions  $E_k(X)$ and  the coupling constants obey the flow equations
 \be
 \label{recursive}
\begin{split}
e_{k+1} =  & L^{\frac12}  e_k \\
\vep_{k+1}   =&  L^3  \B( \vep_k  +  \vep_k(  E_k)  \B)\\
m_{k+1}   =&   L \B( m_k  +   m_k(  E_k) \B)  \\
E_{k+1}   =&    \cL \B( E' _k  +   E^\#_k(  m_k,  E_k) +   E^{\det}_k  \B) \\
 \end{split}
\ee  
Here  $ \vep_k(  E_k) ,  m_k(  E_k)$ are corrections to the vacuum energy and fermion mass obtained by extracting the relevant parts of $E_k$.  The  $E_k'$
 (also called $\cR E_k$ in \cite{Dim15b} )  is $E_k$  with   relevant parts   extracted. The   terms  $E^{\det}_k$ is a correction from a normalizing determinant. 
The term  $E^\#_k(  m_k,  E_k)$  is the result of the operations of the renormalization transformation on the $E_k$ about which we will have much more to say.    The operator  $\cL$ reblocks and rescales.  

These equations are independent the history   and also occur   in the  global small field  version  
 studied in \cite{Dim15b}.   There the following result is proved
 
 \begin{prop} \label{prop}
 Let   $L$  be sufficiently large,  let  $M=L^m $ be sufficiently large (depending on $L$),  and  let $e$  (and hence   $e_k$)   be sufficiently small  (depending on $L,M$).
 Let  $K_0 = N- \log_L M$ and choose a stop point $K \leq K_0$. 
 Then   there is a unique solution  of the flow equations (\ref{recursive})  for $0 \leq k \leq K$ with the boundary 
conditions 
\be   \vep_K  =  0 \hs    m_K =0  \hs E_0 =0
\ee     
This solution satisfies the bounds 
\be  \label{samsung2} 
\begin{split} 
|\vep_{k}|  \leq &  2 e_k^{\frac14- 7 \ep}   \\ 
   |m_{k}|  \leq   &    e_k^{\frac34 - 8 \ep}   \\  
\|    E_k(X, \cA ) \|_{\bh_k}    \leq   &   e_k^{\frac14-7\ep}      e^{-\ka d_M(X)} \\
\end{split} 
\ee 
where $\ep$ is a fixed small positive number.   
 \end{prop}

The norm  here  is an $L^1$ norm on the kernel with fields $\psi^\#$ replaced by weight functions
  \be
    \bh_k  =  ( h_{k},  h'_{k}  )  =  ( e_k^{-\frac14}  , e_k^{-\frac14 + \ep}  )
\ee 
See Appendix \ref{A} 
for the precise definition of this and other norms. The 
 $d_M(X)$ is  defined by stating that $Md_M(X)$ is      the length of the
shortest continuum tree joining cubes in $X$.  If $\sq$ is a single $M$ cube and  $\ka \geq \ka_0 $ with  $\ka_0 = \one$ is   sufficiently large 
 then (see for example the appendix in \cite{Dim13a})
 \be \label{jam} 
 \sum_{X \in \cD_k, X \supset \sq }   e^{ -  \ka   d_M(X  )}  \leq \one
 \ee
 
Proposition 1 also holds  with some variations,    in particular  if  the   function   $E^\#_k(  m_k,  E_k) $ involves a slightly different
 cluster expansion than  \cite{Dim15b}, as is the case in the present paper.    The statement and proof are  the same.    As we will see the extra term 
  $\vep_k^0$ in (\ref{picquant})  now   obeys the flow equation
\be \label{v1}
\vep_{k+1}^0  = L^3 ( \vep_k^0 + \de \vep^0_k)
\ee 
and starts at zero.   The  increment  $\de \vep^0 _k $ is allowed to depend on the history, but   will satisfy the strong bound  $| \de \vep^0 _k | \leq e_k^7  $   (or any power of $e_k$)  and since  $e_k^7  = L^{-\frac72} e_{k+1}^7 $  this implies that for all $k$
 \be \label{v2} 
| \vep_k^0 | \leq e_k^7 
 \ee

\subsection{the theorems}  \label{once} 
We are now almost ready to state our results.   The statement    involves certain analyticity domains defined as follows. 
 Define
\be
\theta_k = \prod_{i=0}^{k-1} ( 1- e_k^{\ep} ) 
\ee
Since $ \sum _{i=0}^{k-1} e_k^{\ep}  = \cO(e^{\ep } ) $  we have for $e$ small (depending on $\ep$) that  $\frac12 \leq \theta_k \leq 2$ for all $k$.

   \bigskip
   
 \begin{defn} \label{defn4}     $  \cR_{k, \bom} $ is  all complex   gauge fields  $  \cA  $   on   $ \tk $   such that       on $\de \Om_j$ 
\be  \label{rk}
 | \cA  | \leq   L^{\frac12(k-j)} \theta_k  e_j^{-\frac34 + \ep}
\hs      |\pa  \cA  | \leq    L^{\frac32(k-j)}   \theta_k    e_j^{-\frac34 + 2\ep}
\hs     | \de_\al  \pa    \cA |  \leq    L^{(\frac32+ \al)(k-j)}   \theta_k  e_j^{-\frac34 + 3\ep}
 \ee
 $\cR_k$ is the restriction to $\de \Om_k = \Om_k$ so $ | \cA  | \leq  \theta_k  e_k^{-\frac34 + \ep}$, etc. 
  \end{defn} 
\bigskip 

\begin{defn}  \label{defn5}   $ \tilde   \cR_{k, \bom} $ is  all  gauge fields of the form    $  \cA=  \cA_0  + \cA_1$
where  
\begin{itemize}
\item    $ \cA_0 $ is real and for  each  $L^{-(k-j)}M $ cube  $\square \in \pi_j(  \de \Om_j) $    is gauge equivalent  in the enlargement  $\square ^{\dagger} \equiv   \sq^{\sim c_0L}$ to a field $\cA' \in   \cR_{k, \bom}$
\item  
 $\cA_1 \in  \cR_{k, \bom}$ is complex. 
 \end{itemize}     
 $\tilde \cR_k$ is the restriction to   $\Om_k$. 
\end{defn}  
 \bigskip
  The  domains  $\cR_{k, \bom}, \tilde \cR_{k, \bom}$ are related to the domains $\fG_{k, \bom}, \tilde \fG_{k, \bom}$ introduced in \cite{Dim20},  which were defined to insure that
 fermi  fields $\psi_{k, \bom} (\cA)$  with background field $\cA$ be   well-defined. 
The domains $\fG_{k, \bom}, \tilde \fG_{k, \bom}$ are defined 
 by conditions similar to $\cR_{k, \bom}$ except   there are no conditions on derivatives and  bounds are weaker by a factor $e_j^{-\frac14}$.   Both conditions are weaker 
so  $\cR_{k, \bom} \subset \fG_{k, \bom}$,  $\tilde \cR_{k, \bom} \subset \tilde \fG_{k, \bom}$.    Moreover if  $ \cA \in   \cR_{k, \bom}$ then $ e_k^{-\frac14} \cA \in \fG_{k, \bom}$ and similarly for  $\tilde \cR_{k, \bom},  \tilde \fG_{k, \bom}$
\bigskip

In the following  theorem we stop the iteration at the first  point $K$ satisfying $K = K_0 -  \log_L r_K = N - \log_L M - \log_L r_K$.    
In section \ref{circus} we show that this uniquely defines $K$ and that  $N-K$ is independent of $N$.

\begin{thm}  \label{maintheorem} 
Let   $L$  be sufficiently large,  let  $M$ be sufficiently large (depending on $L$),  and  let $e$  (hence   $e_k$)   be sufficiently small  (depending on $L,M$).
Let   initial values $\vep_0, m_0$ be chosen as  in Proposition 1 with $K$ as specified above.  
Then  the background field $\tilde \cA$ can be chosen so   that  for all   $0 \leq k \leq K$ the density    $\rho_k(A_{k, \bom}, \Psi_{k, \bom})$  on $\tk$    has the form
(\ref{startrep1})  (after modification) with the following properties:
\begin{enumerate} 
\item $E_{k} (\La)$ has the polymer expansion 
  \be \label{oscar1} 
  E_k(\La_k) = \sum_{X \in \cD_k, X \subset \La_k} E_k(X)
  \ee
   where
  $ 
  E_k(X) = E_k( X, \cA_{k, \bom}, \psi_{k, \bom}^\# (  \cA_{k, \bom} ))
  $
  is independent of the history $\bpi$  and
  depends on the indicated fields only in $X$. 
It is the restriction of a function   
 $ E_k(X,  \cA, \psi^\#(\cA) ) $ to 
 $\cA = \cA_{k, \bom}$. 
The   $ E_k(X ,   \cA, \psi^\# (\cA) ) $ are  gauge invariant, invariant under lattice symmetries,    and    analytic  in  $\cA   \in  \tilde   \cR_k$.
The triple  $(\vep_k, m_k, E_k)$ satisfies the flow equation (\ref{recursive}) with  the stated  bounds 
$|\vep_{k}|  \leq 2 e_k^{\frac14- 7 \ep} $ and $  |m_{k}|  \leq   \one  e_k^{\frac34 - 8 \ep} $ and
\be \label{oscar2} 
 \|    E_k(X, \cA ) \|_{\bh_k}    \leq     e_k^{\frac14-7\ep}     e^{-\ka d_M(X)} 
\ee
The $\vep_k^0$ satisfies the flow equation(\ref{v1}) and the bound  (\ref{v2}).

\item
The boundary term $B_{k, \bpi}$ has the form $
  B_{k, \bpi} = \sum_{j=1}^k  B^{(k)}_{j, \bpi}
$
and with $\cD_j^{(k)}  $ defined to be  connected unions of  $L^{-(k-j)} M $ cubes 
\be   \label{b1}
 B^{(k)}_{j, \bpi} = \sum_{ X \subset \cD_j^{(k)}, X \cap (\La_{j-1} -  \La_j) \neq \emptyset } 
B^{(k)}_{j, \bpi} (X) 
\ee  
 Here
$
B^{(k)}_{j, \bpi} (X) = B^{(k)}_{j, \bpi}(X;   \cA_{k, \bom}  ,  Z_{k, \bpi},  \psi^{\#}_{k, \bom}(\cA_{k, \bom} ), W_{k, \bpi} )
$ 
 depends on the indicated fields only in $X$ and   is the restriction of a  $B^{(k)}_{j, \bpi} (X;   \cA  ,  Z_{k, \bpi},  \psi^\#_{k, \bom}(\cA), W_{k, \bpi}) \ $
which is 
 is  gauge invariant  and 
 analytic in 
  $
   \cA \in   \tilde \cR_k,    |\tilde Z_{j, \Om_j- \La_j} |  \leq    L^{(k-j)/2}   
$ 
  and  satisfies  there  
\be \label{sordid}
\| \cB^{(k)}_{j ,\bpi}  (X,  \cA,  Z_{k, \bpi} ) \|_{ \bh_k \bbI^\#_k,  L\bbI_k}  \leq     e_k^{\frac14-8\ep}    e^{-\ka   d_{L^{-(k-j)}M}(X)} 
\ee
Here the product  $\bh_k\bbI^\#_k$ is elementwise and
$
\bbI^\#_k = (\bbI_k, \bbI'_k)  = (  L^{k-j}, L^{(1+ \al)(k-j) })$  on  $  \de \Om_j $.
\end{enumerate}
\end{thm}
\bigskip

\rem
For the statement of the theorem to make sense we need to know  that
the characteristic functions in (\ref{cinnamon})  restrict the fundamental  gauge fields $A_{k, \bom}$
sufficiently to ensure that the  Landau gauge  field $\cA_{k, \bom} = \cH_{k, \bom} A_{k, \bom}$ is in the domain $\tilde \cR_{k, \bom}$. 
In \cite{Dim20} we showed that on $\sq \subset \de \Om_j$  the field  $\cA_{k, \bom} $  is gauge equivalent in  $ \sq^{\dagger} $  to a field   $\cA'$ satisfying
  \be \label{peachy2}
  \begin{split}
& L^{-\frac12 (k-j)}|\cA'|,\  L^{-\frac32(k-j)}  | \pa \cA'|,\   L^{-(\frac32+ \al)(k-j)} \| \pa  \cA'\|_{(\al)}  \\
 & \leq   
 C M  p_j     \sup_{j'}  L^{-\frac32 (k-j')} p_{j'}^{-1}    \|  d A_{j', \de \Om_j'} \|_{\infty  }  \\
 &\leq  CM p_j  \leq  e_j^{-\ep}  \\
 \end{split}
 \ee
 The last line follows since    $\chi_k(\La), \cC_{k, \bpi} $ enforce $ |d A_{j', \de \Om_{j'}}| \leq L^{\frac32(k-j')}p_{j'}$
 and since $CM p_j  e_j^{\ep} \leq 1$ for $e_j$ sufficiently small.  
   This is much better than the factors  $ e_j^{-\frac34 + \ep}, e_j^{-\frac34 +2 \ep}, e_j^{-\frac34 +3 \ep}$ and so $\cA' \in \cR_{k, \bom}$ and 
  $\cA_{k, \bom} \in \tilde \cR_{k, \bom}$. 
\bigskip

As we will see in section \ref{section4},  theorem 1 yields  a proof  of the stability bound:

\begin{thm}   (stability bound)   \label{theorem2}      Let the coupling constant  $e$ be sufficiently small.  
Then for all $N$
\be
 \frac12  \leq     \B |  \frac{  \sZ(N, e) }{  \sZ(N, 0)} \B|  \leq  \  \frac32
\ee
\end{thm}

\section{proof of  theorem 1}

The proof  incorporates elements from     \cite{Bal82b}, \cite{BIJ88},  \cite{Bal98a}, \cite{Bal98b},     \cite{Dim13b} and especially
Balaban's work on Yang-Mills   \cite{Bal87}, \cite{Bal88a}, \cite{Bal88b}.   
It extends  the global small field  version  in  \cite{Dim15b}.   We recall our convention that $\ga \leq 1$ and $C \geq 1$ are constants which
may depend on $L$  and may change from line to line

 \subsection{extraction}

   We assume it is true for  $k$  and  generate   the representation for $k+1$. 
 To  begin  insert  the expression   (\ref{startrep1}) for  $\rho_k$  into the definition (\ref{basic1})  of   $\tilde  \rho_{k+1}$,   and
 bring the sum outside the integral.
 Then
 \be    \label{rep0} 
 \begin{split}
 &   \tilde \rho_{k+1} ( A_{k+1}, \Psi_{k+1}   )  
=   \sum_{\bpi}
  \int   D\Psi_k  DA_k  \ 
\de_G( \Psi_{k+1} - Q( \tilde \cA  ) \Psi_k )      \de(  A_{k+1} - \cQ A_k  )  \de (\tau A_k ) 
 \\
 &     \sZ'_{k, \bom}( 0 )    \sZ'_{k, \bom}    \       \int   Dm_{k, \bom} (A )Dm_{k, \bpi} (Z )  Dm_{k, \bom}( \Psi)   Dm_{k, \bpi}( W)   \  \cC_{k, \bpi} \     \chi_k( \La_k)  \\
&         \exp \B( -\frac12 \| d \cA\|^2     -         \fS^+_{k, \bom}(\La_k, \cA, \Psi_{k, \bom}, \psi_{k, \bom} (\cA)    )   
  +E_k(\La_k ) +  B_{k,\bpi}         \B) \B|_{\cA =    \cA_{k, \bom} }  \\
\end{split}
\end{equation}

First we  extract the relevant parts  from  $  E_k( \La_k,    \cA,   \psi^\#(\cA)  ) $ where  $\psi^\#(\cA)  = (\psi(\cA), \de_{\al, \cA}\psi(\cA)$) and $\psi(\cA) = \psi_{k, \bom}(\cA)$.  As explained in (327)-(332)  in  \cite{Dim15b}  this has the form    
   \be  \label{renorm3}
 \begin{split}
   E_k(  \La_k,   \cA,   \psi^\# (\cA) )  = &     -  \vep(  E_k)     \Vol( \La_k )
    -      m( E_k)    
  <   \bpsi (\cA),     \psi(\cA) >_{\La_k}  
+      E'_k(\La_k,   \cA,   \psi^\#(\cA) )  +  B^{E} _{k, \bpi} (\psi(\cA))    \\
\end{split}
\ee 
The     term  $E'_k(\La_k)   $  has the relevant parts removed and  a local expansion
  $ E'_k (\La_k) =  \sum_{X \subset \La_k}   E'_k(X) $.
The kernel of  $E'_k(X)  =  E'_k(X,  \cA, \psi^\#(\cA)) $  is independent of the history,  and we have
for $\cA \in \tilde  \cR_{k, \bom}$ and $\bh_k = (e_k^{-\frac14}, e_k^{-\frac14})$  
  \be    \label{retro}
   \|E' _{k}(X, \cA) )\|_{\bh_k}       
      \leq     \one   e_k^{\frac14- 7 \ep}   e^{- \ka  d_M(X)}
\ee
 The  term  $B^{E} _{k, \bpi}$   is a boundary term  and has
a local expansion    $ B^{E} _{k, \bpi}    =   \sum_{X \in \cS, X \# \La_k}  \tilde  B^{E} _{k, \bpi}  (X)$ in small polymers $X$  that intersect both $\La_k, \La_k^c$,
denoted $X \# \La_k$.
It has the form 
\be
 \tilde B^{E} _{k, \bpi}  (X, \psi(\cA) ) =  -  \al_0( E,  X) \    \Vol ( \La \cap X )  -  \int_{X \cap \La}  \bpsi(\cA)  \   \al_2( E,  X) \  \psi(\cA)    
\ee      
for certain $\al_0( E,  X), \al_2( E,  X) $, 
and  has the estimate  from (332) in  \cite{Dim15b} 
  \be    \label{monday}
   \|  \tilde B^{E} _{k, \bpi}(X )\|_{h_k}       
      \leq        \one  e_k^{\frac14- 7 \ep}   e^{- \ka  d_M(X)}
\ee
The effect  in   (\ref{rep0})   is to       adjust  the vacuum energies and coupling constants by  
\be \label{stateside}
     \vep_k'  = \vep_k   +   \vep(  E_k)    \hs   
m'_k    =  m_k +   m( E_k)  
\ee
and so  replace    $  \fS^+_{k, \bom}(\La_k, \cA,  \Psi_{k,\bom}, \psi (\cA) ) $   by  
\be 
       \fS'^{+}_{k, \bom}(\La_k, \cA, \Psi_{k, \bom}, \psi  (\cA)  )   
  \equiv    \fS_{k, \bom}( \cA, \Psi_{k,\bom}, \psi  (\cA) )   
   +   m'_k   \blan  \bar \psi(\cA),  \psi (\cA)  \bran_{\La_k}  + (\vep'_k + \vep^0_k)  \Vol (\La_k) 
\ee
 and to replace     $E_k(\La_k ) $ by   $E'_k(\La_k )$   and add a term    
 $B^E_{k,\bpi}$.

 \subsection{new characteristic functions}  \label{sinbad7} 

We  insert new  characteristic functions.  Let $\sq$ be an $LMr_{k+1}$ cube and define
   \be
     \label{not}
 \chi^0_{k+1}( \sq,  A_{k+1})    =   \chi \B(  \sup_{p \in  \tilde \square}       | d \cA^0_{k+1, \bom^{+}(\sq)  }(p)  |  \leq     p_{k+1}  L^{-\frac32}   \B)  
\ee
where $\tilde \sq$ is an enlargement by a layer of $LMr_{k+1}$ cubes. 
Here   $\bom^{+}(\square) $ is   the decreasing sequence  of cubes
   \be
  \bom^+ (\square) = (\sq_1,  \sq_2,  \dots,  \sq_{k+1}) \hs  \sq _{k+1} \supset  \tilde   \sq
  \ee
  with  $d(\sq_{k+1}^c, \tilde  \sq) =LM  r_{k+1}$ and  $d(\sq_j^c, \sq_{j+1}) = L^{-(k-j)}Mr_{k+1}$, and  the field is 
  the minimizer of $\|d \cA \|^2$ subject to $\cQ_{k+1, \bom^+(\sq) } \cA =\cQ_{k+1, \bom^+(\sq) } \cQ_{k+1}^{s,T} A_{k+1} $
  and either Landau or axial gauge fixing.  We write these respectively as
   \be \label{croc1} 
   \begin{split}
  \cA^0_{k+1, \bom^+(\square)}  = &  \cH^0_{k+1,\bom^+(\sq)} \B( \cQ_{k+1, \bom^+(\sq) } \cQ_{k+1}^{s,T} A_{k+1} \B) \\
   \cA^{0, \sx}_{k+1, \bom^+(\square)}  = &  \cH^{0,\sx}_{k+1,\bom^+(\sq)} \B( \cQ_{k+1, \bom^+(\sq) } \cQ_{k+1}^{s,T} A_{k+1} \B) \\
   \end{split}
  \ee
 This scales to   the function defined for $Mr_{k+1}$ cubes $\sq$ by 
   \be
     \label{not2}
 \chi_{k+1}( \sq,  A_{k+1})    =   \chi \B(  \sup_{p \in \tilde  \square}       | d \cA_{k+1, \bom^{+} (\sq)  }(p)  |  \leq     p_{k+1}   \B)  
\ee
which is (\ref{cinnamon}) for $k+1$. Note that   $d\cA^0_{k+1, \bom^+(\square)} $ only depends on $A_{k+1}$ in the three fold enlargement $ \sq^{\sim 3}$ by
  similar cubes.
  
  In general  
 If $X$ is a union of $Mr_k$  (or $LMr_{k+1}$) cubes then $\tilde X$ is an enlargement by a layer of $Mr_k$  (or $LMr_{k+1}$) cubes and  and $X^{\nat} = ( (X^c)^{\sim} )^c$ is a shrinkage
 by a layer of such  cubes.  Similarly we define $X^{ \sim n  }$ and $X^{n \nat}$, etc.

We shrink  the small field  region $\La_k$ to $\La_k^{5 \nat}$ and break it  up into new  large and small field regions defining  
   $ \zeta^0_{k+1}( \square) = 1-   \chi^0_{k+1}( \square) $  and decomposing the identity as 
\be \label{tiger1}
\begin{split}
1 =&   \prod_{\square  \subset   \La_k^{5 \nat} } \zeta^0_{k+1}( \square) +  \chi^0_{k+1}( \square) \\
= &   \sum_{P_{k+1}  \subset \La_k^{5 \nat} }\prod_{\square  \subset  P_{k+1}} \zeta^0_{k+1}( \square) 
\prod_{\square  \subset \La_k^{5 \nat} - P_{k+1} }  \chi^0_{k+1} (\square)  \\
\equiv &   \sum_{P_{k+1}  \subset  \La_k^{5 \nat}} \zeta^0_{k+1}( P_{k+1})    \chi^0_{k+1} (\La_k^{5 \nat}- P_{k+1})  \\
\end{split}
\ee
Here $P_{k+1}$ is a union of $LMr_{k+1}$ cubes.  We insert this under the integral sign in (\ref{rep0} ). 
\bigskip

Furthermore we  introduce a bound on  fluctuation fields defining  for $LMr_{k+1}$ cubes $\sq$
\be \label{prime}
       \chi'_{k}(\square, A_k, A_{k+1} )  
         =   
\chi  \B( \sup_{b  \in \tilde  \sq }    \B|  ( A_k(b) -   A^{\min }_{k, \bom^{+}(\square) }      (b)\B|    \leq     p_{0,k} ^2   \B)  
\ee
where  $p_{0,k} = \log( -e_k)^{p_0}$ for some $p_0$. We assume $2p_{0} < p$ so  $ p_{0,k}^2 < p_k $.   Let $\bom^{\bullet }(\sq)$ be $\bom^+(\sq)$ with the last entry $\sq_{k+1}$
deleted.
Then  $ A^{\min }_{k, \bom^{+}(\square)} $ is defined as  in  (222) in \cite{Dim20} by   
\be \label{croc2} 
 A^{\min }_{k, \bom^{+}(\square)} \equiv \cQ_{k, \bom^{\bullet}(\sq)} \cA^{0, \sx}_{k+1, \bom^+(\sq)}
 \ee
 The  $A^{\min }_{k, \bom^{+}(\square)}$ only depends on $A_{k+1}$ in $\tilde \sq^2$ and on $\tilde \sq$ it is  $\cQ_k \cA^{0, \sx}_{k+1, \bom^+(\sq)}$.

Now   we have with  $\zeta'_{k} (\square)  =  1- \chi'_{k}(\square)$:
\be    \label{tiger2}
\begin{split}
1 =&   \prod_{\square  \subset   \La_k^{5 \nat}} \zeta'_{k}( \square) +  \chi'_{k}( \square) \\
= &   \sum_{Q_{k+1}  \subset  \La_k^{5 \nat}}\prod_{\square  \subset  Q_{k+1}  } \zeta'_{k}( \square) 
\prod_{\square  \subset  \La_k^{5 \nat} - Q_{k+1} }  \chi'_{k} (\square)  \\
\equiv &   \sum_{Q_{k+1}  \subset   \La_k^{5 \nat}} \zeta'_{k}( Q_{k+1})    \chi'_{k} ( \La_k^{5 \nat} - Q_{k+1})  \\
\end{split}
\ee
where $Q_{k+1}$ is   a union of $LMr_{k+1}$ cubes.    We also  insert this under the integral sign in (\ref{rep0} ).

The new large field regions   $P_{k+1}, Q_{k+1}$   generate  a new small field region   $\Om_{k+1}$, also  a union of  $LMr_{k+1}$  cubes,   defined by 
\be  \label{wordy}
 \Om_{k+1}  =  \La_{k}^{5\nat }  -(  Q^{5 \sim}_{k+1}  \cup  P^{5 \sim}_{k+1})  \hs \textrm{ or  }  \hs
 \Om^c_{k+1}  =  (\La_{k}^c)^{5 \sim}  \cup   Q^{5 \sim}_{k+1}  \cup  P^{5 \sim}_{k+1} 
\ee
(If $\La^{\nat} _k = \emptyset$ then $\Om_{k+1} = \emptyset$ and all subsequent regions are empty.) 
  We  have the required  $d( \La^c_k, \Om_{k+1} ) \geq d(\La^c_k, \La_k^{5 \nat}) \geq 5Mr_k \geq 5Mr_{k+1}$. 
Note that we can also write $ \Om_{k+1}  =  \La_{k}^{5\nat }  \cap  (Q^{c}_{k+1} )^{5 \nat }  \cap  ( P^{c}_{k+1})^{5 \nat}  $ and so
\be  \label{839} 
\Om^{5 \sim}_{k+1}  \subset  \La_{k}  \cap  Q^{c}_{k+1}  \cap   P^{c}_{k+1}  \subset  \La_{k}  -   P_{k+1}  \textrm{ or }
 \La_{k}  -  Q_{k+1}   
 \ee
 Thus the bounds of $ \chi^0_{k+1}, \chi'_{k} $ hold on  $\Om^{5 \sim}_{k+1}$ and in particular $|dA_{k+1}| \leq p_{k+1}L^{-\frac32}$ there.
 
 Next combine the sums (\ref{tiger1}), (\ref{tiger2})  and classify the terms in the double sum by the $\Om_{k+1}$ they generate.
 We abreviate (\ref{wordy}) as $P_{k+1}, Q_{k+1} \to \Om_{k+1}$  and then
\be
\sum_{P_{k+1}, Q_{k+1}}  = \sum_{\Om_{k+1}} \  \sum_{P_{k+1}, Q_{k+1} \to \Om_{k+1}}  
\ee
We also make the split 
 \be 
 \begin{split}
  \chi^0_{k+1} (\La_k^{5 \nat} - P_{k+1}) = &  \chi_{k+1}^0(\Om_{k+1})  \chi^0_{k+1} ( \La_k^{5 \nat} -( P_{k+1} \cup \Om_{k+1} ) )   \\
  \chi'_{k} ( \La_k^{5 \nat}- Q_{k+1})=  &    \chi'_{k}( \Om_{k+1}   )      \chi'_{k}(   \La_k^{5 \nat}-(Q_{k+1} \cup  \Om_{k+1})  )  \\
\end{split}
\ee
The characteristic functions now have the form 
  $  \cC^{\star} _{k+1}\chi_{k+1}^0(\Om_{k+1})  \chi'_{k}( \Om_{k+1}   )  
$
where
  \be  \label{breakup3}
\begin{split} 
\cC^{\star} _{k+1} &  =  \cC^{\star} _{k+1}(\La_k,\Om_{k+1},P_{k+1}, Q_{k+1} )
 \\
&=    
 \zeta^0_{k+1}( P_{k+1} ) 
 \zeta'_{k}(  Q_{k+1}  )   \chi^0_{k+1} (\La_k^{5 \nat} -( P_{k+1} \cup \Om_{k+1} ) )   \chi'_{k}(   \La_k^{5 \nat}-(Q_{k+1} \cup  \Om_{k+1})  )  \\
 \end{split}   
\ee
The only dependence on $A_{k+1}$ in $\Om_{k+1}$ is in the terms $  \chi_{k+1}^0(\Om_{k+1})  \chi'_{k}(\Om_{k+1})$. 

Now     (\ref{rep0}) 
has become
   \be    \label{rep2} 
 \begin{split}
  &  \tilde \rho_{k+1} ( A_{k+1}, \Psi_{k+1}   )  
      \\
=    &
  \sum_{ \bpi,    \Om_{k+1},P_{k+1}, Q_{k+1}  }       \sZ'_{k, \bom}(0)   \sZ'_{k, \bom}   \int   \ 
     Dm_{k, \bom} ( A )     Dm_{k, \bom}( \Psi)\  Dm_{k,\bpi}(Z)  Dm_{k,\bpi}(W)  D\Psi_k  DA_k   
   \\
  & \de_G\B( \Psi_{k+1} - Q(\tilde  \cA   ) \Psi_k \B)      \de(  A_{k+1} - \cQ A_k  )  \de (\tau A_k ) \  \cC_{k, \bpi } \   \cC^{\star} _{k+1}    \chi_k( \La_k)    \chi_{k+1}^0(\Om_{k+1})  \chi'_{k}(\Om_{k+1})          
  \\
&        \exp \B( -\frac12 \| d \cA\|^2      -     \fS'^{+}_{k, \bom}   (\La_k   )   
 +E'_k(\La_k )  +  B_{k,\bpi}       +  B^E_{k,\bpi}  \B) \B|_{\cA =    \cA_{k, \bom} }  \\
\end{split}
\end{equation}

\subsection{redundant characteristic functions}

\label{redundant}

The new characteristic functions have rendered some of the  old characteristic function $\chi_k(\La)$ redundant.  
For an $LMr_{k+1}$ cube $\sq' \subset   \Om^{\sim 5 }_{k+1}$ 
the new characteristic functions enforce on $\tilde \sq'$  
\be \label{new}
\begin{split}
|d \cA^0_{k+1, \bom^+(\sq') } |  \leq &  L^{-\frac32}p_{k+1} \\
| A_k - A^{\min}_{k, \bom^+ (\sq') } |  \leq & p_{0,k}^2\\
\end{split}
\ee

\begin{lem}  \label{crunchy}  The bounds (\ref{new}) enforce on $ \Om^{\sim 5 }_{k+1}$
\be \label{newer} 
|dA_{k+1}| \leq     L^{- \frac32 }p_{k+1}   \hs   |dA_k | \leq   \frac12 p_k
\ee
\end{lem}
\bigskip

\pr  We have
  $ \cQ_{k+1, \bom^+(\sq')}  \cQ^{s,T}_{k+1} A_{k+1} = \cQ_{k+1, \bom^+(\sq')}  \cA^{0} _{k+1, \bom^+(\sq')} $  and on $\tilde \sq'$ 
this says that $A_{k+1} = \cQ_{k+1}  \cA^0_{k+1, \bom^+(\sq')} $ and so  $ dA_{k+1} = \cQ^{(2)}_{k+1} d \cA^0_{k+1, \bom^+(\sq')} $. 
Thus   $|dA_{k+1}| \leq L^{- \frac32 }p_{k+1}$   on any such $\tilde \sq'$ and hence the first bound.
We also have  $A^{\min}_{k, \bom^+ (\sq') }= \cQ_{k,\bom^{\bullet}(\sq')  }  \cA^{0,\sx}_{k+1, \bom^+(\sq')} $ 
and on $\tilde \sq'$ this says   $A^{\min}_{k, \bom^+ (\sq') }= \cQ_k \cA^{0,\sx}_{k+1, \bom^+(\sq')} $ and  $dA^{\min}_{k, \bom^+ (\sq') }= \cQ^{(2)}_k d\cA^{0,\sx}_{k+1, \bom^+(\sq')}$.
  Since  $d\cA^{0,\sx}_{k+1, \bom^+(\sq')} =d\cA^{0}_{k+1, \bom^+(\sq')} $
this gives $|dA^{\min}_{k, \bom^+ (\sq')}| \leq   L^{- \frac32 }p_{k+1}$.
Finally     $dA_k  =  d(A_k-  A^{\min}_{k, \bom^+ (\sq') } ) + dA^{\min}_{k, \bom^+ (\sq')} $.  Since we are on a unit lattice the first term is bounded by $4p_{0,k}^2 $
on $ \textrm{int} ( \tilde \sq' ) $.
Since $p_{0,k}^2p_k^{-1}$ is as small as we like by the choice of $2p_0 < p$   and since $p_{k+1} < p_k$  we get the bound $   |dA_k | \leq   4p_{0,k}^2   +L^{- \frac32 }p_{k+1}  \leq \frac12 p_k $ on $ \textrm{int} ( \tilde \sq' ) $ and hence on all of  $ \Om^{\sim 5  }_{k+1}$. Thus (\ref{newer}) is established.

\bigskip 

\begin{lem} \label{newest} The bounds (\ref{new}), (\ref{newer}) enforce   that  for an $Mr_k$ cube  $  \sq  $   in   $\Om^{\sim 5}_{k+1}$ that  $|d \cA_{k, \bom(\sq) }|   \leq p_k $ on $\tilde \sq$. 
Hence we can take  $\chi_k(\La_k) = \chi_k(\La_k-  \Om^{\sim 5}_{k+1})$.
\end{lem}
\bigskip 

\pr  We  can  work in the axial gauge and show    $|d \cA^{\sx}_{k, \bom(\sq) }|   \leq p_k $ on $\tilde \sq$.
We first claim that for $\sq \subset \sq'$ that  $d\cA^{\sx}_{k,\bom(\sq)} =d \cA^{\sx}_{k,\bom^{\bullet}(\sq')}$. 
To see this we use    $\Om_j(\sq) \subset \Om^{\bullet}_j (\sq') = \sq'_j$.  Then $\{ \cA:  \cQ_{k, \bom(\sq)}\cA  =   \cQ_{k, \bom(\sq)}\cQ^{s,T} A_k \}$
is contained in $\{ \cA:  \cQ_{k, \bom^{\bullet}(\sq')}\cA  =   \cQ_{k, \bom^{\bullet} (\sq')}\cQ^{s,T} A_k \}$ as can be seen by applying  the same extra 
averaging operators on both sides of the equation.  Hence any minimizer of $\| d\cA \|^2$  in the latter set,  like  $ \cA^{\sx}_{k,\bom^{\bullet}(\sq')}$,
is also a minimizer in the former set.  Then by    lemma 13
in \cite{Dim20},    $ \cA^{\sx}_{k,\bom(\sq)}$   and    $   \cA^{\sx}_{k,\bom^{\bullet}(\sq')}$  are on the same restricted orbit and hence the result.

 It now suffices to show  $|d \cA^{\sx}_{k+1, \bom^{\bullet}(\sq') } |  \leq p_k$ on $\tilde \sq'$.
 We write
\be \label{sonny}
d \cA^{\sx}_{k, \bom^{\bullet}(\sq') }  =
 \B( d \cA^{\sx}_{k, \bom^{\bullet}(\sq') }
   -  d \cA^{0,\sx}_{k+1, \bom^+(\sq') } \B) + d \cA^{0,\sx}_{k+1, \bom^+(\sq') } 
\ee
by (\ref{new}) the second   term is bounded by $| d \cA^{0, \sx}_{k+1, \bom^+(\sq') } | \leq L^{-\frac32} p_{k+1} \leq \frac12 p_k$  so it suffices to get the same bound for the first term. 
By (221) in \cite{Dim20} the   first  term  is 
\be \label{six}
\begin{split}
d \cA^{\sx}_{k, \bom^{\bullet}(\sq') }  -  d \cA^{0,\sx}_{k+1, \bom^+(\sq') }
   =  &
d\cH^{\sx}_{k, \bom^{\bullet}(\sq') }\B( \cQ_{k, \bom^{\bullet}(\sq')} \cQ^{s,T}_k A_k -   A^{\min}_{k, \bom^+(\sq')} \B)\\
 \end{split}
\ee
Inside of $\sq'_k$ we have $\cQ_{k, \bom^{\bullet}(\sq')} = \cQ_k$ and so by (\ref{new})
\be  \label{orca1}
|\cQ_{k, \bom^{\bullet}(\sq')} \cQ^{s,T}_k A_k -  A^{\min}_{k, \bom^+(\sq')} |= |A_k -   A^{\min}_{k, \bom^+(\sq')}| \leq  p_{0,k}^2 
\ee
In $\sq'_1 - \sq'_k$ we have $\cQ_{k, \bom^{\bullet}(\sq')}  = \cQ_{k+1, \bom^+(\sq')} $ so  
  $ \cQ_{k+1, \bom^+(\sq')} ( \cQ^{s,T}_{k+1} A_{k+1}  -\cA^{0,\sx}_{k+1, \bom^+(\sq')})=0$ also holds with $\cQ_{k, \bom^{\bullet}(\sq')} $.  Then 
\be
\begin{split}
\cQ_{k, \bom^{\bullet}(\sq')} \cQ^{s,T}_k A_k -   A^{\min}_{k, \bom^+(\sq')} =  & 
\cQ_{k, \bom^{\bullet}(\sq')}  \B( \cQ^{s,T}_k A_k  - \cA^{0, \sx}_{k+1, \bom^+(\sq')}\B) \\
=&\cQ_{k, \bom^{\bullet}(\sq')}  \B(  \cQ^{s,T}_k A_k -   \cQ^{s,T}_{k+1} A_{k+1} \B) \\
\end{split}
\ee
However
\be
\cQ_{k+1} \B(  \cQ^{s,T}_k A_k -   \cQ^{s,T}_{k+1} A_{k+1} \B)
= \cQ A_k - A_{k+1} =0
\ee
and  by (\ref{newer})
\be \label{andyboy}
|d \B(  \cQ^{s,T}_k A_k -   \cQ^{s,T}_{k+1} A_{k+1} \B)| =  |  \cQ^{e,T}_k dA_k -   \cQ^{e,T}_{k+1} dA_{k+1} | \leq  C p_k
\ee
These two bounds plus the fact that $  \cQ^{s,T}_k A_k $ and $  \cQ^{s,T}_{k+1} A_{k+1}$ are axial,  and   lemma 16 in \cite{Dim20}  (and  the identity $d\cQ_k\cA = \cQ^{(2)}_k d\cA$)
imply on $\sq_1'$
\be \label{andyboy2}
 |\cQ_k \B(\cQ^{s,T}_k A_k -   \cQ^{s,T}_{k+1} A_{k+1}\B) | \leq  Cp_k
 \ee
Then (\ref{andyboy}), (\ref{andyboy2}) and lemma 17 in \cite{Dim20} imply  on $\sq_1'$ for $1\leq j \leq k$
\be
 |\cQ_j \B(\cQ^{s,T}_k A_k -   \cQ^{s,T}_{k+1} A_{k+1}\B) | \leq L^{k-j} Cp_k
 \ee
 Therefore on $(\de \Om^{\bullet}_j(\sq')  )^{(j)} = \de \sq_j^{' (j)} =   \sq_j^{'(j)}-\sq_{j+1}^{'(j)} $
 \be  \label{orca2}
 | \cQ_{k, \bom^{\bullet}(\sq')}  \B(  \cQ^{s,T}_k A_k -   \cQ^{s,T}_{k+1} A_{k+1} \B)| \leq   L^{k-j} Cp_k
 \ee

Using the bounds  (\ref{orca1}) and (\ref{orca2})  and the estimate on $d\cH^{\sx}_{k, \bom^{\bullet}(\sq') }= d\cH_{k, \bom^{\bullet}(\sq') } $ from (302) in \cite{Dim20},   we  estimate   (\ref{six})  on a unit cube $\De_x \subset  \tilde \sq' $ by 
\be \label{contradiction}
\begin{split}
&
\B|d\cH^{\sx}_{k, \bom^{\bullet}(\sq') } \B( \cQ_{k, \bom^{\bullet}(\sq')} \cQ^{s,T}_k A_k -   A^{\min}_{k, \bom^+(\sq')}  \B)\B|\\
&
 \leq  C p_{0, k+1}  \sum_{b \in \sq'^{(k)}_k }    e^{- \ga d_{\bom^{\bullet}(\sq')}(x, b)} 
 +  C   p_k   \sum_{j=1}^{k-1} \sum_{b \in \de \sq_j^{' (j)}} e^{- \ga d_{\bom^{\bullet}(\sq')}(x, b)}  L^{k-j}  \\
  & \leq    C p_{0, k}^2 + 
 Cp_k e^{-\ga LM r_{k+1}}\leq \frac12 p_k\\
\end{split}
\ee
Here in the second sum  we used  $d_{\bom^{\bullet}(\sq)}(x, b)  \geq d_{\bom^{\bullet}(\sq)}( (\sq'_k)^c, \tilde \sq' ) \leq (L+1)M r_{k+1}$ to extract a factor $e^{-\ga M r_{k+1}}$  from
the decay factor  $e^{- \ga d_{\bom^{\bullet}(\sq')}(x, b)} $.    
The decay factor also kills the $L^{k-j}$ which is less than   $Ce^{-\one (Mr_{k+1})^{-1}d_{\bom^{\bullet}(\sq')}(x, b)}$ and gives the convergence of the sum as in \cite{Dim20}. 
The last step follows since $Cp_{0,k}^2p_{k}^{-1}$ and $  C e^{-\ga LM r_{k+1}}$ 
can be as small as we like.  This completes the proof.

  \subsection{gauge  field  translation}  \label{gfm}

In   the exponent in (\ref{rep2})  we   have    $ \frac12 \| d     \cA_{k,\bom}  \|^2 $, a function of $A_{k,\bom}$.    We expand around the minimizer in $A_{k, \bom}$  subject to the constraints $\cQ A_k =A_{k+1}$   and  $\tau A_k = 0$ on $\Om_{k+1}$ and fixed at $A_{k, \bom}$ on the complement.  The minimizer  is denoted  $ A^{\min} _{k,  \bom^+}$  where  again 
$  \bom^+   =    (  \bom , \Om_{k+1}  )$.
We make the change of variables on $\Om_{k+1}$
\be  \label{lager1} 
  A_{k,\bom}     =  A^{\min} _{k,  \bom^+}  + Z_k  
\ee
where $Z_k$ is a function on  the unit lattice $\Om^{(k)}_{k+1} \subset \tz$.
Then    $\cA_{k, \bom}  = \cA_{k, \bom}(A_{k, \bom})= \cH_{k, \bom} A_{k, \bom}$ becomes
 \be
 \cA_{k, \bom} =
 \cA_{k,  \bom}(A^{\min} _{k, \bom^+} )      + \cZ_{k,  \bom } 
\ee 
 where
\be
 \cZ_{k,  \bom } 
= \cH_{k,  \bom}Z_k 
\ee
The  term $ \cA_{k,  \bom}(A^{\min} _{k, \bom^+} )$   is mixed axial and Landau gauge.
But we have the equivalence  to all axial gauge  $\cA_{k,  \bom}(A^{\min} _{k, \bom^+} )      \sim    \cA^{\sx}_{k,  \bom}(A^{\min} _{k, \bom^+} )  $
and  the identification   $ \cA^{\sx}_{k,  \bom}(A^{\min} _{k, \bom^+} ) =  \cA^{0, \sx}_{k+1, \bom^+ }$.   Furthermore    $ \cA^{0, \sx}_{k+1, \bom^+ } 
\sim   \cA^{0}_{k+1, \bom^+ } $  takes us back to Landau gauge.   Overall we have for some scalar  $\om$ on $\tk$
\be  
 \cA_{k,  \bom}(A^{\min} _{k, \bom^+} )     =    \cA^{0}_{k+1, \bom^+ }  - d \om
\ee
In gauge invariant positions the change of variables is now
\be   \label{lager2} 
  \cA_{k, \bom}  =      \cA^{0}_{k+1, \bom^+ }     + \cZ_{k,  \bom } 
\ee
In particular 
we have 
    as in section 3.5  in \cite{Dim20}      
\be        \label{underwood} 
    \frac12    \|  d  \cA _{k, \bom}  \|^2 
=  \frac12 \|  d  \cA^0_{k+1, \bom^+ }   \|^2     + \frac12 \blan Z_k,  [\De_{k, \bom} ]_{\Om_{k+1} }Z_k \bran     
\ee

For fermions  besides the change in the background field we make the change of variables $\Psi_j \to e^{ie_k \om^{(k-j)} }\Psi_j $  (and
$\bPsi_j \to e^{-ie_k \om^{(k-j)} }\bPsi_j $)
where $\om^{(k-j)}$ is the restriction of $\om$ to $\bbT^{-(k-j)}_{N-k}$.  This has  
Jacobian one.  Then  $\Psi_{k, \bom} \to   e^{ie_k \om(\bom) }\Psi_{k, \bom}$   where $\om(\bom)$ is the restriction of $\om$ to $\cup_j \de \Om_j^{(j)}$.
 The fermi field     $  \psi_{k, \bom} ( \cA_{k, \bom },  \Psi_{k, \bom}  )$     becomes 
 \be
     \psi_{k, \bom} \B( \cA^{0}_{k+1, \bom^+ }  + \cZ_{k,  \bom }   - d \om ,    e^{ie_k \om(\bom) } \Psi_{k, \bom}\B)
       =  e^{ie_k \om}   \psi_{k, \bom}  \B( \cA^{0}_{k+1, \bom^+ }   + \cZ_{k,  \bom }, \Psi_{k, \bom} \B)
 \ee
  (We now include $\psi_{1, \Om_1^c}$ in $\Psi_{k, \bom}$.)
The free fermi action is gauge invariant and so 
\be
\begin{split} 
   &   \fS'^{+}_{k, \bom}\B(\La_k,  \cA^{0}_{k+1, \bom^+ }
    +   \cZ_{k, \bom}  - d \om,\  e^{ie_k \om(\bom)} \Psi_{k, \bom},\  e^{ie_k \om}   \psi _{k, \bom} ( \cA^{0}_{k+1, \bom^+ }+ \cZ_{k,  \bom } )    \B)    \\
 & =     \fS'^{+}_{k, \bom}\B(\La_k,  \cA^{0}_{k+1, \bom^+ } 
 +   \cZ_{k, \bom},\  \Psi_{k,\bom},\  \psi _{k, \bom} ( \cA^{0}_{k+1, \bom^+ } + \cZ_{k,  \bom })     \B)   \\
\end{split} 
\ee
The   Gaussian delta function  $  \de_{G}\B( \Psi_{k+1} - Q(\tilde  \cA   ) \Psi_k \B)  $  becomes
\be   \de_G  \B( \Psi_{k+1} -  Q( \tilde \cA  )  e^{ie_k \om^{(0)} }   \Psi_k \B)  =   
 \de_G \B(  \Psi_{k+1} -  e^{ie_k \om^{(-1)} }Q( \tilde \cA'   )    \Psi_k \B) 
   \ee
   where $\tilde \cA' = \tilde \cA - d \om$ is still arbitrary.  
We     replace  $\Psi_{k+1} $ by    $ e^{ie_k \om(-1) } \Psi_{k+1}$ so the phase factor here
disappears   as well.  This change  means we have made a modification of the original renormalization group transformation,
 but   subsequent integrals over  $\Psi_{k+1}$ are not affected.

Some of the characteristic functions are also affected by the translation, but
we postpone the discussion of this  until the next section.

 Let us collect the changes so far.  Identify 
 \be
    Dm^0_{k+1, \bom^+} (A ) 
    =    \de_{\Om^c_{k+1} } (A_{k+1} -   \cQ  A_k )   \de_{\Om^c_{k+1} }  (\tau A_k ) \  DA_{k, \Om^c_{k+1}} Dm_{k, \bom} (A ) 
 \ee
 and then  (\ref{rep2}) has become
   \be    \label{rep3} 
 \begin{split}
&  \tilde \rho_{k+1} ( A_{k+1}, \Psi_{k+1}   )      
=    
   \\
  &   \sum_{ \bpi,    \Om_{k+1},P_{k+1}, Q_{k+1}  }  \sZ'_{k,\bom}(0)   \sZ'_{k, \bom}    \int     Dm^0_{k+1, \bom^+}  ( A )      Dm_{k, \bpi}  (Z)\     Dm_{k,\bom}(\Psi)   Dm_{k,\Pi}(W) \     D\Psi_k\ DZ_k\ \\
  &
   \de_{G}\B( \Psi_{k+1} - Q( \tilde \cA'  ) \Psi_k \B)      \de_{ \Om_{k+1}}(    \cQ Z_k  )  \de_{ \Om_{k+1}} (\tau Z_k ) 
\  \cC_{k, \bpi } \     \cC^{\star}_{k+1}     \chi_k( \La_k- \Om_{k+1}^{\sim 5})       \chi_{k+1}^0(\Om_{k+1})  \chi'_{k+1}(\Om_{k+1})    
  \\
&  
      \exp \B( -\frac12 \|  d  \cA^0_{k+1, \bom^+ }  \|^2  - \frac12  \blan   Z_k, [ \De_{k, \bom}]_{\Om_{k+1} }Z_k \bran
     -    \fS'^{+}_{k, \bom}  (\La_k) +   E'_k(\La_k  )  
 +  B_{k,\bpi}       +  B^E_{k,\bpi}        \B) \\   
\end{split}
\end{equation}
with the last four terms in the  exponential evaluated at ${\cA =     \cA^{0} _{k+1, \bom^+ } + \cZ_{k,  \bom}  }$.

\subsection{translated  characteristic functions}

 Now we consider  the    effect  of the gauge field   translation in $A_k$ on $\Om_{k+1}$  on the characteristic functions.  
 There is no effect on $\cC_{k,\bpi}$ (which  does not  depend  on $A_k$ on $\La_k^{3 \nat} \supset \Om_{k+1}$) or on $\cC^{\star} $. There is also  no effect
  on 
 $  \chi_k( \La_k-  \Om^{\sim 5}_{k+1})  $ or $\chi^0_{k+1}(\Om_{k+1})$. 
 The only effect is through    $\chi'_{k+1}(\Om_{k+1}) $.
 This also the only characteristic function affected by the change of variables. 
 It  has now become  
 \be  \label{picata}
\chi'_{k}(\Om_{k+1}) = \prod_{\sq \subset \Om_{k+1}}   \chi'_{k}(\sq) 
       \hs       \chi'_{k}(\sq) =  \chi'_{k}\B(\square, Z_k +  (  A^{\min}_{k, \bom^+}   - A^{\min}_{k, \bom^+  (\sq)}  ) \B)   
        \ee

Our immediate task is to get bounds on  $ A^{\min}_{k, \bom^+}   - A^{\min}_{k, \bom^+  (\sq)}  $  so that these characteristic functions
generate bounds on $Z_k$.   To this end  we  use the representation   $A^{\min}_{k, \bom^+}= \cQ_{k, \bom} \cA^{0, \sx}_{k+1, \bom^+}$.       
 As in (\ref{peachy2}) the bounds  on $dA_j$ (including (\ref{newer})) and our basic regularity results from \cite{Dim20}  imply that for   $\sq \subset \de \Om_j$, $j=1, \dots, k+1$,
the field $\cA^{0,\sx}_{k+1, \bom^+} \sim \cA'$  on $\sq^{\dagger}$ satisfies
  \be
   \label{william}
 L^{-\frac12 (k-j)}|\cA'|,\  L^{-\frac32(k-j)}  | \pa \cA'|,\   L^{-(\frac32+ \al)(k-j)} \| \pa  \cA'\|_{(\al)}    
 \leq  CM p_j  
 \ee
It follows that $d\cA^{0,\sx}_{k+1, \bom^+}=d\cA^{0}_{k+1, \bom^+} $ satisfies  on $\de \Om_j$ 
\be \label{exeter}
  \hs  |d\cA^{0,\sx}_{k+1, \bom^+}| \leq L^{\frac32(k-j)} CMp_{j}
\ee

 \begin{lem} \label{peanut0}
Let $\sq$  be an  $LMr_{k+1}$  cube    in    $\Om_{k+1} $. Then  on $\tilde \sq$  
\be
\label{pecan1}
| A^{\min}_{k, \bom^+} -  A^{\min}_{k, \bom^+(\sq)}|  \leq  Cp_{k+1}
\ee
and  hence  $\chi'_{k+1}(\Om_{k+1}) $  enforces on $\Om_{k+1}$
\be
\label{pecan2}
 |  Z_k| \leq  CMp_{k+1}
\ee
\end{lem} 
 \bigskip

 \pr   We have on $\tilde \sq$ that  $dA^{\min}_{k, \bom^+}= \cQ^{(2)}_{k} d\cA^{0, \sx}_{k+1, \bom^+}$.  
 Then the bound (\ref{exeter}) gives 
 $|dA^{\min}_{k, \bom^+}| \leq CMp_{k+1}$.   Similarly  $dA^{\min}_{k, \bom^+(\sq)}= \cQ^{(2)}_{k}  d\cA^{0, \sx}_{k+1, \bom^+(\sq)}$
 and 
 $| d\cA^{0, \sx}_{k+1, \bom^+(\sq)}| \leq L^{-\frac32}p_{k+1}$ from $\chi^0_{k+1}(\Om_{k+1})$
 imply that   $|dA^{\min}_{k, \bom^+(\sq)}| \leq Cp_{k+1}$.
 Altogether then  
 \be \label{bacon1}
|d (A^{\min}_{k, \bom^+} -  A^{\min}_{k, \bom^+(\sq)} )| \leq  | d(\cA^{0, \sx}_{k+1, \bom^+(\sq)} -\cA^{0, \sx}_{k+1, \bom^+})| \leq  CMp_{k+1}
 \ee
 We also have on $\tilde \sq$ 
 \be  \label{bacon2}
  \cQ (A^{\min}_{k, \bom^+} -  A^{\min}_{k, \bom^+(\sq)} ) = A_{k+1} - A_{k+1} =0 
  \ee
 Since $A^{\min}_{k, \bom^+} -  A^{\min}_{k, \bom^+(\sq)}$ is axial the result (\ref{pecan1})  follows by (\ref{bacon1}), (\ref{bacon2}) and  lemma 16 in \cite{Dim20}. 
 
 The second result follows from the bound on $\sq \subset \Om_{k+1} $
 \be
 \begin{split}
| Z_k| \leq  &   | Z_k+  (  A^{\min}_{k, \bom^+}   - A^{\min}_{k, \bom^+  (\sq)} ) |  +  | A^{\min}_{k, \bom^+}   - A^{\min}_{k, \bom^+  (\sq)}| \\
 \leq &  p_{0,k} ^2+  CMp_{k+1}   \leq CMp_{k+1}  \\
 \end{split} 
 \ee
This completes the proof.
\bigskip

If we shrink $\Om_{k+1}$  we can do better.

\begin{lem} \label{peanut1}
Let $\sq$  be an  $LMr_{k+1}$  cube    in    $\Om^{3\nat}  _{k+1} $. Then
 on $\tilde \sq $
\be \label{walnut1}
 |d \cA^{0,\sx}_{k+1, \bom^+}  - d \cA^{0, \sx}_{k+1, \bom^+(\sq)} |   \leq   e^{- r_{k+1}}
\ee
\be
\label{walnut2}
| A^{\min}_{k, \bom^+} -  A^{\min}_{k, \bom^+(\sq)}|  \leq  Ce^{-r_{k+1} }
\ee
\end{lem} 
\bigskip

\pr 
We only need prove the first; the second follows as in the previous lemma.
 The proof is similar to the proof in lemma  \ref{newest}.  Since we take $\sq \subset \Om^{3 \nat}_{k+1}$ we  have   $\Om_j(\sq ) \subset \Om_j$ for $j=1 \dots k+1$.  
  Then   since   $\cA^{0, \sx}_{k+1, \bom^+}$ is a minimizer of $\|d\cA\|^2$ subject to $\cQ_{k+1, \bom^+} \cA = \cQ_{k+1, \bom^+ } \cA^{0,\sx}_{k+1, \bom^+}$
   it is also a minimizer of  $\|d\cA\|^2$ subject to $\cQ_{k+1, \bom^+(\sq)} \cA = \cQ_{k+1, \bom^+(\sq) } \cA^{0,\sx}_{k+1, \bom^+}$
   since the latter is more restrictive.    Then by lemma 13 in \cite{Dim20}    
   $\cA^{0, \sx}_{k+1, \bom^+}$ is on the same restricted orbit as  $ \cH^{0,\sx}_{k+1, \bom^+(\sq) }( \cQ_{k+1, \bom^+(\sq) } \cA^{0,\sx}_{k+1, \bom^+} )$
   and so  $d\cA^{0, \sx}_{k+1, \bom^+} = d\cH^{0, \sx}_{k+1, \bom^+(\sq) } (\cQ_{k+1, \bom^+(\sq) } \cA^{0,\sx}_{k+1, \bom^+} )$.
   Therefore  it suffices to show 
   \be \label{boxer}
 \B|  d\cH^{0,\sx}_{k+1, \bom^+(\sq) } \cQ_{k+1, \bom^+(\sq) }\B( \cA^{0,\sx}_{k+1, \bom^+}   - \cQ_{k+1}^{s,T} A_{k+1} \B) \B| \leq e^{- r_{k+1}}
  \ee

Note that  on  $\sq_1$ (actually $\sq_1^{(k+1)}$) 
\be   \label{x1}
\cQ_{k+1 } \B(\cA^{0, \sx} _{k+1, \bom^+}  - \cQ_{k+1}^{s,T} A_{k+1} \B) = A_{k+1} - A_{k+1}  =0
\ee
We also have  that   on $\sq_1$ from our bounds on $dA_{k+1}$ and  $d \cA^{0, \sx} _{k+1, \bom^+} $ 
\be \label{x2}
|d \B(\cA^{0, \sx} _{k+1, \bom^+}  - \cQ^{s,T} A_{k+1} \B)| = |d \cA^{0, \sx} _{k+1, \bom^+}  - \cQ^{e,T} dA_{k+1} |  \leq CMp_{k+1}
\ee 
Furthermore both $\cA^{0, \sx} _{k+1, \bom^+} $ and $ \cQ^{s,T} A_{k+1}$ are axial, so  by   (\ref{x1}), (\ref{x2}) and  lemma 16 in \cite{Dim20} 
\be  \label{x3}
\B|\cQ_{k} \B(\cA^{0, \sx} _{k+1, \bom^+}  - \cQ_{k+1}^{s,T} A_{k+1} \B) \B| \leq CMp_{k+1}
\ee
Then by   (\ref{x2}), (\ref{x3}) and  lemma 17  in \cite{Dim20}
 \be
 |\cQ_j \B( \cA^{0, \sx} _{k+1, \bom^+}  - \cQ^{s,T} A_{k+1} \B)| \leq L^{k-j}  CMp_{k+1}
 \ee
Hence  on $\de \sq_j^{(j)}$
\be   \label{corndog}
| \cQ_{k+1, \bom^+(\sq) }\B( \cA^{0,\sx}_{k+1, \bom^+}   - \cQ_{k+1}^{s,T} A_{k+1} \B)| \leq L^{k-j} CMp_{k+1}
\ee

We use this bound in  (\ref{boxer}).
First note that since     $\cQ_{k+1 }  = \cQ_{k+1, \bom^+(\sq) }$ on $\sq_{k+1}$ the identity (\ref{x1}) implies that 
$\cQ_{k+1,\bom^+(\sq) } (\cA^{0, \sx} _{k+1, \bom^+}  - \cQ^{s,T} A_{k+1} )  =0$  on   $\sq_{k+1}^{(k+1)}$.
Thus we can exclude such points from our analysis of (\ref{boxer}).
Then by (\ref{corndog})  and the basic bound on $d \cH^{0,\sx}_{k+1, \bom^+(\sq)}=d \cH^0_{k+1, \bom^+(\sq)}$ from (302) in \cite{Dim20}
to get for $p$ in a unit cube  $\De_x \subset \tilde \sq $
\be   \label{quonset}
| d \cA' (p) - d \cA^0_{k+1, \bom^+(\sq)}(p) | \leq \sum_{j=1}^k \sum_{p \in \de \sq_j^{(j)} } e^{- \ga d(x,  p ) } L^{k-j} CMp_{k+1}
\leq  e^{-\ga LMr_{k+1} } CMp_{k+1}
  \leq  e^{ -  r_{k+1}}
\ee
Here we used $d( \sq^c_{k+1}, \tilde \sq ) \geq LMr_{k+1}$ to extract the factor  $e^{-\ga LMr_{k+1} }$ and estimated the sum as in (\ref{contradiction}). 
This completes the proof of (\ref{walnut1}) and the lemma.

\subsection{another small field expansion}  \label{another}

The $\chi'_{k}(\Om_{k+1})$ in (\ref{picata})   depends on $A_{k+1}$  in a  nonlocal way due to the term  $A^{\min}_{k, \bom^+}$.    This   is unsatisfactory for subsequent steps.   
Instead     we  introduce  sharper     bounds on the fluctuation variable  $Z_k$   which will  help remove this dependence.

  For an $LMr_{k+1}$ cube $\sq \subset  \Om^{3\nat}_{k+1}$   define the characteristic function
\be  \label{sister1}
 \chi^{\da}_k(\square)   =   \chi  \B(  \sup_{b \subset \tilde  \square}  |Z_k(b) | \leq    p^{\frac43} _{0,k} \B)
\ee
and then with $ \zeta^{\da}_k(\sq)   = 1 -  \chi^{\da}_k(\sq) $ write
\be \label{suds} 
\begin{split}
1 = &  \prod_{\sq \subset \Om^{3\nat}_{k+1} } (   \zeta^{\da}_k(\sq)   +  \chi^{\da}_k(\sq)  )  \\
 = & \sum_{R_{k+1} \subset \Om^{3\nat}_{k+1} }\prod_{\sq \subset R_{k+1}}  \zeta^{\da}_k(\sq)  \prod_{\sq \subset (\Om^{\nat}_{k+1} -  R_{k+1})}  \chi^{\da}_k(\sq) \\
 \equiv   &  \sum_{R_{k+1}   \subset  \Om^{3\nat}_{k+1} }   \zeta^{\da}_k  (R_{k+1})   \chi^{\da}_k  ( \Om^{3\nat}_{k+1} - R_{k+1} ) \\
\end{split}
\ee
where $R_{k+1}$ is a union of $LMr_{k+1}$ cubes.

For  $\sq \subset \Om_{k+1}^{3 \nat}-R_{k+1}$ the bound  $|Z_k| \leq  p^{\frac43} _{0,k}$   and the bound of
 lemma  \ref{peanut1} yield on $\tilde \sq$
 \be
     |  Z_k +  (  A^{\min}_{k, \bom^+}   - A^{\min}_{k, \bom^+(\sq)} )|  \leq  p^{\frac43} _{0,k}  +   e^{-r_k} CMp_k  \leq    p_{0,k}^2
\ee
and therefore  
\be  \label{one}
  \chi'_{k}\B(\square , Z_k +  (  A^{\min}_{k, \bom^+}   - A^{\min}_{k, \bom^+(\sq)}  ) \B)=1
  \ee
 Since $R_{k+1} \subset \Om_{k+1}^{3 \nat}$  we  can now shrink the offending function by
\be \label{bogus}
 \chi'_{k}(\Om_{k+1} ) = \chi'_{k}((\Om_{k+1}- \Om^{3\nat}_{k+1} ) \cup R_{k+1} ) \chi'_{k}( \Om^{3\nat}_{k+1} - R_{k+1} ) = \chi'_{k}(S_{k+1})
\ee
where 
\be S_{k+1} =  (\Om_{k+1}- \Om^{3\nat}_{k+1} ) \cup R_{k+1} 
\ee
 Note that $R_{k+1}$ and $S_{k+1}$ determine each other  ($R_{k+1} = S_{k+1} \cap \Om^{3\nat}_{k+1}$) so 
the sum over $R_{k+1}$ can be regarded as a sum over $S_{k+1}$.  

We make a further shrinkage and introduce a tentative new small field region
\be \label{wordy2} 
T_{k+1}  = \Om_{k+1}^{5 \nat} - R_{k+1}^{ \sim 5} 
\ee 
The characteristic functions   only depends on $A_{k+1}$ in $\La^{3 \nat}_{k+1}$  through  the function $\chi'_{k+1}(S_{k+1})$, but now it is weakened since $S_{k+1}$ is a considerable
distance  from
from $\La_{k+1}$.   
It has the form 
\be    \label{lax2}
\chi'_k(S_{k+1}) = \prod_{\square \subset  S_{k+1}}   \chi'_{k}(\square) 
       \hs       \chi' _{k}\B(\square ,   Z_k +  (   A^{\min}_{k, \bom^+   }  - A^{\min}_{k, \bom^+(\square)}  ) \B) 
 \ee
  We  want  to replace $A^{\min}_{k, \bom^+   }$ by something better localized, namely a function
$A^{\min}_{k, \bom'   }$ where $\bom'$ is 
is   approximately   localized in  $\La_{k} - T_{k+1}  $ and will now be defined.

First  for any $X$ a union of $Mr_k$ cubes  with enlargement  $\tilde X$    define $\bom(X) = (\Om_1(X), \dots \Om_k(X))$ by specifying that
$\Om_k(X)$  adds a  layer of $Mr_k$ cubes to $\tilde X$,  then $\Om_{k-1}(X)$ adds another layer of $L^{-1}Mr_k$ cubes,  then a layer of $L^{-2}Mr_k$ cubes, etc.
This generalizes the construction of $\bom(\sq)$.
Similarly if $X$ is specified as  a union of $LMr_{k+1}$ cubes  with enlargement  $\tilde X$    define $\bom(X) = (\Om_1(X), \dots \Om_{k+1}(X))$ by adding successively
smaller layers as before.  
Now define  $\bom'=  \bom( \La_{k},  \Om_{k+1},  T_{k+1} )$ by specifying  $\bom'=(\Om'_1, \dots, \Om'_{k+1})$
where
\be
\Om'_j = \begin{cases} \Om_j(\La_k) \cap \Om_j(T^c_{k+1}) & \hs j=1,\dots, k \\
\Om_{k+1}  \cap \Om_{k+1}(T^c_{k+1}) & \hs  j= k+1 \\
\end{cases}
\ee
Then define the field  
  \be    \label{piggy}
  \begin{split}
   \cA^{0, \sx}_{k+1,  \bom'}   =  &    \cA^{0,\sx}_{k+1,  \bom'} (  \tilde A_{k+1,\bom'}  )\\
       \tilde A_{k+1,\bom'}  = &
   \B(  [\cQ_{k, \bom(\La_k)} \cQ_k^{s,T} A_k  ]_{\de \Om_k} ,  [\cQ_{k+1, \bom(\La^c_{k+1})} \cQ_{k+1}^{s,T} A_{k+1}  ]_{ \Om_{k+1}} \B)
     \\
     \end{split}
 \ee
 Note that inside $\Om'_{k+1}$  we have 
 \be
 \begin{split}
[ \cQ_{k, \bom(\La_k)} \cQ_k^{s,T} A_k  ]_{ \de \Om_k}=  & [\cQ_{k} \cQ_k^{s,T} A_k  ]_{\de \Om_k} = A_{k, \de \Om_k} \\
 [\cQ_{k+1, \bom(\La^c_{k+1})} \cQ_{k+1}^{s,T} A_{k+1}  ]_{ \Om_{k+1}} =  &   [\cQ_{k+1} \cQ_{k+1}^{s,T} A_{k+1}  ]_{ \Om_{k+1}} = A_{k+1,  \Om_{k+1}}  \\
 \end{split}
 \ee
 Thus in this region  $ \tilde A_{k+1,\bom'}  = A_{k+1, \bom^+}$ which is  the  the argument of $\cA^{0,\sx}_{k+1, \bom^+}$.
Also note that inside $\Om_1'$  the field  $  \cA^{0, \sx}_{k+1,  \bom'}  $ only depends on $A_k,A_{k+1}$  on a small neighborhood of $\Om' _1$.  We define   $A^{\min}_{k, \bom'   }= \cQ_{k, \bom'^{\bullet}} \cA^{0, \sx}_{k+1,  \bom'} $ where $\bom'^{\bullet}$ deletes the last entry of $\bom'$. 
We  want to   compare it  with $A^{\min}_{k, \bom^+  }= \cQ_{k, \bom} \cA^{0, \sx}_{k+1,  \bom^+} $.
\bigskip
 
 \begin{lem}  \label{peanut2}  On $\tilde S_{k+1} $
 \be \label{walnut3}
|d\cA^{0,\sx}_{k+1, \bom^+}    - d \cA^{0,\sx}_{k+1, \bom'} |   \leq e^{-r_{k+1}}
\ee
\be \label{walnut4} 
|A^{\min}_{k, \bom^+}-  A^{\min}_{k, \bom' } | \leq  C e^{-r_{k+1} }
\ee
\end{lem}
\bigskip

 \pr The  proof  is   similar to the proof of lemma \ref{peanut1}. 
  First note  that  $\Om'_j \subset  \Om_j$   implies that $d\cA^{0,\sx}_{k+1, \bom^+}  = d \cH^{0, \sx}_{k+1, \bom'} (\cQ_{k+1, \bom'} \cA^{0,\sx}_{k+1, \bom^+} )$.
 So  it suffices to  bound $d\cH^{0,\sx}_{k+1, \bom' }( \cQ_{k+1, \bom' }  \cA^{0,\sx}_{k+1, \bom^+}   - \tilde A_{k+1, \bom'} )$.
 However on $\tilde S_{k+1}$ (or even $\Om'_{k+1} $)  we have  $ \cQ_{k+1, \bom' }  \cA^{0,\sx}_{k+1, \bom^+} =  \cQ_{k+1, \bom^+}  \cA^{0,\sx}_{k+1, \bom^+} = A_{k+1, \bom^+}$ 
 and $\tilde A_{k+1, \bom'} = A_{k+1, \bom^+}$. They agree and so  
  \be \label{puff1}
  \cQ_{k+1, \bom' }  \cA^{0,\sx}_{k+1, \bom^+}   - \tilde A_{k+1, \bom'} =0  \textrm{ on } \tilde S_{k+1}
 \ee
  Outside of $\tilde  S_{k+1}$ we can treat the piece in $\de \Om_k$ and the piece in $\Om_{k+1}$ separately.    In $\de \Om_k \cap \tilde S^c_{k+1}$ we have
 $ \cQ_{k+1, \bom' }  \cA^{0,\sx}_{k+1, \bom^+}   - \tilde A_{k+1, \bom'}  =   \cQ_{k, \bom(\La_k) } (  \cA^{0,\sx}_{k+1, \bom^+}   -  \cQ_k^{s,T} A_k  )$.
 But $\cQ_k(  \cA^{0,\sx}_{k+1, \bom^+}   -  \cQ_k^{s,T} A_k  ) = A_k - A_k =0$ and 
 $|d(  \cA^{0,\sx}_{k+1, \bom^+}   -  \cQ_k^{s,T} A_k  )| \leq CMp_{k+1}$ as in lemma \ref{peanut1}.  By lemma 17 in \cite{Dim20} it follows that  
 $|\cQ_j(  \cA^{0,\sx}_{k+1, \bom^+}   -  \cQ_k^{s,T} A_k  )| \leq  L^{k-j}CMp_{k+1}$ and hence   $|\cQ_{k, \bom(\La_k)} (  \cA^{0,\sx}_{k+1, \bom^+}   -  \cQ_k^{s,T} A_k  )| \leq  L^{k-j}CMp_{k+1}$ on $\de \Om_j(\La_k)$.
   Similarly we establish the same bound
  on   $ \Om_{k+1} \cap \tilde S^c_{k+1}$.    altogether then we have the bound  on $\de \Om'_j \cap \tilde S^c_{k+1} $
  \be \label{puff2}
  |  \cQ_{k+1, \bom' }  \cA^{0,\sx}_{k+1, \bom^+}   - \tilde A_{k+1, \bom'}| \leq     L^{k-j}CMp_{k+1}
 \ee
  Now   using (\ref{puff1}),(\ref{puff2}) and  the bound  on $d\cH^{0,\sx}_{k+1, \bom' }=d\cH^{0}_{k+1, \bom' }$ from (302) in \cite{Dim20} we find on $\tilde S_{k+1}$
  \be \label{boxer2}
 \B|  d\cH^{0,\sx}_{k+1, \bom' } \B(  \cQ_{k+1, \bom' }  \cA^{0,\sx}_{k+1, \bom^+}   - \tilde A_{k+1, \bom'} \B) \B| 
 \leq   e^{-\ga LMr_{k+1} } CMp_{k+1}
  \leq  e^{ -  r_{k+1}}  
  \ee
 Here we used $d(\tilde S^c, S) \geq LMr_{k+1}$ to extract a factor $  e^{-\ga LMr_{k+1} }$. 
 This  gives the bound (\ref{walnut3}).  
 
 For the  bound (\ref{walnut4})  we have $\cQ A^{\min}_{k, \bom^+} = A_{k+1}$ on $\Om_{k+1}$ and          $\cQ  A^{\min}_{k, \bom' } = A_{k+1}$ on $\Om'_{k+1}$.
 Then since on $\tilde S_{k+1} \subset  \Om_{k+1}\cap \Om'_{k+1}$ we have
 \be \label{jones1}
 \cQ ( A^{\min}_{k, \bom^+} - A^{\min}_{k, \bom'}) =A_{k+1} -A_{k+1} =  0 
 \ee
 Also $A^{\min}_{k, \bom^+} = \cQ_{k, \bom} \cA^{0,\sx}_{k+1, \bom^+}$ and   $A^{\min}_{k, \bom'} = \cQ_{k, \bom'^{\bullet} } \cA^{0,\sx}_{k+1, \bom'}$. 
 But on $\tilde S_{k+1}$ we have  $\cQ_{k, \bom}  = \cQ_k = \cQ_{k, \bom'^{\bullet} } $. 
 Hence  on $\tilde S_{k+1}$ by (\ref{walnut3})
 \be \label{jones2}
| d   ( A^{\min}_{k, \bom^+} - A^{\min}_{k, \bom'}) | = |\cQ^{(2)}_k (  d\cA^{0,\sx}_{k+1, \bom^+} - d\cA^{0,\sx}_{k+1, \bom'} )| \leq e^{-r_{k+1}}
\ee
  The bound (\ref{walnut4})   now follows  by (\ref{jones1}) and (\ref{jones2}) and lemma 16 in \cite{Dim20}.    This completes the proof.

Now define      with  $\bpi^+  = (\bpi, \Om_{k+1}, P_{k+1}, Q_{k+1}, R_{k+1} )$
\be
 \fH_{k, \bpi^+}  = 
     \B[ A^{\min}_{k, \bom' }   - A^{\min}_{k, \bom^+}  \B]_{S_{k+1} }
\ee
We  have just seen that   $|\fH_{k, \bpi^+}  | \leq  C e^{-r_{k+1} }$.
We would like to make   the change of variables   $
  Z_k \to   Z_k  +    \fH_{k, \bpi^+}  \textrm{ in } S_{k+1}
 $
in which case we would have 
 \be   
\chi'_{k}(S_{k+1}) = \prod_{\square \subset  S_{k+1}}   \chi'_k(\square) 
       \hs       \chi' _{k}\B(\square , Z_k+  (   A^{\min}_{k, \bom'   }  - A^{\min}_{k, \bom^+(\sq)}  ) \B) 
 \ee
 and then    there  would be    no dependence on $A_{k+1} $  or on $Z_k$ in the new small field region  $T^{3 \nat}_{k+1}$.
 However  this would have an unpleasant effect on $ \zeta^{\da}(R_{k+1} ) $.   This is not a problem in $ (\Om_{k+1}- \Om^{3\nat}_{k+1} )$ which has no $\hat \zeta$.    As in \cite{Bal88b}  in $R_{k+1}$   we only make the translation at points
satisfying a large field condition intermediate  between  $p_{0,k}^2$ ( for $\chi'$) and $p_{0,k} ^{\frac43} $ ( for $ \chi^{\da},   \zeta^{\da} $).  
 Define  $g$ on $ S_{k+1} =  (\Om_{k+1}- \Om^{3\nat}_{k+1} ) \cup R_{k+1} $
 by 
 \be
 g_b(Z)  = \begin{cases} 
  1 &       b \in ( \Om_{k+1}- \Om^{3\nat}_{k+1}) \textrm{ or }    b \in R_k \textrm{ and }   |Z(b)| \geq   p_{0,k}^{\frac53}   \\ 
  0 &    b \in R_k \textrm{ and }  |Z(b) | <   p_{0,k}^{\frac53} \\
 \end{cases}
 \ee
 and make the translation
 \be \label{alibaba} 
  Z_k \to   Z_k  + g(Z_k)  \   \fH_{k, \bpi^+}  \textrm{ in } S_{k+1}
  \ee
 The  fluctuation action $ \frac12  \blan   Z_k, [ \De_{k, \bom}]_{\Om_{k+1} }Z_k \bran$
becomes $
 \frac12  \blan   Z_k, [ \De_{k, \bom}]_{\Om_{k+1} }Z_k \bran +  R^{(1)} _{k, \bpi^+ }
$
where $ R^{(1)} _{k, \bpi^+ }$ is tiny, see lemma \ref{outsize5}.
  There is also the following:

\begin{lem}    For $\square \subset S_{k+1} $ after the translation (\ref{alibaba}) the new characteristic functions are 
\begin{enumerate}
\item  
 $   \chi' _{k} (\square, Z_k  + g  (   A^{\min}_{k, \bom'     }  - A^{\min}_{k,\bom^+}   )
   +  (   A^{\min}_{k, \bom^+    }  - A^{\min}_{k, \bom^+(\sq)}   )\B) $
which  is independent of $A_{k+1} $ in $T_{k+1} ^{3 \nat }$.  
\item
  $
 \zeta^{\da}_k \B(\square , Z_k+  g (   A^{\min}_{k, \bom'   }  - A^{\min}_{k, \bom^+(\sq)}  ) \B) = \zeta^{\da} _{k} (\square,  Z_k)
  $
\end{enumerate} 
 \end{lem}

 \pr  It suffices to consider $\sq \subset R_{k+1} $ 
 We write 
 \be
 \begin{split}
 &
   \chi' _{k} (\square, Z_k  + g  (   A^{\min}_{k, \bom'     }  - A^{\min}_{k,\bom^+}   )
   +  (   A^{\min}_{k, \bom^+    }  - A^{\min}_{k, \bom^+(\sq)}   )\B) \\
 &  =  \prod_{ b \in  \tilde \square}  \chi \B(\B |Z_k(b)   + g_b(Z_k)   (   A^{\min}_{k, \bom'     }(b)   - A^{\min}_{k,\bom^+}(b)    )
   +  (   A^{\min}_{k, \bom^+    } (b)  - A^{\min}_{k, \bom^+(\sq)}(b) ) \B| \leq p^2_{0,k}     \B) \\
 \end{split}
 \ee
At points $b$  where  $| Z_k(b) | \geq p_{0,k} ^{\frac53} $ we have $g_b = 1$ and the factor 
is
\be
\chi  \B(\B |Z_k(b)   +   \ (   A^{\min}_{k, \bom'     }(b)   - A^{\min}_{k, \bom^+(\sq)}(b) ) \B| \leq p^2_{0,k}     \B)
 \ee
 At points where $| Z_k(b) | <  p_{0,k} ^{\frac53} $ we have $g_b  = 0 $
  and the factor is   
 \be
   \chi \B(\B |Z_k(b)     +  (   A^{\min}_{k, \bom^+    } (b)  - A^{\min}_{k, \bom^+(\sq)}(b) ) \B| \leq p^2_{0,k}     \B)   =1  
   \ee
   Here we use the bound $| A^{\min}_{k, \bom^+} -  A^{\min}_{k, \bom^+(\sq)}|  \leq  Ce^{-r_{k+1} }$ from lemma \ref{peanut1}
   and $p_{0,k} ^{\frac53} +  Ce^{-r_{k+1} } \leq p_{0,k}^2$. 
   In both cases the factor is independent of  $A_{k+1} $ in $T_{k+1} ^{3 \nat }$.
  
For the second point  suppose  $ |Z_k(b)| \geq   p_{0,k}^{\frac53} $ for some   $b \in  \tilde \sq $.   Then  since $|A^{\min}_{k, \bom^+}-  A^{\min}_{k, \bom' } | \leq  C e^{-r_{k+1} }$ by lemma \ref{peanut2} and since $p_{0,k}^{\frac53} - Ce^{-r_{k+1} } \geq p_{0,k}^{\frac43}$,   both sides are one and hence equal.
 Otherwise  $g_b=0$ for all $b \in \tilde \sq $,  and the result follows.

 \bigskip
 
No other characteristic functions are affected, but   the   translation  
forces adjustments   in   the action.  This is tolerable since the dependence is analytic and we can  deal with it. 
The translation means we replace $  \cZ_{k, \bom}  =\cH_{k, \bom}  Z_k $  by 
\be \label{bullet} 
   \cZ_{k, \bpi^+}  \equiv    \cH_{k, \bom} \B(Z_k + g \    \fH_{k, \bpi^+} \B)
\ee

\bigskip
 
 \rem  We claim  that $ \fH_{k, \bpi^+}  $ can be regarded as a function of  $d \cA^0_{k+1, \bom^+} $.     
 Indeed we used lemma 16 in \cite{Dim20}  to  express  $ \fH_{k, \bpi^+}  $ as a local function of  $d\cH^{0}_{k+1, \bom' }( \cQ_{k+1, \bom' }  \cA^{0,\sx}_{k+1, \bom^+}   - \tilde A_{k+1, \bom'} )$.  Then we used lemma 17 in \cite{Dim20} to express $( \cQ_{k+1, \bom' }  \cA^{0,\sx}_{k+1, \bom^+}   - \tilde A_{k+1, \bom'} )$ as a local function of either
 $d(  \cA^{0,\sx}_{k+1, \bom^+}   -  \cQ_k^{s,T} A_k  ) =  d\cA^{0}_{k+1, \bom^+}   -  \cQ_k^{e,T} dA_k  $ in $\de \Om_k$ or   $d(  \cA^{0,\sx}_{k+1, \bom^+}   -  \cQ_{k+1}^{s,T} A_{k+1}  ) = d  \cA^{0}_{k+1, \bom^+}   -  \cQ_{k+1}^{e,T} dA_{k+1}  $ in $\Om_{k+1}$.  Since $dA_{k+1, \bom^+} = \cQ^{(2)}_{k+1, \bom^+} d \cA^0_{k+1, \bom^+}$ this  proves the claim.
 
 The  only non-local part of this construction comes with the operator $d\cH^{0}_{k+1, \bom' }$.  Thus when we introduce weakening parameters later on
 it will suffice to weaken this operator to   $d\cH^{0}_{k+1, \bom' }(s)$.

\bigskip 

\noindent
\textbf{Summary:}. We collect inactive characteristic functions by  defining
\be
\cC^{\star \star}_{k+1}  =  \cC^{\star} _{k+1}      \chi_k( \La_k- \Om_{k+1}^{\sim 5})    \chi'_{k}(S_{k+1})  \zeta^{\da}_k  (R_{k+1}) 
\ee
Now (\ref{rep3}) has become
  \be    \label{rep4} 
 \begin{split}
&  \tilde \rho_{k+1} ( A_{k+1}, \Psi_{k+1} , Q_{k+1}   ) =  \sum_{ \bpi, \Om_{k+1}, P_{k+1}, Q_{k+1},   R_{k+1} }   \sZ'_{k,\bom}(0)   \sZ'_{k, \bom}  
     \\
= &   
       \int      Dm^0_{k+1, \bom^+}  ( A )      Dm_{k, \bpi}  (Z)\     Dm_{k,\bom}(\Psi)   Dm_{k,\bpi}(W) \   D\Psi_k \   DZ_k  \\
  & 
   \de_{G}\B( \Psi_{k+1} - Q( \tilde \cA'  ) \Psi_k \B)      
    \de_{ \Om_{k+1}}(    \cQ Z_k  )  \de_{ \Om_{k+1}} (\tau Z_k ) \  
     \cC_{k, \bpi }\  \cC^{\star \star}_{k+1}      \chi_{k+1}^0(\Om_{k+1})    \chi^{\da}_k  ( \Om^{3\nat}_{k+1} - R_{k+1} )    
 \\
 &      \exp \B(  -\frac12 \|  d  \cA^0_{k+1, \bom^+ }  \|^2   - \frac12  \blan   Z_k, [ \De_{k, \bom}]_{\Om_{k+1} }Z_k \bran    -    \fS'^{+}_{k, \bom} (\La_k)  +   E'_k(\La_k  ) +  B_{k,\bpi}       +  B^E_{k,\bpi}     + R^{(1)}_{k, \bpi^+}     \B) \\   
\end{split}
\end{equation}
with the last four terms in the exponential  evaluated at $\cA =     \cA^{0} _{k+1, \bom^+ } + \cZ_{k, \bpi^+}  $.

\subsection{ultralocal fluctuation integral} \label{ultralocal}

We       parametrize    the integral  over  $Z_k$  on $\Om_{k+1}$ with the constraints $\tau Z=0, \cQ Z=0$      
   by  $Z_k = C \tilde Z_k$.           Here $\tilde Z_k $ belongs to  a vector space $V = V(\Om_{k+1})  $ consisting  of pairs  
  $\tilde Z_k = (\tilde Z_1, \tilde Z_2) $ where   $\tilde Z_1$ satisfies the axial gauge condition  $\tau Z  =0$ in each $L$-block
  and $\tilde Z_2$ is  defined  on  bonds joining $L$ blocks, except for the central bond in each face.  The map $Z_k = C \tilde Z_k$ assigns 
  values to the central bond so that $\cQ Z =0$.   See  \cite{Dim15}  for more details

 Then we  identify a  Gaussian measure with the identities 
\be   \label{stark}
 \int f( Z_k)   \de_{\Om_{k+1}} ( \cQ  Z_k )     \de_{\Om_{k+1}}( \tau  Z_k ) \exp\B( - \frac12 \blan Z_k, [ \De_{k, \bom} ]_{\Om_{k+1}} Z_k \bran \B)   D  Z_k  
    =     \de \sZ_{k, \bom^+}    \    \int f(C \tilde Z_k)    d \mu_{C_{k, \bom^+}}(  \tilde Z_k   )    
  \ee
 Here  the covariance is   
 \be
 C_{k, \bom^+} =   C_{k, \bom}(\Om_{k+1} ) =( C^T [ \De_{k, \bom} ]_{\Om_{k+1}}C)^{-1}
 \ee

We would like  to make the change of variables $\tilde Z_k \to  C _{k, \bom^+} ^{\frac12} \tilde Z_k$ to change the last integral
to  the ultralocal version  $  \int f( C C_{k, \bom^+}^{\frac12}  \tilde Z_k)    d \mu_{I}(  \tilde Z_k   ) $ where now  the Gaussian measure has identity 
covariance.  But the non-locality $C_{k, \bom^+} ^{\frac12} $ causes problems elsewhere,  particularly in the characteristic functions. 
So before we do this we make a local approximation.

First  we study modifications of   $  C_{k, \bom^+ }$.    It suffices to consider the    operator $CC_{k, \bom^+} C^T$ 
which is defined on  functions on all bonds in $\Om_{k+ 1} $ and agrees with $ C_{k, \bom^+} $
as quadratic forms on $V \times V$ ( see the discussion in  \cite{Dim15b} ).     This    has the representation
   \cite{Bal85b}, \cite{Dim15},  \cite{Dim20}.
\be
C C_{k, \bom^+}  C^T = (1 + \pa \cM) [\cQ_k  \tilde \cG_{k+1, \bom^+ } \cQ_k^T]_{\Om_{k+1} }( 1 + \pa \cM)^T
\ee
where $ (1 + \pa \cM) $ is a  local operator and  $\tilde \cG_{k+1, \bom^+ }$  is a certain Green's function. 
The Greens function has a multiscale  random walk expansion and corresponding  exponential decay estimates, and this 
carries over to    $C C_{k, \bom^+}  C^T  $ and $C_{k, \bom^+} $.   If   $\{\Up \} $ is an orthonormal basis for the space $V$
and if  we define $C_{k, \bom^+} (\Up, \Up' ) = < \Up, C_{k, \bom^+} \Up'> $ then 
\be \label{tingle}  
 | C_{k, \bom^+ } (\Up,\Up') |   \leq   Ce^{ - \ga d(\Up,\Up') }
\ee
The random walk expansion enables the introduction of weakening parameters and defines  $\tilde \cG_{k+1, \bom^+ }(s)$ 
and thereby $ C_{k, \bom^+ } (s) $  which also satisfies the bound (\ref{tingle}). 

Again consider an   $Mr_k$ cube $\sq$  with   $\tilde \sq$ an enlargement  by a layer of $M r_k$ cubes.  We define
 a local approximation to   $C _{k, \bom^+}  $ by defining  as in \cite{Bal98a},  \cite{Dim13b} 
\be
C^{\loc} _{k, \bom^+} = \sum_{\sq \subset \Om_{k+1}}  1_{\sq}  C_{k, \bom^+,\sq}  \hs \hs   C_{k, \bom^+,\sq} = C _{k, \bom} \B(s_{\tilde \sq} =1, s_{\tilde \sq^c} =0 \B) 
\ee 
This is  more local since it only connects points in the same $\tilde \sq$  (and in $\Om_{k+1}$) .   
The difference is  
\be
\de C_{k, \bom^+} \equiv  C _{k, \bom^+}-C^{  \loc} _{k, \bom^+} =
\sum_{\sq \subset \Om_{k+1}}  1_{\sq}  \B(  C _{k, \bom^+}  - C_{k, \bom^+,\sq}  \B) 
\ee
We claim that this is very small and  satisfies the bound
\be
\de C_{k, \bom^+} (\Up, \Up')  \leq  Ce^{-r_k } e^{ - \ga d(\Up,\Up') }
\ee
Indeed suppose $\Up \in \sq$ so the only term that contributes is  $ C _{k, \bom^+}(\Up, \Up')   - C_{k, \bom^+,\sq} (\Up, \Up')$.
In the random walk expansion only paths which start in $\sq$ and leave $\tilde  \sq$ contribute,  since otherwise the contribution
of the two terms cancels.    These walks are based on $M$ cubes rather than the $Mr_k$ cube $\sq$ considered here.    Thus these walks must have
$\cO(r_k)$  steps and $r_k$ inverse powers of $M$  can  generate the factor  $e^{ -r_k} $ and still give  convergence of the walk.

Now  consider the square root which can be represented as 
\be
 C^{\frac12}_{k, \bom^+ }  = \frac{1}{\pi} \int_0^{\infty}  \frac{dx}{ \sqrt{x}}   C_{k, \bom^+,x}
 \hs     C_{k, \bom^+,x} =   \B(  C^T [ \De_{k, \bom} ]_{\Om_{k+1}}C + x \B)^{-1}
\ee
It  suffices to establish the expansions and estimates for $ C_{k, \bom^+,x} $ or  $C C_{k, \bom^+,x}  C^T$ which  has the representation
   \cite{Bal85b}, \cite{Dim15},  \cite{Dim20}.
\be
C C_{k, \bom^+,x }  C^T = (1 + \pa \cM) [\cQ_k  \tilde \cG_{k+1, \bom^+,x } \cQ_k^T]_{\Om_{k+1} }( 1 + \pa \cM)^T
\ee
for a modified Green's function  $\tilde \cG_{k+1, \bom^+,x } $. 
The Greens function has a multiscale  random walk expansion and corresponding  exponential decay estimates, and this 
carries over to    $C C_{k, \bom^+x}  C^T  $ and $C_{k, \bom^+,x} $ and $ C^{\frac12}_{k, \bom^+ } $. 
In particular
\be \label{tingle2}  
 | C^{\frac12}_{k, \bom^+ } (\Up,\Up') |   \leq   Ce^{ - \ga d(\Up,\Up') }
\ee
 There are also
weakened versions  such as   $C^{\frac12}_{k, \bom^+ } (s) $ which satisfy the same  bound.

As before we define
 a local approximation to   $C _{k, \bom^+}  $ by defining 
\be
\begin{split} 
C^{\frac12, \loc} _{k, \bom^+} = & \sum_{\sq \subset \Om_{k+1}}  1_{\sq}  C^{\frac12}_{k, \bom^+,\sq}  \hs \hs   C^{\frac12}_{k, \bom^+,\sq} = C^{\frac12} _{k, \bom} \B(s_{ \tilde\sq} =1, s_{\tilde \sq^c} =0 \B) 
\\
\de C^{\frac12}_{k, \bom^+} \equiv &  C^{\frac12} _{k, \bom^+}-C^{ \frac12, \loc} _{k, \bom^+} =
\sum_{\sq \subset \Om_{k+1}}  1_{\sq}  \B(  C^{\frac12} _{k, \bom^+}  - C^{\frac12}_{k, \bom^+,\sq}  \B) \\
\end{split}
\ee
The latter satisfies
\be \label{pipsqueak}
|\de C^{\frac12}   _{k, \bom^+ } (\Up,\Up') |   \leq   C  e^{-r_k}  e^{ - \ga d(\Up,\Up') } 
\ee
\bigskip

Now return to    the integral in (\ref{stark}).
With $\tilde \De_{k, \bom^+ }  =   C^T [ \De_{k, \bom} ]_{\Om_{k+1}}C$  and $C_{k, \bom^+}   = \tilde \De_{k, \bom^+ } ^{-1} $  it is 
\be
   \int f(C \tilde Z_k)    d \mu_{C_{k, \bom^+}}(  \tilde Z_k   )    = 
\frac{  \int f( C \tilde Z_k)    \exp\B( - \frac12 \blan  \tilde Z_k, \tilde  \De_{k, \bom^+ }  \tilde Z_k \bran \B)   D  \tilde Z_k  }
{ \int    \exp\B( - \frac12 \blan  \tilde Z_k, \tilde  \De_{k, \bom^+ }  \tilde Z_k \bran \B)   D  \tilde Z_k   }
\ee
In the numerator   make the change of variables $\tilde  Z_k \to   C_{k, \bom^+} ^{\frac12,\loc}\tilde Z_k$
This changes the quadratic form to 
\be  \label{c2}
  \frac12 \blan  C_{k, \bom^+} ^{\frac12,\loc}\tilde Z_k,  \tilde \De_{k, \bom^+ } C_{k, \bom^+} ^{\frac12,\loc}\tilde Z_k \bran 
 =   \frac12 \| \tilde Z_k\|^2  -  R^{(2)}_{k, \bom^+ }    
 \ee
 which defines $ R^{(2)}_{k, \bom^+ }   $. 
It also  introduces a factor  $\det (  C_{k, \bom^+} ^{\frac12,\loc} ) $.  In the denominator   make the change of variables $\tilde  Z_k \to   C_{k, \bom^+} ^{\frac12}\tilde Z_k$.     This changes the quadratic form to $ - \frac12 \| \tilde Z_k \|^2$  and introduces a factor  $\det (  C_{k, \bom^+} ^{\frac12} ) $.
We identify the Gaussian measure with identity covariance $dm_{I, \Om_{k+1}} (\tilde Z_k)$ and define $   R^{(3)}_{k, \bom^+ }  $ by 
\be \label{c3} 
\det (  C_{k, \bom^+} ^{\frac12,\loc} )     = \det  (  C_{k, \bom^+} ^{\frac12} ) \exp (     R^{(3)}_{k, \bom^+ }    )
\ee
Then we have 
\be \label{stark2} 
   \int f(C \tilde Z_k)    d \mu_{C_{k, \bom^+}}(  \tilde Z_k   )    = 
  \int f(C   C_{k, \bom^+} ^{\frac12,\loc} \tilde Z_k)  \exp \B(     R^{( 2)}_{k, \bom^+ }  +    R^{( 3)}_{k, \bom^+ }  \B)
 dm_{I, \Om_{k+1}} (\tilde Z_k)
 \ee

Making these changes in  (\ref{rep4}) we now have with $R^{(\leq 3)} = R^{(1)} + R^{(2)} + R^{(3)} $
 \be    \label{rep5} 
 \begin{split}
&  \tilde \rho_{k+1} ( A_{k+1}, \Psi_{k+1}   )  
=   
  \sum_{ \bpi, \Om_{k+1},P_{k+1} , Q_{k+1}, R_{k+1}   }  \sZ'_{k,\bom}(0)   \sZ'_{k, \bom}   \de \sZ_{k, \bom^+}  \\
  &
      \int      Dm^0_{k+1, \bom^+}  ( A )      Dm_{k, \bpi}  (Z)\     Dm_{k,\bom}(\Psi)   Dm_{k,\bpi}(W) \   D\Psi_k\    d \mu_{I, \Om_{k+1}} (\tilde Z_{k}) \ \\
  &
   \de_{G}\B( \Psi_{k+1} - Q( \tilde \cA'  ) \Psi_k \B)    \cC_{k, \bpi }\  \cC^{\star \star}_{k+1}      \chi_{k+1}^0(\Om_{k+1})    \chi^{\da}_k  ( \Om^{3\nat}_{k+1} - R_{k+1} )  
   \\
   &  \exp \B(  -\frac12 \|  d  \cA^0_{k+1, \bom^+ }  \|^2 
     -    \fS'^{+}_{k, \bom}(\La_k)   +     E'_k(\La_k  )  
 +  B_{k,\bpi}       +  B^E_{k,\bpi}     + R^{(\leq 3 )}_{k, \bpi^+ }     \B) \B|_{\cA =     \cA^{0} _{k+1, \bom^+ } + \cZ^{\bullet} _{k,  \bpi^+ }  } \\   
\end{split}
\end{equation}
Here  $\cZ_{k, \bpi^+} $ has been replaced by 
\be 
\cZ^{\bullet}_{k, \bpi^+ } = \cH_{k, \bom} \B( C C_{k, \bom^+} ^{\frac12,\loc} \tilde  Z_k   + g \    \fH_{k, \bpi^+} \B)
\ee

\subsection{a final  small field expansion} 

The characteristic functions  $\chi'_{k+1}(S_{k+1})$, $    \zeta^{\da}_k  (R_{k+1})$ in $\cC^{\star \star} _{k+1, \bom^+}$  as well as  $ \chi^{\da}_k  ( \Om^{3\nat}_{k+1} - R_{k+1} ) $ 
all have had $\tilde Z_k$ replaced by $ C_{k, \bom^+} ^{\frac12,\loc}  \tilde Z_k$. 
This enlarges the domain of dependence by $3Mr_k$.    So   $    \zeta^{\da}_k  (R_{k+1})$ which formerly depended on $\tilde Z_k$  in $\tilde R_{k+1}$
now depends on $\tilde Z_k$ in  $R_{k+1} ^{ \sim 4} $.  This is outside the new small field region $T_{k+1}$ and will not affect our subsequent analysis.  
Similarly   $\chi'_{k+1}(S_{k+1})$  which formerly depended on  $\tilde Z_k$ in   $\tilde S_{k+1}$ now depends on  $\tilde Z_k$  in
$S_{k+1}^{\sim 4} $.  But $S_{k+1} \subset \Om^{3 \nat,c } _{k+1} \cup R_{k+1}  = \Om^{c, \sim 3 }_{k+1} \cup R_{k+1} $ and so
$
S_{k+1}^{\sim 4}  \subset  \Om^{c, \sim 7 }_{k+1} \cup R^{\sim 4} _{k+1} 
$.
 This is outside $T^{2 \nat}_{k+1}$ and also will not affect our subsequent analysis. 
But for     $  \chi^{\da}_k  ( \Om^{3\nat}_{k+1} - R_{k+1} ,     C_{k, \bom^+} ^{\frac12,\loc} Z  ) $ the $ C_{k, \bom^+} ^{\frac12,\loc} $ is a obstacle to strict  localization.  
  To avoid this  we
 introduce a new small field expansion.
  \bigskip

 For an  $LMr_{k+1}$ cube $\sq \subset  T_{k+1}$   define the  strictly local characteristic function
\be  \label{sister2}
\hat  \chi_k(\square)   =   \chi  \B(    \sup_{\Up   \in \square}  |\tilde Z_k(\Up)  | \leq    p _{0,k} \B)
\ee
Then with $ \hat \zeta_k(\sq)   = 1 -  \hat \chi_k(\sq) $ we insert inside the integral over $
\tilde Z_k$
 \be 
\begin{split}
1 = &  \prod_{\sq \subset T_{k+1}} ( \hat  \zeta_k(\sq)   +  \hat \chi_k(\sq)  )  \\
 = & \sum_{U_{k+1} \subset T_{k+1} }\prod_{\sq \subset U_{k+1}} \hat  \zeta_k(\sq)  \prod_{\sq \subset ( T_{k+1} -  U_{k+1})}  \hat \chi_k(\sq) \\ \equiv   &  \sum_{U_{k+1}   \subset T_{k+1} } \hat \zeta_k  (U _{k+1})  \hat  \chi_k  (T_{k+1}- U _{k+1} ) \\
\end{split}
\ee
where $U _{k+1}$ is a union of $LMr_{k+1}$ cubes. 
Then  define
\be \label{1781}
\La_{k+1} = T^{8 \nat}_{k+1}  - U_{k+1}^{\sim 8} 
\ee
We  classify the terms in the sums on $R_{k+1}, U_{k+1} $ by the $\La_{k+1}$ they generate (with intermediary $T_{k+1} $)
\be
\sum_{R_{k+1}, U_{k+1}   }
= \sum_{\La_{k+1} }\ \   \sum_{R_{k+1}, U_{k+1}  \to   \La_{k+1} } 
\ee
We have $ \La^{\sim 8} _{k+1}  \subset T_{k+1} - U_{k+1} $ and so  on $ \La^{\sim 8} _{k+1} $
\be  \label{exact} 
|\tilde Z_k| \leq p_{0,k} 
\ee

Since $\La_{k+1}^{\sim 4} \subset T_{k+1}  \subset \Om^{3\nat}_{k+1} - R_{k+1}$
the old small field function can   be split
\be
 \label{parsley} 
  \chi^{\da}_k \B ( \Om^{3\nat}_{k+1} - R_{k+1},C^{\frac12,\loc} _{k, \bom^+} \tilde Z_k \B) =   \chi^{\da} \B ( \Om^{3 \nat}_{k+1} - (R_{k+1} \cup \La^{\sim 4}_{k+1} ),C^{\frac12,\loc} _{k, \bom^+} \tilde Z_k\B )  \chi^{\da}\B(\La^{\sim4}_{k+1},C^{\frac12,\loc} _{k, \bom^+} \tilde Z_k \B)
\ee
Now $C^{\frac12,\loc} _{k, \bom^+} \tilde Z_k $ only depends on $\tilde Z_{k,\La_{k+1}} $   in  $\La_{k+1} ^{\sim 3}$,  so 
$\chi^{\dagger} (\sq , C^{\frac12,\loc} _{k, \bom^+} \tilde Z_k ) $  depends on $\tilde Z_{k,\La_{k+1}} $  only for $\sq \subset \La^{\sim 4} _{k+1} $.  Thus the first factor in  (\ref{parsley}) does not depend on on $\tilde Z_{k,\La_{k+1}} $, and will eventually come outside our fluctuation integral.    Furthermore the bound (\ref{exact}) implies that 
$|C^{\frac12,\loc} _{k, \bom^+} \tilde Z_k | \leq   Cp_{0,k} \leq  p_{0,k}^{\frac43} $ in $\La^{\sim 5} _{k+1}$.  Hence   $\chi^{\dagger} (\sq , C^{\frac12,\loc} _{k, \bom^+} \tilde Z_k ) =1$ for $\sq \subset  \La_k^{\sim 4} $.   
 Thus    $ \chi^{\da}(\La^{\sim 4} _{k+1},C^{\frac12,\loc} _{k, \bom^+} \tilde Z_k ) =1$ 
and we can omit it. This was our goal.

\bigskip

We make the splits
\be
\begin{split}
\hat \chi (T _{k+1}- U_{k+1}  ) = &  \hat \chi  ( T_{k+1}- (R_{k+1} \cup  \La_{k+1}  ) )\hat \chi( \La_{k+1} )\\
\chi_{k+1}^0(\Om_{k+1}) = &\chi_{k+1}^0(\Om_{k+1}- \La_{k+1}) \chi_{k+1}^0(\La_{k+1}) \\
\end{split} 
\ee 
The characteristic functions now have the form
$
 \cC^0_{k+1, \bpi^+ } \chi_{k+1}^0(\La_{k+1})   \hat  \chi_k(\La_{k+1} )    
 $
 where 
 \be
  \cC^0_{k+1, \bpi^+ } =   \cC_{k, \bpi } \     \cC^{\star \star }_{k+1}   \chi_{k+1}^0(\Om_{k+1}- \La_{k+1} )    \chi^{\da}_k  (  \Om^{3 \nat}_{k+1} - (U_{k+1} \cup \La^{\sim 4}_{k+1})    \   \hat \zeta_k  (U _{k+1}) \hat \chi  ( T _{k+1}- (U_{k+1} \cup  \La_{k+1}  ) )
\ee
Also  $ d \mu_{I, \Om_{k+1 } }  =  d \mu_{I, \Om_{k+1} - \La_{k+1 } }   d \mu_{I, \La_{k+1 } }  $ and we  identify
$   Dm^0_{k, \bom^+}  (Z) =   Dm_{k, \bom}  (Z)\  d \mu_{I, \Om_{k+1} - \La_{k+1 }} (\tilde Z_k)$. 

The representation is now with  $\bpi^+ = ( \bpi, \Om_{k+1}, \La_{k+1}; P_{k+1} , Q_{k+1}, R_{k+1}, U_{k+1}) $ 
 \be    \label{rep6} 
 \begin{split}
&  \tilde \rho_{k+1} ( A_{k+1}, \Psi_{k+1}   )  = \sum_{ \bpi^+ }    \sZ'_{k,\bom}(0)   \sZ'_{k, \bom}   \de \sZ_{k, \bom^+}  
     \\
= &   
     \int      Dm^0_{k+1, \bom^+}  ( A )      Dm^0_{k+1, \bpi}  (Z)\     Dm_{k,\bom}(\Psi)   Dm_{k,\bpi}(W) \   D\Psi_k\  d \mu_{I, \La_{k+1} } (\tilde Z_k)
   \\
  & 
  \cC^0_{k+1, \bpi^+ } \chi_{k+1}^0(\La_{k+1})   \hat  \chi_k(\La_{k+1} ) \   \de_{G}\B( \Psi_{k+1} - Q( \tilde \cA'  ) \Psi_k \B)   
       \\
   &           \exp \B( -\frac12 \|  d  \cA^0_{k+1, \bom^+ }  \|^2   -    \fS'^{+}_{k, \bom} (\La_k)   +   E'_k(\La_k  )  
 +  B_{k,\bpi}       +  B^E_{k,\bpi}     + R^{(\leq 3)}_{k, \bom}     \B) \B|_{\cA =     \cA^{0} _{k+1, \bom^+ } + \cZ^{\bullet} _{k,  \bpi^+ }  } \\   
\end{split}
\end{equation}

\subsection{first   localization} 
 
 \label{localization}

In  the effective action we have  $ \fS'^{+}  _{k,\bom} (\La_k, \cA+ \cZ , \Psi_{k, \bom},   \psi_{k, \bom}(  \cA+ \cZ ))$
as well as
$ E'_k(\La_k,    \cA + \cZ,     \psi^\#_{k,  \bom}(  \cA+ \cZ)  )$
evaluated at    $\cA =  \cA^{0} _{k+1, \bom^+}$ and $\cZ= \cZ^{\bullet}_{k,  \bom^+ }$.    We   want to  isolate the   behavior in  $  \cZ$.
Breaking up $ \fS'^{+}  _{k,\bom} (\La_k)$ as in (\ref{picquant}) we  define   $E^{(1)}, E^{(2)}, E^{(3)}$ by  
\be  
\begin{split}
\label{sanibel}
 \fS _{k,\bom} \B(\cA+ \cZ,  \Psi_{k, \bom}, \psi_{k, \bom}\B(  \cA  + \cZ  )\B)     
            =  & \fS _{k,\bom} \B( \cA, \Psi_{k, \bom} , \psi_{k, \bom}(  \cA  )   \B)    -  E^{(1)}_{k, \bom} \B (  \cA ,   \cZ,  \psi_{k, \bom}(\cA),  \Psi_{k, \bom}        \B )         \\
 E'_k\B(\La_k,     \cA + \cZ,     \psi^\#_{k,  \bom}(  \cA  + \cZ  )  \B) 
           =   &    E'_{k} \B( \La_k,  \cA,  \psi^\#_{k,\bom}(  \cA   )\B)   + E^{(2)}_{k, \bom}\B( \La_k,  \cA ,   \cZ,  \psi^\#_{k, \bom}(\cA),  \Psi_{k, \bom}   \B)  
           \\
m_k' \blan \bpsi_{k, \bom}(\cA+\cZ), \bpsi_{k, \bom}(\cA+ \cZ) \bran_{\La_k}      = &
 m_k' \blan \bpsi_{k, \bom}(\cA), \bpsi_{k, \bom}(\cA) \bran_{\La_k}  +E^{(3)}_{k, \bom}\B( \La_k,  \cA ,   \cZ, \psi^\#_{k, \bom}(\cA),  \Psi_{k, \bom}     \B)  \\
\end{split}
\ee 
These are analytic in  say $\cA \in (1-e_k^{\frac12} )  \tilde \cR_{k,,\bom},  \cZ \in  e_k^{\frac12 }\tilde \cR_{k, \bom}$
 It  will be convenient to consider a smaller domain, namely  $\cA \in (1-e_k^{\frac12} )  \tilde \cR_{k, \bom}$ and   
 \be  \label{d1}
 |\cZ| \leq  L^{\frac12(k-j)}e_k^{-\frac52 \ep} \hs   |\pa \cZ| \leq  L^{\frac32(k-j)}e_k^{-\frac32\ep} \hs  |\de_{\al} \pa \cZ| \leq  L^{(\frac32+ \al)(k-j)}e_k^{-\frac12 \ep} \hs \textrm{ on } \de \Om_j
 \ee 
 Note that if $\cZ$ is in this domain then $\frac12 e_k^{-\frac34 + \frac92 \ep} \cZ \in e_k^{\ep}\cR_{k, \bom}  $.  
  Indeed in $\de \Om_j$ for $j \leq k$ we have
 \be
 | \frac12 e_k^{-\frac34 + \frac92  \ep} \cZ| \leq L^{\frac12(k-j) } \theta_k e_k^{- \frac34 + 2 \ep}  \leq e_k^{\ep} [ L^{\frac12(k-j) } \theta_k e_k^{- \frac34 + \ep}]
 \ee
 The derivatives are similar.

 Also note  that  due to our characteristic functions   $\cA = \cA^0_{k+1, \bom^+}$  is easily in  domain (\ref{d1})  by  the estimates (\ref{william}).   Furthermore
  $\cZ =   \cZ^{\bullet} _{k, \bpi^+}  =  \cH_{k, \bom} ( C C_{k, \bom^+} ^{\frac12,\loc} \tilde  Z_k   + g \    \fH_{k, \bpi^+} ) $ is in the domain.
   Indeed suppose $|\tilde Z_k| \leq e_k^{-\frac14 \ep} $ which is better than the characteristic function bound (\ref{pecan2}).  Then by
  (\ref{tingle2}) 
   $| CC^{\frac12,\loc} _{k, \bom^+} \tilde Z_k| \leq CM  e_k^{-\frac14 \ep} $. Then  on $\de \Om_j$ by (303) in \cite{Dim20}
  \be \label{lumbar} 
    | \cH_{k, \bom} CC_{k, \bom^+} ^{\frac12,\loc} \tilde Z_k|        \leq     L^{\frac12(k-j)} CM   e_k^{-\frac14 \ep}   \leq    \frac12   L^{\frac12(k-j)} e_k^{- \frac12 \ep} 
\ee
 The correction   $ \cH_{k, \bom} \  g  \fH_{k, \bpi^+}$  is even smaller,     hence     $|\cZ^{\bullet} _{k, \bpi^+}| \leq   L^{\frac12(k-j)} e_k^{- \frac 12 \ep} $ which suffices. 
Derivatives are estimated similarly.

 The $ E^{(i)}_{k, \bom}$  have some dependence on the bare fields $\Psi_{k, \bom}  $  but we will eventually change this to dependence on $ \psi_{k, \bom}(\cA)$
as follows.    First define  for any   $\bom$ define  a multiscale  operator
\be  
T_{k, \bom}(\cA)  = ( \bb^{(k)})^{-1}  Q_{k, \bom}(\cA)  \B( \fD_{\cA}  + \bar m_k   +  P_{k, \bom} (\cA )\B)  
\ee
This satisfies  
\be \label{tprime}
  \Psi_{k, \bom}  =     [ T_{k, \bom}(\cA) ]_{\Om_1}   \psi _{k, \bom}(\cA)  +( \bb^{(k)})^{-1}  Q_{k, \bom}(\cA)1_{\Om_1} \fD_{\cA} \psi_{\Om^c_1}
  \equiv   T'_{k, \bom}(\cA)   \psi _{k, \bom}(\cA)  
\ee
The operator $ T'_{k, \bom}(\cA) $   is a local.  We show in appendix \ref{B}  that $ | T_{k, \bom }(\cA) f | \leq C \|f\|_{\infty}$.  We also have  $|( \bb^{(k)})^{-1}  Q_{k, \bom}(\cA)1_{\Om_1} \fD_{\cA}f|
\leq  C \|f\|_{\infty}$. Thus for some constant $C_T$
\be \label{llama0}
 | T'_{k, \bom }(\cA) f | \leq C_T \|f\|_{\infty}
\ee

In the expressions  (\ref{sanibel}) for 
 $  E^{(i)}_{k, \bom } $ 
we have
  $ \cZ = \cZ^{\bullet} _{k, \bpi^+}  =  \cH_{k, \bom} ( C C_{k, \bom^+} ^{\frac12,\loc} \tilde  Z_k   + g \    \fH_{k, \bpi^+} )  $
   with $ \fH_{k, \bpi^+} = \fH_{k, \bpi^+} (d\cA^0_{k+1,  \bom^+})$.
The next few results    localize  the dependence  of $\cZ^{\bullet} _{k, \bpi^+}=$ in $\tilde Z_k$ and $d\cA^0_{k+1,  \bom^+} $ and replace
$\Psi_{k, \bom}  $ by $\psi_{k, \bom}( \cA^0_{k+1, \bom^+} ) $ using (\ref{tprime})

\bigskip

\begin{lem}  \label{outsize1}   
\be    \label{serendipity}
\begin{split}
   &   E^{(1)}_{k, \bom}\B(\La_k,  \cA^{0} _{k+1, \bom^+},     \cZ^{\bullet}_{k, \bpi^+}, \psi_{k, \bom}(\cA^{0} _{k+1, \bom^+}) ,    \Psi_{k, \bom})     \B) \\
   &   =   \sum_{  X \subset \La_{k+1}    }   E^{(1)}_{k}\B(X, \cA^{0} _{k+1, \bom^+},     \tilde Z_k ,   \psi_{k, \bom}(\cA^{0} _{k+1, \bom^+}) \B)  \\
   & +  \sum_{  X \cap (\Om_{k+1} - \La_{k+1})   \neq \emptyset  }   B^{(1)}_{k, \bpi^+}\B(X, \cA^{0} _{k+1, \bom^+},     \tilde Z_k ,   \psi_{k, \bom}(\cA^{0} _{k+1, \bom^+}) \B)    \\
\end{split}
\ee
where the sum is over $X \in \cD_k$ and 
where with a constant $\ka = \one$ fixed and large
\begin{enumerate} 
\item  $  E^{(1)}_{k}(X, \cA^{0} _{k+1, \bom^+},     \tilde Z_k ,     \psi_{k, \bom}(\cA^{0} _{k+1, \bom^+}) ) $   are the restrictions 
 of  $ E^{(1)}_{k}(X, \cA,   \tilde Z_k,   \psi_{k, \bom}( \cA))$
 depending on the indicated fields only in $X$,  are analytic in  $\cA \in (1-e_k^{\frac12} )  \tilde  \cR_{k,\bom},  \  |\tilde Z_k| \leq e_k^{-\frac14 \ep}$  and satisfying   there
 with $h_k = e_k^{- \frac14}$
  \be    \label{gumball} 
  \|  E^{(1)}_{k} \B(X,  \cA  ,    \tilde Z_k   \B) \| _{ h_k}  \leq   \one   e_k^{\frac14 - 5 \ep}  e^{- \ka d_{M}(X)}
  \ee
The kernel  is independent of the history $\bpi^+$.
 \item 
$ B^{(1)}_{k, \bpi^+}\B(X, \cA^{0} _{k+1, \bom^+},     \tilde Z_k ,   \psi_{k, \bom}(\cA^{0} _{k+1, \bom^+})  \B)  $ are the restrictions of 
 $ B^{(1)}_{k, \bpi^+}\B(X, \cA ,     \tilde Z_k ,   \psi_{k, \bom}(\cA)  \B)  $ depending on the indicated fields in $X$,  analytic in  $\cA \in (1-e_k^{\frac12} )  \tilde  \cR_{k,\bom},  \ |\tilde Z_k| \leq e_k^{-\frac14 \ep}$
and satisfying  there
\be
\| B^{(1)}_{k, \bpi^+ }\B(X, \cA ,     \tilde Z_k    \B)  \| _{  h_k \bbI_k}  \leq   \one   e_k^{\frac14 - 5 \ep}  e^{- \ka d_{M}(X)}
\ee
\end{enumerate} 
\end{lem}
\bigskip   

\pr 
\textbf{A.} We first  consider     $ E^{(1)}_{k, \bom} ( \cA, \cZ ,  \Psi_{k, \bom},\psi_{k, \bom}(\cA))  $  for $\cA \in (1-e_k^{\frac12} )  \tilde  \cR_{k,\bom} $ and $\cZ$  in the domain (\ref{d1}). It is the change in 
$\fS _{k,\bom} ( \cA, \Psi_{k, \bom}, \psi_{k, \bom}(  \cA  ) ) $
and we use the representation 
\be \label{succotash} 
\begin{split}
& \fS _{k,\bom} \B( \cA, \Psi_{k,\bom}, \psi_{k, \bom}(  \cA  )   \B)  =  \   \blan \bPsi_{k, \bom}, D_{k, \bom}(\cA)\Psi_{k, \bom} \bran 
  +  \blan  \bPsi_{k, \bom} , \bb^{(k)}Q_{k, \bom}(\cA)  S_{k, \bom}(\cA) \fD_{\cA}
 \psi_{\Om_1^c}   \bran
  \\
 & + \blan  \bpsi_{\Om_1^c}  , \fD_{\cA} S_{k, \bom}(\cA) 
Q^T_{k, \bom}(-\cA) \bb^{(k)} \Psi_{k, \bom}  \bran 
+ \blan   \bpsi_{\Om_1^c} \B(  (\fD_{\cA} + \bar m_k) - \fD_{\cA}S_{k, \bom}(\cA) \fD_{\cA}\B) \psi_{\Om_1^c}   \bran
\end{split}
\ee

We suppose initially that the leading term $ \blan \bPsi_{k, \bom}, D_{k, \bom}(\cA)\Psi_{k, \bom} \bran $ is the only contribution. 
We have  $D_{k, \bom}(\cA) = \bb^{(k)}  - M_{k, \bom} (\cA)$
where    
\be \label{runrun}
  M_{k, \bom} (\cA)
 =   \bb^{(k)} Q_{k, \bom}(\cA) S_{k, \bom}(\cA)Q^T_{k, \bom}(-\cA) \bb^{(k)}  
\ee
Then
\be
\begin{split}
 E^{(1)}_{k, \bom} ( \cA, \cZ ,  \Psi_{k, \bom})   
=   &  \blan \bPsi_{k,\bom} ,\B( M_{k,\bom}( \cA)  - M_{k,\bom}( \cA+ \cZ)  \B)  \Psi_{k,\bom} \bran  \\
\end{split}
\ee

In this first part of the proof we study  $M_{k, \bom}(\cA)$ and its  variation.   We will  need an explicit expression for $Q_{k, \bom} (\cA)$.  On $\de \Om_j$ it is $Q_j(\cA)$  which is given for $f, \cA$ on $\tk$
and $y \in \bbT^{-(k-j)}_{N-k}$ by
\be
\label{studly}
(Q_j(\cA) f ) (y)= L^{-3j}  \sum_{x \in B_j(y)} \exp\B(  ie_k L^{-j}(\tau_j\cA)(y,x) \B)  f(x) 
\ee
Here for a sequence $x=x_0, x_1, \dots x_j = y$ with $x_{i} \in B(x_{i+1})$ 
\be
(\tau_j\cA)(y,x)  = \sum_{i=0}^{j-1} (\tau \cA) (x_{i+1}, x_i)
\ee
If $j=k$ this is established for example in lemma 2 in \cite{Dim15a}.  For $j<k$ it follows by scaling the result for $\cQ_j(\cA)$ on $\bbT^{-j}_{N-j}$ down to $\tk$. 
It follows that for $h$ on $ \bbT^{-(k-j)}_{N-k}$ and $x \in \tk$ 
\be
(Q^T_j(-\cA)h ) (x)  = \exp\B( - ie_k L^{-j}(\tau_j\cA)(y,x) \B)h(y) \hs x \in B_j(y)
\ee

Now consider  the kernel $M_{k, \bom} ( \cA; y,y')  =  <\de_y, M_{k, \bom}(\cA) \de_{y'} >$.   Here for $y \in \de \Om_j^{(j)} \subset \bbT^{-(k-j)}_{N-k}$ and  we define  $\de_{y} (z) =L^{3(k-j)} \de_{y,z}$. 
We also have for  $y \in \de \Om_j^{(j)}$   and $\De_y = B_j(y)$ the $L^{- (k-j)}$ cube  centered on $y$
\be
\begin{split}
Q^T_{k, \bom}(-\cA)  \bb^{(k)}  \de_y
=  &    b_j L^{k-j}  Q^T_j(-\cA)\de_y    = b_j L^{4(k-j)}  e^{ - i  \si_j}  1_{\De_y }    \\
\si_j(x) =  &e_k  L^{-j}  (\tau\cA)(y,x)    \hs   x \in B_j(y) \\ 
\end{split}
\ee
This  gives  for $y \in \de \Om_j^{(j)},  y' \in \de \Om_{j'}^{(j')}   $
\be
M_{k , \bom}(\cA;  y,y')  = b_j b_{j'}    L^{4(k-j)} L^{4(k-j')} \blan e^{i\si_j} 1_{\De_y}, S_{k, \bom} (\cA  )e^{-i\si_{j'} }1_{\De_{y'} }\bran
\ee

To estimate this we use  $| L^{-j}(\tau_j\cA)(y,x)| \leq 2 \|\cA\|_{\infty} $ so $|\si_j| \leq 2 e_k\|\cA\|_{\infty} $.
For  $\cA= \cA_0 + \cA_1 \in   \tilde \cR_{k, \bom}$      only the complex part $\cA_1$ contributes to  $\im \si_j$  .  We have in $\de \Om_j$ that
 $|\cA_1| \leq   L^{\frac12(k-j)} e_j^{-\frac34+ \ep}$ and so 
 \be
|\im \si_j |  \leq 2e_k \|\cA_1\|_{\infty}   \leq  2e_k  L^{\frac12(k-j)} e_j^{-\frac34+ \ep} = 2 e_j^{\frac14+ \ep} 
\ee
Hence $|e^{i \si_j }| = e^{- \im \si_j } \leq  2 $

Now   estimate  $S_{k, \bom} (\cA  )$ by    (95) in \cite{Dim20} and obtain
\be \label{forty-five}
\begin{split}
|  \blan e^{i \si_j} 1_{\De_y}, S_{k, \bom} (\cA  )e^{-i \si_{j'} }1_{\De_{y'} }\bran |
 \leq & 
   \| e^{i \si_j}  1_{\De_y} \|_1 \| 1_{\De_y }S_{k, \bom} (\cA  )e^{-i \si_{j'}}  1_{\De_{y'} }  \|_{\infty}  \\
\leq   &
C L^{- 3(k-j)}L^{-(k-j')} e^{- \ga d_{\bom} (y,y') } \\
 \leq   &
 CL^{- 2(k-j)}L^{-2(k-j')} e^{- \ga d_{\bom} (y,y') } \\
\end{split}
\ee
The last step follows  from our freedom to switch powers of $L^j$ and $L^{j'}$ in the presence of the $e^{- \ga d_{\bom} (y,y') } $.  
Thus we have
\be  \label{squash0}
|M_{k , \bom}( \cA; y,y')|  \leq  
C L^{2(k-j)} L^{2(k-j')}e^{- \ga d_{\bom} (y,y') } 
\ee

\bigskip

We can completely   localize in $\cA$ by  making a polymer expansion for $S_{k, \bom} (\cA )$  as in lemma 9 in \cite{Dim20}.
We have for $\cA \in  \tilde  \cR_{k,\bom}$
 \be 
S_{k, \bom} (\cA ) = \sum_{X \in \cD_{k,\bom}}  \hat  S_{k, \bom} (X, \cA )
\ee 
Here the sum is over multiscale polymers $\cD_{k, \bom}$ as defined in section II.G in \cite{Dim20}. 
The $S_{k, \bom} (X, \cA )$ only depend on $\cA$ in $X$ and   satisfy the same bounds as $S_{k, \bom} (\cA ) $ but with an extra factor 
$e^{-( \hat \ka +1) |X|_{\bom}}$ where $|X|_{\bom}$ is the number of cubes in $X$ and $\hat \ka = \one$ is as large as we like for $M$ sufficiently large. 
This generates an expansion   $M_{k, \bom} (\cA ) = \sum_{X \in \cD_{k, \bom}}  \hat M_{k, \bom} (X, \cA )$ again with $X \in  \cD_{k, \bom} $.  The $\hat  M_{k, \bom} (X, \cA )$ only depend on $\cA$ in $X$ and have   kernels
$\hat M_{k, \bom} (X, \cA, y, y' )$  as above with  support in $y, y' \in X$ which satisfy
\be
  \label{squash1}
|\hat M_{k , \bom}( X, \cA; y, y' )|  \leq  
C L^{2(k-j)} L^{2(k-j')}e^{- \ga d_{\bom} (y,y') }  e^{- (\hat \ka +1) |X|_{\bom}}
\ee
Now 
\be 
\blan \bPsi_{k, \bom},  \hat M_{k,\bom}(\cA) \Psi_{k, \bom} \bran 
= \sum_{X} \blan \bPsi_{k, \bom},  \hat M_{k,\bom}(X,\cA) \Psi_{k, \bom} \bran
\ee 
where  (with $y \in \de \Om^{(j)}_j$ abbreviated as  $y \in \de \Om_j$)
 \be \label{sss}
\begin{split}
\blan \bPsi_{k, \bom}, \hat M_{k,\bom}(X, \cA) \Psi_{k, \bom} \bran
 = &
\sum_ {j, y \in  \de \Om_j } \sum_ {j', y' \in  \de \Om_{j'} }   \bPsi_{k, \bom}(y) L^{-3(k-j)}\hat M_{k,\bom}(X,  \cA; y, y')  L^{-3(k-j')} \Psi_{k, \bom} (y') \\
\end{split}
\ee
The $\bbI_k$ supplies a factor $L^{k-j}$ in $\de \Om_j$ for each field and so by (\ref{squash1})
\be  \label{slumbering}
\begin{split} 
&\|\blan \bPsi_{k, \bom}, \hat M_{k,\bom}(X, \cA) \Psi_{k, \bom} \bran \|_{ C_Th_k\bbI_k} \\
\leq 
 & \frac12(C_Th_k)^2  \sum_ {j, y \in  \de \Om_j } \sum_ {j', y' \in  \de \Om_{j'} } 
 L^{-2(k-j)} |M_{k, \bom}(X, \cA, y,y') |   L^{-2(k-j')} \\
\leq 
 &  C(C_Th_k)^2e^{- (\hat \ka +1)  |X|_{\bom}}\sum_ {j, y \in  \de \Om_j \cap X } \sum_ {j', y' \in  \de \Om_{j'} \cap X} e^{- \ga d_{\bom} (y,y') }
\\
\leq  &    CM^3(  C_Th_k)^2 e^{- \hat \ka   |X|_{\bom}}         \\
\end{split}
\ee
Here the   sum over $j',y' \in\de \Om^{j'}_{j'} \cap X$ was bounded by a constant (see \cite{Bal84b}) and the number of elements in 
the sum over $j,y \in\de \Om^{j}_{j} \cap X$ was identified as  $M^3 |X|_{\bom}$.

Now we can write
  \be  \label{tutu}
  E^{(1)}_{k, \bom} = \sum_{X \in \cD_{k, \bom}  } \tilde E^{(1)}_{k, \bom} (X) 
  \ee
     where
\be
\tilde  E^{(1)}_{k, \bom} (X, \cA, \cZ, \Psi_{k, \bom}) 
=    \blan \bPsi_{k,\bom} ,\B(\hat  M_{k,\bom}(X, \cA)  - \hat M_{k,\bom}(X, \cA+ \cZ)  \B)  \Psi_{k,\bom} \bran
\ee
\bigskip

\noindent
\textbf{B.}  At first  we restrict ourselves  in (\ref{tutu}) to the sum over $X \in \La_k$, in which case $X \in \cD_k$.
In this case we have $\Psi_{k, \bom} = \Psi_k$ and  $Q_{k, \bom}(\cA) = Q_k (\cA )$ and $S_{k, \bom}(X)  =  S_k(X)$.
Therefore 
\be
 \blan \bPsi_{k,\bom}, M_{k,\bom}(X, \cA) \Psi_{k, \bom} \bran 
 = \blan \bPsi_{k}, M_{k}(X, \cA), \Psi_{k} \bran  
\equiv  b_k^2 \blan  \bPsi_{k}, Q_{k}(\cA) S_{k}(X,\cA)Q^T_{k}(-\cA) \Psi_{k} \bran 
\ee
Since we assume $\cZ$ is in the domain (\ref{d1}) we have $t\cZ \in e_k^{\ep}  \cR_{k,\bom}$ for complex $t$ with $|t| \leq  \frac12 e_k^{-\frac34 + \frac92 \ep}$ 
(use $e_k^{-1} \leq e_j^{-1}$ for $j \leq k$). Then $\cA +t \cZ \in \tilde \cR_{k, \bom}$ for such $t$ and  $ M_{k}(X, \cA+ t \cZ)$ is analytic in such $t$.  
Hence we can write
\be \label{musty} 
\tilde  E^{(1)}_{k, \bom} (X, \cA, \cZ, \Psi_{k, \bom}) 
=   \frac{- 1}{2 \pi i} \int_{|t| =\frac12 e_k^{-\frac34 + \frac92 \ep} } \frac{dt}{t(t-1)}    \blan \bPsi_{k}, M_{k}(X, \cA+ t \cZ), \Psi_{k} \bran 
\ee
Since $|X|_{\bom} = |X|_M \geq d_M(X)$  the bound (\ref{slumbering}) gives (here $\bbI_k = 1$)
\be
 \| \blan \bPsi_{k}, M_{k}(X, \cA+ t \cZ), \Psi_{k} \bran  \|_{ C_Th_k \bbI_k}
\leq  CM^3(  C_Th_k)^2 e^{-\hat \ka d_M(X)}
\ee
With $h _k = e_k^{-\frac14}$  and   $CM^3 C_T^2e_k^{\frac12 \ep}  \leq 1$ for $e_k$ sufficiently small this gives 
\be \label{tinker} 
  \|  \tilde  E^{(1)}_{k, \bom} (X, \cA, \cZ)  \|_{ C_Th_k\bbI_k}
\leq  CM^3(  C_Th_k)^2   e_k^{\frac34 -\frac 92  \ep} e^{-\hat \ka d_M(X)} \leq   e_k^{\frac14 - 5  \ep} e^{- \hat \ka d_M(X)} 
\ee  
\bigskip

Now specialize to $ \cZ =\cZ^{\bullet} _{k, \bpi^+}  =  \cH_{k, \bom} ( C C_{k, \bom^+} ^{\frac12,\loc} \tilde  Z_k   + g \    \fH_{k, \bpi^+} ) $ with   $\tilde Z_k$ and $\fH_{k, \bpi^+}=\fH_{k, \bpi^+}(d\cA)$ supported in
$\Om_{k+1}$.
 We seek to localize  $ \tilde E^{(1)}_{k, \bom}(X, \cA,  \cZ^{\bullet}_{k, \bpi^+}  , \Psi_{k}) $ in  $\tilde Z_k$ and  $ d\cA$  (still for $X \subset \Om_{k+1}$). 
Accordingly    we    introduce weakening parameters    $s_{\bom}= \{s_{\sq} \}_{\sq \in \pi'(\bom)}$ , i.e.  $\sq$ ranges over  $2L^{-(k-j)}M$ cubes in $\de \Om_j$, and 
 define
\be    \label{saturn2}
  \cZ^{\bullet} _{k, \bpi^+}(s) =    \cH_{k, \bom} (s) \B( C C_{k, \bom^+} ^{\frac12,\loc} (s) \tilde  Z_k   + g \    \fH_{k, \bpi^+}(s)  \B)
 \ee
The definition   of $\cH_{k, \bom}(s)$ depends on random walk expansions and   is discussed in \cite{Dim20}.  The definition of  $C_{k, \bom^+} ^{\frac12,\loc} (s)$ is discussed in section \ref{ultralocal}.    The definition of   $\fH_{k, \bpi^+}(s)$ is discussed  in section \ref{another}.    
We have  $\cZ^{\bullet} _{k, \bpi^+}(1) = \cZ^{\bullet} _{k, \bpi^+}$.  Instead of  (\ref{musty}) we consider  $ \tilde E^{(1)}_{k, \bom}(X, \cA,  \cZ^{\bullet}_{k, \bpi^+}(s)  , \Psi_{k}) $
which again satisfies a bound (\ref{tinker}). 

Now in each variable $\sq \in \pi'(\bom)$  except those in $X$  we   interpolate between  $s_{\sq} =1$  and   $s_{\sq} =0$
with the identity
\be 
  f(s_{\sq} =1  )     =  f(s_{\sq} =0  )     +   \int_0^1      d  s_{\square}   \frac{\pa  f}{  \pa  s_{\sq}}  
 \ee
Then we find
 \be \label{spumoni0}
\sum_{X \subset   \La_{k}  }  \tilde  E^{ (1)}_{k, \bom}  (X)
=  \sum_{X \subset   \La_{k},   Y \supset X   }    \breve   E^{(1)}_{k, \bpi^+}  (X,   Y) 
\ee 
The sum over $Y$ is a sum  over unions of cubes   $\sq \in \pi'(\bom) $ in $X^c$ 
and 
 \be  \label{winter}
\begin{split}
& \breve     E^{(1)}_{k, \bpi^+}(X,Y, \cA,  \tilde Z_k,  \Psi_{k})   
=      \int   ds_{Y- X} 
 \frac { \pa  }{ \pa s_{Y-X}}   \left[ \tilde E^{(1)}_{k, \bom}(X, \cA,  \cZ^{\bullet}_{k, \bpi^+}(s)  , \Psi_{k})  \right]_{s_{Y^c} = 0, s_{X}=1}\\
 \end{split}
 \ee
 With $s_{Y^c} =0$ only walks that stay in $Y$ contribute so  $\cZ^{\bullet}_{k, \bpi^+}(s) =0$ on $Y^c$, and if  $Y$ has connected components $\{ Y_{\al } \}$ then
  \be
  [\cZ^{\bullet}_{k,\bpi^+}(s) ]_Y = \bigoplus\   [\cZ^{\bullet}_{k, \bpi^+}(s_{Y_{\al}}) ]_{Y_{\al}}
 \ee
However only  the connected component containing $X$  gives a non-zero contribution.   So the sum in (\ref{spumoni0})  is restricted   to connected   $Y$ satisfying $Y \supset  X $, i.e.
$Y \in \cD_{k, \bom} $.   Also   $Y$ must intersect $\Om_{k+1}$ else $Y \subset \Om_{k+1}^c$ and $[ \cH_{k, \bom}(s_Y)]_Y $ vanishes on functions with support in $\Om_{k+1}$
like $\tilde Z_k,   \fH_{k, \bpi^+}(s) $.   Now $  \breve     E^{(1)}_{k, \bom}(X,Y,\cA,  \tilde Z_k,   \Psi_{k}  )    $
 only depends  on  the indicated fields    in $Y$.   
 \bigskip

Note that if $Y \subset \La_{k+1}$, then with  $s_{Y^c} =0$ the random walk expansion for  $\cH_{k, \bom}(s)$  only involves units in $\La_{k+1}$
and hence agrees with the global $\cH_{k}(s)$.  The second term in (\ref{saturn2}) does not contribute since $  \fH_{k, \bpi^+}(s)  $ is localized in $S_{k+1} \subset \La^c_{k+1}$.
Hence in   (\ref{winter})  we can replace $\cZ^{\bullet} _{k, \bpi^+}(s)$ by $\cZ_k(s) = \cH_k(s)C \tilde Z_k$.  Then $\breve     E^{(1)}_{k, \bpi^+}(X,Y)$ is independent 
of $\bpi^+$.

Now  $\tilde E^{(1)}_{k, \bom^+}(s, X  )$ is analytic  in  $|s_{\square} |  \leq   M^{\frac12} $ and satisfies a bound of the form 
(\ref{tinker}) there.
By  Cauchy inequalities  each derivative introduces a factor   $M^{- \frac12}  \leq   e^{- \hat \ka} $   so we have
 \be  
 \|  \breve    E^{(1)}_{k, \bom^+}(X,  Y, \cA, \tilde Z_k )    \|_{ C_T  h_k \bbI_k }
\leq   e_k^{\frac14 - 5 \ep} e^{- \hat  \ka|Y- X|_{\bom}  }       e^{-  \hat \ka  d_M(X)}   
\ee

We do a partial resummation of  (\ref{spumoni0}).     Let $Y' =  \bar Y \in \cD_k $ be the union  of all $M$-cubes intersecting $Y \in \cD_{k, \bom}$.
We classify the terms in the sum by the $Y'$ they determine   
\be \label{spumoni4}
\sum_{X \subset   \La_{k}  }  \tilde  E^{ (1)}_{k, \bom}  (X)
=  \sum_{ Y'  \cap  \Om_{k+1} \neq \emptyset}      E^{(1),\st }_{k, \bpi^+}  (Y') 
\ee 
 where
\be   
   E^{(1),\st}_{k, \bpi^+}  (Y', \cA,  \tilde Z_k,   \Psi_{k} )    =    \sum_{X \subset   \La_{k},   Y \supset X,  \bar Y = Y'    }   \breve   E^{(1)}_{k, \bpi^+}  (X,   Y,\cA,  \tilde Z_k,   \Psi_{k} )    
\ee
We have 
 \be  \label{catty}
 \|    E^{(1), \st}_{k, \bpi^+}(Y',\cA,  \tilde Z_k  )    \|_{ C_Th_k\bbI_k }
\leq    e_k^{\frac14 - 5  \ep} \sum_{X \subset   \La_{k},   Y \supset X,  \bar Y = Y'    }    e^{- \hat \ka|Y- X |_{\bom}  }   e^{- \hat \ka d_M(X ) }         
\ee
But 
$|Y- X|_{\bom} \geq | \bar Y  -X|_{M}$
and then 
\be \label{sinful}
|\bar Y  - X|_{M}+   d_M(X)  \geq  d_{M} (Y' ) 
\ee
Thus we can extract a factor $e^{- (\hat \ka- \ka_0)   d_{M} (Y' ) }$.  The sum over $X$ can be relaxed to $X  \subset Y' $ and the condition $\bar Y = Y'$ dropped.  
 Then 
 \be  \label{catty1}
 \|    E^{(1), \st}_{k, \bpi^+}(Y',\cA,  \tilde Z_k)    \|_{ C_Th_k \bbI_k}
\leq    e_k^{\frac14 - 5  \ep} e^{- (\hat \ka- \ka_0)   d_{M} (Y' ) } 
\sum_{X \subset Y' }  e^{-\ka_0 d_M(X ) }\sum_{ Y : Y \supset  X  }      e^{-\ka_0|Y- X |_{\bom}  }     
\ee
But by (\ref{jam} ) 
 \be
 \sum_{ Y : Y \supset  X  }      e^{-\ka_0|Y- X |_{\bom}  }      \leq  \one |X|_{\bom}   =  \one |X|_M   \leq  \one |Y'|_M 
 \leq  \one   \exp ( d_M(Y') + 1  ) 
 \ee
and 
\be 
   \sum_{X: X  \subset  Y' } e^{-\ka_0 d_M(X ) }         
\leq \one |Y' |_M  \leq \one ( d_M(Y') + 1)
\ee
These factors can be absorbed by the exponential and  assuming $\hat \ka - \ka_0 -1 > \ka$
  \be  \label{catty2}
 \|    E^{(1), \st}_{k, \bpi^+}(Y',\cA,  \tilde Z_k)    \|_{ C_Th_k \bbI_k}
\leq  \one    e_k^{\frac14 - 5  \ep}     e^{- (\hat \ka- \ka_0 -1)   d_{M} (Y' ) }  \leq  \one    e_k^{\frac14 - 5  \ep}     e^{- \ka d_{M} (Y' ) }  
\ee

The identity  $\Psi_{k,\bom}  = T'_{k, \bom} (\cA)  \psi_{k,\bom}(\cA)$
 becomes on $\La_{k} \subset \Om_{k}$ the identity $\Psi_k  = T_k (\cA)  \psi_{k,\bom}(\cA)$.
 We use it to define    
\be
  E^{(1)}_{k, \bpi^+}\B(Y' , \cA, \tilde Z_k, \psi_{k, \bom}(\cA )\B) =  E^{ (1),\st}_{k, \bpi^+}\B(Y', \cA, \tilde Z_k,  T_k(\cA)   \psi_{k,\bom}(\cA) \B) 
\ee
Then  since   $|T'_{k}(\cA) f|  \leq  C_T  \|f\|_{\infty}$   we have
\be \label{orthogonal} 
   \|  E^{(1)}_{k, \bpi^+}(Y', \cA, \tilde Z_k) \|_{h_k \bbI_k }
   \leq 
 \|   E^{ (1), \st}_{k,\bpi^+}(Y', \cA, \tilde Z_k )  \|_{C_T h_k\bbI_k} 
\leq \one 
 e_k^{\frac14 - 5 \ep}   e^{- \ka   d_{M}(Y') } 
\ee

The sum (\ref{spumoni4}) can now be split into terms with $Y' \subset \La_{k+1}$ and terms with $Y' \cap \Om_{k+1} \neq \emptyset$ and $Y' \cap \La^c_{k+1} \neq \emptyset$.
Since $\La_{k+1} \subset \Om_{k+1}$ the latter condition is the same as $Y'  \cap ( \Om_{k+1} -   \La_{k+1}) \neq \emptyset$ and we have
\be 
 \sum_{ Y'  \cap \Om_{k+1} \neq \emptyset}      E^{(1) }_{k, \bpi^+}  (Y') 
=  \sum_{ Y'  \subset \La_{k+1} }      E^{(1) }_{k, \bpi^+}  (Y') 
+  \sum_{ Y'  \cap ( \Om_{k+1} -   \La_{k+1}) \neq \emptyset}      E^{(1) }_{k, \bpi^+}  (Y') 
\ee
By our earlier remarks the terms in the first sum  have kernels independent of $\bpi^+$.  
This is the first sum in the lemma and $\| \cdot \|_{h_k \bbI_k} = \| \cdot \|_{h_k}$ in this case.  The terms in the second sum are contributions to $B^{(1)}_{k, \bpi^+}(Y')$. 
\bigskip

\noindent
\textbf{C.}
Now consider terms in the sum (\ref{tutu})  with the sum over $X \cap \La^c_k \neq \emptyset$.
The previous analysis can be characterized as anchoring the expansion around $X \subset \La_k$.
If $X$ is far from $\La_k$ this does not work so well.   We still would like to anchor around something in $\La_k$ or perhaps  $\Om_{k+1}$  
which we accomplish by "pre-localizing" following Balaban \cite{Bal88b}.   The following analysis works for any $X$ but we only use it for  $X \cap \La^c_k \neq \emptyset$. 

Instead of (\ref{musty}) we now  take for $X \in \cD_{k, \bom}$
\be
\begin{split}
\tilde  E^{(1)}_{k, \bom} (X, \cA, \cZ, \Psi_{k, \bom}) 
= &  \int_0^1 \frac{d}{dt} \blan \bPsi_{k,\bom} , M_{k,\bom}(X, \cA+ t \cZ) )  \Psi_{k,\bom} \bran \ dt \\
= &    \int_0^1  \blan \frac{ \pa }{ \pa \cA}\B[ \blan \bPsi_{k,\bom} ,   M_{k, \bom} (X,  \cA+ t \cZ)  \Psi_{k,\bom} \bran \ \B], \cZ\bran \ dt\\
\end{split}
\ee
We specialize to   $\cZ = \cZ^{\bullet}_{k, \bpi^+}$  and use this as a basis for further localization. 
Note that  because we are taking $\cA \in  \tilde \cR_{k, \bom}$ which involves derivatives of $\cA$,  we cannot vary  $\cA(b)$  independently at one point
and so cannot use a Cauchy bound to bound the derivatives in $\cA$ here.

 Instead we introduce    a sharp partition of unity $\chi(\sq_0)$ on $\tz$  with support in $M$-cubes.
 Replace  $\cZ^{\bullet}_{k, \bpi^+} $ by  a sum over $\sq_0$ of
 \be    \label{saturn3}
  \cZ^{\bullet} _{k, \bpi^+} (\sq_0) =    \cH_{k, \bom}  1_{\sq_0} \B( C C_{k, \bom^+} ^{\frac12,\loc}  \tilde  Z_k   + g \    \fH_{k, \bpi^+}  \B)
 \ee
 Then
 \be
\tilde  E^{(1)}_{k, \bom} (X, \cA, \cZ^{\bullet}_{k, \bpi^+}, \Psi_{k, \bom}) 
= \sum_{\sq_0 \in \Om_{k+1} }  \tilde E^{(1)}_{k, \bom} \B (X, \cA,  \cZ^{\bullet}_{k, \bpi^+},   \cZ^{\bullet}_{k, \bpi^+} ( \sq_0) , \Psi_{k,\bom}  \B) 
\ee
where  
\be \label{pollyanna} 
\begin{split}
&\tilde E^{(1)}_{k, \bom} \B (X,  \cA,  \cZ^{\bullet}_{k, \bpi^+},   \cZ^{\bullet}_{k, \bpi^+} ( \sq_0) , \Psi_{k,\bom}  \B)  \\
= &    \int_0^1  \ dt \blan \frac{ \pa }{ \pa \cA}\B[ \blan \bPsi_{k,\bom} , 
\hat  M_{k, \bom} (X,  \cA+ t\ \cZ^{\bullet}_{k,\bpi^+} )  \Psi_{k,\bom} \bran \ \B],  \cZ^{\bullet}_{k, \bpi^+} ( \sq_0) \bran 
  \\
= &   \int_0^1  \ dt \frac{ \pa }{ \pa u }\B[ \blan \bPsi_{k,\bom} , 
\hat  M_{k, \bom} \B(X,   \cA+ t\ \cZ^{\bullet}_{k, \bpi^+} +  u   \cZ^{\bullet}_{k, \bpi^+} ( \sq_0)    \B)  \Psi_{k,\bom} \bran \B]_{u =0} 
   \\
= &\int_0^1  \ dt
\frac{1} { 2\pi i} \int_{|u| = \frac14 e_k^{-\frac34 + 5 \ep} }   \frac{ du}{u^2}
\blan \bPsi_{k,\bom} ,  \hat M_{k, \bom}\B (X,   \cA+ t\ \cZ^{\bullet}_{k, \bpi^+} +  u \cZ^{\bullet}_{k, \bpi^+} ( \sq_0) \B  )  \Psi_{k,\bom} \bran \\
\end{split}
\ee
As before $\cZ^{\bullet}_{k, \bpi^+} ( \sq_0) $ is in the domain (\ref{d1}). Hence  $ \frac14 e_k^{- \frac34 + \frac92 \ep}  \cZ^{\bullet}_{k, \bpi^+} ( \sq_0)
\in \frac 12 e_k^{\ep} \tilde \cR_{k, \bom} $.   The $ \cZ^{\bullet}_{k, \bpi^+} $ is even smaller.   Thus  for $0 \leq t \leq 1$ and  $|u| \leq  \frac14 e_k^{-\frac34 + \frac92 \ep} $
we have
$ t\ \cZ^{\bullet}_{k, \bpi^+} +  u \cZ^{\bullet}_{k, \bpi^+} ( \sq_0) \in e_k^{\ep}   \cR_{k, \bom} $. 
 The integrand in  (\ref{pollyanna})  is analytic in  $u$ and the representation 
 holds.

\bigskip

 We  again seek to localize  in  $\tilde Z_k,   d\cA$. 
Accordingly    we    introduce weakening parameters    $s_{\bom}= \{s_{\sq} \}_{\sq \in \pi'(\bom)}$
and replace $\cZ^{\bullet} _{k, \bpi^+} $ by $\cZ^{\bullet} _{k, \bpi^+}(s) $ as in (\ref{saturn2}), and we replace
$ \cZ^{\bullet}_{k, \bpi^+} ( \sq_0)$ by 
\be   
  \cZ^{\bullet} _{k, \bpi^+} (s,\sq_0) =    \cH_{k, \bom} (s) 1_{\sq_0} \B( C C_{k, \bom^+} ^{\frac12,\loc} (s) \tilde  Z_k   + g \    \fH_{k, \bpi^+}(s)  \B)
 \ee
This again satisfies the bound (\ref{tinker}) : 
\be
 \label{possible}
 \| \tilde E^{(1)}_{k, \bom} \B (X,   \cA, \cZ^{\bullet}_{k, \bpi^+}(s),   \cZ^{\bullet}_{k, \bpi^+} (s,  \sq_0) , \Psi_{k,\bom}  \B) \|_{ C_Th_k\bbI_k} 
\leq    e_k^{\frac14 - 5 \ep}     e^{- \hat \ka |X|_{\bom}}
\ee
  Now we expand around $\sq_0 \subset \Om_{k+1}$. In
 each variable $\sq \in \pi'(\bom)$  except $\sq_0$ we   interpolate between  $s_{\square} =1$  and   $s_{\square} =0$
and find 
 \be 
\tilde E^{(1)}_{k, \bom} \B (X;   \cA , \cZ^{\bullet}_{k, \bpi^+},   \cZ^{\bullet}_{k, \bpi^+} ( \sq_0) , \Psi_{k,\bom}  \B)
=  \sum_{Y \supset \sq_0  }    \breve   E^{(1)}_{k, \bpi^+}  (X,  Y, \sq_0,   \cA,   \tilde Z_k , \Psi_{k,\bom} ) 
\ee 
where the  sum is over    multiscale  polymers   $Y \in \cD_{k, \bom} $,
and where
 \be  \label{summer1}
\begin{split}
&  \breve   E^{(1)}_{k, \bpi^+} \B (X,  Y,  \sq_0, \cA,   \tilde Z_k , \Psi_{k,\bom} \B) \\
&
=      \int   ds_{Y- \sq_0} 
 \frac { \pa  }{ \pa s_{Y-\sq_0 }}   \left[  \tilde E^{(1)}_{k, \bom} \B (X,   \cA ,  \cZ^{\bullet}_{k, \bpi^+}(s),   \cZ^{\bullet}_{k, \bpi^+} (s,  \sq_0) , \Psi_{k,\bom}  \B) \right]_{s_{Y^c} = 0, s_{\sq_0}=1}\\
 \end{split}
 \ee
  This vanishes unless $X \cap Y \neq \emptyset$.   The  $ \breve     E^{(1)}_{k, \bpi^+}(X, Y, \sq_0)   $
 only depend  on  the indicated fields    in $ X \cup Y$.  
Furthermore  $ \tilde E^{(1)}_{k, \bom} \B (X,   \cA ,  \cZ^{\bullet}_{k, \bpi^+}(s),   \cZ^{\bullet}_{k, \bpi^+} (s,  \sq_0) , \Psi_{k,\bom}  \B) $ is analytic  in  $|s_{\square} |  \leq   M^{\frac12} $  and satisfies the bound
(\ref{possible}) there.
Again by   Cauchy inequalities  each derivative introduces a factor   $M^{- \frac12 }  \leq   e^{- \hat \ka} $   so we have
 \be  
 \|  \breve   E^{(1)}_{k, \bpi^+}  (X,  Y, \sq_0;  \cA,   \tilde Z_k)   \|_{ C_T  h_k \bbI_k }
\leq 
      e_k^{\frac14 - 5 \ep}   e^{- \hat \ka |Y- \sq_0|_{\bom}  }   e^{-\hat \ka |X|_{\bom}}
\ee

 At this point we have for $X,Y \in \cD_{k, \bom}$
 \be \label{spumoni}
\sum_{X \cap \La_k^c \neq \emptyset }  \tilde  E^{ (1)}_{k, \bom} (X) 
=\sum_{X \cap \La_k^c \neq \emptyset } \  \sum_{\sq_0 \in \Om_{k+1} } \ \sum_{ Y \supset \sq_0, \  Y \cap X \neq \emptyset  }  
 \breve   E^{(1)}_{k, \bpi^+}  \B(X,  Y, \sq_0;  \cA,   \tilde Z_k , \Psi_{k,\bom} \B)
\ee 
We do a partial resummation here.     Let $Y^+ = X \cup Y \in \cD_{k, \bom}$.
Then   (\ref{spumoni}) can be written
 \be
  \sum_{\sq_0 \subset \Om_{k+1}} \  \  \sum_{Y^+}      E^{ (1),+}_{k, \bpi^+}  (Y^+, \sq_0) 
\ee
 where 
\be   
      E^{ (1),  + }_{k, \bpi^+}  (Y^+,\sq_0; \cA,   \tilde Z_k , \Psi_{k,\bom} )  = \sum_{ X \cup  Y = Y^+, X \cap Y \neq \emptyset } \ \     1_{ Y \supset \sq_0   }  \ \
   1_{X \cap \La_k^c \neq \emptyset}     \ 
     \breve   E^{(1)}_{k, \bpi^+}  \B(X, Y;  \cA,   \tilde Z_k , \Psi_{k,\bom} \B)
\ee
 vanishes unless $Y^+ \supset \sq_0$ and $ Y^+ \cap \La_k^c \neq \emptyset$.
With $|Y- \sq_0|_{\bom} = |Y|_{\bom} +1$    we  have the estimate
\be
 \| E^{ (1), +}_{k, \bpi^+}  (Y^+,   \sq_0 ,\cA, \tilde Z_k)  \|_{ C_T  h_k \bbI_k }
 \leq  \one  e_k^{\frac14 - 5 \ep}  \sum_{X, Y: X \cup  Y = Y^+, X \cap Y \neq \emptyset } \ \    1_{ Y \supset \sq_0   }      e^{- \hat  \ka |Y|_{\bom}  }   e^{- \hat \ka |X|_{\bom}} 
\ee
But 
$|Y |_{\bom} + |X|_{\bom}  \geq | Y^+|_{\bom}$
so we can extract a factor $e^{- ( \hat \ka - \ka_0) |Y^+|_{\bom} }$.    Relaxing the remaining sums 
we have 
\be
 \| E^{ (1), + }_{k, \bpi^+}  (Y^+,  \sq_0 ,\cA,   \tilde Z_k  ) \|_{ C_Th_k \bbI_k} \leq  \one e_k^{\frac14 - 5 \ep}   e^{- ( \hat \ka - \ka_0) |Y^+|_{\bom} } \ \  \sum_{Y \supset \sq_0}  \ \       e^{- \ka_0|Y|_{\bom}  }
 \sum_{X \subset Y^+}    e^{- \ka_0 |X|_{\bom}} 
\ee
The sum over $Y$ is bounded by one (see Appendix in \cite{Dim13b}). The sum over $X$ is bounded by $|Y^+|_{\bom}$ which  is absorbed by the exponential.  
Thus
\be 
 \|    E^{  (1),+}_{k, \bpi^+}(Y^+, \sq_0 ,\cA,   \tilde Z_k )    \|_{ C_Th_k \bbI_k}
\leq    
  \one    e_k^{\frac14- 5\ep}  e^{- ( \hat \ka - \ka_0 - 1) |Y^+|_{\bom} }
      \ee

Now we do another partial resummation.  Let $ Y' = \bar Y^+$ be the smallest element of $\cD_k$  containing $Y^+$.
Then we have   
\be \label{conjugal}
 \sum_{ X \cap \La^c_k \neq \emptyset} \tilde  E^{ (1)}_{k, \bom}(X)  =   \sum_{Y' \in \cD_k}      E^{ (1), \st}_{k,\bpi^+}  (Y') 
 \ee
where
\be
  E^{ (1), \st}_{k, \bpi^+}  (Y', \cA,   \tilde Z_k , \Psi_{k,\bom} ) = \sum_{\bar  Y^+ = Y'} \ \sum_{\sq_0   \subset (Y \cap \Om_{k+1} ) } 
       E^   { (1),  +}_{k, \bpi^+}  (Y^+,\sq_0; \cA,   \tilde Z_k , \Psi_{k,\bom} )  
 \ee 
 This vanishes unless $Y'  \cap \Om_{k+1} \neq \emptyset$ and $ Y'  \cap \La_k^c \neq \emptyset$.
We have $ |Y^+|_{\bom} \geq | \bar Y^+|_M = |Y'|_M \geq  d_M(Y') $ so we can extract a factor $e^{- (\ka + 1) d_M(Y') }$. Relaxing the remaining
sums we have  
\be
 \|    E^{ (1), \st }_{k, \bpi^+}  (Y' ,\cA,   \tilde Z_k )    \|_{ C_Th_k \bbI_k}
\leq     \one e_k^{\frac14- 5\ep}   e^{- (\ka+1)  d_M(Y') } \sum_{\sq_0 \subset (Y' \cap \Om_{k+1})}   \sum_{Y^+ \supset \sq_0}   e^{-(\hat \ka - \ka  - \ka_0-2) |Y^+|_{\bom} }  
\ee
For $\hat \ka$ large enough the  sum over $Y^+$ gives one.  The sum over $\sq_0$ gives $|Y' \cap \Om_{k+1}|_M \leq |Y'|_M \leq \one (d_M(Y') +1) $ which can be absorbed by the exponential.
Thus 
\be
 \|    E^{ (1),\st }_{k, \bpi^+}  (Y' ,\cA,   \tilde Z_k )    \|_{ C_Th_k \bbI_k }
\leq     \one e_k^{\frac14- 5\ep}   e^{-  \ka d_M(Y') }
\ee

Finally   express  $\Psi_k$ in terms of         $ \psi_{k, \bom}(\cA)$   $\Psi_{k,\bom}(\cA)  = T'_{k, \bom} (\cA)  \psi_{k,\bom}(\cA)$
defining  
\be
  E^{(1)}_{k, \bpi^+}\B(Y', \cA,   \tilde Z_k, \psi_{k, \bom}(\cA )\B) =  E^{ (1),\st}_{k, \bpi^+}\B(Y', \cA,   \tilde Z_k,   T'_{k, \bom} (\cA)  \psi_{k,\bom}(\cA)\B) 
\ee
Again   since   $\|T'_{k, \bom}(\cA) f\|  \leq  C_T  \|f\|_{\infty}$   we have   
\be 
   \|  E^{(1)}_{k, \bpi^+}(Y',\cA,   \tilde Z_k) \|_{h_k   \bbI_k }
    \leq \|   E^{ (1), \st}_{k, \bpi^+}(Y', \cA,   \tilde Z_k )  \|_{C_T h_k \bbI_k} 
    \leq \one    e_k^{\frac14- 5\ep}  e^{- \ka    d_{M} (Y' ) }  
\ee

Now (\ref{conjugal}) is written  $\sum_{Y'}  E^{(1)}_{k, \bpi^+}(Y') $.  
These are  contributions to $\sum_{Y'}  B^{(1)}_{k, \bpi^+}(Y') $.  Note also   that $Y' \cap \Om_{k+1} \neq \emptyset$
and $ Y' \cap \La_k^c \neq \emptyset$ imply  $Y' \cap (\Om_{k+1}- \La_{k+1} ) \neq \emptyset$.
\bigskip

 \noindent
 \textbf{D.}   Finally  we discuss   the omitted terms in (\ref{succotash}). 
 Another contribution to  (\ref{succotash}) is  the expression $< \bPsi_{k, \bom} ,N_{k, \bom}(\cA) \psi_{\Om_1^c}   >
$ where
\be   
N_{k, \bom}(\cA)  =   \bb^{(k)}  Q_{k, \bom}(\cA) S_{k, \bom}(\cA) \fD_{\cA}
\ee
with kernel   $ N_{k, \bom}(\cA;y,y' ) =  <\de_y,  N_{k, \bom}(\cA)   \de_{y'}>$. 
For $y \in \de \Om_j$ we have as before $ Q^T_{k, \bom}(-\cA) \bb^{(k)}  \de_y = b_j L^{4(k-j)}  e^{ - i  \si_j}  1_{\De_y } $.
For $y' \in \Om_1^c$ we have from the definition of $\fD_{\cA} $ with $\eta = L^{-k}$
\be
(\fD_{\cA}  \de_{y'})(x)  = L^{4k} f_{y'}(x) 
= L^{4k}  \begin{cases} 
\frac12 ( 1  \pm \ga_{\mu} )e^{ie_k \eta \cA(y',x)}   &  x= y' \mp \eta e_{\mu} \\
1  &  x =y' \\
0 & \textrm{otherwise} \\
\end{cases} 
\ee
which we write as
\be
\fD_{\cA}  \de_{y'} =
 L^{4k}  \sum_{|x'-y'| \leq \eta}  f_{y'} \ 1_{\De_ {x'}}  
\ee
where   $1_{\De_{x'}} $ selects the single point $x'$.
So   for $y\in \de \Om_{j}^{(j)}$ and  $y'  \in \Om_1^c$ 
\be
 N_{k, \bom}(\cA;y,y'  )  = b_j  L^{4(k-j)}L^{4k}  \sum_{|x'-y'| \leq \eta}  \blan e^{i \si_j }1_{\De_y},  S_{k, \bom}(\cA)  f_{y'} 1_{\De_{x'}}    \bran 
\ee

The $\cA$ dependence  in $ f_{y'}$   comes from  expression  $\exp( ie_k \eta \cA(b ))$ where $b$ crosses from $\Om^c_1$ to  $\Om_1$. 
On such a bond we have  $|\im \cA(b)| \leq  L^k e_k^{-\frac34 + \ep}$ and so $  |\exp( ie_kL^{-k} \cA(b ))| \leq
\exp (e_k^{\frac14+ \ep} ) \leq 2$.  We  then have $| f_{y'} | \leq \one $.
 This gives the estimate as in (\ref{forty-five})
\be
\begin{split}
| \blan e^{i \si_j }1_{\De_y},  S_{k, \bom}(\cA)  f_{y'} 1_{\De_{x'}}    \bran  |
 \leq &\| e^{i \si_j }1_{\De_y}  \|_1 
\| 1_{\De_y} S_{k, \bom}(\cA) f_{y'}1_{\De_ {y'}}\|_{\infty} \\
\leq  &    C L^{-3(k-j)} L^{ - k}  e^{-\ga d_{\bom}(y,y') } \\
\leq   & C  L^{-2(k-j)} L^{-2k}  e^{-\ga d_{\bom}(y,y') }\\
\\
\end{split} 
\ee
Then we have just as for $M_{k , \bom}( \cA; y,y')$ with $j'=0$
\be
  \label{squash}
|N_{k , \bom}( \cA; y,y')|  \leq  
C L^{2(k-j)} L^{2k}e^{- \ga d_{\bom} (y,y') } 
\ee
The rest of the analysis goes as before.
We have  a polymer expansion $N_{k , \bom}= \sum_X N_{k , \bom}(X)$ where now $X$ must intersect $\Om_1^c$.
The  $N_{k , \bom}(X)$ satisfy an estimate like (\ref{slumbering}). Continuing as
in (\textbf{B.}), (\textbf{C.}) we end with contributions to $  B^{(1)}_{k, \bom}(Y') $ where now $Y'$ must intersect $\Om_{k+1}$ and $\Om_1^c$.
These terms are very small (except in the first step).

The other terms in (\ref{succotash}) are treated similarly, except $< \bpsi_{\Om_1^c}, ( \fD_{\cA} + \bar m ) \psi_{\Om_1^c}>$ which anyway  does
not contribute when we vary $\cA$ in $\Om_1$. This completes the proof. 
\bigskip

\begin{lem}   \label{outsize2}   
\be    \label{serendipity2}
\begin{split}
   &   E^{(2)}_{k, \bom}\B(\La_k,  \cA^{0} _{k+1, \bom^+},     \cZ^{\bullet}_{k, \bpi^+},    \psi^\#_{k, \bom}(\cA^{0} _{k+1, \bom^+}),    \Psi_{k, \bom}   \B) \\
   &   =   \sum_{X \subset \La_{k+1}    }   E^{(2)}_{k}\B(X, \cA^{0} _{k+1, \bom^+},     \tilde Z_k ,   \psi^\#_{k, \bom}(\cA^{0} _{k+1, \bom^+}) \B)  \\
   & +  \sum_{  X \cap (\Om_{k+1} -\La_{k+1})  \neq \emptyset  }   B^{(2)}_{k, \bpi^+}\B(X, \cA^{0} _{k+1, \bom^+},     \tilde Z_k ,   \psi^\#_{k, \bom}(\cA^{0} _{k+1, \bom^+})  \B)    \\
\end{split}
\ee
where  the sums  are over $X \in \cD_k$ and
\begin{enumerate} 
\item  $  E^{(2)}_{k}(X, \cA^{0} _{k+1, \bom^+},     \tilde Z_k ,     \psi^\#_{k, \bom}(\cA^{0} _{k+1, \bom^+}) ) $   are the restrictions 
 of  $ E^{(2)}_{k}(X, \cA,   \tilde Z_k,   \psi^\#_{k, \bom}( \cA))$
 depending on the indicated fields only in $X$,  are analytic in  $\cA \in (1-e_k^{\ep} ) \tilde  \cR_{k,\bom} , \ |\tilde Z_k| \leq e_k^{-\frac14 \ep}$  and satisfy  there
  \be   
  \|  E^{(2)}_{k} \B(X, \cA  ,    \tilde Z_k   \B) \| _{ \frac12  \bh_k}  \leq   \one   e_k^{\frac34 - 7 \ep}  e^{- ( \ka - \ka_0-1) d_{M}(X)}
  \ee
The kernel is independent of the  history $\bpi^+$
\item 
$ B^{(2)}_{k, \bpi^+}\B(X, \cA^{0} _{k+1, \bom^+},     \tilde Z_k ,   \psi^\#_{k, \bom}(\cA^{0} _{k+1, \bom^+})  \B)  $ are the restrictions of 
 $ B^{(2)}_{k, \bpi^+ }\B(X, \cA ,     \tilde Z_k ,   \psi^\#_{k, \bom}(\cA)  \B)  $ depending on the indicated fields in $X$,  analytic in  $\cA \in (1-e_k^{\ep} ) \tilde  \cR_k, \ |\tilde Z_k| \leq e_k^{-\frac14 \ep}$
and satisfying  there
\be
\| B^{(2)}_{k,\bpi^+}\B(X, \cA ,     \tilde Z_k    \B)  \| _{ \frac12 \bh_k   \bbI^\#_k}  \leq   \one   e_k^{\frac34 - 7 \ep}  e^{- ( \ka - \ka_0-1) d_{M}(X)}
\ee
\end{enumerate} 
\end{lem}
\bigskip

\rem
In the same way we have the splitting
of $E^{(3)}(\La_k)$ which in an abbreviated notation is
\be    \label{serendipity3}
    E^{(3)}_{k, \bom}(\La_k)   =   \sum_{ X \subset \La_{k+1}    }   E^{(3)}_{k}(X) 
   +  \sum_{ X \cap (\Om_{k+1}- \La_{k+1})  \neq \emptyset  }   B^{(3)}_{k, \bpi^+}( X)    \
\ee
and there are similar bounds on $ E^{(3)}_{k}(X) $ and  $ B^{(3)}_{k, \bpi^+}( X)   $. 
\bigskip

\pr 
We first study the operators   
 $
   \psi^\#_{k, \bom}(\cA)    =     (  \psi_{k, \bom}(\cA) , \de_{\al, \cA}   \psi_{k, \bom}(\cA))
 $
and $ \cJ^\#_{k, \bom}(\cA, \cZ) \Psi_{k, \bom}   \equiv        \psi^\#_{k, \bom}(\cA+ \cZ )  -    \psi^\#_{k, \bom}(\cA)  $
in the domain (\ref{d1}).
Since $
   \psi_{k, \bom}(\cA)    =    \cH_{k, \bom}(\cA)  \Psi_{k} $ we   have 
\be 
 \label{bugsbunny} 
\cJ^{\#}_{k, \bom} (\cA)   \Psi_{k, \bom} \equiv  \B (  \cJ_{k, \bom} ( \cA, \cZ )   \Psi_{k, \bom}  , \  \cJ^{(\al) }_{k, \bom} ( \cA, \cZ )    \Psi_{k, \bom} \B )   
\ee
where
\be   
\begin{split}
   \cJ_{k, \bom} ( \cA, \cZ )    \Psi_{k, \bom} 
 = & \B(  \cH_{k, \bom}(\cA+ \cZ) - \cH_{k, \bom}(\cA) \B)    \Psi_{k, \bom}   \\
   \cJ^{(\al)}_{k, \bom} ( \cA, \cZ )  \Psi_{k, \bom} 
 = &  \B(\de_{\al, \cA+ \cZ}    \cH_{k, \bom}(\cA+ \cZ) -\de_{\al, \cA}     \cH_{k, \bom}(\cA) \B)   \Psi_{k, \bom}    \\
\end{split} 
\ee
As before if    we  take complex    $|t|  \leq  \frac12   e_k^{-\frac34+\frac92 \ep}  $
 then    $t \cZ  \in  e_k^{\ep}  \cR_k$,   
 hence  $\cA  + t \cZ$ is  in  $\tilde   \cR_k$  and hence  well within the analyticity domain of $t \to   \cH_{k, \bom}(\cA+ t \cZ)$.  
 Thus we  can write
\be  \label{leak1}
   \B( \cH_{k, \bom}(\cA+ \cZ) - \cH_{k, \bom}(\cA) \B)  f  =   
  \frac{1}{2 \pi i}
 \int_{|t|  = \frac12 e_k^{-\frac34+ \frac92 \ep}   }      \frac{dt}{t(t-1)}     \cH_{k, \bom}(\cA+t \cZ)    f
\ee
and similarly for the Holder derivative.  Using the bounds  (147) from \cite{Dim20}   on  $ \cH_{k, \bom}(\cA)$
we get  for $f = \{ f_{j, \de \Om_j} \}$
 on $\de \Om_j$   
\be  
  \label{log}
L^{-(k-j)}  |  \cJ_{k, \bom} ( \cA, \cZ )  f|,   \  L^{-(1+ \al)(k-j)}  |      \cJ^{(\al)}_{k, \bom} ( \cA, \cZ ) f |  \leq  Ce_k^{\frac34 -\frac92 \ep} \sup_{j'} L^{-(k-j')}  \|f\|_{\infty, \de \Om_{j'}}  
   \ee
Define
  \be
  \bbI_k^\# = ( \bbI_k, \bbI_k')  =( L^{k-j}, L^{(1 + \al)(k-j)} )
  \  \textrm{ on } \de \Om_j
  \ee
     Then replacing $f$ by $\bbI_k f = \{ L^{k-j} f_{j, \de \Om_j } \}$  (\ref{log})  can be written
\be  
  \label{log2}
   |  \cJ^\#_{k, \bom} ( \cA, \cZ )\bbI_k   f  | \leq  C\  \bbI^\#_k  e_k^{\frac34 -\frac92 \ep}   \|f\|_{\infty}  
   \ee

\bigskip

 Now  consider   first  $ E^{(2)}_{k \bom} (\La_k, \cA, \cZ , \psi_{k, \bom}(\cA ), \Psi_{k, \bom}     )  $ on the domain (\ref{d1}).     
We  write 
 $
 E^{(2)} (\La_k  ) =   \sum_{X \subset  \La_k}   \tilde  E^{(2)}_{k, \bom}(X ) 
 $
where
\be   \label{tingly}
\begin{split} 
 &  \tilde  E^{(2)}_{k, \bom}\B(X, \cA, \cZ,    \psi^\#_{k, \bom}(\cA ), \Psi_{k, \bom}  \B ) \\  & =      E'_k\B(  X,  \cA + \cZ,    \psi^\#_{k, \bom}( \cA + \cZ  )  \B) 
     -       E'_k\B( X,   \cA,   \psi^\#_{k, \bom}(  \cA   )  \B)     \\
 &=      E'_k\B(  X,  \cA + \cZ,    \psi^\#_{k, \bom}( \cA  ) 
  +  \cJ^\#_{k, \bom}(\cA, \cZ) \Psi_{k, \bom}) \B)      -       E'_k\B( X,   \cA,   \psi^\#_{k, \bom}(  \cA   )  \B)  \\
&=       \frac{1}{2 \pi i}
 \int_{|t|  = e_k^{-\frac34+6 \ep}   }      \frac{dt}{t(t-1)}    E'_k\B(   X, \cA  + t \cZ,      \psi^\#_{k, \bom}(\cA)  + t  \cJ^\#_{k, \bom}(\cA, \cZ) \Psi_{k, \bom})\B )  \\
 \end{split}
\ee 
Let 
$
\cE(    \psi^\#_{k, \bom}(\cA)  ,  \Psi_{k, \bom})  ) =   E'_k(   X, \cA  + t \cZ,      \psi^\#_{k, \bom}(\cA)  + t  \cJ^\#_{k, \bom}(\cA, \cZ) \Psi_{k, \bom} )
$
from the last line.
For $|t|   = e_k^{-\frac34+6 \ep} $ we have  
by (\ref{log2})   $|t|  \   | \cJ^{\#}_{k, \bom}  f \bbI_k |   \leq   \bbI^\#_k C e^{\frac32 \ep}_k   \|f\|_{\infty}$ and so    (see (A44)  in  \cite{Dim15b} )     
\be  
\begin{split}  \label{lingo} 
  \| \cE \|_{  \frac12   \bh_k \bbI^\#_k , C_Th_k   \bbI_k }  &  
    \leq   \| E'_k( X, \cA + t \cZ)  \|_{ \frac12   \bh_k \bbI_k^\#   +(\bbI_k^\# C e^{\frac32 \ep}_k )( C_Th_k) }\\
  &   \leq   \| E'_k( X, \cA+ t \cZ)  \|_{  \bh_k  \bbI_k^\#  }   \  \leq  \    \one  e_k^{\frac14-7 \ep}   e^{-\ka d_M(X) } \\
 \end{split}
 \ee
 Here we used  
 \be 
  (\bbI_k^\# C e^{\frac 32 \ep}_k )( C_Th_k ) \leq   \frac12 e^{  \ep}_k (h_k, h_k )  \bbI^\#_k  \leq  \frac12 \bh_k \bbI_k^\# 
 \ee 
 and the bound (\ref{retro}) on $ E'_k$  (here with $\bh_k \bbI_k = \bh_k$). 
 Then the representation (\ref{tingly})
 gives
 \be \label{lingo2}
\| \tilde  E^{(2)}_{k, \bom}(X, \cA, \cZ ) \|_{ \frac12   \bh_k \bbI^\#_k , C_Th_k   \bbI_k  }
\leq \one  e_k^{1 - 13 \ep }   e^{-\ka d_M(X) } 
\ee
 
 \bigskip

  Now we  specialize to $\cZ =  \cZ^{\bullet}_{k, \bpi^+}$ and further localize
 $ \tilde  E^{(2)}_{k, \bom}(X, \cA,\cZ^{\bullet}_{k, \bpi^+},    \psi^\#_{k, \bom}(\cA ), \Psi_{k, \bom}   ) $.
Accordingly    we  again   introduce   weakening parameters    $s_{\bom}= \{s_{\sq} \}_{\sq \in \pi'(\bom)}$ and replace $ \cZ^{\bullet} _{k, \bpi^+}$ by
$ \cZ^{\bullet} _{k, \bpi^+}(s)$ as  in (\ref{saturn2}).
   We  also  introduce weakening parameters $\cH_{k, \bom} \to \cH_{k, \bom}(s)$   in the fermion operators  as explained in \cite{Dim20}  and define
  \be
  \begin{split}
   \cJ^{\#}_{k, \bom}(s,\cA, \cZ) \Psi_{k, \bom} 
= & \B(  \cJ_{k, \bom} (s,  \cA, \cZ ) \Psi_{k, \bom}   , \  \cJ^{(\al)}_{k, \bom} (s,  \cA, \cZ ) \Psi_{k, \bom}    \B)  \\ 
  \cJ_{k, \bom}(s,  \cA, \cZ)   =  &    \cH_{k, \bom}(s,\cA+ \cZ)  -   \cH_{k, \bom}(s,\cA)  \\
    \cJ^{(\al)}_{k, \bom}(s,  \cA, \cZ)   = &     \de_{\al, \cA + \cZ} \cH_{k, \bom}(s,\cA+ \cZ)  - \de_{\al, \cA}  \cH_{k, \bom}(s,\cA) \\
  \end{split}
    \ee 
Make these    replacements  in   (\ref{tingly}) 
and  define 
\be   \label{tingly2}
\begin{split} 
 &  \tilde  E^{(2)}_{k, \bpi^+}\B(s, X,   \cA, \tilde Z_k ,  \psi^\#_{k, \bom}(\cA),     \Psi_{k, \bom}   \B )  \\
 =    &    \frac{1}{2 \pi i}
 \int_{|t|  = e_k^{-\frac34+7 \ep}   }      \frac{dt}{t(t-1)}    E'_k\B(   X, \cA  + t  \cZ^{\bullet}_{k,\bpi^+}(s) ,    
   \psi^\#_{k, \bom}(\cA)  + t  \cJ^{\#}_{k, \bom}(s,\cA,\cZ^{\bullet}_{k, \bpi^+}(s)) \Psi_{k, \bom}  \B )  \\
 \end{split}
\ee 
which is not yet local in $\tilde Z_k$. 
 We  can repeat the  estimates above for  complex  $|s_{\square}| \leq   M^{-\frac12}$   and   get
    instead of  (\ref{lingo2}) 
   \be       \label{lingo3} 
\|  \tilde  E^{(2)}_{k, \bpi^+}(s, X,   \cA, \tilde Z_k) \|_{ \frac12   \bh_k \bbI^\#_k , C_Th_k   \bbI_k  }  
\leq     \one    e_k^{1-13 \ep} e^{-\ka d_M(X) } 
\ee

Again in  each variable  $s_{\sq}$   with $\sq \in \pi'(\bom) $ except those in $X$ we  interpolate between  $s_{\sq} =1$  and   $s_{\sq} =0$  and 
find    
   $E^{ (2)}_{k, \bom}  (\La_k)  
=  \sum_{ X \subset  \La_{k}, Y \supset X  }    \breve   E^{(2)}_{k, \bpi^+}  (X, Y) $
 where the sum is over $Y \in \cD_{k, \bom}$ and 
 \be  \label{summer2}
\begin{split}
& \breve    E^{(2)}_{k, \bpi^+}(X, Y,    \cA,  \tilde Z_k,     \psi^\#_k(\cA) , \Psi_{k, \bom} )   \\
= &        \int   ds_{Y-X} 
 \frac { \pa  }{ \pa s_{Y-X}}   \left[  \tilde  E^{(2)}_{k, \bpi^+}(s,X,   \cA,  \tilde Z_k,     \psi^\#_{k, \bom}(\cA)  , \Psi_{k, \bom} )   \right]_{s_{Y^c} = 0, s_X=1}\\
 \end{split}
 \ee
 This  depends on the indicated fields only   in $Y$.  Also   $Y$ must intersect $\Om_{k+1}$ since if not $[\cZ_{k,\bpi^+} (s)]_Y =0$
then $  \cJ^{\#}_{k, \bom}(s,\cA, 0) =0$ and the integral (\ref{tingly2}) vanishes.
 
The rest of   proof  more or less follows the proof in part B of lemma  \ref{outsize1}.  The  differences are
 (i) we have   $ e_k^{\frac34- 7\ep} $
 instead of $ e_k^{\frac14- 6\ep}$
(ii.) there is the extra  explicit field $ \psi^\#_{k, \bom}(\cA) $
 (iii.) our decay starts with $\ka$ rather than the larger $ \hat \ka$.  
      We estimate $ \breve    E^{(2)}_{k, \bpi^+}(X, Y)  $ by Cauchy inequalities and find
 \be
  \|  E^{(2)}_{k, \bpi^+}(X, Y)     \|_{  \frac12   \bh_k \bbI^\#_k , C_Th_k   \bbI_k  }   \leq 
  \one     e_k^{1-13 \ep} e^{- \hat \ka|Y-X|_{\bom} } e^{-\ka d_M(X) } 
 \ee     
We resum to $Y' \in \cD_k$  and  so  $E^{ (2)}_{k, \bom }  (\La_k)   = \sum_{Y' \cap \Om_{k+1} \neq \emptyset}  E^{(2), \st}_{k, \bpi^+}(Y') $
where
\be   
      E^{(2), \st}_{k, \bpi^+}(Y', \cA,  \tilde Z_k,     \psi^\#_k(\cA) , \Psi_{k, \bom})    =  \sum_{ X \subset \La_k, Y \supset X,    \bar  Y   =  Y' }     \breve     E^{(2)}_{k, \bpi^+}(X, Y, \cA,  \tilde Z_k,     \psi^\#_k(\cA) , \Psi_{k, \bom})   
\ee
As in lemma \ref{outsize1}
\be
 \| E^{(2), \st}_{k, \bpi^+}(Y', \cA,  \tilde Z_k)   \||_{  \frac12    \bh_k \bbI^\#_k , C_Th_k   \bbI_k  }  
\leq     \one     e_k^{1-13 \ep} e^{-(\ka- \ka_0 -1) d_M(Y') } 
\ee
(The  $\bbI^\#_k$ is optional here since the function only depends on $ \psi^\#_k(\cA) $ in $X \subset \La_k \subset \Om_k$.) 
Then we define 
\be  E^{(2)}_{k, \bpi^+}\B(Y', \cA,  \tilde Z_k, \psi^\#_{k, \bom}(\cA ) \B)  
  =  E^{(2), \st }_{k,\bpi^+}\B(Y' , \cA,  \tilde Z_k, \psi^\#_{k, \bom}(\cA ), T_{k, \bom}(\cA)  \psi_{k, \bom}(\cA)\B)
  \ee
and since  $|T'_{k, \bom}(\cA) f |  \leq  C_T  \|f\|_{\infty}$  
\be 
\begin{split}
\|  E^{(2)}_{k, \bpi^+}(Y', \cA,   \tilde Z_k )  \|_{  \frac12  \bh_k \bbI^\#_k }
\leq    \|   E^{(2), \st }_{k,\bpi^+}(Y' , \cA,  \tilde Z_k) \|_{  \frac12   \bh_k \bbI^\#_k ,   C_Th_k   \bbI_k }  
\leq   &  \one    e_k^{1-13 \ep}e^{-(\ka- \ka_0-1) d_M(Y') }  \\
\end{split}
\ee

Finally we split the sum over $Y' \cap \Om_{k+1} \neq \emptyset$  into  $Y' \subset  \La_{k+1}$ and $Y' \cap  ( \Om_{k+1} -\La_{k+1}) \neq \emptyset$ 
as before to generate 
the leading term  $ E^{(2)}_{k}(Y' )$ independent of $\bpi^+$  with the norm  $\| \cdot \|_{  \frac12  \bh_k \bbI^\#_k } =  \| \cdot \|_{  \frac12   \bh_k} $ and the boundary term $ B^{(2)}_{k, \bpi^+}(Y')$. 
   This completes the proof, indeed with a much better power of $e_k$ than claimed. 
\bigskip

Now we turn to the localization of the existing   boundary terms.
First define $\de B^E_{k,  \bpi} $ by 
\be
    B^E_{k,  \bpi} (  \psi_{k, \bom} (\cA + \cZ^{\bullet} _{k,  \bpi^+} )  ) 
=   B^E_{k,  \bpi} (  \psi_{k, \bom} (\cA )  )  + \de B^E_{k,  \bpi} (  \cA,  \tilde Z_k,   \psi_{k, \bom} (\cA ) )     
\ee

\begin{lem}    \label{outsize3}  
\be
\de B^E_{k,  \bpi} \B( X,  \cA^0_{k+1, \bom^+},  \tilde Z_k,   \psi_{k, \bom} ( \cA^0_{k+1, \bom^+} ) \B)   
    = \sum_{X \in \cD_k,  X \# \La_k  }    \de  B^E _{k,  \bpi^+} \B(X,   \cA^0_{k+1, \bom^+},   \tilde Z_k,   \psi_{k, \bom} (\cA^0_{k+1,\bom^+} )  \B)  
    \ee
The $ \de  B^E_{k,  \bpi^+} (X )$
  are the restrictions of functions  $\de  B^E _{k,  \bpi^+} (X,   \cA,   \tilde Z_k, \psi_{k, \bom} (\cA )  ) $ depending on the indicated fields in $X$,  
   analytic in   $\cA \in \frac12 \tilde \cR_k, |\tilde Z_k| \leq  e_k^{-\frac14 \ep} $
  and satisfying  there
 \be  \label{slurp2}
 \| \de B^E _{k,  \bpi^+} (X,   \cA,   \tilde Z_k )  \|_{\frac12 h_k\bbI_k}  \leq     \one e_k^{\frac14 - 5 \ep} e^{- (\ka- \ka_0 -1)  d_M(X)} 
\ee
 \end{lem} 
\bigskip

 The proof follows the proof of lemma \ref{outsize2}.   The main difference is that our initial expansion $  \de  B^E_{k,  \bpi} = \sum_X \de \tilde B^E_{k,  \bpi} (X)$
 is over $X \# \La_k$ (that is $X \cap \La_k \neq \emptyset, X \cap \La_k^c \neq \emptyset) $ rather than $X \subset \La_k$.
 \bigskip

 Now  consider    $B_{k, \bpi} = \sum_j   B^{(k)}_{j, \bpi} $.
and define $\de  B^{(k)}_{j, \bpi}$ by 
 \be
 \begin{split}
&B^{(k)}_{j, \bpi}
 \B( \cA+ \cZ,Z_{k, \bpi}, \psi^\#_{k, \bom}(\cA+ \cZ)  ,W_{k, \bpi}  \B)  \\
  = &
B^{(k)}_{j, \bpi} \B(  \cA,Z_{k, \bpi}, \psi^\#_{k, \bom}(\cA)  ,W_{k,\bpi}  \B)  
+ \de  B^{(k)}_{j, \bpi} \B(  \cA, \cZ, Z_{k, \bpi}, \psi^\#_{k, \bom}(\cA)  , \Psi_{k, \bom}, W_{k,\bpi}  \B)  \\
\end{split} 
\ee   
\bigskip

 \begin{lem}  \label{outsize4}  For   $1 \leq j \leq k$
 \be 
 \begin{split}
&
 \de  B^{(k)}_{j, \bpi}\B ( \cA^0_{k+1,\bom^+ }, \cZ^{\bullet} _{k, \bpi^+}, Z_{k, \bpi}, \psi^\#_{k, \bom}(\cA^0_{k+1,\bom^+ }), \Psi_{k, \bom},  W_{k,\bpi} \B) \\
&
 = \sum_{X \in \cD_k, X \cap \Om_{k+1} \neq \emptyset,  X \cap \La_j^c \neq  \emptyset } \de  B^{(k)}_{j,  \bpi^+} (X, \cA^0_{k+1,\bom^+ }, 
 Z_{k, \bpi}, \tilde Z_k,\psi^\#_{k, \bom}(\cA^0_{k+1,\bom^+ }),   W_{k,\bpi})\\
\end{split}
 \ee
 where the $ \de  B^{(k)}_{j,  \bpi^+} (X)$ are restrictions of functions $ \de  B^{(k)}_{j,  \bpi^+} (X, \cA,  Z_{k, \bpi}, \tilde Z_k, \psi^\#_{k, \bom}(\cA),  W_{k,\bpi})$
which depend on the indicated fields only in $X$,  are analytic in  $\cA \in  (1- e_k^{\ep} )  \tilde  \cR_{k, \bom}, , \ |\tilde Z_k| \leq e_k^{-\frac14 \ep}$ 
and  $ |\tilde Z_j|  \leq L^{\frac12(k-j)}$  on $\Om_{j+1}- \La_{j+1}$  and satisfy   there
\be
 \label{sordid2}
\| \de  B^{(k)}_{j,  \bpi^+}  (X,\cA,  Z_{k, \bpi},\tilde Z_k) \|_{\frac12 \bh_k \bbI^\#_k,  L\bbI_k}  
 \leq  
\one  e_k^{\frac14 -5\ep}e^{-(k-j)}  e^{- (\ka - \ka_0 -2) d_{M}(X)  }
\ee
  and so  
   \be
\| \sum_{j=1}^k      \de B^{(k)}_{j, \bpi} (X) \|_{\frac12 \bh_k \bbI^\#_k,L \bbI_k}  
  \leq  \one   e_k ^{\frac14-5\ep} e^{ -( \ka - \ka_0 -2)  d_{M}(X)}
\ee
\end{lem}
 \bigskip

\pr For the proof drop  the superscript $k$  in   $B^{(k)}_{j, \bpi} $. Also we 
suppress $Z_{k, \bpi}$ and $W_{k,\bpi}$ from the notation since  they are spectators throughout the proof. 
The proof combines elements of the previous two  lemmas.     

Now $B_{j, \bpi}$ has a local expansion   $B_{j, \bpi} = \sum_X B_{j, \bpi}(X)$ and this generates the expansion   
\be 
\de  B_{j, \bpi}  =  \sum_{X \in \cD_j, X \cap (\La_{j-1} - \La_j) \neq \emptyset }  \de \tilde B_{j, \bpi} (X)
 \ee
 where
    \be \label{bisquit}
\begin{split}
 \de  \tilde  B_{j, \bpi} \B(  X, \cA, \cZ,  \psi^\#_{k, \bom}(\cA) , \Psi_{k, \bom} \B)  & =    B_{j, \bpi} \B(  X, \cA + \cZ,  \psi^\#_{k, \bom}(\cA+ \cZ)  \B) 
-B_{j, \bpi} \B(  X, \cA,  \psi^\#_{k, \bom}(\cA)  \B) 
\\
& =  \int_0^1  dt  \frac{d}{dt}   B_{j, \bpi}
 \B( X,\cA+ t\cZ, \psi^\#_{k, \bom}(\cA+ t  \cZ)\B)  \\
\end{split} 
\ee

Specialize to  $\cZ =\cZ^{\bullet} _{k,  \bpi^+}$
Now  as in part (\textbf{C.)} of lemma \ref{outsize1} we evaluate the derivative as a derivative in $\cA$ gaining an expression with 
$\cZ^{\bullet} _{k, \bom^+}$ in a linear position.    Replace this $\cZ^{\bullet} _{k,  \bpi^+}$ by a sum over  $\cZ^{\bullet} _{k,  \bpi^+}(\sq_0)$
and identify the result as a derivative in a new parameter $u$ which we express as a contour integral.
The effect is to replace  $t\cZ^{\bullet} _{k,  \bpi^+}$ by 
\be
\cZ(t,u)  \equiv  t\cZ^{\bullet} _{k,  \bpi^+}+ u \cZ^{\bullet} _{k, \bpi^+}(\sq_0)
\ee
and we have
$ \de  \tilde  B_{j, \bpi} ( X) = \sum_{\sq_0 \subset \Om_{k+1}}   \de  \tilde B_{j,  \bpi^+} ( X, \sq_0)$
where  
\be \label{dumdum}
\begin{split}
& \de  \tilde  B_{j,  \bpi^+} ( X, \sq_0,\cA, \tilde Z_k, \psi^\#_{k,\bom}(\cA),
  \Psi_{k, \bom} )  \\
= & 
 \int_0^1  \ dt
\frac{1} { 2\pi i} \int_{|u| =e_k^{-\frac34+ \frac92\ep}  } \frac{ du}{u^2}  B_{j, \bpi} \B( X,\cA+\cZ(t,u),  \psi^\#_{k, \bom}(\cA+\cZ(t,u)) \B) \\
= &   
 \int_0^1  \ dt
\frac{1} { 2\pi i} \int_{|u| =e_k^{-\frac34+ \frac92\ep}}   \frac{ du}{u^2} B_{j, \bpi} \B( X,\cA+\cZ(t,u),  \psi^\#_{k, \bom}(\cA) 
+ \cJ^\#_{k, \bom}( \cA, \cZ(t,u)) \Psi_{k, \bom}\B) \\
\end{split}
\ee
This is not local  in $ \cA, \tilde Z_k,  \Psi_{k, \bom}$.    Note that for $0 \leq t \leq 1, |u| \leq e_k^{-\frac34 + \frac92 \ep}$ we have  as before that $\cZ(t, u) \in e_k^{\ep}   \cR_{k, \bom}$
so we are within the domain of analyticity.   In fact $\cZ(t, u) \in \frac12 e_k^{\ep}  \cR_{k, \bom}$ is still true and we can write 
\be
\cH_{k,\bom} (\cA +  \cZ(t, u)) - \cH_{k,\bom}( \cA ) = \frac{1} { 2 \pi i} \int_{|z| = 2 }\frac{ dz} { z(z-1)}  \cH_{k, \bom}(\cA +  z\cZ(t, u))
\ee
As in (\ref{log2}) this leads to  the bound
  $ |  \cJ^\#_{k, \bom} ( \cA, \cZ )\bbI_k   f  | \leq  C\  \bbI^\#_k    \|f\|_{\infty}$.  Then from  our  assumed estimate (\ref{sordid}) on  $B_{j, \bpi}(X) $
we have as in (\ref{lingo})
 \be \label{boxcars}
 \begin{split}
&
 \|  B_{j, \bpi} \B( X,\cA+\cZ(t,u),  \psi^\#_{k, \bom}(\cA) 
+ \cJ^\#_{k, \bom}( \cA, \cZ(t,u)) \Psi_{k, \bom}\B)\|_{ \frac12 \bh_k  \bbI^\#_k , C_T \bbI_k, L \bbI_k }  \\
&
 \leq   \|  B_{j, \bpi} ( X,\cA+\cZ(t,u) ) \| _{\frac12 \bh_k  \bbI^\#_k +   CC_T \bbI^\#_k, \ L \bbI_k  } \\
 &
   \leq   \|  B_{j, \bpi} ( X,\cA+\cZ(t,u)  \| _{\bh_k  \bbI^\#_k , L\bbI_k}  \leq    \one   e_k^{\frac 14 - 8 \ep}  e^{- \ka d_{L^{-(k-j)}M}(X)  }\\
\end{split}  
 \ee
 Here the weighting  $( \frac12 \bh_j  \bbI^\#_k , C_T \bbI_k,  \bbI_k ) $  refers to 
$\psi^\#_{k, \bom}(\cA)  , \Psi_{k, \bom}, W_{k,\bpi}  $ respectively and we used $CC_T\bbI^\#_k \leq \frac12 \bh_j \bbI^\#_k$.
Use  this  bound in  (\ref{dumdum}) and find
 \be \label{bisquit6}
 \begin{split}
& \|  \de \tilde B_{j,  \bpi^+} \B( X,  \sq_0 ,\cA, \tilde Z_k \B) \|_{ \frac12 \bh_k  \bbI^\#_k , C_T \bbI_k, L \bbI_k }
 \leq   \one  e_k^{1 - \frac72 \ep}   e^{- \ka d_{L^{-(k-j)}M}(X)  }
\end{split}  
 \ee

 We need to localize this in $ \cA, \tilde Z_k, \Psi_k$. As in lemma \ref{outsize1} we 
 replace the fields  $\cZ^{\bullet} _{k,  \bpi^+}, \cZ^{\bullet} _{k,  \bpi^+}(\sq_0)$
by  the  weakened version $\cZ^{\bullet}_{k,  \bpi^+}(s) , \cZ^{\bullet} _{k, \bpi^+}(s,\sq_0)$, replace $\cZ(t,u)$ by a weakened version $\cZ(s,t,u)$,   and replace
 $\cJ_{k, \bom}(\cA,\cZ(t,u))$ by   $\cJ_{k, \bom}(s,\cA,\cZ(s,t,u))$. 
This yields  instead of  (\ref{dumdum})
\be \label{dumdum2}
\begin{split}
& \de  \tilde  B_{j, \bpi^+} (s, X, \sq_0,\cA, \tilde Z_k, \psi^\#_{k,\bom}(\cA),  \Psi_{k, \bom} )  \\
= &   
 \int_0^1  \ dt
\frac{1} { 2\pi i} \int_{|u| =e_k^{-\frac34+ \frac92 \ep} }   \frac{ du}{u^2} B_{j, \bpi} \B( X,\cA+\cZ(s,t,u),  \psi^\#_{k, \bom}(\cA) 
+ \cJ^\#_{k, \bom}( \cA, \cZ(s, t,u)) \Psi_{k, \bom}\B) \\
\end{split}
\ee
This   still satisfies a bound of the form  (\ref{boxcars}) even for $|s_{\sq}| \leq M^{-\frac12}$. 
Now expand around $s=1$ outside the fixed cube $\sq_0$.  
Then  $ \de \tilde   B_{j,  \bpi^+} (X, \sq_0)=   \sum_{ Y \supset \sq_0} \de    \breve  B_{j,  \bpi^+}(X, Y, \sq_0  )$   with the sum over $Y \in \cD_{k,\bom}$
and
 \be  
\begin{split}
& \de    \breve  B_{j,  \bpi^+}(X, Y, \sq_0   ) =
  \int   ds_{Y- \sq_0}  
 \frac { \pa  }{ \pa s_{Y-\sq_0}}   \left[  \de  \tilde B_{j ,  \bpi^+}(s, X, \sq_0)  \right]_{s_{Y^c} = 0, s_{\sq_0}=1}\\
 \end{split}
 \ee
 Now  $
  \de    \breve  B_{j,  \bpi^+}(X, Y, \sq_0  )=  \de    \breve  B_{j,  \bpi^+}\B(X, Y,  \sq_0,   \cA, \tilde Z_k,  \psi^\#_{k, \bom}(\cA)  , \Psi_{k, \bom}    \B )
 $
 depends on the indicated fields only in $X \cup Y$.   Furthermore we  must have $X \cap Y \neq \emptyset$. 
 Each derivative   gives a factor $M^{-\frac12} \leq e^{- \hat \ka}$ and we have 
\be
\|    \de    \breve  B_{j,  \bpi^+}(X, Y, \sq_0, \cA, \tilde Z_k  )\|_{ \frac12 \bh_k  \bbI^\#_k , C_T \bbI_k, L \bbI_k }\leq 
 \one    e_k^{1 - \frac72 \ep}  e^{- \hat \ka|Y |_{\bom}  }  e^{- \ka d_{L^{-(k-j)}M}(X)  }
\ee

At this point we have
\be   \label{kumquat}
 \de B_{j, \bpi}  =   \sum_{X \in \cD_j,  X \cap (\La_{j-1} - \La_j) \neq \emptyset } \sum_{\sq_0 \subset \Om_{k+1}} \ \  
 \sum_{ Y \supset \sq_0, X \cap Y \neq \emptyset}   \     \de    \breve  B_{j,  \bpi^+}(X, Y, \sq_0  )
  \ee 
Next let  $\bar Y$ be the smallest element of $\cD_k$ containing $Y$.   We classify the terms in the last   sum by the $Y' \equiv \bar Y \cup \bar X $ they generate.
Then 
\be 
 \de  B_{j, \bpi}  = \sum_{ \sq_0 \subset \Om_{k+1}} \sum_{Y' \supset  \sq_0 } \de B^{\st} _{j,  \bpi^+}(Y', \sq_0)
\ee
where
\be
 \de B^{\st} _{j,  \bpi^+}(Y', \sq_0)   = \sum_{ \bar X \cup \bar Y = Y' ,  X \cap Y \neq \emptyset   }\ \  1_{ Y \supset \sq_0} \ 
\ 1_{ X \cap (\La_{j-1}- \La_j) \neq \emptyset } \ \   \de    \breve  B_{j,  \bpi^+}(X, Y, \sq_0) 
\ee
and this vanishes unless $Y'  \cap \Om_{k+1} \neq \emptyset$ and $Y' \cap   (\Om_j - \La_j) \neq \emptyset $. 
We have the estimate
\be \label{pippin}
\begin{split}
&\| \de B^{\st}_{j,  \bpi^+}(Y', \sq_0, \cA, \tilde Z_k) \|_{ \frac12 \bh_k  \bbI^\#_k , C_T \bbI_k, L \bbI_k } \\ 
&  \leq   \one    e_k^{1 - \frac72 \ep}   \sum_{ \bar X \cup \bar Y = Y' ,  X \cap Y \neq \emptyset   }\ \  1_{ Y \supset \sq_0}  1_{ X \cap (\La_{j-1} - \La_j) \neq \emptyset }
 \  e^{- \hat \ka |Y|_{\bom}  }  e^{- \ka d_{L^{-(k-j)}M}(X)  }  \\
 \end{split}   
 \ee
Next note that  $d_{L^{-(k-j)}M}(X) \geq L^{k-j} d_M (\bar X) \geq d_M(\bar X)$ and so 
\be
|Y|_{\bom} +  d_{L^{-(k-j)}M}(X)   \geq  |\bar  Y|_M +  d_{M}(\bar X)   \geq  d_M(Y')
\ee
Hence we can extract a factor $ e^{- (\ka - \ka_0  - 1) d_M(Y')} $  leaving decay factors $   e^{- (\hat \ka -\ka_0 - 1)  |Y|_{\bom}  } \leq e^{- (\ka_0+2)  |Y|_{\bom}  }$ and  $  e^{- (\ka_0 +1) d_{L^{-(k-j)}M}(X)  } $.
Also since $X \cap Y \neq \emptyset $ there must be an $L^{-(k-i)}M$ cube $\sq$ in some $ \de \Om_i$ in both of them  ($j \leq i \leq k$). 
This means we can replace the restriction $X \cap Y \neq \emptyset$ by   conditions $X \supset \sq$ and $Y \supset \sq$ if we also sum over $i$
and $\sq \subset \de \Om_i$.  
 Also drop the 
condition $Y' = \bar Y \cup  \bar X$
and  we have 
\be \label{cunning}
\begin{split}
&\| \de B^{\st}_{j,  \bpi^+}(Y', \sq_0, \cA, \tilde Z_k) \|_{ \frac12 \bh_k \bbI^\#_k , C_T \bbI_k, L \bbI_k } 
 \leq     \one     e_k^{1 - \frac72 \ep}  e^{- (\ka - \ka_0 -1 ) d_M(Y')}  \\
&
 \sum_{i=j}^k  \sum_{\sq \subset \de \Om_i}
  \sum_{   Y \supset \sq, Y \supset \sq_0 } e^{-(\ka_0 +2)|Y|_{\bom}  }
  \sum_{  X \supset \sq, X \cap (\La_{j-1} - \La_j) \neq \emptyset }  e^{-( \ka_0 +1)  d_{L^{-(k-j)}M}(X)  }\\
 \end{split}
 \ee

We estimate the sum over $X$.  If $i=j$ then this is $\one$ by a standard estimate.  If $i>j$ then $ \La^c_j \subset \La^c_{i-1}$. 
and since $X$ has points in both $\La_j^c$ and $\Om_i$ 
\be
\diam (X) \geq d(\La_j^c, \Om_i) \geq  d( \La^c_{i-1}, \Om_i) \geq 5L^{-(k-i)}M
\ee
On the other hand 
\be  
L^{-(k-j)}Md_{L^{-(k-j)}M}(X)  + 2 L^{-(k-j)}M \geq   \diam(X)
\ee
These combine to give 
\be  
d_{L^{-(k-j)}M}(X) + 2  \geq    M^{-1}L^{k-j} \diam(X)  \geq  5L^{(i-j))}
\ee
hence we can extract a factor $e^{- 5L^{(i-j)}}$ from the exponential $e^{- (\ka_0 + 1) d_{L^{-(k-j)}M}(X)  }$. Now drop the condition
$X \cap (\La_{j-1}- \La_j) \neq \emptyset$  and estimate. 
 \be
 \sum_{ X \supset \sq} e^{- \ka_0 d_{L^{-(k-j)}M}(X)  }
\leq  \one | \sq|_{L^{-j} M}
= \one  L^{3(i-j) }
\ee
Now use $ L^{3(i-j)}e^{- 5L^{(i-j)}} \leq   e^{- L^{(i-j)}}  \leq e^{-(i-j)}$  and obtain
\be \label{cunning2}
\begin{split}
&\| \de B^{\st}_{j,  \bpi^+}(Y', \sq_0, \cA, \tilde Z_k) \|_{ \frac12 \bh_k  \bbI^\#_k , C_T \bbI_k, L \bbI_k } \\ &  \leq     \one      e_k^{1 - \frac72 \ep}  e^{- (\ka - \ka_0 -1 ) d_M(Y')}  \sum_{i=j}^k e^{-(i-j)} \sum_{\sq \subset \de \Om_i} 
 \sum_{   Y \supset \sq, Y \supset \sq_0 } e^{-(\ka_0 +2)|Y|_{\bom}  }
 \\
 \end{split}
 \ee

Note that  since $Y$ intersects $\de \Om_i$ and $ \Om_{k+1} $ it must have at least $(k-i)$ cubes and we use $e^{- | Y|_{\bom} } \leq e^{-(k-i)}$.
Then identify  $e^{-(k-i)}  e^{-(i-j)}  = e^{-(k-j)}$ which  comes outside the sum over $i$.
  Next bring the sum over $i, \sq$ inside the sum over $Y$ where it becomes a sum over $ \sq \subset (Y \cap \de \Om_i)$
But  
\be
\sum_{i=j}^k  \sum_{\sq \subset (Y \cap \de \Om_i) } =  \sum_{i=j}^k  |Y \cap \de \Om_i|_{L^{-(k-i)}M}  \leq |Y|_{\bom}
\ee
and this can be absorbed by a factor $ e^{-|Y|_{\bom}} $.  The
sum over $Y$ is estimated   by
 $
   \sum_{ Y \supset \sq_0}   e^{-\ka_0|Y|_{\bom}  }  \leq  \one
 $
 and we have  
\be 
\label{cunning3}
\| \de B^{\st}_{j,  \bpi^+}(Y', \sq_0, \cA, \tilde Z_k) \|_{ \frac12 \bh_k  \bbI^\#_k , C_T \bbI_k, L \bbI_k } \leq 
  \one    e_k^{1 - \frac72 \ep} e^{-(k-j)}e^{- (\ka - \ka_0 -1 ) d_M(Y')}  
   \ee

Now (\ref{kumquat}) can be written $ \de B^{\st}_{j, \bpi} = \sum_{Y'} \de B^{\st}_{j,  \bpi^+}(Y') $
where
\be 
 \de B^{\st}_{j, \bpi^+}(Y') = \sum_{\sq_0 \subset (Y' \cap \Om_{k+1}) }\de B^{\st}_{j, \bpi^+}(Y', \sq_0 )
 \ee
 The sum over $\sq_0$ is bounded by $|Y' \cap \Om_{k+1} |_M \leq   |Y'|_M \leq \one(d_M(Y') +1)$ and can be absorbed by the exponential
 and we have
   \be 
\label{cunning4}
\| \de B^{\st}_{j,  \bpi^+}(Y',  \cA, \tilde Z_k) \|_{ \frac12 \bh_k  \bbI^\#_k , C_T \bbI_k, L \bbI_k }   \leq 
  \one      e_k^{1 - \frac72 \ep} e^{-(k-j)} e^{- (\ka - \ka_0 -2 ) d_M(Y')}  
   \ee
Finally  we make the change $\Psi_{k, \bom} \to \psi_{k,\bom} (\cA)$ as before
defining 
\be 
 \de  B_{j,  \bpi^+}\B(Y', \cA,  \tilde Z_k, \psi^\#_{k, \bom}(\cA) \B)
= \de B^{\st }_{k,  \bpi^+}\B(Y', \cA,  \tilde Z_k, \psi^\#_{k, \bom}(\cA), T_{k, \bom}(\cA) \psi_{k, \bom}(\cA) \B)
\ee
and then 
\be \label{kumquat2}
\| \de  B_{j,  \bpi^+}(Y' ) \|_{\frac12 \bh_k  \bbI^\#_k , L \bbI_k}
\leq   \| \de B^{\st} _{k ,  \bpi^+}(Y')\|_{ \frac12 \bh_k  \bbI^\#_k , C_T \bbI_k, L \bbI_k }
\leq  \one     e_k^{1 - \frac72 \ep} e^{-(k-j)}e^{ - (\ka - \ka_0 -2) d_M(Y') }
\ee
This the desired bound, indeed with a higher power of $e_k$. 
\bigskip

  Next we study the tiny terms  $R^{(\leq 3) }_{k, \bpi^+} =  R^{(1) }_{k, \bpi^+} + R^{(2)}_{k, \bom^+}  + R^{(3)}_{k, \bom^+}  $

 \begin{lem}  \label{outsize5} {  \  }
 \begin{enumerate}
 \item  
 $R^{(1)}_{k, \bpi^+ }  = \sum_{X \cap (\Om_{k+1} - \La_{k+1})  \neq \emptyset}  R^{(1 )}_{k, \bpi^+ } (X) $
 where $R^{(1 )}_{k, \bpi^+} (X, \cA^0_{k+1, \bom^+ }, \tilde Z_k) $ is the restriction of a function    $R^{(1 )}_{k, \bpi^+ } (X, \cA, \tilde Z_k) $   analytic in  $\cA \in \frac12 \tilde  \cR_{k,\bom},  \  |\tilde Z_k| \leq e_k^{-\frac12 \ep}$   and  satisfies
 \be
| R^{(1)}_{k, \bpi^+ } (X) | \leq Ce^{-r_k} e^{- \ka d_M(X) }
\ee
\item  
 $R^{(2)}_{k, \bom^+ }  = \sum_{X \cap \Om_{k+1}  \neq \emptyset}  R^{(2 )}_{k, \bom^+ } (X) $
 where $R^{(2)}_{k, \bom^+ } (X,  \tilde Z_k) $ is   analytic in  $  |\tilde Z_k| \leq e_k^{-\frac12 \ep}$   and  satisfies  
 \be
| R^{(2)}_{k, \bom^+ } (X) | \leq Ce^{-r_k} e^{- \ka d_M(X) } \ee
If $X \subset \La_{k+1} $ then   $ R^{(2)}_{k, \bom^+ } (X) \equiv  R^{(2)}_{k} (X)$ is independent of $\bom^+ $.
\item 
 $ R^{(3)}_{k, \bom^+ }  = \sum_{\sq_0 \subset \Om_{k+1} } R^{(3)}_{k, \bom^+ } (\sq_0)$
 where for $M$-cubes $\sq_0$:
 $
| R^{(3)}_{k, \bom^+ } (\sq_0) | \leq Ce^{-r_k} 
$
\end{enumerate} 
\end{lem}  
 \bigskip
 
 \pr 
 Start with  $ R^{(2)}_{k, \bom^+}$ which by (\ref{c2})  is given by 
 \be \label{cindy1}
   R^{(2)}_{k, \bom^+}= 
   \blan    C^{\frac12} _{k, \bom^+}  \tilde Z_k,   \tilde \De_{k, \bom^+ } \de C^{\frac12} _{k, \bom^+} \tilde Z_k \bran
 + \frac12   \blan   \de C^{\frac12} _{k, \bom^+}  \tilde Z_k,   \tilde \De_{k, \bom^+ } \de C^{\frac12} _{k, \bom^+} \tilde Z_k \bran 
 \ee
The first term  can be written   with $\cZ_{k, \bom^+}  = \cH_{k, \bom} C C^{\frac12} _{k, \bom^+} \tilde Z_k$ and
$ \cZ' _{k, \bom^+ }  = \cH_{k, \bom} C \de C^{\frac12} _{k, \bom^+} \tilde Z_k$ as
\be \label{cindy2} 
 \blan  d \cZ_{k, \bom^+},  d \cZ'_{k, \bom^+ } \bran
= 
\sum_{ \sq_0 \subset \Om_{k+1} }  \blan   d \cZ_{k, \bom^+} (\sq_0)  , d \cZ'_{k, \bom^+}    \bran
\ee
Here we have localized by  introducing $ d \cZ_{k, \bom^+} (\sq_0)  = \cH_{k, \bom}1_{\sq_0}  C C^{\frac12} _{k, \bom^+} \tilde Z_k$. 
As in lemma \ref{outsize1} we  use random walk expansions to introduce weakened versions $\cZ_{k, \bom^+} (s, \sq_0),\cZ'_{k, \bom^+ } (s)  $ and expand around $s_{\sq} =1$ in
$\sq_0 ^c$.  Following lemma  \ref{outsize1}   we find   (\ref{cindy2}) can be written 
\be
\sum_{ \sq_0 \Om_{k+1} } \sum_{ X \supset \sq_0}   R'_{k, \bom^+}(X, \sq_0  )
 = \sum_{ X  \cap  \Om_{k+1}  \neq 0 }    R'_{k, \bom^+ }(X)
 \ee
 Then   $   R'_{k, \bom^+ }(X)  $ only depends on $\tilde Z_k$ in $X$
and   satisfies  the announced  bound with    $ \de C^{\frac12} _{k, \bom^+}$ supplying the tiny factor $e^{-r_k}$  by (\ref{pipsqueak}). 
Similarly the second  term  $\frac12 \| d \cZ' _{k, \bom^+}  \|^2$ in (\ref{cindy1})  has a polymer expansion with $  R''_{k, \bom^+ }(X)$.  The result follows
with $ R^{(2)} = R' + R''$.  
If $X \subset \La_{k+1} $ then  only random walks which  stay inside $\La_{k+1} $ contribute.  These are independent of the
history  so  $R^{(2)} _{k, \bom^+}(X) $ is independent of $\bom^+ $. 

The term $R^{(1)} _{k, \bpi^+} $ defined just below (\ref{alibaba}) 
is given by 
 \be
 \begin{split}
   R^{(1)}_{k, \bpi^+ }=  &
   \blan   C C^{\frac12} _{k, \bom^+}  \tilde Z_k,   \De_{k, \bom^+ } \  g  \fH_{k, \bpi^+}  \bran
 + \frac12   \blan    \  g  \fH_{k, \bpi^+} ,   \ \De_{k, \bom^+ } \  g  \fH_{k, \bpi^+}  \bran  \\
 = &   \blan  d \cZ_{k,\bom},  d \cH _{k, \bom} \  g  \fH_{k, \bpi^+}  \bran
 + \frac12   \blan     d \cH _{k, \bom}\   g  \fH_{k, \bpi^+} ,    d \cH _{k, \bom}\   g  \fH_{k, \bpi^+} \bran  \\ 
\end{split} 
 \ee
 This is treated in the same way as  $ R^{(2)}_{k, \bom^+ }$  
 starting by replacing    $\fH_{k, \bpi^+} $ by $\sum_{\sq_0 \subset S_{k+1}} 1_{\sq_0} \fH_{k, \bpi^+} $.
 The  $\fH_{k, \bpi^+} $ supplies the tiny factor $e^{- r_k}$.    We end with a sum over $X  \cap S_{k+1} \neq \emptyset$
which is included in the restriction $X \cap (\Om_{k+1} - \La_{k+1})   \neq \emptyset$

Finally  consider  $  R^{(3)}_{k, \bom^+ } $which by (\ref{c3}) is given by 
\be 
 R^{(3)}_{k, \bom^+ }=  \log \det \B( C^{\frac12, \loc} _{k, \bom^+}  C^{-\frac12 } _{k, \bom^+}  \B) 
 =  \log  \det \B( I  - \de C^{\frac12} _{k, \bom^+}  C^{-\frac12 } _{k, \bom^+ } \B) 
  =  - \sum_{n=1}^{\infty}\frac{1}{n} \tr 
  ( \de C_{k, \bom^+} ^{ \frac12}   C_{k, \bom^+} ^{- \frac12}  )^n   
\ee
We can write this as    $ R^{(3)}_{k, \bom^+ } = \sum_{\sq_0 \subset \Om_{k+1} } R^{(3)}_{k, \bom^+  } (\sq_0) $
where
\be
  R^{(3)}_{k, \bom^+ }(\sq_0) =  -  \sum_{n=1}^{\infty}\frac{1}{n} \tr \B(  1_{\sq_0} 
  ( \de C_{k, \bom^+} ^{ \frac12}   C_{k, \bom^+} ^{- \frac12}  )^n  \B) 
 \ee
 Now $ C_{k, \bom^+} ^{- \frac12}  = ( C^T[\De_{k, \bom}]_{\Om_{k+1}} C )  C_{k, \bom^+} ^{ \frac12}  $
 has exponential decay since both factors have exponential decay.   The same is true for $ ( \de C_{k, \bom^+} ^{ \frac12}   C_{k, \bom^+} ^{- \frac12}  )^n $
 and it supplies a factor  $( e^{-r_k} )^n$.  Thus the series converges and we get the bound  $| R^{(3)}_{k, \bom^+ } (\sq_0) | \leq Ce^{-r_k} $.
\bigskip

 \noindent
\textbf{Summary:} 
 We collect all the changes. 
 For $i=1,2,3$  write
 $  R^{(i )}   =  R^{( i  )a}     + R^{(i  )b }$.
 The first term is the sum over $X \subset \La_{k+1} $  and the second term is the rest, so  $ R^{( 1  )a}_{ k, \bpi^+ }  =0$.
 Then define
   \be 
   E^{( \leq 3)}_k(\La_{k+1} )= \sum_{\al =1}^3 E^{( \al )}_k(\La_{k+1} ) +  R^{( 2 )a}_k   
   \ee
 Then   $E^{( \leq 3)}_k(\La_{k+1} )
 = \sum_{X \subset \La_{k+1} }E_k^{( \leq 3)}(X) $ and   $E_k^{( \leq 3)}(X)$ is analytic in  $\cA \in \frac12 \tilde \cR_k, | \tilde Z_k| \leq e_k^{-\frac14 \ep}$ 
 and satisfies there
\be \label{urgent}
\| E_k^{( \leq 3)}(X, \cA, \tilde Z_k) \|_{ \frac12 \bh_k}
\leq  \one e_k^{\frac14 - 5 \ep} e^{- (\ka - \ka_0 -1) d_M(X)} 
\ee
We collect all  the new boundary terms and get an expression which will eventually contribute to $B_{k+1, \bpi^+}$.
These are 
  $    B^{\bullet}_{k, \bpi^+}  =   B^{\bullet}_{k, \bpi^+} ( \cA,    Z_{k, \bpi}, \tilde Z_k, \psi^\#_{k, \bom}(\cA), W_{k, \bpi} )
$ 
given by
\be 
\label{tootsie2}
 B^{\bullet}_{k, \bpi^+}   =  B^E_{k, \bpi^+}
+ \de  B^E_{k, \bpi^+} +
 \sum_{\al=1}^3 B^{(\al)}_{k, \bpi^+} +  \sum_{j=1}^k  \de   B^{(k)}_{j, \bpi^+} +  R^{( \leq 3) b}_{k , \bpi^+ } 
\ee
We then have  
$
 B^{\bullet}_{k, \bpi^+} =  \sum_X     B^{\bullet}_{k, \bpi^+} (X)
$
where  the sum is over $X \in \cD_k$ such that    $X  \cap ( \Om_{k+1}  - \La_{k+1}) \neq \emptyset $.   A bound on $ B^{\bullet}_{k, \bpi^+} (X)$ follows from  the bounds
of lemmas  \ref{outsize1}, \ref{outsize2}, \ref{outsize3}, \ref{outsize4}, and also from lemma \ref{outsize5} where we use that $Ce^{-r_k}$ is smaller than any 
power of $e_k$. 
  The
 $ B^{\bullet}_{k, \bpi^+} (X)$ are 
 are analytic in  $\cA \in \frac12  \tilde \cR_k,   |Z_{j,\Om_{j+1}- \La_{j+1}}|  \leq L^{\frac12(k-j)}, | \tilde Z_k|\leq e_k^{-\frac12 \ep}$
  and  satisfy there
\be  
\begin{split}
\label{sonnyboy}
\|   B^{\bullet}_{k, \bpi^+} (X, \cA,    Z_{k, \bpi}, \tilde Z_k) \|_{\frac12 \bh_k \bbI^\#_k, L \bbI_k}  &  \leq  \one e_k^{\frac14 - 5 \ep}   e^{- (\ka - \ka_0 -2) d_M(X)} \\
\end{split}
\ee

Now (\ref{rep6}) becomes
\be    \label{rep7} 
 \begin{split}
&  \tilde \rho_{k+1} ( A_{k+1}, \Psi_{k+1}   )  = \sum_{ \bpi^+ }  \sZ'_{k,\bom}(0))   \sZ'_{k, \bom} \ \de \sZ_{k, \bom^+}    
     \\
  &       
     \int      Dm^0_{k+1, \bom^+}  ( A )    Dm^0_{k+1, \bpi^+ }(Z )\      Dm_{k,\bom}(\Psi)   Dm_{k,\bpi}(W) \ D\Psi_k\  d \mu_{I, \La_{k+ 1}} (\tilde Z_k) \   \\
  &
    \de_G\B( \Psi_{k+1} -  Q( \cA ) \Psi_k \B)      
\  \cC^0_{k+1, \bpi^+ } \chi_{k+1}^0(\La_{k+1})    \hat  \chi_k(\La_{k+1} )      \\
&     \exp \B(     -\frac12 \|  d  \cA   \|^2       -    \fS'^{+}_{k, \bom}(\La_k)     +     E'_k(\La_{k} ) 
     +    E^{( \leq  3)}_{k}(\La_{k+1})    +  B_{k,\bpi^+ }  +    B^{\bullet}_{k, \bpi^+}    +   R^{3a}_{k,  \bom^+}     \B) |_{\cA =     \cA^{0} _{k+1, \bom^+} }  \\   
\end{split}
\end{equation}

\subsection{fermion translation} 

In the last expression the Gaussian   approximate   delta function  $  \de_G( \Psi_{k+1} -  Q(\tilde  \cA'  ) \Psi_k ) $ still 
has an arbitrary gauge field $\tilde \cA'$.  We now make the choice  $\tilde \cA' = (0, \cA^0_{k+1, \bom^+})$ on $(\Om^c_{k+1}, \Om_{k+1})$. 
By allowing $\tilde \cA'$ to depend on $\Om_{k+1}$ we have made another modification of the original renormalization group transformation  (namely
different  averaging operators under the sum over $\Om_{k+1}$).  We allow it because  integrals over $\Psi_{k+1}$ are not affected. 
With this choice we have the split 
\be
   \de_G( \Psi_{k+1} -  Q( \tilde   \cA'  ) \Psi_k )= \de_{G, \Om^c_{k+1}}( \Psi_{k+1} -  Q( 0 ) \Psi_k )  \de_{G, \Om_{k+1}}( \Psi_{k+1} -  Q( \cA^0_{k+1, \bom^+} ) \Psi_k )
\ee
Here   by definition
\be \label{deltafunction} 
\de_{G, \Om_{k+1}}( \Psi_{k+1} -  Q( \cA) \Psi_k )
= N_{k+1, \Om_{k+1} }  \exp \B( -   bL^{-1}  \blan  \bPsi_{k+1} -  Q(  -   \cA) \bPsi_k,   \Psi_{k+1} -  Q(  \cA) \Psi_k  \bran _{\Om_{k+1} } 
\B)
\ee

Temporarily drop  the counterterms in   $  \fS'^{+}_{k, \bom}(\La_{k+1})$ leaving just  $ \fS_{k, \bom} $.  With $\cA = \cA^0_{k+1, \bom^+}$ the    $  \de_{G, \Om_{k+1}}( \Psi_{k+1} -  Q( \cA ) \Psi_k )$ and $\exp (  -  \fS_k(\cA ,\Psi_k, \psi_{k, \bom} (\cA    ) )$  taken together form the exponential 
of a quadratic form which  
is  minus 
\be   \label{laugh}
   bL^{-1}  \blan  \bPsi_{k+1} -  Q(  -   \cA) \bPsi_k,   \Psi_{k+1} -  Q(  \cA) \Psi_k  \bran _{\Om_{k+1} } 
+          \fS_{k, \bom}\B( \cA , \Psi_{k, \bom}, \psi_{k, \bom} ( \cA )      \B)
\ee
Now  we diagonalize  this expression    
as   in section  II.D in \cite{Dim20}   by the transformation on  $\Om_{k+1}$
 \be
     \Psi_{k, \Om_{k+1}}   =\Psi^{\crit}_{k, \bom^+}(\cA) + W'_k
 \ee
 where again    $\bom^+ =  (\bom, \Om_{k+1} ) $.
  This induces  the transformation    
 \be
  \psi_{k, \bom} (\cA)   =  \psi^{0}_{k+1, \bom^+}(\cA)  +     \cW_{k, \bom}( \cA)  
 \ee
where    
 \be 
 \begin{split}
   \psi^{0}_{k+1, \bom^+ }(\cA) =   &   \psi_{k,\bom}(\cA,  \Psi^{\crit}_{k, \bom^+}(\cA) )  \\
      \cW_{k, \bom} (\cA)  =  &   \cH_{k, \bom} (\cA)  W'_k   \\
 \end{split}      
 \ee
   For the pair  $\psi^\#_{k,   \bom}(\cA) =   ( \psi_{k,\bom} (\cA)  ,  \de_{\al, \cA} \psi_{k,\bom}(\cA)  )$
we   define    
  \be
  \begin{split}
  \psi^{0,\#}_{k+1, \bom^+ }(\cA) 
  =   & ( \psi^{0}_{k+1,\bom^+ }(\cA)  ,  \de_{\al, \cA} \psi^{0}_{k+1,\bom^+}(\cA)  \B)   \\
 \cW_{k,\bom }^{\#} =    & 
  ( \cW_{k,  \bom}(\cA)  ,  \de_{\al, \cA} \cW_{k, \bom}(\cA)  )\\
  \end{split}
  \ee
   and then  $     \psi^{ \#}_{k, \bom}(\cA)      =       \psi^{0, \#}_{k+1, \bom^+ }(\cA)  +   \cW_{k,\bom }^{\#} (\cA)$.
  By  lemma 4 in \cite{Dim20}       the quadratic form     (\ref{laugh})    becomes   
  \be   
  \begin{split}
     \label{laugh2}
      &    \fS^0_{k+1, \bom^+}\B(\cA, \Psi_{k, \bom^+},  \psi^{0}_{k+1, \bom^+ }(\cA)     \B)  
  +         \blan    \bar W'_k,  \B[ D_{k,\bom }(  \cA)+   bL^{-1} P(\cA) \B]_{\Om_{k+1}  } W'_k  \bran      \\
\end{split}  
\ee

The integral over $\Psi_k$ becomes an integral over $\Psi_{k, \Om^c_{k+1}}$ and $W_k'$. 
We  identify the Gaussian integral    
 \be 
\int [ \cdots ]   \exp \B( -  \blan \bar  W'_k, \B[ D_{k, \bom }(  \cA) +  bL^{-1} P(\cA)\B]_{\Om_{k+1}}  W'_k  \bran  \B)   D W'_k      
   =     \de \sZ_{k, \bom^+} (\cA)   \  \int [ \cdots ]   d \mu_{\Ga}  (W' _k)  
\ee
where
\be  
\Ga  =    \Ga _{k, \bom^+}  (\cA)   =  \B[ D_{k,\bom }(  \cA)+   bL^{-1} P(\cA) \B]_{\Om_{k+1}}  ^{-1}
\ee
and
\be
 \de \sZ_{k, \bom^+} (\cA)  =  \det \B(  \Ga _{k, \bom^+}  (\cA) \B)^{-1}
\ee
Furthermore we make the change of variables  on $\Om_{k+1}$
\be   W'_k  =     \Ga _{k, \bom^+}  (\cA) W_k   \hs    \bar  W'_k  =  \bar  W_k  
\ee
which we  abbreviate   as   $ W' _k  =    \tilde   \Ga _{k, \bom^+}  (\cA) W_k  $.
This changes     $ d \mu_{\Ga } (W' _k) $ to the ultra-local   $ d \mu_{I, \Om_{k+1}}  (W_k )  $
and the translations are now   
     \be
      \begin{split} 
     \Psi_k   =  &  \Psi^{\crit}_{k, \bom^+}(\cA)  +   \tilde   \Ga _{k, \bom^+}  (\cA) W_k   \\
   \psi_{k, \bom} (\cA)   =   &     \psi^{0}_{k+1, \bom^+}(\cA)  +     \cW_{k, \bom}( \cA)   \\
   \end{split}   
 \ee
where we replace $\cW_{k, \bom} (\cA)$ by  
\be
   \cW_{k, \bom^+} (\cA)  =     \cH_{k, \bom} (\cA)   \tilde   \Ga _{k, \bom^+}  (\cA) W_k
\ee
We  split   
$  
 d \mu_{I, \Om_{k+1} }  (W_k )   = d \mu_{I,\Om_{k+1}- \La_{k+1}}  (W_k )    d \mu_{I, \La_{k+1} }  (W_k )  
$
and identify     
\be 
\begin{split}
   Dm^0_{k+1,\bom^+}(\Psi) = &  Dm_{k,\bom}(\Psi) \de_{G, \Om^c_{k+1}}\B( \Psi_{k+1} -  Q( 0 ) \Psi_k \B)   D\Psi_{k, \Om^c_{k+1}} \\
   Dm^0_{k+1,\bpi^+}(W) =  &  Dm_{k,\bpi}(W)  d \mu_{I, \Om_{k+1}- \La_{k+1} }  (W_k )         \\
\end{split}
\ee

With these changes   (\ref{rep7}) becomes  
\be    \label{rep8} 
 \begin{split}
&  \tilde \rho_{k+1} ( A_{k+1}, \Psi_{k+1}   )  
= 
  \sum_{ \bpi^+ }(   N_{k+1, \Om_{k+1} }  \sZ'_{k,\bom}(0)    \de \sZ_{k, \bom^+} (\cA)  )( \sZ'_{k, \bom}  \de \sZ_{k, \bom^+})  \\
& 
  \int  \   Dm^0_{k+1, \bom^+}  ( A )    Dm^0_{k+1, \bpi^+ }(Z) \   Dm^0_{k+1, \bom^+}  ( \Psi)\   Dm^0_{k+1,\bpi^+}(W)  \ 
 d \mu_{I ,\La_{k+1}}  (W_k ) d \mu_{I,\La_{k+1}} (\tilde Z_k)   \\
  &  
   \cC^0_{k+1, \bpi^+ } \chi_{k+1}^0(\La_{k+1}) 
 \hat \chi_k(\La_{k+1} )    \exp \B( -\frac12 \|  d  \cA   \|^2 
     -     \fS^{0}_{k+1, \bom^+}( \cA, \Psi_{k+1,\bom^+}, \psi^0_{k+1, \bom^+} (\cA))    \B)   
  \\
&       
      \exp \B( - m'_k < \bpsi   ,   \psi >_{\La_{k}} 
- \vep_k' \Vol ( \La_{k})  +   E'_k(\La_{k} ) +  E^{( \leq  3)}_{k}(\La_{k+1})    +  B_{k,\bpi^+ }  +    B^{\bullet}_{k, \bpi}      +   R^{(3)a}_{k,  \bom^+}      \B)   \\   
\end{split}
\ee
all evaluated at $ \cA =     \cA^{0} _{k+1, \bom^+}$,  with the last exponential  evaluated  at  $ \psi  = \psi^{0}_{k+1, \bom^+ }(\cA)  +   \cW_{k,\bom^+ }(\cA)  $.

  \subsection{second localization}

We  need to study the effects of the fermion  translation on the various terms in the action.  
We         define  for $\cA \in \frac12   \tilde   \cR_k$   functions   $E_{k}^{(4)},  E_{k} ^{(5)}$  analytic in this domain   by 
 \be  \label{a}
\begin{split}
& E'_k\B(\La_{k}, \cA, \psi^{0, \#}_{k+1, \bom^+}(\cA)  +     \cW^{\#}_{k, \bom^+}( \cA)    \B) \\ 
  &   \hs  =
 E'_k\B(\La_{k+1}, \cA,    \psi^{0, \#}_{k+1, \bom^+}(\cA)    \B)
   +     E^{(4)} _{k, \bom^+ }\B(\La_{k}, \cA,     \psi^{0, \#}_{k+1, \bom^+}(\cA)  ,   \cW^{\#}_{k, \bom^+}( \cA)        \B )\\
&   m'_k \blan  \B( \bpsi^{0}_{k+1, \bom^+}(\cA)  +  \bar   \cW_{k, \bom^+}( \cA) \B)   ,  \B( \psi^{0}_{k+1, \bom^+}(\cA) 
 +     \cW_{k, \bom^+}( \cA) \B)  \bran_{\La_{k}}  + (\vep'_k + \vep^{0}_k ) \Vol(\La_k) \\
  &    =
    m'_k \blan   \bpsi^{0}_{k+1, \bom^+}(\cA)     ,   \psi^{0}_{k+1, \bom^+}(\cA)   \bran_{\La_{k+1}} + (\vep'_k + \vep^{0}_k ) \Vol(\La_{k+1} )
-        E^{(5)}_{k, \bom^+ }\B(   \La_{k}, \cA,      \psi^{0}_{k+1, \bom^+}(\cA) , \cW_{k, \bom^+}( \cA)        \B)\\
\end{split}
\ee 
The  counterterms  here are  reunited with   $  \fS^0_{k+1, \bom^+}$  to give 
 \be 
\begin{split}  \label{miller}
& \fS^{0,+}_{k+1, \bom^+}(\La_{k+1}, \cA, \Psi_{k+1, \bom^+ },  \psi^{0}_{k+1, \bom^+}(\cA) ) \\
& \equiv  \fS^{0}_{k+1, \bom^+}( \cA, \Psi_{k+1, \bom^+},  \psi^{0}_{k+1, \bom^+}(\cA) )  +m'_k \blan   \bpsi^{0}_{k+1, \bom^+}(\cA)     ,   \psi^{0}_{k+1, \bom^+}(\cA)   \bran_{\La_{k+1}} 
+ ( \vep_k'+\vep^0_k) \Vol ( \La_{k+1}) \\
\end{split}
\ee

We  localize the various   polymer functions in the fundamental fermion  fluctuation variable $W_k$.

  \begin{lem}  \label{outsize6}  
For    
$ \cA \in (1-e_k^{\ep} )   \tilde \cR_{k, \bom}$  
\be    \label{serendipity4}
\begin{split}
 E^{(4)}_{k, \bom^+ }\B(\La_k,   \cA,     \psi^{0, \#}_{k+1, \bom^+}(\cA) ,    \cW^{\#}_{k, \bom^+}(\cA)   \B) 
  =  &  \sum_{X \subset \La_{k+1}}    E^{(4)}_{k}\B(X; \cA,     \psi^{0, \#}_{k+1, \bom^+}(\cA) , W_k  \B)  \\
   &   +   \sum_{ X \cap  (\Om_{k+1} - \La_{k+1} ) \neq \emptyset  }   B^{(4)}_{k, \bom^+ }\B(X; \cA,      \psi^{0, \#}_{k+1, \bom^+}(\cA) ,W_k  \B)    \\
\end{split}
\ee
where the  sums are over $X \in \cD_k$ and
\begin{enumerate}
\item
The   $  E^{(4)}_{k, \bom^+ }(X, \cA,        \psi^{0, \#}_{k+1, \bom^+}(\cA)  , W_k) $  depend on the indicated fields only  in $X$ 
 and    have kernels  independent of the history  which   satisfy 
  \be   
  \|  E^{(4)}_{k} (X, \cA ) \| _{ \frac12 \bh_k, L }  \leq   \one   e_k^{\frac12 - 10 \ep}  e^{- ( \ka - \ka_0-1) d_M(X)}
\ee 
\item The   $  B^{(4)}_{k, \bom^+ }(X, \cA,     \psi^\#_{k, \bom}(\cA), W_k) $  depend on the indicated fields only    in $X$,  
  and satisfy
\be    
  \|  B^{(4)}_{k,  \bom^+ } (X,  \cA ) \| _{ \frac12 \bh_k \bbI^\#_k, L}  \leq   \one   e_k^{\frac12 - 10 \ep}  e^{- ( \ka - \ka_0-1) d_M(X)}
\ee
\end{enumerate} 
\end{lem}
\bigskip

\pr  The proof is similar to the proof of lemma \ref{outsize2}. 
  We have $ E^{(4)}_{k, \bom^+}(\La_k) = \sum_{X \subset \La_k} \tilde E^{(4)}_{k,\bom^+}(X)$ where
\be   \label{step1}
\begin{split}
&\tilde  E^{(4)} _{k, \bom^+}\B(X, \cA,     \psi^{0, \#}_{k+1, \bom^+}(\cA )  ,   W_k    \B)  \\
=   &   \frac{1}{2 \pi i}
 \int_{|t|  = e_k^{-\frac14 + 3\ep}   }      \frac{dt}{t(t-1)}  
 E' _{k}\B(X, \cA ,   \psi^{0, \#}_{k+1, \bom^+}(\cA ) + t   \cW^{\#}_{k, \bom^+}( \cA )       \B ) \\
 \end{split} 
\ee
This is not yet local in $W_k$.    The function $E'_{k}(X)$ under the integral is regarded  as a function of 
$\psi^{0, \#}_{k+1, \bom^+}(\cA ) $ and $ W_k $   through   $ \cW_{k, \bom^+} (\cA)  =     \cH_{k, \bom} (\cA)   \tilde   \Ga _{k, \bom^+}  (\cA) W_k$.
From \cite{Dim20} we have the estimates  in $\Om_k$  
\be \label{cinnamon2}
  | \cH_{k, \bom} (\cA)   \tilde   \Ga _{k, \bom^+}  (\cA) f |, \  \| \de_{\al, \cA}\cH_{k, \bom} (\cA)   \tilde   \Ga _{k, \bom^+}  (\cA) f \| \leq C \|f\|_{\infty}
 \ee
 Indeed the estimates on $\cH_{k,\bom} (\cA)$  are given in (146) in \cite{Dim20}. The estimate on $\tilde   \Ga _{k, \bom^+}  (\cA)$
follows from the estimate (96) in \cite{Dim20} on $S_{k,\bom} (\cA)$ or $\cS^0_{k+1, \bom^+}(\cA)$ and the representation from  (A8) in \cite{Dim20}  of  $ \Ga _{k, \bom^+}  (\cA)$ in terms of $\cS^0_{k+1, \bom^+}(\cA)$.
Then we have   for $|t| \leq e_k^{-\frac14 + 3\ep} $
\be
\begin{split}
& \|  E' _{k}\B(X, \cA ,   \psi^{0, \#}_{k+1, \bom^+}(\cA ) + t    \cW^{\#}_{k, \bom^+}( \cA )   \B)\|_{ \frac12 \bh_k, L} \\
& \leq   \| E' _{k}(X, \cA  ) \|_{ \frac12 \bh_k+ |t| C (L,L)  } \leq  \one   \|E' (X, \cA)\|_{\bh_k} \leq \one e_k^{\frac14-7\ep}  e^{-\ka d_M(X)} \\
\end{split} 
\ee
Here we used (A44) from \cite{Dim15b} and
\be \label{less}
 |t| C (L,L) \leq  e_k^{-\frac14 + 3\ep} C(L,L)\  \leq  e_k^{-\frac14 + 2\ep}(1,1) \leq  \frac12 \bh_k
\ee
  and the estimate (\ref{retro}) on $E' (X, \cA)$.
 Then   (\ref{step1}) gives
 \be \label{superman}
\|\tilde   E^{(4)}_{k, \bom^+ }(X, \cA)\|_{\frac12  \bh_k, L}  \leq  \one  e_k^{\frac12 - 10\ep}  e^{-\ka d_M(X)}
\ee

Next     we    introduce  weakened fluctuation    fields.
As explained in \cite{Dim20} the random walk expansions for $S_{k,\bom} (\cA), \cS^0_{k+1, \bom^+}(\cA)$
give weakened versions  $S_{k,\bom} (s, \cA), \cS^0_{k+1, \bom^+}(s, \cA)$
and these give weakened versions  $\cH_{k, \bom} (s, \cA)$ and    $ \tilde   \Ga _{k, \bom^+}  (s, \cA) $.
Hence we can define
    $   \cW_{k, \bom^+}(s,  \cA)  =  \cH_{k, \bom} (s, \cA)   \tilde   \Ga _{k, \bom^+}  (s, \cA) W_k$.
Then  define 
$\tilde  E^{(4)} _{k, \bom^+  }(s, X, \cA,     \psi^{0, \#}_{k+1, \bom^+}(\cA )  ,   W_k)$ by replacing   $\cW_{k, \bom^+}( \cA)$ by $ \cW_{k, \bom^+}(s,  \cA)$
in (\ref{step1}).  These again satisfy the bound (\ref{superman}).     Again interpolate between $s_{\sq}=1$ and $s_{\sq}=0$ outside of $X$ 
and find   
$
  E^{(4)}_k( \La_k) =  \sum_{X \subset \La_k, Y  \supset X } \breve E^{(4)}_{k, \bom^+} (X, Y)
$
 where the sum is over $Y \in \cD_{k, \bom}$ and 
\be \label{sudden}
\begin{split} 
 &
  \breve E^{(4)}_{k,  \bom^+ }\B(X, Y, \cA ,       \psi^{0, \#}_{k+1, \bom^+}(\cA),   W_k        \B ) \\
  & =
  \int   ds_{Y-X} 
 \frac { \pa  }{ \pa s_{Y-X}}   \left[ \tilde   E^{(4)} _{k, \bom^+}(s, X, \cA,     \psi^{0, \#}_{k+1, \bom^+}(\cA ), W_k) 
  \right]_{s_{Y^c} = 0, s_X=1}\\
\end{split}
\ee
The expression is now local in $Y$ in the indicated variables. 
We resum  to  $\cD_k$ polymers $Y'$  and have 
\be \label{pinsky} 
E^{(4)}_{k} (\La_k) = \sum_{Y' \cap \Om_{k+1} \neq \emptyset} E^{(4)}_{k, \bom^+}(Y')
\ee
  where
\be
      E^{(4)}_{k, \bom^+}(Y' \cA ,       \psi^{0, \#}_{k+1, \bom^+}(\cA),   W_k )    
      =  \sum_{X,Y:  X \subset \La_k, Y \supset X,   \bar  Y  = Y'}     \breve      E^{(4)}_{k, \bom^+}(X,Y, \cA ,       \psi^{0, \#}_{k+1, \bom^+}(\cA),   W_k )   
\ee  
The derivatives in (\ref{sudden})  are  estimated by a Cauchy bounds as before and   as in lemma \ref{outsize2} this leads to
 \be \label{superman2}
\|  E^{(4)}_{k, \bom^+}(Y' , \cA)\|_{\frac12 \bh_k \bbI_k^\#, L}  \leq  \one  e_k^{\frac12 - 10\ep}  e^{-(\ka- \ka_0-1) d_M(X)}
\ee
where again the $\bbI_k^\#$ is optional. 
Split the sum (\ref{pinsky})  into $Y' \subset \La_{k+1}$ and $Y' \cap (\Om_{k+1} - \La_{k+1}) \neq \emptyset$
to generate the two sums on the right side of the lemma.  For  $Y' \subset \La_{k+1}$ the $ E^{(4)}_{k, \bom^+}(Y')$  are independent of $\bom^+$
since this is true of the random walks that generate it. 
 The estimates follow by relaxing the parameters in the norms. 
 This completes the proof.
\bigskip

\noindent
\textbf{Variations:} The same result holds for  $  E^{(5)}_{k}(   \La_{k} )$.  With the same bounds we have
\be
 E^{(5)}_{k}(\La_k ) 
  =    \sum_{X \subset \La_{k+1}}    E^{(5)}_{k}(X )  
      +   \sum_{ X \cap  (\Om_{k+1} - \La_{k+1} ) \neq \emptyset  }   B^{(5)}_{k, \bom^+ }(X)
\ee

  We can also consider 
   \be  
   \tilde E^{(\leq 3) }_{k, \bom^+} \B( \La_{k+1}, \cA,   \tilde Z_k,   \psi^{0, \#}_{k+1, \bom^+}(\cA),   W_k  \B) 
  \equiv      E^{(\leq 3)}_{k}\B(\La_{k+1}, \cA,   \tilde Z_k,         \psi^{0, \#}_{k+1, \bom^+}(\cA)  +     \cW^{\#}_{k, \bom^+}( \cA)  \B) 
  \ee
  This has a polymer expansion  $\tilde E^{(\leq 3)}_{k, \bom^+}(\La_{k+1}) = \sum_{X \subset \La_{k+1}} \tilde E^{(\leq 3)}_{k, \bom^+}(X)$ , which satisfies  by (\ref{urgent})
  and a variation of  (\ref{less})
  \be
    \| \tilde E^{(\leq 3) }_{k, \bom^+}(  X, \cA,     \tilde Z_k)  \|_{ \frac14  \bh_k, L} \leq    \| E^{(\leq 3) }_k (  X, \cA,     \tilde Z_k)  \|_{ \frac12 \bh_k }
\leq  \one e_k^{\frac14 - 5 \ep}e^{-(\ka- \ka_0 -1) d_M(X)}
\ee
The $\tilde E^{(\leq 3)}_{k, \bom^+}(X) $ is not local  in $W_k$, but 
 we  proceed with the localization as  for $E^{(4)}_k$  and obtain 
  \be
 E^{(\leq 3)}_{k, \bom^+}(\La_k ) 
  =    \sum_{X \subset \La_{k+1}}    E^{(\leq 3)}_{k}(X )  
      +   \sum_{ X \cap  (\Om_{k+1} - \La_{k+1} ) \neq \emptyset  }   B^{(\leq 3)}_{k, \bom^+ }(X)
\ee
where $ E^{(\leq 3)}_{k}(X ) $   
and $ B^{(\leq 3)}_{k, \bpi^+}(X ) $ 
are functions of $( \cA, \tilde Z_k,  \psi^{0, \#}_{k+1, \bom^+}(\cA) ,W_k ) $ localized in $X$ and
 satisfy the bounds
\be
\begin{split} 
    \label{gumball2} 
  \|  E^{(\leq 3) }_{k} \B(X,  \cA,     \tilde Z_k   \B) \| _{ \frac14 \bh_k , L } &  \leq   \one   e_k^{\frac14 - 5 \ep}  e^{- ( \ka - 2\ka_0-2) d_M(X)}\\
  \|  B^{(\leq 3)}_{k, \bom^+} \B(X  \cA  ,     \tilde Z_k   \B) \| _{ \frac14 \bh_k \bbI^\#_k,  L} &  \leq   \one   e_k^{\frac14 - 5 \ep}  e^{- ( \ka -2 \ka_0-2) d_M(X)}\\
\end{split}
\ee

Combine all these terms and define 
  \be 
  \begin{split}
   E^{(\leq  5) }_k (X)  =   &   E^{(\leq  3) }_{k} (X) 
+        E^{(4)} _{k } (X)  +    E^{(5)}_{k} (X) \\
  B^{(\leq  5) }_{k, \ \bom^+ }(X)   =   &   B^{(\leq  3) }_{k, \bom^+ }(X)
+        B^{(4)} _{k,  \bom^+ } (X)  +   B^{(5)}_{k,  \bom^+ }(X) \\
\end{split}
\ee
which again satisfy the bounds (\ref{gumball2}). 
  \bigskip

 The  previous  boundary terms  $B_{k, \bpi} = \sum_{j=1}^k B^{(k)}_{j, \bpi}$  and $B^{\bullet}_{k, \bpi^+}$  are    also modified by the fermion  translation.    
  After the fermion  translation (with a new definition of $ \de B^{(k)}_{j, \bpi} $) 
 \be
 \begin{split}
B^{(k)}_{j, \bpi}
= &
B^{(k)}_{j, \bpi} \B( \cA, Z_{k, \bpi}, \psi^{0, \#}_{k+1, \bom^+}(\cA) +   \cW^{\#}_{k+1, \bom^+}(\cA)   ,W_{k, \bpi}  \B)  \\
 = & 
 B^{(k)}_{j, \bpi}  \B( \cA, Z_{k, \bpi}, \psi^{0, \#}_{k+1, \bom^+}(\cA)   ,W_{k, \bpi}  \B)  
 + \de B^{(k)}_{j, \bpi} \B( \cA, Z_{k, \bpi}, \psi^{0, \#}_{k+1, \bom^+}(\cA) ,   \cW^\#_{k+1, \bom^+}(\cA)   ,W_{k,  \bpi}  \B) \\
 \end{split} 
\ee

\begin{lem}  
For   $1 \leq j \leq k$   
\be
\begin{split} 
   \de B^{(k)}_{j, \bpi} &   =      \sum_{  X \cap \Om_{k+1} \neq \emptyset, X \cap \La^c_j \neq \emptyset  } 
   \de B^{(k)}_{j, \bpi^+} \B(X,   \cA,  Z_{k, \bpi}, \psi^{0, \#}_{k+1, \bom^+}(\cA) , W_{k, \bpi}, W_k  \B)         \\
\end{split}
\ee
where the sum is over $X \in \cD_k $.  The 
  $ \de B^{(k)}_{j, \bpi} (X)$  are analytic in $\cA \in (1-e_k^{\ep} )   \tilde \cR_{k, \bom},  |Z_{j,\Om_{j+1}- \La_{j+1}}|  \leq L^{\frac12(k-j)}$   and  depend on the  indicated variables
  in X.
They  satisfy
\be    \label{ollie}
  \|  \de   B^{(k)}_{j, \bpi^+} \B(X;   \cA,    Z_{k, \bpi}  \B)   \|_{\frac12 \bh_K \bbI_k^\#, L\bbI_k,1 }  
\leq    \one   e_k^{\frac14-5\ep} e^{-(k-j)} e^{ -( \ka - \ka_0 -2)  d_{M}(X)}
\ee
\end{lem}   
\bigskip

\pr  The proof is similar to the proof of lemma \ref{outsize4}.
 Suppress the superscript $k$    and the  $ \cA,  Z_{k, \bpi}$ which are spectators for this proof.

 At first we have     $\de B_{j, \bpi}  = \sum_{X \in \cD_j}  \de  \tilde B_{j,\bpi^+} (X)$ where
 \be
\begin{split}
&\de  \tilde B_{j, \bpi^+}\B(X,\psi^{0, \#}_{k+1, \bom^+}(\cA)    ,W_{k, \bpi} ,W_k  \B ) \\
& = \int_0^1 \frac{d}{dt} B_{j, \bpi} \B(  \psi^{0, \#}_{k+1, \bom^+}(\cA) + t  \cW^{\#}_{k, \bom^+}(\cA)   ,W_{k, \bpi}  \B)\ dt
\\
&     = \int_0^1 \blan    \cW^{\#}_{k, \bom^+}(\cA), \   \frac{\pa}{\pa \psi } B_{j, \bpi} \B(  \psi^{0, \#}_{k+1, \bom^+}(\cA) + t  \cW^{\#}_{k, \bom^+}(\cA)   ,W_{k, \bpi}  \B)
 \bran \ dt\\
 \end{split} 
\ee
As in lemma \ref{outsize1}, part \textbf{(C.)},   we replace the first $ \cW^{\#}_{k, \bom^+}(\cA)$ by 
\be
\cW^{\#}_{k, \bom^+}(\sq_0, \cA)  \equiv  \cH^\#_{k, \bom}(\cA) \tilde \Ga_{k, \bom^+}(\cA) \chi(\sq_0) W_k
\ee
and sum over  $M$-cubes $\sq_0 \in \Om_{k+1}$.   (Here   $\cH^\#_{k, \bom}(\cA)=(\cH_{k, \bom}(\cA), \de_{\al \cA}\cH_{k, \bom}(\cA)$.)  Then as in (\ref{dumdum}) we have 
$
  \de  \tilde B_{j, \bpi^+}(X ) = \sum_{\sq_0 \subset \Om_{k+1} }
\de  \tilde B_{j, \bpi^+}(X, \sq_0 )
$
where
\be \label{wally}
\begin{split}
&\de  \tilde B_{j, \bpi^+}\B(X, \sq_0 , \psi^{0, \#}_{k+1, \bom^+}(\cA), W_{k, \bpi} ,W_k  \B ) \\
= &   
 \int_0^1  \ dt
\frac{1} { 2\pi i} \int_{|u| =e_k^{-\frac14+ 3 \ep}} 
  \frac{ du}{u^2}  B_{j, \bpi}\B(X,\psi^{0, \#}_{k+1, \bom^+}(\cA) +  t \cW^\#_{k, \bom^+}(\cA) + u  \cW^{\#}_{k, \bom^+}(\sq_0, \cA),  W_{k , \bpi}   \B ) \\
\end{split}
\ee
This is not yet local in $W_k$.  The bounds (147) in \cite{Dim20} imply for $f$ on $\Om_{k+1}$  that  $ |\cH^\#_{k, \bom}(\cA) \tilde \Ga_{k, \bom^+}(\cA) f| \leq \bbI_k^\# \|f \|_{\infty}$.
Using this and   our fundamental bound (\ref{sordid})  on $B_{j, \bpi}(X) $ 
\be \label{endless}
\begin{split}
\|\de \tilde B_{j, \bpi^+}(X  ) \|_{ \frac12 \bh_k \bbI_k^\#,  L \bbI_k,1 }\
& \leq  \one e_k^{\frac14 -3 \ep} \| B_{j, \bpi}(X  ) \|_{  \frac12 \bh_k \bbI_k^\#+ Ce_k^{-\frac14 +3 \ep}\bbI^\#_k, \  L \bbI_k} \\
&\leq   \one e_k^{\frac14 -3 \ep} \| B_{j, \bpi}(X  ) \|_{   \bh_k \bbI_k^\#, L \bbI_k} \\
& \leq  \one e_k^{\frac12 -11 \ep}   e^{- \ka d_{L^{-(k-j)}M}(X)} \\
\end{split}
\ee
Here we used $Ce_k^{-\frac14 +3 \ep}\bbI^\#_k  \leq e_k^{-\frac14 +2 \ep}\bbI^\#_k 
\leq  \frac12 \bh_k \bbI^\#_k $.

Now we  localize  in $W_k$.   
 Replace the fields  $\cW^{\bullet} _{k,  \bom^+}$, $\cW^{\bullet} _{k,  \bom^+}(\sq_0)$
by    weakened versions  $\cW^{\bullet} _{k,  \bom^+}(s)$, $\cW^{\bullet} _{k,  \bom^+}(s, \sq_0)$.
This yields  instead of  (\ref{wally})
\be \label{dumdum3}
\begin{split}
&\de  \tilde B_{j, \bpi^+}\B(s,X, \sq_0 , \psi^{0, \#}_{k+1, \bom^+}(\cA), W_{k, \bpi}, W_k   \B ) \\
= &   
 \int_0^1  \ dt
\frac{1} { 2\pi i} \int_{|u| =e_k^{-\frac14+ 2 \ep}} 
  \frac{ du}{u^2}  B_{j, \bpi}\B(X,\psi^{0, \#}_{k+1, \bom^+}(\cA) +  t \cW^\#_{k, \bom^+}(s) + u  \cW^{\#}_{k, \bom^+}(s,\sq_0 )    ,W_{k , \bpi}   \B ) \\
\end{split}
\ee
This   still satisfies a bound of the form  (\ref{endless}) even for $|s_{\sq}| \leq M^{-\frac12}$. 
Now expand around $s=1$ outside the fixed cube $\sq_0$.  
Then  $\de  \tilde   B_{j,  \bpi^+} (X, \sq_0)=   \sum_{ Y \supset \sq_0} \de    \breve  B_{j,  \bpi^+}(X, Y, \sq_0  )$   with the sum over $Y \in \cD_{k,\bom}$
and
 \be 
\begin{split}
& \de    \breve  B_{j,  \bpi^+}(X, Y, \sq_0   ) =
  \int   ds_{Y- \sq_0}  
 \frac { \pa  }{ \pa s_{Y-\sq_0}}   \left[  \de  \tilde B_{j ,  \bpi^+}(s, X, \sq_0)  \right]_{s_{Y^c} = 0, s_{\sq_0}=1}\\
 \end{split}
 \ee
 Now  
 $
  \de    \breve  B_{j,  \bpi^+}(X, Y, \sq_0  )=  \de    \breve  B_{j,  \bpi^+}\B(X, Y,  \sq_0, \psi^{0, \#}_{k+1, \bom^+}(\cA),  W_{k, \bpi}, W_k  \B )
 $
 depends on the indicated fields   only in $X \cup Y$.   Furthermore we  must have $X \cap Y \neq \emptyset$. 
 Each derivative   gives a factor $M^{-\frac12} \leq e^{- \hat \ka}$ and we have 
\be
\|    \de    \breve  B_{j,  \bpi^+}(X, Y, \sq_0 )\|_{ \frac12 \bh_k \bbI_k^\#, L \bbI_k, 1 } 
\leq 
 \one     e_k^{\frac12 -11 \ep}   e^{- \hat \ka|Y |_{\bom}  }  e^{- \ka d_{L^{-(k-j)}M}(X)  }
\ee 
Now the  proof  follows (\ref{kumquat}) - (\ref{cunning4}) in  the proof of lemma \ref{outsize4} and gives the result with a better power of $e_k$.

\bigskip

Now define  $\de B^{\bullet}_{k, \bpi^+}$ by
\be 
\begin{split}
 & B^{\bullet}_{k, \bpi^+}\B( \cA,  Z_{k, \bpi},\tilde Z_k,\psi^{0,\#}_{k+1, \bom^+}(\cA) + \cW_{k, \bom^+}(\cA), W_{k, \bpi} \B)\\
 & =   B^{\bullet}_{k, \bpi^+}\B(\cA,  Z_{k, \bpi},\tilde Z_k,\psi^{0,\#}_{k+1, \bom^+}(\cA), W_{k, \bpi}\B)  
+\de  B^{\bullet}_{k, \bpi^+}\B(\cA,  Z_{k, \bpi},\tilde Z_k,\psi^{0,\#}_{k+1, \bom^+}(\cA), W_{k, \bpi}, W_k\B)  \\
\end{split}
\ee

\begin{lem} 
$
\de B^{\bullet}_{k, \bpi^+}  =   \sum_{X}  \de B^{\bullet}_{k, \bpi^+}(X)$ where the sum is over $X \in \cD_k$ with $X \cap (\Om_{k+1} - \La_{k+1}) \neq \emptyset$.
The  $\de B^{\bullet}_{k, \bpi^+}(X)$
 are analytic in the same domain,  depend only on the fields in $X$,  and satisfy
\be
\| \de  B^{\bullet}_{k, \bpi^+} (X,\cA,  Z_{k, \bpi},\tilde Z_k)  \|_{\frac14 \bh_k \bbI_k^\#, L \bbI_k,1} \leq  \one e_k^{\frac14 - 5 \ep}    e^{ - (\ka - 2 \ka_0 - 3)   d_{M}(X)}
\ee
\end{lem}
\bigskip

\pr Suppress  $(\cA,  Z_{k, \bpi},\tilde Z_k)$ from the notation.    We have   $\de  B^{\bullet}_{k, \bpi^+}  = \sum_{X} \de   \tilde  B^{\bullet}_{k, \bpi^+} (X)$
where
   \be 
\begin{split}
 & \de \tilde B^{\bullet}_{k, \bpi^+}\B(X,\psi^{0, \#}_{k+1, \bom^+}(\cA), W_{k, \bpi}, W_k \B)\\
 & = \frac{1}{2 \pi i} \int_{|t| = e_k^{-\frac14 + 3 \ep}}   \frac{dt}{t(t-1)}    B^{\bullet}_{k, \bpi^+}\B(X,\psi^{0, \#}_{k+1, \bom^+}(\cA) + t\cW^{\#}_{k, \bom^+}(\cA), W_{k, \bpi}\B)  \\
\end{split}
\ee
is not local in $W_k$.  The analyticity in $t$ follows  as in (\ref{endless}).  Using the    bound (\ref{sonnyboy}) on $ B^{\bullet}_{k, \bpi^+} (X)$ 
we find as in (\ref{bisquit6}), (\ref{endless})
\be
\| \de \tilde   B^{\bullet}_{k, \bpi^+}(X) \|_{ \frac14 \bh_k \bbI_k^\#, L \bbI_k,1 } \leq 
\one e_k^{\frac12 - 8 \ep} \|  B^{\bullet}_{k, \bpi^+}(X) \|_{ \frac12 \bh_k \bbI_k^\#, L \bbI_k }
\leq \one e_k^{\frac12 - 8 \ep}  e^{ - (\ka -  \ka_0 - 2)   d_{M}(X)}
\ee

As in the previous lemma introduce weakening parameters replacing $\cW^\#_{k, \bom^+}(\cA)$ by $\cW^\#_{k, \bom^+}(s, \cA) $.
Expand around $s = 1$ outside $X$  and get a sum of  expressions  local in $W_k $.  After some rearrangement we    gain the announced result.  Details are much the same as  in previous lemmas and complete the proof with a stronger bound.  
\bigskip

\noindent
\textbf{Summary:} 
 In the last exponent in (\ref{rep8}) we now  have $E'_k(\La_{k} )+ E^{(\leq 5 )}_k ( \La_{k+1} ) + B_{k, \bpi } + B^{\star}_{k, \bpi^+}   +   R^{(3) a}_{k,  \bom^+}  $. 
Here   the new  boundary terms are collected into  
 $ B^{\star}_{k, \bpi^+}  =   B^{\star}_{k, \bpi^+} \B( \cA,  Z_{k, \bpi},\tilde Z_k, \psi^{0,\#}_{k+1, \bom^+}(\cA), W_{k, \bpi }, W_k \B)
$  defined   by
\be \label{tootsie3}
 B^{\star}_{k, \bpi^+}   = B^{(\leq 5)}_{k, \bpi^+} + \sum_{j=1}^k  \de   B^{(k)}_{j, \bpi^+}  +   B^{\bullet}_{k, \bpi^+}  + \de    B^{\bullet}_{k, \bpi^+}  
\ee
Then $  B^{\star }_{k, \bpi^+} =  \sum_{X }   B^{\star  }_{k, \bpi^+} (X)$
where the sum is over $X \in \cD_k$ such that $X \cap ( \Om_{k+1}  - \La_{k+1}) \neq \emptyset$. The   $ B^{\star} _{k, \bpi^+} (X)$ are 
 analytic in  $\cA \in (1-e_k^{\ep} )   \tilde \cR_{k, \bom},   |Z_{j,\Om_{j+1}- \La_{j+1}}|  \leq L^{\frac12(k-j)}, | \tilde Z_k|\leq e_k^{-\frac14 \ep}$, are local in the indicated fields 
  and satisfy  there
\be  \label{sonnyboy2}
\|   B^{\star }_{k, \bpi^+} (X, \cA,  Z_{k, \bpi},\tilde Z_k) \|_{\frac14 \bh_k \bbI_k^\#,  L\bbI_k,1} \leq  \one e_k^{\frac14 - 5 \ep}  e^{- (\ka - 2\ka_0 - 3)  d_M(X)}
\ee
 We split the sum over $ X $ into  
terms intersecting $\La_{k+1}$ and terms contained in $\La^c_{k+1} $ and correspondingly write
 \be  B^{\star}_{k, \bpi^+}   = B^{\star a}_{k, \bpi^+}   + B^{\star b }_{k, \bpi^+}   \ee 
 Now the terms   $E'_k(\La_{k} )  + B_{k, \bpi}  +  B^{\star b }_{k, \bpi^+}  +   R^{(3) a}_{k,  \bom^+}     $
come outside the integral  over $\tilde Z,   W_k $ on $\La_{k+1}$. Left inside is
\be
 V_{k, \bpi^+} ( \La_{k+1})  \equiv E^{( \leq  5)}_{k}(\La _{k+1})  +    B^{\star a }_{k,\bpi^+}  
 \ee
 The  fluctuation integral is now   
\be 
\label{fluctuation}
   \int  \   d \mu_{I, \La_{k+1}   } (W_k)d \mu_{I, \La_{k+1} }(\tilde  Z_k)  \hat \chi_k(  \La,  \tilde Z_k )     
 \exp \B(   V_{k, \bpi^+} ( \La_{k+1}) \B)   =  \Xi_{k,  \bpi^+} \int  \hat \chi_k(  \La_{k+1} ,  \tilde Z_k )  \ d \mu_{I, \La_{k+1}  }(\tilde  Z_k)  
   \ee 
 Here we defined a probability measure 
 \be
 d  \hat \mu_{\La}   ( \tilde Z_k )  = \frac {  \hat  \chi_k(  \La,  \tilde Z_k )  \ d \mu_{I, \La }(\tilde  Z_k)   }
{  \int   \hat\chi_k(  \La,  \tilde Z_k )  \ d \mu_{I, \La }(\tilde  Z_k)   }   
 \ee
 and 
 \be \label{fluctuation2} 
 \Xi_{k,  \bpi^+}  =     \int   \exp \B(   V_{k, \bpi^+} ( \La_{k+1}) \B)   \   d \mu_{I, \La_{k+1}   } (W_k)d \hat \mu_{I, \La_{k+1} }(\tilde  Z_k)   
   \ee 
   \bigskip

We combine the normalization factor with $\exp( R^{(3)a}_{k+1, \bom^+ } )
 = \exp \B(\sum_{\sq_0 \subset \La_{k+1}} R^{(3)a }_{k, \bom^+ }(\sq_0) \B)   $ from lemma \ref{outsize5}  to get $\exp (\de \vep_k^0 |\La_{k+1} |  )  $ 
defining
\be \label{kiwi} 
\de  \vep_k^0   |\La_{k+1} |   =\sum_{\sq_0 \subset \La_{k+1}} R_{k, \bom^+ }^{(3)a }(\sq_0)    + \log  \int  \hat \chi_k(  \La,  \tilde Z_k )  \ d \mu_{I, \La }(\tilde  Z_k) 
   \ee
 Then   $\de  \vep_k^0  = \cO(e^{-r_k}) + \cO(e^{-p_{0,k}^2} ) $ is tiny  (see \cite{Dim13b} for the latter). 
 Add the $\de \vep_k^0   |\La_{k+1} |  $ onto $ \fS^{0,+}_{k+1, \bom^+}$ defined in (\ref{miller}).  This  changes
 the $ (\vep' _k  + \vep^0_k) |\La_{k+1} | $ to   $ (\vep' _k  + (\vep^0_k)') |\La_{k+1} | $ where  $ (\vep^0_k)'=  \vep^0_k + \de  \vep^0_k$.  
 We make these changes but keep the notation  $ \fS^{0,+}_{k+1, \bom^+} $. 
\bigskip

Now   (\ref{rep8}) becomes
\be  
\begin{split}    \label{rep9}
&
\tilde   \rho_{k+1} (A_{k+1},  \Psi_{k+1} )  
=  \sum_{\bpi^+} (  N_{k+1, \Om_{k+1} }   \sZ'_{k,\bom}(0)    \de \sZ_{k, \bom^+} (\cA)  )( \sZ'_{k, \bom}  \de \sZ_{k, \bom^+})   \\
&
   \int  \   Dm^0_{k+1, \bom^+}  ( A )    Dm^0_{k+1, \bpi^+ }(Z) \   Dm^0_{k+1, \bom^+}  ( \Psi)\   Dm^0_{k+1,\bpi^+}(W)    
  \cC^0_{k+1, \bpi^+}    \chi^0_{k+1} (  \La_{k+1}  )  \\ 
 &   \exp \B( -\frac12 \|  d  \cA  \|^2  -    \fS^{0,+ }_{k+1, \bom^+}\B( \La_{k+1}, \cA, \Psi_{k,\bom^+},  \psi \B)  
       +   E'_k(\La_{k} )  + B_{k, \bpi}  +  B^{ \star b }_{k, \bpi^+}   \B)  \   \Xi_{k,  \bpi^+} \\   
 \end{split}
\ee
with the last line evaluated  at $\cA =     \cA^{0} _{k+1, \bom^+}$ and  $ \psi = \psi^0_{k+1, \bom^+} (\cA)$.

\subsection{cluster expansions}

We analyze the fluctuation integral  (\ref{fluctuation2}).

\begin{lem}  (cluster expansion)  
 \be
 \begin{split}  
  &  \Xi _{k,  \bpi^+}  =   
 \exp   \B(  V^\#_{k,\bpi^+ }   \B)  =  \exp   \B(  E^\#_{k} (\La_{k+1})  + B^{\#} _{k, \bpi^+}  \B)     \\
  \end{split}  
 \ee 
where      
 \be   
 \begin{split}
    &   E^\#_{k} (\La_{k+1})
    =       \sum_{  X \subset  \La_{k+1} } E^{\#}_{k} (X) \hs 
B^{\#} _{k, \bpi^+}  =  \sum_{   X  \cap  \La_{k+1} \neq \emptyset }   B^{\#} _{k, \bpi^+}  (X )    \\
  \end{split}  
  \ee
  with the following properties
  \begin{enumerate}
  \item
   $ E^{\#}_k (X) =  E^{\#}_{k} \B(X,  \cA,   \psi^{0, \#}_{k+1, \bom^+}(\cA)\B ) $  is analytic in $\cA \in (1-e_k^{\ep} )   \tilde \cR_{k, \bom}$ and has a kernel  independent of the history.    It satisfies
 \be \label{good}
     \| E^\#_k(X,  \cA)\|_{\frac14 \bh_k}  \leq    \one e_k^{\frac14-5\ep}e^{-(\ka - 6\ka_0 -6)d_M(X)} 
 \ee  
   \item 
   $ B^\#_{k, \bpi^+}(X) =  B^\#_{k, \bpi^+}\B(X;    \cA, Z_{k+1, \bpi^+};    \psi^{0, \#}_{k+1, \bom^+}(\cA), W_{k+1, \bpi^+} \B)$  is analytic in 
     $\cA \in (1-e_k^{\ep} )   \tilde \cR_{k, \bom}$, and $ |\tilde Z_{j, \Om_{j+1} - \La_{j+1} }| \leq L^{\frac12(k-j)}$ for $1 \leq j \leq k-1$
      and  $|\tilde Z_{k, \Om_{k+1}- \La_{k+1}}| \leq \frac12e_k^{- \ep/2}$ . It satisfies  
\be   \label{better}
 \|  B^\#_{k, \bpi^+}(X;   \cA, Z_{k+1, \bpi^+}) \|_{\frac14 \bh_k, L \bbI_k}  \leq   \ \one e_k^{\frac14-5\ep}e^{-(\ka - 6\ka_0 -6)d_M(X)}  
\ee
\end{enumerate} 
\end{lem} 
\bigskip

\rem   The following proof follows along the lines of  \cite{Bal98a}, \cite{Bal98b}, \cite{Dim13b}  where the cluster expansion starts with an ultralocal measure, 
rather than  \cite{Bal88a},     \cite{Dim15b}   where establishing the ultralocal measure is part of the cluster expansion. The latter  strategy  did not seem to work so well in the multiscale setting.

 \bigskip
\pr   
   We add a parameters $0 \leq t   \leq 1$ and define
\be
V_{t}   =  E^{( \leq  5)}_{k}(\La_{k+1}  )  +   t B^{\star a }_{k,\bpi^+}  \ee
so that $V_{1}  = V_{k, \bpi^+} ( \La_{k+1} )$. Then   $V_{t} = \sum_X V_{t}(X)$ where the sum is over $X \in \cD_k$ and $X \cap \La_{k+1} \neq \emptyset$ 
and where
\be
V_{t} (X) =V_{t}  \B(X, \cA, Z_{k+1, \bpi^+ },\tilde Z_{k, \La_{k+1} } , \psi^{0, \#} _{k+1, \bom^+} (\cA), W_{k+1, \bpi^+ }, W_{k, \La_{k+1} }  \B)
\ee
Here we have regrouped variables by
\be
\begin{split}
(Z_{k, \bpi} , \tilde Z_k)  = &  (Z_{k, \bpi} , \tilde Z_{k,\Om_{k+1} - \La_{k+1}} ,\tilde Z_{k, \La_{k+1} }  )  = (Z_{k+1, \bpi^+ },\tilde Z_{k, \La_{k+1} }  )\\
(W_{k, \bpi} ,  W_k)  = &  (W_{k, \bpi} , W_{k,\Om_{k+1} - \La_{k+1}} , W_{k, \La_{k+1} }  )  = (W_{k+1, \bpi^+ }, W_{k. \La_{k+1} }  )\\
\end{split} 
\ee
The weight for  $ W_{k+1, \bpi^+ }$ is now $(L \bbI_k, 1) = \bbI_{k+1}$. 
 From    the bounds (\ref{gumball2}) on  $ E^{( \leq  5)}_{k}(\La_{k+1}  ) $,  and the bound (\ref{sonnyboy2}) on $  B^{\star  a }_{k, \bpi^+} (X)$   we have in the stated region
  \be \label{ticino1}
\|  V_{t}   (X,\cA, Z_{k+1, \bpi^+ },\tilde Z_{k, \La_{k+1} } ) \|_{\frac14 \bh_k \bbI_k^\#, \bbI_{k+1}, 1} \leq 
\one e_k^{\frac14-5\ep} e^{-(\ka - 3\ka_0 -3)d_M(X)}  
  \ee
Here the weighting   $\bbI_k$ is still one in $\Om_{k+1} $. 
  We define
 \be \label{fluctuation1.5}
\Xi_{t} 
 =     \int  \ e^{V_t}    d \mu_{I, \La_{k+1}   } (W_k) d  \hat \mu_{I, \La_{k+1}  }(\tilde Z_k)   
   \ee 
Then   $ \Xi _{k,  \bpi^+}   = \Xi_{1}$ is the object we want to study.

We now make a standard cluster expansion,  see for example the appendix in \cite{Dim13a}. 
 Start with    a Mayer expansion   
   \be  \label{startpoint} 
 \begin{split}
     \exp \B(   \sum_{X  \cap  \La_{k+1}  \neq \emptyset }  V_{t} (X )  \B) 
  =    \sum_{X \cap  \La_{k+1}  \neq \emptyset}    K_{t} (X)
 \end{split}
 \ee
 where with distinct $X_i \cap \La_{k+1} \neq \emptyset$
  \be
 K_{t} (X)   =   \sum_{\{ X_i \} :  \cup X_i  = X  }   \prod_i   \B( e^{ V_t (X_i )  } -1 \B)  
 \ee
Then    $ K_{t}( X)  $ factors over its connected components   i.e if  $X = \bigcup_{\al} X_{\al} $ then $K_{t} (X) = \prod_{\al} K_{t} (X_{\al})$.
If  $X$ is connected   $ K_{t} (X) $ again  satisfies a bound of the form (\ref{ticino1}) but now with a  weaker decay factor $\exp( - (\ka- 4 \ka_0 -5)d_M(X) )$.
Now $\Xi_t = \sum_X H_t(X) $ with
\be
H_{t}  (X) =  \int      K_{t} (X) \  d \mu_{I, \La_{k+1}   } (W_k) d  \hat \mu_{I, \La_{k+1}  }(\tilde Z_k)     
\ee
and $H_t(X)  = H_t(X, \cA, Z_{k+1, \bpi^+ } , \psi^{0, \#} _{k+1, \bom^+} (\cA), W_{k+1, \bpi^+ })$. 
Thanks to the ultralocal measures this also factors over its connected components and if  $X$ is  connected satisfies  
 \be \label{ticino3 }
\|  H_{t} (X)  \|_{\frac14 \bh_k \bbI_k^\#, \bbI_{k+1} }\leq 
\one  e_k^{\frac14- 5\ep}   e^{-(\ka - 5\ka_0 -5)d_M(Y)}  
  \ee
  If $\{ X_{\al}\}$ are the connected components of $X$  with  $X_{\al} \cap \La_{k+1} \neq \emptyset$ then 
    \be \label{fluctuation3}
\Xi _{t} 
 =     \sum_{ \{ X_{\al} \}  }  \prod_{\al} H_t(X_{\al} )
\ee
We  can now take the 
 logarithm in the standard fashion  and  find 
\be  
  \Xi_{t}    = \exp \B( \sum_{X \cap \La_{k+1}   \neq \emptyset}      V^{\#} _{t} (X) \B)
\ee
where
\be   \label{hstar}
   V^{\#}_{t} (X) =    \sum_{n=1}^{\infty}
 \frac{1}{n!}  \sum_{(X_1, \dots,  X_n): \cup_i X_i  =X}  \rho^T(X_1, \dots,  X_n)  \prod_{\al}  H_{t}(X_{\al} )  
 \ee
 Here  $\rho^T(X_1, \dots,  X_n) $ is a standard function enforcing that $X$ is connected. 
Then  $V^\#_{t} $ satisfies
\be \label{pikes}
\|   V^{ \#} _{t} (X)  \|_{\frac14 \bh_k \bbI_k^\#, \bbI_{k+1} }  \leq    \one  e_k^{\frac14- 5 \ep}   e^{-(\ka - 6\ka_0 -6)d_M(X)}  
\ee

Next we remove the boundary terms from   $\Xi_{1} $ writing
\be \label{sinister1}
 \Xi_{1}    = \exp \B( \sum_{X \cap \La_{k+1}   \neq \emptyset}     B^{\#} (X)        \B) \Xi_{0}  
\ee 
where 
\be 
\label{astro} 
 B^\#_{k, \bpi^+}(X) =    V^{\#} _{1} (X) -   V^{\#} _{0} (X) 
\ee
again satisfies (\ref{pikes}). 
These really are boundary terms.   That is $B^{\#}_{k, \bpi^+} (X)$ is zero if $X \subset \La_{k+1} $ and so the sum is over $X \# \La_{k+1}$.   This  follows since if $X \subset \La_{k+1}  $ then only terms $V_{t} (X') $ with
$X' \subset \La_{k+1} $ can contribute to  $V^{\#} _{t} (X) $  and this  excludes any term $tB_{k,\bpi^+}(X) $.   Thus in this case  $ V^{\#} _{t} (X) $ is 
independent of $t$ and  $B^\#_{k, \bpi^+}(X)$  vanishes.

We are left with 
 \be \label{fluctuation1.7}
\Xi_{0} 
 =     \int   \exp \B(    \sum_{  X \subset  \La_{k+1} } E_{k, \leq 5} (X) \B) 
 \   d \mu_{I, \La_{k+1}   } (W_k)d   \hat \mu _{I, \La_{k+1}  }(\tilde Z_k)    = \exp \B( \sum_{X \subset \La_{k+1} } E_k^\#(X) \B)
   \ee 
where  $E_k^\#(X) = V_0^\#(X) $ again satisfies  (\ref{pikes}) and is independent of the history.  
Combining  this with (\ref{sinister1})    completes the proof.

 \subsection{normalization factor} 
 
The $\cA$ dependence in the normalization factor is handled as  follows.     We continue with the the notation  $\bom^+ =   (\bom, \Om_{k+1} )$. 

\begin{lem} For $\cA \in  \tilde \cR_{k, \bom} $ 
\be
\begin{split} 
 \de  Z_{k ,\bom^+ } (\cA)   =   & \de  Z_{k ,\bom^+ } (0 )          \exp  \B(  E^{\det}_k (\La_{k+1},\cA  )   
      +  B_{k, \bpi^+}^{\det}(\La_{k+1}, \cA )   \B) \\
 \end{split}       
\ee
where  $ E^{\det}_k (\La_{k+1})= \sum_{ X   } E^{\det}_{k} (X)$  with the sum over   $X \in \cD_k, X \subset  \La_{k+1} $.  The  $ E^{\det}_{k} (X)$
are independent of history  and satisfy
\be
 |E^{\det}_{k} (X, \cA  )|  \leq  e_k^{\frac14-\ep} e^{- \ka d_M(X)}
\ee
Furthermore   $ B_{k, \bpi^+}^{\det}  =  \sum_{X }  B_{k, \bpi^+}^{\det}  (X ) $
with the sum over $X \in \cD_k, X  \cap  (\Om_{k+1} - \La_{k+1}) \neq \emptyset $
and
\be \label{best}
 | B_{k, \bpi^+}^{\det}  (X,  \cA   ) | \leq  e_k^{\frac14-\ep} e^{- \ka d_M(X )}
\ee
 \end{lem}
\bigskip

\pr     By  lemma 10  in \cite{Dim20}   for $\cA$ in $\tilde \fG_{k, \bom}$
 \be 
 \de  Z_{k ,\bom^+  } (\cA )    =   \de  Z_{k ,\bom^+  } (0 )  
     \exp  \B(   \sum_{ X \cap \Om_{k+1} \neq \emptyset } E^{\det}_{k,\bom^+  } (X, \cA  )   \B) 
  \ee
 where 
$E^{\det}_{k, \bom^+  } (X, \cA  )  =  E^d_{k, \bom^+  }  (X, \cA  )  -  E^d_{k, \bom^+  }(X, 0  )$.
From a representation of  $ E^d_{k, \bom^+  }  (X, \cA  ) $ in terms of certain Greens functions and
the random walk expansions of the same,  one finds that  it is analytic in $\cA \in   \tilde \fG_{k, \bom^+} $ and satisfies     $  |E^d_{k, \bom^+ }  (X, \cA  )|  \leq CM^3e^{- \ka d_M(X)}$. 
If    $\cA \in \tilde \cR_{k, \bom^+}$   then $e_k^{-\frac14  } \cA \in \tilde \fG_{k, \bom^+}$ as mentioned in section \ref{once}.
Hence  we can write
\be E^{\det}_{k, \bom^+} (X, \cA  ) = \frac{1}{2 \pi i} \int_{|t| = e_k^{-\frac14}} \frac{dt}{t(t-1)}   E^d_{k, \bom^+}  (X, t\cA  ) 
\ee
which leads to the estimate
\be 
 |E^{\det}_{k, \bom^+  } (X, \cA  )| \leq C M^3e_k^{\frac14} e^{- \ka d_M(X)} \leq  e_k^{\frac14- \ep} e^{- \ka d_M(X)}
\ee
Now split the sum over $X \cap \Om_{k+1} \neq \emptyset $ into  $X \subset  \La_{k+1}$ and $X  \cap  (\Om_{k+1} - \La_{k+1}) \neq \emptyset $.  
If $X \subset  \La_{k+1}$ then  $E^{\det}_{k, \bom^+} (X, \cA  )  = E^{\det}_{k} (X, \cA  ) $ is independent of the history since this is true of the random
walk expansions that generate it.
 On the other hand if   $X  \cap  (\Om_{k+1} - \La_{k+1}) \neq \emptyset $ we define   $B^{\det}_{k, \bom^+} (X, \cA  )=E^{\det}_{k, \bom^+} (X, \cA  )$.
 This completes the proof. 
  \bigskip

\subsection{reblocking} 

We recall the  reblocking operation.   Consider an  expression $\sum_{X \subset \La} E(X) $ with $X \in \cD_k$, a connected union of $M$ blocks, and 
$\La $ a union of $LM$ blocks.  We   rewrite it as  $\sum_{Y \subset \La} (\cB E)(Y) $ where the sum is now over connected unions of $LM$ blocks $Y$
and
\be
(\cB E)(Y) = \sum_{X: \bar X =Y} E(X) 
\ee 
where $\bar X$ is the union of all $LM$ blocks intersecting $X$. 
Then one can show \cite{Dim13a} that if in some norm $\| E(X) \| \leq C e^{ -\ka d_M(X) }$ then
\be \label{reblock}
\|  (\cB E)(Y) \| \leq \one L^3 C  e^{- L(\ka- \ka_0 -1)  d_{LM}( Y) }
\ee
If $L$ is large enough, one can improve the decay constant back to $\ka$ or even better. 

Now after the cluster expansion and  the analysis of the normalization factor we have the leading terms 
$ E'_k +  E^{\#}_{k}  + E^{\det}_k      $ which we reblock to 
$\cB (  E'_k +  E^{\#}_{k}  + E^{\det}_k      ) $.   We estimate these quantities using  $h_{k+1} = L^{-\frac18} h_k$
to replace   $\bh_k$ or $\frac14 \bh_k$ by the smaller $\bh_{k+1} $. 
The bound  $ \|  E'_k(X)   \|_{\bh_{k+1} } \leq \one e_k^{\frac14- 7 \ep}  e^{-\ka d_M(X) } $ then becomes
\be \label{sinbad}
  \|( \cB  E'_k)(Y)   \|_{\bh_{k+1} }  \leq \one  L^3  e_k^{\frac14- 7 \ep} e^{-\ka d_{LM} (Y ) } 
\ee
The bound  $ \|  E^\#_k(X)   \|_{\bh_{k+1} } \leq  e_k^{\frac14 - 5 \ep}  e^{-(\ka- 6 \ka_0 - 6)  d_M(X) } $  becomes (using $L(\ka- 7 \ka_0 - 7) \geq \ka$)  
\be
\| (\cB E^\#_k) (Y)   \|_{\bh_{k+1} } \leq  \one L^3  e_k^{\frac14 - 5 \ep}  e^{-\ka d_{LM} (Y ) }  \leq  e_k^{\frac14 - 6 \ep}  e^{-\ka d_{LM} (Y ) }
\ee
The bound $  |  E^{\det} _k(X)  | \leq  e_k^{\frac14 -  \ep}  e^{- \ka d_M(X) } $  becomes
\be
|  (\cB E^{\det} _k) (Y)   | \leq   \one L^3  e_k^{\frac14 -  \ep}  e^{- \ka d_{LM} (Y) }  \leq   e_k^{\frac14 -  2\ep}  e^{- \ka d_{LM}(Y) }  
\ee

For the boundary terms 
 we have     $B_{k, \bpi}  =\sum_{j=1}^k B^{(k)}_{j, \bpi^+}$  to which we add   new terms 
$ B^{ \star b }_{k, \bpi^+}  + B_{k, \bpi^+}^{\#}+  B_{k, \bpi^+}^{\det}$, as well as $B'_{k, \bpi^+ } \equiv E'_k(\La_k ) - E'_k(\La_{k+1}) $.
After reblocking this new term  has the form    $  B^{(k)}_{k+1, \bpi^+}  = \sum_Y   B^{(k)}_{k+1, \bpi^+} (Y)$ summed on
$Y \cap (\La_k- \La_{k+1}) \neq \emptyset$   
with  $ B^{(k)}_{k+1, \bpi^+} (Y)$ is given by 
\be
  B^{(k)}_{k+1, \bpi^+} =    \cB \B( B'_{k, \bpi^+ }+  B^{ \star b }_{k, \bpi^+}  + B_{k, \bpi^+}^{\#}+  B_{k, \bpi^+}^{\det} \B) 
\ee
(If $k=0$ there is no $B'_{0, \bpi^+} $ and the sum can be taken over $Y \cap (\Om_1 - \La_1) \neq \emptyset$.) 
From (\ref{retro}),  (\ref{sonnyboy2}), (\ref{better}), (\ref{best}),  the reblocking estimate (\ref{reblock}),   and  replacing  $\bh_k$ or $ \frac14 \bh_k$ by  $\bh_{k+1} $ we have
\be \label{397}
\| B^{(k)}_{k+1, \bpi^+}  (Y )\|_{ \bh_{k+1} \bbI^{0,\#}_{k+1} , \bbI_{k+1}} \leq \one L^3 e_k^{\frac14- 7 \ep}  e^{- \ka d_{LM} (Y)} 
\leq  e_k^{\frac14- 8\ep}  e^{- \ka d_{LM} (Y)} 
\ee
Here we also  replaced $\bbI^\# =(\bbI_k, \bbI' _k) $ which have $1$ in
$\Om_{k+1} $ by  $\bbI^{0,\#}_{k+1} = ( \bbI^0_{k+1}, \bbI'^0_{k+1}) $ which are  the same but  with   $(L^{-1}, L^{-(1+ \al)} )$  in  $\Om_{k+1} $.

  All boundary terms are now  $
 B^0_{k+1, \bpi^+}  \equiv  \sum_{j=1}^{k+1}B^{(k)}_{j, \bpi^+}
  $
  and we have  
\be  
 \begin{split}    \label{rep10}
&
\tilde   \rho_{k+1} (A_{k+1},  \Psi_{k+1} )  
=  \sum_{\bpi^+}   ( \sZ'_{k, \bom}(0) N_{k+1, \Om_{k+1}}  \de \sZ_{k, \bom^+} (0) ) \  (  \sZ'_{k, \bom}   \   \de \sZ_{k, \bom^+} ) \\
&   \int  \   Dm^0_{k+1, \bom^+}  ( A )    Dm^0_{k+1, \bpi^+ }(Z) \   Dm^0_{k+1, \bom^+}  ( \Psi)\   Dm^0_{k+1,\bpi^+}(W)  
 \cC^0_{k+1, \bpi^+}    \chi^0_{k+1} (  \La_{k+1}  )     \\
&
    \exp \B( -\frac12 \|  d  \cA  \|^2   -    \fS^{0,+}_{k+1}\B( \La_{k+1}, \cA, \Psi_{k+1, \bom^+},  \psi^0_{k+1, \bom^+} (\cA)\B)  
      \\   
       &
    \exp \B(\B( \cB(  E'_k +  E^{\#}_{k}   + E^{\det}_k  )    \B)  (\La_{k+1}) +  B^0_{k+1, \bpi^+} )   \B)      \B|_{\cA =     \cA^{0} _{k+1, \bom^+} } \\   
 \end{split}
\ee

\subsection{scaling}

We  scale  according to  (\ref{scaleit}) replacing $A_{k+1}, \Psi_{k+1}$ on $\bbT^1_{N-k}$   by $A_{k+1,L}, \Psi_{k+1,L}$ now with $A_{k+1}, \Psi_{k+1}$
on $\bbT^0_{N-k-1} $ and  $A_{k+1,L}(x) = L^{-\frac12}  A_{k+1}(x/L)$ and  $\Psi_{k+1,L}(x) = L^{-1}  \Psi_{k+1}(x/l)$. 
This makes the following changes,  many of which we have already noted.  

\begin{itemize}

\item   The sum over    regions     $\bom^+ = ( \Om_1,  \dots,  \Om_{k+1})$  with   $\Om_j$  a  union of  $L^{-(k-j)}M$ blocks 
in   $\bbT^{-k}_{N -k}$   is    relabeled    as     $L \bom^+ =   ( L\Om_1,  \dots, L \Om_{k+1})$  
where  now   $\bom^+ = ( \Om_1,  \dots,  \Om_{k+1})$  with   $\Om_j$  a  union of  $L^{-(k+1-j)}M$ blocks 
in   $\bbT^{-k-1}_{N -k-1}$.  Similarly $\bpi^+$ is relabeled as  $L\bpi^+$.

\item  The fields    $A_{k+1,  \bom^+} =  ( A_{1, \de  \Om_1},  \dots,  A_{k+1, \de   \Om_{k+1}})$  defined
on subsets of   $\bbT^{-k}_{N-k}$  have become  with the relabeling    $A_{k+1, L \bom^+} =  ( A_{1,L\de   \Om_1},  \dots,  A_{k+1,L \de  \Om_{k+1}})$.   Now  replace   $A_{j,L\de \Om_j}$    by   $[A_{j,L}  ]_{L\de \Om_j}  =    [A_{j,\de \Om_j}]_L$.  Then      $A_{k+1, L \bom^+} $  becomes
 $A_{k+1,  \bom^+,L} =  ( A_{1,\de   \Om_1,L},  \dots,  A_{k+1, \de   \Om_{k+1},L})$.   
 
 \item
The relabeled  
 $ Dm^0_{k+1, L\bom^+}  ( A )$ scales to   $ Dm_{k+1, \bom^+}  ( A )$ by definition, and similarly we generate  
   $ Dm_{k+1, \bpi^+ }(Z) \   Dm_{k+1, \bom^+}  ( \Psi)\   Dm_{k+1, \bpi^+}(W)  $.

\item  
The gauge field      $\cA^0_{k+1,  \bom^+}$  becomes   $\cA_{k+1, \bom^+, L}$   and   
$  \| d \cA^0_{k+1,  \bom^+} \| ^2$ remains  $ \| d \cA_{k+1,  \bom^+} \|^2$.

 The characteristic function     $\chi^0_{k+1}(\Om_{k+1})$  becomes  $\chi_{k+1}(\Om_{k+1})$. 
 Also    $ \cC^0_{k+1,  \bpi^+}  $  becomes    $ \cC_{k+1,  \bpi^+}   $  by definition. Thus
 \be
  \cC_{k+1,  \bpi^+}  (A_{k+1}, \dots) =   (\cC^0_{k+1, L \bpi^+} ) (A_{k+1,  L}, \dots) 
\ee
Note that (\ref{newer}) which scales to $|dA_{k+1}| \leq    p_{k+1}  $ gives (\ref{simplesimon}) on $\Om_{k+1} $.

\item  The fermi field  $  \psi^{0}_{k+1, \bom^+} (\cA^0_{k+1,  \bom^+})  $  becomes   $[ \psi_{k+1, \bom^+} (\cA_{k+1,  \bom^+})]_L $. 
The   fermion action     formerly      $  \fS^{0,+}_{k+1}(\La_{k+1}, \cA^0_{k+1,  \bom^+}, \Psi_{k+1, \bom^+ },   \psi^0_{k+1, \bom^+} (\cA^0_{k+1,  \bom^+}))  $
now scales to  \be
  \fS^{+}_{k+1}(\La_{k+1}, \cA_{k+1,  \bom^+}, \Psi_{k+1, \bom^+ },  \psi_{k+1, \bom^+} (\cA_{k+1,  \bom^+}))    
  \ee
 Here we  have    identified  $e_{k+1} = L^{\frac12} e_k $   and
 \be
 m_{k+1}= L m'_k =L( m_k +   m( E_k)  ) \hs  \vep_{k+1}=  L^3 \vep_k' =L^3( \vep_k   +   \vep(  E_k)  )   
 \ee   
These are the first  three flow equations in Proposition \ref{prop}, and the bounds (\ref{samsung2}) are satisfied.   Similarly the equation (\ref{v1}) for  $ \vep^0_k$ is established together with the bound (\ref{v2}).

\item
The  $(\cB E')(\La_{k+1}, \cA, \psi^{0,\#} _{k+1,\bom^+ } (\cA)  )$ with  $\cA = \cA^0_{k+1, \bom^+ }$   becomes
  $(\cB E')(L\La_{k+1}, \cA_L,  [\psi^{\#} _{k+1,\bom^+ } (\cA)]_L  )$ with  $\cA  =   \cA_{k+1, \bom^+ }$.  The latter
  is a sum over polymers  $X \subset L \La_{k+1}$ with   $X \in \cD_{k+1} $.   We write this instead as a sum over $ LX$ where $X \subset \La_{k+1}  $
and $X \in \cD_k$.   Then the expression becomes
\be
\sum_{X \subset \La_{k+1} } (\cL E_k' ) \B(X , \cA,  \psi^{\#}_{k+1,\bom^+ } (\cA)\B) 
\ee
where
\be
\begin{split}
(\cL E_k' ) \B(X , \cA,  \psi^{\#}_{k+1,\bom^+ } (\cA)\B)  \equiv&   (\cB E_k' )_{L^{-1} } \B(X , \cA,  \psi^{\#}_{k+1,\bom^+ } (\cA)\B) \\
\equiv  &    (\cB E_k' ) \B(LX , \cA_L, [ \psi^{\#}_{k+1,\bom^+ } (\cA) ]_L\B ) \\
\end{split} 
\ee
Similarly for  $E^\#_k, E^{\det} _k$.  Thus we identify $E_{k+1} (\La_{k+1} ) = \sum _{X \subset \La_k+1}E_{k+1} (X )$
with polymer functions $E_{k+1} (X )$  given by 
\be     E_{k+1}    =  \cL  \B(      E'_k  +   E^\#_{k}  +   E^{\det}_k     \B)   
\ee
This  is the last  flow equation in Proposition 1 and so  $E_{k+1}$ satisfies the bound
\be
\| E_{k+1}  \|_{\bh_{k+1} } \leq e_{k+1}^{\frac14 - 7 \ep} e^{ - \kappa d_M(X) }
\ee   
An important point  here is the estimate on $\cL E'_k $.  Because it recycles the previous step  the factor $L^3$ in (\ref{sinbad}) might lead to  
exponential growth.  This does not occur because $E_k'$ has the relevant parts of $E_k$ removed. ($E_k^\#$ also recycles   previous terms, but here a higher power of the coupling constant cancels the $L^3$.) 

 Note that $E_{k+1}$  is  analytic in   $\cA \in \tilde \cR_{k+1, \bom^+ }$  since then  $ \cA_L  \in (1- e_k^{\ep} )  \tilde  \cR_{k, \bom} $.     Indeed   on $\de \Om_j$, with $j\leq k$,  the bound  $|\cA| \leq  L^{\frac12(k+1-j)} \theta_{k+1} e_j^{-\frac34 + \ep} $ and $\theta_{k+1} = (1- \ep_k^{\ep} ) \theta_k$
imply  $
|\cA_L| \leq (1-e_k^{\ep} )\B[ L^{\frac12(k-j)} \theta_{k} e_j^{-\frac34 + \ep}\B]
$.
In $\Om_{k+1}$ the bound $|\cA| \leq  \theta_{k+1} e_{k+1} ^{-\frac34 + \ep} $   and $e_{k+1} ^{-1} \leq e_k^{-1} $
imply $  |\cA_L| \leq (1-e_k^{\ep} )\B[ \theta_{k} e_k^{-\frac34 + \ep}\B]$. Derivatives are similar.

\item      Consider   the boundary term 
  $B^{(k)}_{j, \bpi^+} = \sum_X   B^{(k)}_{j, \bpi^+} (X) $ summed on $X\in \cD_j^{(k) } $ and $X \cap (\La_{j-1}  - \La_j)  \neq \emptyset$.   
  After relabeling and scaling this becomes
   $B_{j, \bpi^+} = \sum_X   B^{(k+1)}_{j, \bpi^+} (X) $ summed on $X\in \cD_j^{(k+1) }  $ and $X \cap (\La_{j-1}  - \La_j)  \neq \emptyset$  
 where  
\be \label{fourohfour}
\begin{split}
& B^{(k+1)} _{j,\bpi^+} (X , \cA, Z_{k+1, \bpi^+ }, \psi_{k+1, \bom^+} (\cA),W_{k, \bpi} ) \\
& \equiv B^{(k)} _{j,L\bpi^+} (LX , \cA_L,  Z_{k+1, \bpi^+,L }, [\psi_{k+1, \bom^+ } (\cA)]_L,W_{k, \bpi, L} )\\
\end{split} 
\ee
This   is analytic in  $\cA \in \tilde \cR_{k+1, \bom^+ }$ and  $ |\tilde Z_j |  \leq L^{ \frac12(k+1-j) }$ on $\Om_j- \La_j$.
In general if $B_{L^{-1} } (\psi^\# )  = B(\psi^\#_L) $ then $\| B_{L^{-1} } \|_{h,h'}  = \| B \|_{L^{-1} h, L^{ -(1+ \al)} h'}  $.  We apply this to (\ref{fourohfour})
and use $L^{-1}  \bbI_{k+1}  = \bbI^0_{k+1} $ and $L^{-(1+ \al)  }  \bbI'_{k+1}  = \bbI' {}^0_{k+1} $ and $d_{LM}(LX) = d_M(X) $ 
and  have by (\ref{397}) 
\be
\begin{split}
 \|  B^{(k+1)} _{j,\bpi^+} (X , \cA,  Z_{k+1, \bpi^+ } ) \|_{ \bh_{k+1} \bbI^\#_{k+1},L \bbI_{k+1}} 
 & \leq   \|  B^{(k)} _{j,\bpi^+} (LX , \cA_L,  Z_{k+1, \bpi^+,L }) \|_{\bh_{k+1}  \bbI^{0,\#} _{k+1}, \bbI_{k+1}} \\
  & \leq  \one e_k^{-\frac14 - 8 \ep}  e^{-\ka   d_{L^{-(k-j)}M}(LX)}           \\ 
   & \leq  e_{k+1} ^{-\frac14 - 8 \ep} e^{-\ka   d_{L^{-(k+1-j)}M}(X)}           \\ 
\end{split}
\ee
which is the required bound.  Overall $B^0_{k+1, \bpi^+}$ has scaled to $B_{k+1, \bpi^+}$.

\item After including the scaling factors from (\ref{scaleit})  we identify \be
\begin{split}
   \sZ'_{k+1, \bom^+}(0))  =  & L^{-8(s_N -s_{N-k-1})   }    N_{k+1, L\Om_{k+1}}  \sZ'_{k, L\bom}(0) \de \sZ_{k,L \bom^+} (0) \  \\
\hs  \sZ'_{k+1, \bom^+} = &     L^{\frac12 (b_N- b_{N-k-1})   -\frac12 (s_N- s_{N-k-1}) }    \sZ'_{k, L \bom}   \   \de \sZ_{k, L\bom^+} \\
\end{split}
 \ee

\end{itemize}
 Combining all the above we have the required
 \be  
 \begin{split}    \label{rep11}
&  \rho_{k+1} (A_{k+1},  \Psi_{k+1} )  
=   \sum_{\bpi^+}  \sZ'_{k+1, \bom^+}(0)     \sZ'_{k+1, \bom^+}  \\
&  \int  \  \   Dm_{k+1, \bom^+}  ( A )    Dm_{k+1, \bom^+}  ( \Psi)\    Dm_{k+1, \bpi^+ }(Z)  Dm_{k+1,\bpi^+ }(W)    
  \cC_{k+1, \bpi^+}    \chi_{k+1} (  \La_{k+1}  )  \\
&   \exp \Big( - \frac12 \| d   \cA   \|^2 -     \fS^+_{k+1, \bom^+}\B(\La _{k+1}, \cA, \Psi_{k+1, \bom^+},  \psi_{k+1, \bom^+ } (\cA) \B) 
    +E_{k+1}(\La_{k+1} )  +  B_{k+1,\bpi^+})     \B)   \B|_{\  \cA =     \cA _{k+1, \bom^+  } }  \\   
 \end{split}
\ee
This completes the proof of theorem \ref{maintheorem}.

% \newpage

  \section{Convergence}  \label{section4}

\subsection{the last  step}  \label{circus} 

The main theorem has generated a sequence of densities $\rho_0, \rho_1, \dots, \rho_k$.  We  stop the iteration when 
 our $Mr_k$ cubes no longer give a partition of the torus $\bbT^0_{N-k} $.   They do give a partition if $Mr_k \leq L^{N-k} $
(or with $M = L^m$ if  $r_k \leq L^{N-k-m} $).   In fact there  are indices  $k=K$ such that $Mr_K= L^{N - K}$ exactly, and we take the  first such.    The existence of $K$  is demonstrated in lemma  \ref{sync} 
to follow, and  it is shown that $N- K$ is bounded in $N$.

 Stopping at $k=K$ we  are on the  torus  $\bbT^{-K}_{N-K}$  
with dimension $Mr_K $.   The density      $\rho_K  = \rho_K (A_K, \Psi_K ) $  is  a function of  fundamental  fields   
 $A_K, \Psi_K $ on $\bbT^0_{N-K} $ and has 
the form    (\ref{startrep1}).  We   now  want to integrate it to get the partition function.  But this  last   step requires special treatment.

     In the integral  over  $A_K  $   we  still   impose    axial gauge fixing  by  inserting 
    $  \de  ( \tau^*A_K)  $
  where    $\tau^*$ is a tree in  $\bbT^0_{N-K}$  rather  than an $L$-cube.
   This gauge  fixing is not enough to give convergent integrals.     
     At this point we allow ourselves to suppress the contribution from    torons (Wilson lines ),  which come from integrals
around the torus.   They   are an artifact of the topology of the torus and not  fundamental to the model.     Thus we could   impose that       $A_K(   \Ga_{x, \mu} )  =0  $
  where     $\Ga_{x, \mu} $ is  the global    circle thru $x$ in the direction $e_{\mu}$. 
In fact we impose the weaker condition that      $  L^{-3}  \sum_{x \in \tz}      A_K (   \Ga_{x, \mu} )  =0  $.    But this    is just  the condition that the single vector
    $ \cQ^* A_K    $  satisfies    $ \cQ^*A_K =0 $.     Here  $\cQ^*$  is the single averaging operator  $\cQ$ but with  $L$ replaced by  $Mr_K $.      Contours  in   the averaging operation   join    the   the origin back to itself     instead of joining neighboring  centers.
 See  \cite{Dim15} for further discussion of  this operator.

This modification  of the gauge field     could   be avoided if  we took circle valued    gauge fields as in \cite{BIJ88}.
It could also  be avoided if we worked in a rectangular box rather than a torus.    Then the axial gauge fixing  alone    would be 
sufficient to enable convergence of the last integral.     In the latter case the overall translation invariance of the model would be spoiled ,   but
it could be retained for polymers  separated from the boundary.     This should be sufficient to carry out the analysis.   A similar problem is addressed in   \cite{Dim09}.

In any case  the  representation of the partition function is now.
\be
\sZ(N,e)   =      \int   \de (  \cQ^*A_K  )    \de  ( \tau^* A_K)    \rho_K (A_K, \Psi_K )   D\Psi_K DA_K  
\ee      
Insert  the   expression for  $\rho_K$ from  (\ref{startrep1}).  
 Taking account also that   $M_K =0,  \vep_K =0$ so that  $\fS^+_K =\fS_K + \vep_K^0 \Vol(\La_K) $    we  find
\begin{equation}     \label{z}
\begin{split}
&  \sZ(N,e)   =       \sum_{\bpi}   \sZ'_{K, \bom}(0)   \sZ'_{K, \bom} \\
&  \int   \de (  \cQ^* A_K  )    \de  ( \tau^*A_K) 
DA_K    D\Psi_K  Dm_{K, \bom} (A )Dm_{K, \bpi} (Z )  Dm_{K, \bom}( \Psi)   Dm_{K, \bpi}( W)    \     \cC_{K, \bpi} \     \chi_{K} (\La_K)    \\
&        \exp \B( -\frac12 \| d \cA_{K, \bom}\|^2      -         \fS_K\B( \cA_{K, \bom},  \Psi_{K, \bom},   \psi_{K, \bom} (\cA_{K, \bom})   \B)   
  +E_K(\La_K ) +  B_{K,\bpi}  + \vep_K^0 \Vol(\La_K)     \B)  \\
\end{split}
\end{equation}

The final torus $\bbT^{-K}_{N-K}$ is a single $Mr_K$  cube.  For the last region $\La_K$      the only  possibilities are   $\La_K = \bbT^{-K}_{N-K} $ or $\La_K  = \emptyset$.   (The latter case
includes the possibility that $\La_k = \emptyset$  at some earlier stage  in which case all subsequent small field regions are defined to  be empty.)  
    We separate the two cases writing  
 \be 
\sZ(N,e)  =    \sZ^{\cL}(N,e)  +    \sZ^{\cS}(N,e) 
\ee
Here   $ \sZ^{\cL}(N,e)$ is the sum of terms with  $\La_K  = \emptyset$.    In this case the  $   \chi_{K} (\La_K) $, $E_k(\La_k)$ and $\vep_K^0 \Vol(\La_K)   $ are all absent.
  We have     
  \begin{equation}     \label{zL}
\begin{split}
&    \sZ^{\cL} (N,e)    =      \sum_{\bpi:  \La_K = \emptyset  }    \sZ'_{K, \bom}(0)   \sZ'_{K, \bom} \\
&   \int   \de (  \cQ^*A_K  )    \de  ( \tau^*A_K)  
DA_K    D\Psi_K  Dm_{K, \bom} (A )Dm_{K, \bpi} (Z )  Dm_{K, \bom}( \Psi)   Dm_{K, \bpi}( W)  \       \\
&  \     \cC_{K, \bpi} \        \exp \B( -\frac12 \| d \cA_{K, \bom}\|^2      -         \fS_{K, \bom}\B( \cA_{K, \bom},  \Psi_{K, \bom},   \psi_{K, \bom} (\cA_{K, \bom})   \B)   
  +  B_{K,\bpi}       \B)  \\
\end{split}
\end{equation}
The term  $ \sZ^{\cS}(N,e) $ is the sum of terms  with  $\La_K  =  \bbT^0_{N-K}  $.  There is only one such term.  It  is the case where  every small field 
region is maximal:    $ \Om_j, \La_j =\bbT^{-k}_{N-K}$.     
 The large field integrals $Dm_{K, \bom} (A), \dots $ are absent, as are $\cC_{K, \bpi},  B_{K, \bpi} $.
The  normalization factors $  \sZ'_{K, \bom}(0)   \sZ'_{K, \bpi }  $   become the global   $   \sZ'_{K}(0)   \sZ'_{K}  $. 
Thus it is 
\be  \label{snuffit}
\begin{split}
&   \sZ^{\cS}   (N,e)  =    \sZ'_{K}(0)   \sZ'_{K}   \int    \de (  \cQ^*A_K  )    \de  ( \tau^*A_K)    DA_K D\Psi_K  \   \chi_{K}  \\&
\exp  \B(   -  \frac12   \| d \cA_K \|^2  -           \fS_K\B( \cA_{K},  \Psi_{K},   \psi_{K} (\cA_{K})   \B)   +    E_K(\La_K,    \cA_K,   \psi^\#_K(\cA_K)) +
\vep_K^0 \Vol( \La_K ) \B) \\
\end{split}  
\ee
with fields
$    \cA_K = \cH_K A_k $   and $    \psi_K(\cA)  =  \cH_K(\cA) \Psi_K   
$
and
$  \psi^\#_K(\cA) = ( \psi_K(\cA), \de_{\al, \cA} \psi_K(\cA)) $.
 \bigskip

 We are going to compare $\sZ(N,e)$ to the free partition function $\sZ(N,0)$ which is a product of the free fermion partition function
 and the free boson  partition function.  We have
 \be 
 \sZ(N,0) =  \sZ^f(N,0) \sZ^b(N,0)
 \ee
 Here the free fermion partition function is 
\be \label{sounder}
\begin{split}
 \sZ^f(N, 0) 
 \equiv &  \int \exp \B( - \blan \bPsi_0 ( \fD_{0} + \bar m_0 ) \Psi_0 \bran   \B) D\Psi_0 \\
=& \B[ \int   \exp \B( - \blan \bPsi_K, D_K(0) \Psi_K  \bran  \B)  \  D\Psi_K \B]  \  \sZ'_K(0)\\
=&    \det (D_K(0) ) \ \prod_{j=0}^{K-1}      L^{-8 (s_N - s_{N-j-1} )  }  N_{j} \de \sZ_{j}(0)   \\
\end{split}
  \ee
The second line is the result of repeated global  block averaging, see (25)-(27) in \cite{Dim20} where  $ D_K(0)$
is defined.

 The free boson partition function is a gauge fixed version of $\int \rho_0 (A_0)  DA_0$ with  density $\rho_0 (A_0)  = \exp ( - \frac12 \| dA_0 \|^2 )$.        
   As in   (\ref{basic1}),(\ref{scaleit})  one  generates a 
 a sequence   $\rho_0, \rho_1, \dots, \rho_K$  by 
 \be  
 \begin{split}
 &   \tilde \rho_{k+1} ( A_{k+1}  )  
   =      
 \int   \      \de(  A_{k+1} - \cQ A_k  )  \de (\tau A_k ) 
  \rho_{k}(A_k )  \ DA_k    \\
  &   \rho_{k+1} ( A_{k+1}  )   =   L^{-8 (s_N - s_{N-k-1} )  } \   \tilde \rho_{k+1} ( A_{k+1,L}   )  \\
\end{split} 
\ee 
 Then the free   partition function is $Z^b(N,0)  =      \int       \de (  \cQ^*A_K  )    \de  ( \tau^*A_K)  \rho_K (A_K)     dA_K $
 This has a nice global expression \cite{Dim15b}.   Here we give an alternate expression adapted to our proof.   Fix $\bom = (\Om_1, \dots, \Om_K)$
 and repeat theorem \ref{maintheorem} with no fermions,  with $\La_k = \Om_k$ for all $k$, and with  no characteristic functions.    There are  also  no $E_K, B_{k, \bpi}, \vep_K^0$. 
This yields 
 \be     \label{twice}
 \sZ^b(N,0)  =    \sZ'_{K, \bom}\   \int    \de (  \cQ^*A_K  )    \de  ( \tau^*A_K)   \ DA_K   Dm_{K, \bom}(A)
     \exp  \B( - \frac12  \| d \cA_{K , \bom } \|^2  \B) 
\ee
\bigskip

\begin{lem} { \ } \label{sync}  Let $r_k$ be the smallest power of $L$  greater than or equal to $r^0_k  = ( - \log e_k)^r$.
\begin{enumerate}
\item  $r_{k+1} =L^{-1} r_k$ or $r_k$.  
\item  There is a $K<N$ so $Mr_K = L^{N-K} $ so $|\bbT^0_{N-K}| = (Mr_K)^3$
\item $N-K$ is bounded by a constant independent of $N$, as are $|\bbT^0_{N-K}| $ and $r_K$. 
\end{enumerate} 
\end{lem} 
\bigskip

\pr
\begin{enumerate} 
\item  
$r_k^0$ is decreasing in $k$ so
$r_k$ is non-increasing. Hence   $r_{k+1} = L^{-n} r_k$ for some integer  $n \geq 0 $.  On the other hand  since $e_{k+1} = L^{\frac12} e_k$ we  have $r^0_{k+1}  = (- \log e_k -\frac12 \log L ) ^r$.  Hence
for  $e_k$ sufficiently small
$
 \frac12 r^0_k  \leq   r^0_{k+1} \leq r^0_k
$. 
Also  $r^0_k \leq r_k \leq  Lr^0_k$ and so 
\be \label{orange}
  \frac{1}{2L}  r_k  \leq  \frac12 r^0_k \  \leq   r^0_{k+1}  \leq  r_{k+1} \ee
This inequality excludes  $r_{k+1} = L^{-n} r_k$ for $n \geq 2$.

\item 
Define integer-valued $d_k$ by 
\be 
d_k = (N-k) -  \log_L  Mr_k = (N-k)  - \log_L r_k -m
\ee
This starts out positive for $N$ sufficiently large (since $\log_L r_k$    grows logarithmically in $N$)   but eventually goes negative
at $k = N$.  Since $\log_L r_k$ only decreases by $0,1$  there must be a points  $k = K$  where $d_K=0$

\item
  We have $\ N-K=   \log_L r_K + m$  and $ r_K \leq Lr^0_K$ and hence  
$
 N-K   \leq  \log_L r^0_K  +m +1
$.
But 
\be
\log_L r^0_K  = r \log_L (- \log e_K)  =  r \log_L \B(  - \log e + \frac12(N-K) \log L \B)  
\ee
Thus $x= N-K$ satisfies an inequality of the form $ x \leq  r\log_L (a + bx ) + c$ with positive constants $a,b,c$ independent of $N$.
Equivalently $L^x \leq  L^c (a+bx) ^r$.
But  $ ( L^c (a+bx) ^r ) L^{-x/2} $ is bounded by a constant independent of $N$, hence the same is true for $L^{x/2}$ and $x$. 
This completes the proof. 
\end{enumerate} 
\bigskip

\subsection{estimates}

 Our goal is to get good estimates on $ \sZ^{\cL}(N,e)$ and    $ \sZ^{\cS}(N,e)$ and in particular to show that the sum in (\ref{zL})
converges.  This requires good estimates  on the integrals  for each $\bpi$.

 \subsubsection{fermion integral - large fields region}  

For each  fixed   $\bpi $  we first consider the fermion integral  in   the $\bpi$ term in      $ \sZ^{\cL}(N,e) $ which is 
\be 
\begin{split} 
\label{super} 
\cJ_{K, \bpi}   \equiv  &    \int  \cF_K (  \Psi_{K, \bom},  W_{K, \bpi}) \    D\Psi_{K}  \   Dm_{K, \bom } ( \Psi)\  \    Dm_{K, \bpi} ( W)\\
\cF_K (  \Psi_{K, \bom},  W_{K, \bpi})   =   &    \exp \B(-\fS_K( \cA_{k, \bom}, \Psi_K, \psi_{K, \bom}(\cA_{K, \bom} ))  +   B_{K, \bpi}  \B)\\
 \end{split}  
\ee

To estimate this we write it as a sequence of integrals at different levels.   
First we scale up replacing $\Psi_j(\sx) $ on $\bbT^{-(K-j) } _{N-K} $ by $ \Psi_{j,L^{-1} }(\sx) = L  \Psi_j ( L  \sx) $ now with  $\Psi_j$    on $\bbT^{-(K-j)+1 } _{N-K+1} $.
Then    $\Psi_{j,  \de \Om_j}  $   becomes  $ ( \Psi_{j, L^{-1}} )_{ \de \Om_j }  =  (  \Psi_{j,  L \de \Om_j })_{L^{-1} }  $. 
The $\Psi_K$  on $\bbT^0_{N- K}$ is replaced by   $\Psi_{K, L^{-1} } $ with $\Psi_K$ on $\bbT^1_{N- K+1}$   and    $D \Psi_ k  $ is replaced by 
 $L^{ - |\bbT^0_{N- K}|} D \Psi_K $.  Similarly   $W_j $ on $\bbT^{-(K-j) } _{N-K} $  is replaced by   $W_j$    on $\bbT^{-(K-j)+1 } _{N-K+1} $.     The integral  $   Dm_{K, \bom }$ scales to  $ ( Dm_{K, \bom })_L  =   Dm^0_{K, L \bom }$, etc. 
Thus we obtain 
\be 
\label{super4} 
 \cJ_{ K, \bpi}    =  L^{ - |\bbT^0_{N- K}|}   \int    \cF_{K}  ( ( \Psi_{K, L\bom} )_{L ^{ -1}}  ,  (W_{K, L\bpi})_{L ^{ -1}}) 
\   D\Psi_{K }    Dm^0_{K, L\bom} ( \Psi)   Dm^0_{K,L \bpi} ( W)\
\ee
Now by (\ref{post3}), (\ref{post4}) with $\La_K = \emptyset$  
\be    \label{twotwo}
\begin{split}
 Dm^0_{K,L \bom} ( \Psi)
  =  &\de_{G,L \Om_{K}^c}\B(\Psi_{K} - Q(0) \Psi_{K-1}\B) \     D \Psi_{K-1,L \Om^c_{K} }   Dm_{K-1, L\bom} ( \Psi) \\
Dm^0_{K, L\bpi} ( W)\ 
=  & D \mu _{ I,L \Om_{K}  } (W_{K-1})  Dm_{K-1,L \bpi} (W) \\
\end{split}
\ee
Split the integral     by $  D \Psi_{K }  = D \Psi_{K, L\Om^c_{K} }   D \Psi_{K, L \Om_{K}  } $ 
Nothing depends   on $\Psi_{K, L\Om^c_K } $  and we can use the normalization
$
\int   \de_{G, L\Om_{K}^c}\B(\Psi_{K} - Q(0) \Psi_{K-1} \B)   D \Psi_{K,L \Om^c_K}  = 1
$.
Then define 
\be \label{prince}
  \cF_{K-1}(   \Psi_{K-1, L\bom} , W_{K-1, L\bpi}) =   \int     \cF_{K}  ( ( \Psi_{K, L\bom} )_{L ^{ -1}}  ,  (W_{K, L\bpi})_{L ^{ -1}}) 
  D \Psi_{K, L \Om_K }   d \mu _{ I, L\Om_{K}} (W_{K-1})
\ee
where $   \Psi_{K-1,L \bom}=  (\Psi_{1,L \de \Om_1}, \dots,  \Psi_{K-1, L\de \Om_{K-1}}  )$, etc. 
This gives the representation
\be 
\label{super5} 
 \cJ_{ K, \bpi}    =   L^{ - |\bbT^0_{N- K}|}    \int    \cF_{K-1}(   \Psi_{K-1,L \bom} , W_{K-1, L\bpi})    D\Psi_{K-1, L\Om^c_K }    Dm_{K-1,L \bom} ( \Psi)\      Dm_{K-1, L\bpi} ( W)
\ee

 Now repeat this procedure to eliminate more fields.       To state the result    define $\Om_j(k)   = L^{K-k} \Om_j$ and  $\La_j(k)  = L^{K-k} \La_j$.   These are what  the regions were  at level $k$,  i.e. on $\tk$,
 before they were relabeled $K-k$ times  to  their present life in $\bbT^{-K} _{N-K} $.   Also define $\bom(k)  = ( \de \Om_1(k) , \dots , \de \Om_k(k) )$ and similarly
 $\bpi(k)$.   And we define for fields $\Psi_j, W_j$  on $\bbT^{-(k-j)}_{N-k} $, $1 \leq j \leq k$,
 \be
 \begin{split}
  \Psi_{k, \bom(k) } = &\B ( \Psi_{0, \Om_1^c(k)},\Psi_{1, \de \Om_1(k)},  \cdots, \Psi_{k-1, \de \Om_{k-1}(k)},  \Psi_{k,  \Om_k(k)}\B ) \\
   W_{k, \bpi(k) }  =  &  \B( W_{0, \Om_1(k) - \La_1(k)}, \dots, W_{k-1, \Om_k(k) - \La_k(k)} \B ) \\
 \end{split}
 \ee
 which is what they were at level $k$.  
\bigskip

 \begin{lem}  \label{tucson} 
  Define a sequence    $\cF_{K}, \cF_{K} \dots, \cF_{1}, \cF_0 $     
 by  
 \be \label{alpaca}
 \begin{split}
&   \cF_{k}(   \Psi_{k, \bom(k)} , W_{k, \bpi(k) })  \\
&=   \int   \cF_{k+1,L}\B (    \Psi_{k+1, \bom(k)} ,   W_{k+1, \bpi(k) } \B) \ 
  D \Psi_{k+1, \de \Om_{k+1} (k)}  d \mu_{I,\Om_{k+1}(k) - \La_{k+1}(k) } (W_k) \\
 \end{split} 
\ee
where    $ \cF_{k+1,L}\B (    \Psi_{k+1, \bom(k)} ,   W_{k+1, \bpi(k) } \B) = \cF_{k+1}\B (  (  \Psi_{k+1, \bom(k)}  )_{ L^{-1} },  ( W_{k+1, \bpi(k) })_{  L^{-1} }  \B) $.
Then for any $k$ 
 \be 
\label{llama}
\cJ_{ K, \bpi}   = J_k \int \cF_{k} \B(  \Psi_{k, \bom(k)},    W_{k, \bpi(k)} \B)\   \ D \Psi_{k, \Om^c_{k+1} (k)  } \  Dm_{k, \bom(k) } ( \Psi)   Dm_{k, \bpi(k) } (W)
\ee
where 
\be
   J_k = \prod_{ j=k+1}^{K} L^{ - {|   \Om_{j+1}^{{(j)},c} |} }
\ee
\bigskip
\end{lem}  
 
  \pr  
 We have seen it  is true for $k= K-1$  (since   $\La_K,  \Om_{K+1} = \emptyset$) .   We assume it is true for  $k+1$ and prove it for $k$. We have
  \be 
\label{llama1}
\begin{split}
&\cJ_{ K, \bpi} = J_{k+1} \\
& \int \cF_{k+1} \B(  \Psi_{k+1, \bom(k+1)},    W_{k+1, \bpi(k+1) } \B)\   \ D \Psi_{k+1, \Om^c_{k+2} (k+1)  } \  Dm_{k+1, \bom(k+1)} ( \Psi)   Dm_{k+1, \bpi(k+1) } (W)\\
\end{split} 
\ee
We scale up replacing  $\Psi_{j , \de \Om_j(k+1) }$ by $(\Psi_ {j , L^{-1}} )_{ \de \Om_j(k+1)    }  =(\Psi_ {j ,   \de \Om_j(k) }  )_{L^{-1} }  $ 
(since $L \Om_j(k+1) = \Om_j(k)$). 
Then   $ \Psi_{k+1, \bom(k+1)}$ is replaced by $ (\Psi_{k+1, \bom(k)} )_{L^{-1}  } $ 
and the  integral $  D \Psi_{k+1, \Om^c_{k+2} (k+1)  } $ is replaced by   $ L^{ -  | \Om^{(k+1) ,c} _{k+2} | }  D \Psi_{k+1, \Om^c_{k+2} (k+1)  }$. 
We identify $  L^{ -  | \Om^{(k+1) ,c} _{k+2} | } J_{k+1} =J_k$.   The integral $ Dm_{k+1, \bom(k+1)}  $ scales to  $ (Dm_{k+1, \bom(k+1)})_L =  Dm^0_{k+1, \bom(k)} ( \Psi) $, etc.    
Thus we have     
 \be 
\label{llama2}
\cJ_{ K, \bpi}   = J_k  \int  \cF_{k+1}\B (  (  \Psi_{k+1, \bom(k)}  )_{ L^{-1} },  ( W_{k+1, \bpi(k) })_{  L^{-1} }  \B)    \ D \Psi_{k+1, \Om^c_{k+2} (k)  } \  Dm^0_{k+1, \bom(k) } ( \Psi)   Dm^0_{k+1, \bpi(k) } (W)
\ee
 However 
 \be
 \begin{split} 
 Dm^0_{k+1, \bom(k)}  (\Psi) =  & Dm_{k, \bom(k)}(\Psi)\ \de_{G, \Om^c_{k+1}(k)  } (\Psi_{k+1} - Q(0) \Psi_k)  D\Psi_{k, \Om^c_{k+1}(k)  }  \\
   Dm^0_{k+1, \bpi (k) } (W) = &  Dm_{k, \bpi(k) }(W)\  d \mu_{I, \Om_{k+1}(k) - \La_{k+1}(k) } ( W_k ) \\
 \end{split} 
 \ee
 Split the integral   $ D \Psi_{k+1, \Om^c_{k+2} (k)  } =  D \Psi_{k+1, \Om^c_{k+1} (k)  }  D \Psi_{k+1,  \de \Om_{k+1} (k)  } $.
 and   integrate out 
 \be
\int    \de_{G, \Om^c_{k+1}(k)  } (\Psi_{k+1} - \cQ(0) \Psi_k)  D \Psi_{k+1, \Om^c_{k+1} (k)  } =1
 \ee
 So now we have 
  \be 
  \begin{split}
\label{llama3}
& \cJ_{ K, \bpi}   =  J_k \int    \ D\Psi_{k, \Om^c_{k+1} (k) }    Dm_{k, \bom(k)}(\Psi)       Dm_{k, \bpi(k) }(W)\\\
&  \B[  \int   \cF_{k+1} \B(  (  \Psi_{k+1, \bom(k)}  )_{ L^{-1} },  ( W_{k+1, \bpi(k) })_{  L^{-1} }  \B)  \ D \Psi_{k+1,   \de \Om_{k+1} (k)  } \ 
 d \mu_{I, \Om_{k+1} (k)- \La_{k+1}(k)} ( W_k )  \B]   \ \\
 \end{split} 
\ee
The expression in brackets is identified as $  \cF_{k}(  \Psi_{k, \bom(k)} ,W_{k, \bpi(k)}  ) $ to complete the proof. 
 \bigskip

Now we estimate the final density $\cF_K$ for a fixed history $\bpi$.

\begin{lem}  \label{tucson2}  With  $ \La_{K} = \emptyset$, $\La_0 \equiv \Om_1 $ and $\de \La_{j-1} = \La_{j-1} - \La_j$
\be \label{sometimes} 
\| \cF_{K} \|_{\bbI_K, L \bbI_K}
 \leq    \exp \B(  C \sum_{j=1 }^{K}        |  \de  \Om^{(j)}_j  |   +  \sum_{j=1}^{K}   |\de \La_{j-1}^{(j)}  |\B)    
\ee
\end{lem} 
 \bigskip

\pr    From (\ref{super})    $\cF_K =  \exp (-\fS_{K, \bom}    +   B_{K, \bpi} ) $.     For each of  $\fS_{K, \bom} $  and  $B_{K, \bpi} $ we express the object in the fundamental fields $\Psi_{k, \bom}, W_{K, \bom} $  and then estimate the norm of the 
kernel. 
\bigskip

\noindent
\textbf{  A. }   
For $\fS_{K, \bom}$  we  have from (\ref{succotash}) 
 \be \label{tootootoo}
 \fS_{K, \bom} \B(\cA, \Psi_{K, \bom},  \psi_{K, \bom} (\cA)  \B) = \blan \bPsi_{K, \bom}, D_{K, \bom} (\cA) \Psi_{K, \bom} \bran + \dots 
 \ee
 where  
 \be   \label{keykey}
  D_{K, \bom} (\cA) = \bb^{(K)} - \bb^{(K)}\  Q_{K, \bom}(\cA) S_{K, \bom}(\cA) Q^T_{K, \bom}(-\cA)\ \bb^{(K)}
 \ee
First consider the term 
\be
\begin{split}
 \blan \bPsi_{K, \bom}, \bb^{(K)} \  \Psi_{K, \bom} \bran 
= \sum_{j}  \sum_{y \in \de \Om_j^{(j)} }   L^{-3(K-j)}   b^{( K)} _j \bPsi_j(y) \Psi_j (y)
\end{split} 
\ee
Since $\bbI_K$ is $L^{K-j} $ in  $\de \Om_j^{(j)} $  and since  $b_j^{ (K) } = b_j L^{K-j}$ with $b_j$ bounded
   we have
\be \label{lemon1} 
 \|  \blan \bPsi_{K, \bom}, \bb \  \Psi_{K, \bom} \bran  \|_{\bbI_K}  \leq   \sum_{j=1}^K \sum_{y \in \de \Om_j^{(j)} }  L^{-3(K-j)} b^{(K)} _j L^{2(K-j)} 
 =   \sum_j b_j    |\de \Om_j^{(j)}  |  \leq C   \sum_j   |\de \Om_j^{(j)}  |
 \ee
 The second term in (\ref{keykey}) is called $M_{K, \bom}(\cA) $  as in (\ref{runrun}) and we have
\be
\blan \bPsi_{K, \bom}, M_{K, \bom} (\cA) \Psi_{K, \bom} \bran 
 = \sum_{j,j'} \sum_{y \in \de \Om_j^{(j)},  y' \in \de \Om_{j'}^{(j')} }\bPsi_j (y)
 L^{-3(k-j)} M_{K, \bom} (\cA;  y,y')   L^{-3(k-j')}  \Psi_{j'} (y')
\ee
Using the estimate  (\ref{squash0})  on $M_{K, \bom} (\cA, y,y') $ we have  
\be \label{lemon2}
\begin{split}
\| \blan \bPsi_{K, \bom}, M_{K, \bom} (\cA) \Psi_{K, \bom} \bran \|_{\bbI_K}
\leq &   \sum_{j,j'} \sum_{y \in \de \Om_j^{(j)},  y' \in \de \Om_{j'}^{(j')} }\Psi_j (y)
 L^{-2(k-j)} | M_{K, \bom} (\cA;  y,y') |   L^{-2(k-j')}  \\
\leq & 
C \sum_{j,j'} \sum_{y \in \de \Om_j^{(j)},  y' \in \de \Om_{j'}^{(j')} }
 e^{- \ga d_{\bom} (y,y') }  \leq C \sum_{j } \sum_{y \in \de \Om_j^{(j)} }
   \leq      C \sum_{j = 1}^K |\de \Om_j^{(j)}  |  \\
   \end{split} 
\ee 
The other terms in (\ref{tootootoo}) are estimated similarly,  and so $\| \fS_{k, \bom}(\cA) \|_{\bbI_K,L \bbI_K}   \leq   C \sum_j |\de \Om_j^{(j)}  | $
\bigskip

\noindent
\textbf{  B. }   
For the boundary terms let   $\hat B_{K, \bpi } ( \cA, Z_{K, \bom},   \Psi_{K, \bom} , W_{K, \bom} )
$ be   $ B_{k, \bpi } ( \cA,  Z_{K, \bom},  \psi^\# _{K, \bom}(\cA)  ,W_{K, \bom} )$
 with the evaluation  from (54) in \cite{Dim20} 
\be
\psi^\# _{K, \bom}(\cA)  = \hat  \cH^{\#} _{K,  \bom }  (\cA)  \Psi_{K, \bom} \equiv 
  \cH^{\#} _{K,  \bom } (\cA) 1_{\Om_1}  \Psi_{K, \bom}
- S^\#_{K, \bom} (\cA)  \fD_{\cA} \Psi_{0, \Om_1^c} 
\ee
where  $\cH^\#_{k, \bom} (\cA) = ( \cH _{k, \bom}(\cA) , \de_{\al, \cA} \cH_{k, \bom}(\cA) )$, etc. 
 We   suppress the gauge fields $\cA, Z_{K, \bom}$ from the notation   and  write 
 \be
 \begin{split}
\|  \hat B_{K, \bpi } ( \Psi_{K, \bom} , W_{K, \bom} ) \|_{\bbI_K,L \bbI_K} 
= &  \|  \hat B_{K, \bpi }( \bbI_K   \Psi_{K, \bom} , W_{K, \bom} )  \|_{1, L\bbI_K} \\
= &  \|   B_{K, \bpi }( \hat \cH^\#_{K, \bom}  \bbI_K    \Psi_{K, \bom} , W_{K, \bom} )  \|_{1, L\bbI_K } \\
= &  \|   B_{K, \bpi }( (\bbI^\#_K) ^{-1}  \hat \cH^\#_{K, \bom}  \bbI_K    \Psi_{K, \bom} , W_{K, \bom} )  \|_{\bbI^\#_K,L \bbI_K } \\
\end{split} 
\ee
But from the estimate (147) in \cite{Dim20}  we have
\be
|\bbI^\#_K {}^{-1}  \cH^\#_{K, \bom}  \bbI_K f  | \leq   C \| f \|_{\infty} 
\ee
and the same holds for $ \hat \cH^\#_{K, \bom} $. 
Then  from the  appendix in \cite{Dim15b}    
\be \label{toasty} 
\|  \hat B_{K, \bpi }  \|_{\bbI_K, L\bbI_K} 
\leq     \|   B_{K, \bpi }  \|_{C\bbI^\#_K, L\bbI_K } 
\leq     \|   B_{K, \bpi }  \|_{\bh_K \bbI^\#_K,L \bbI_K } 
\ee
We need a bound on the latter quantity. 
We have our basic bound (\ref{sordid}) 
\be
\| B_{j, \bpi} ^{(K)} (X) \|_{\bh_K \bbI^\#_K, L\bbI_K }  \leq     e_K^{\frac14 - 8\ep} \exp\B( - \ka d_{L^{-(K-j)}M} (X  )\B)
\ee
We need to sum this  over connected unions of   $L^{-(K-j)}M$ cubes $X$   with $X \cap  \de \La_{j-1}  \neq \emptyset$.  Instead we let $X=L^{K-j} Y$
and sum over    $M$ cubes.   This gives
\be
\| B_{j, \bpi} ^{(K)}  \|_{\bh_K \bbI^\#_K, L \bbI_K }  \leq     e_K^{\frac14 - 8\ep} \sum_{Y \cap L^{K-j}  \de \La_{j-1}  \neq \emptyset}  \exp\B( - \ka d_M (Y  )\B)
\leq   \one  e_K^{\frac14 - 8\ep} |L^{K-j} \de \La_{j-1} |_M
\ee
Now $L^{K-j} \de \La_{j-1}$  is a subset of  $\bbT^{-j}_{N-j}$,  The number of unit cubes in this set is the number of centers of $L^j$ cubes, which in our notation
 is  $|(L^{K-j} \de \La_{j-1})^{(j)} | $.  This is the
same as $ | \de \La_{j-1}^{ (j)} | $ by the scale invariance of the latter.   Hence the number of $M$ cubes is $|L^{K-j} \de \La_{j-1} |_M = M^{-3}   | \de \La_{j-1}^{( j)} |$.
Now sum over $j$ and get
\be
\| B_{K, \bpi}  \|_{\bh_K \bbI^\#_K, L\bbI_K }  \leq   \one  \sum_{j=1}^K  e_K^{\frac14 - 8 \ep} M^{-3}   |\de \La_{j-1}^{(j)}  |
 \leq  \one     e_K^{\frac14 - 8 \ep}  M^{-3}    \sum_{j=1}^{K} |\de \La_{j-1}^{(j)}  |
\ee
We throw away the small factor $\one    e_K^{\frac14 - 8 \ep}  M^{-3}    \leq 1$.
The resulting bound  combined  with (\ref{toasty})    completes the proof. 
 \bigskip

 \begin{lem} \label{apple1}
  \be \label{english}
  \B |  \cJ_{ K, \bpi}    \B|  \leq     \exp \B(  C \sum_{j=0 }^{K}        |  \de  \Om^{(j)}_j  |   + \sum_{j=1}^{K}    |\de \La_{j-1}^{(j)}  |\B)  
  \ee  
  \end{lem}
  \bigskip

\pr
Start with (\ref{prince}) 
\be
  \cF_{K-1}(   \Psi_{K-1, L\bom} , W_{K-1, L\bpi}) =   \int     \cF_{K }  ( (\Psi_{K,L\bom })_{L^{-1} }  , ( W_{K, L\bpi})_{L^{-1} } ) 
  D \Psi_{K, L \Om_K }   d \mu _{ I, L\Om_{K}} (W_{K-1})
\ee
In this formula      $\Psi_{j} $ is on $\bbT^{-(K-1-j) } _{N-(K-1)} $ 
so $\Psi_{j,L^{-1}} $  is on $\bbT^{-(K-j) } _{N-K} $.  In particular    $\Psi_K $ is on $\bbT^1 _{N-K+1} $ and 
$\Psi_{K-1}  $ is on $\bbT^0 _{N-K+1}$ and so forth.  
Furthermore  $W_{K-1} $ is on $\bbT^0_{N-K}$  so that     $W_{K-1, L^{-1} } $ is on $\bbT^{-1}_{N-K-1}$

 In the $\Psi_K$ integral we change to $\Psi_K =  \Psi'_{K,L} $ with $\Psi'_K$  back  on $\bbT^0 _{N-K} $. 
Then $D \Psi_{K, L \Om_K }  = L^{ |\Om^{(K)} _K | }   D \Psi'_{K,  \Om_K }   $.   To facilitate estimates we  artificially introduce a Gaussian integral by 
$
  D \Psi'_{K,  \Om_K }  = e^{ < \bPsi'_K, \Psi'_K>_{\Om_K}  }  d\mu_{I} ( \Psi'_K) 
 $.
 Then we have
 \be
 \begin{split}
&   \cF_{K-1}(   \Psi_{K-1, L\bom} , W_{K-1, L\bpi})  \\
& =   L^{ |\Om^{(K)} _K | }  \int     \cF_{K } \B ( (\Psi_{K-1,L\bom })_{L^{-1} }, \Psi'_{K, \Om_K}   , ( W_{K-2, L\bpi})_{L^{-1} } ,  (W_{K-1, L\Om_K } )_{L^{-1}}    \B) \\
& \hs \hs \hs 
e^{< \bPsi'_K, \Psi'_K>_{\Om_K } }  d\mu_{I, \Om_K} ( \Psi'_K)  d \mu _{ I, L\Om_{K}} (W_{K-1})\\
& =   L^{ |\Om^{(K)} _K | }   \int     \cF'_{K,L  } \B ( \Psi_{K-1,L \bom  }, \Psi'_{K, \Om_K}   ,  W_{K-2, L\bpi}  ,  W_{K-1, L\Om_K }    \B) \\
& \hs \hs \hs e^{ < \bPsi'_K, \Psi'_K>_{\Om_K } }  d\mu_{I,\Om_K} ( \Psi'_K)  d \mu _{ I, L\Om_{K}} (W_{K-1})\\
\end{split} 
\ee
 where the prime means    the scaling in $ \cF'_{K,L  } $  is in all fields except  $ \Psi'_{K, \Om_K}  $ which is already scaled. 
 
 In general on a unit lattice $ |  \int   f(\Psi )   d \mu_I (\Psi )  | \leq \|f\|_1$, see the appendix in \cite{Dim15b}.  
So  with  $\bbI_{K} = ( L^{K} , \dots , 1)  $ we have $(L\bbI_{K-1},1)= \bbI_K$    and hence  
\be
\begin{split} 
\|  \cF_{K-1} \|_{\bbI_{K-1}, L\bbI_{K-1}}
\leq & \
  L^{ |\Om_K | }  \| \cF'_{K,L} e^ {  <\bPsi'_K, \Psi'_K  >_{\Om_K}}   \|_{  \bbI_{K-1},1, L\bbI_{K-1}, 1}  \\
 \leq  & \   L^{ |\Om^{(K)} _K | }   \| \cF'_{K,L }  \|_{  \bbI_{K-1}, 1,  \bbI_K}   \exp \B(  \|  <\bPsi'_K, \Psi'_K  >_{\Om_K}    \|_1 \B) \\
 \leq  & \  L^{ |\Om^{(K)} _K | }   \| \cF_{K }  \|_{L  \bbI_{K-1}, 1,L\bbI_K}   \exp \B( |  \Om_K^{(K)}  | \B)   \\
 =   & \   \exp \B( C |  \Om_K^{(K)}  | \B)   \| \cF_{K }  \|_{  \bbI_{K},L \bbI_{K} }  \\
 \end{split}
\ee

Now  repeat this estimate.    For the general step start with equation (\ref{alpaca}) expressing $\cF_k$ in terms of  $\cF_{k+1} $.
Here there is an integral over $\Psi_{k+ 1, \de \Om_{k+1} (k)  }$   a field on $\bbT^1_{N-k +1 } $.   We  scale down  to $\Psi_{k+1}  =  \Psi'_{k+ 1, L} $
now with  $\Psi'_{k+ 1}$ 
on  $\bbT^0_{N-k } $.
Then   $(\Psi _{k+1, \de \Om_{k+1} (k) } )_{L^{-1} }   =   \Psi'_{k+ 1, \de \Om_{k+1}(k+1)} $.
Thus we have 
 \be \label{alpaca3}
 \begin{split}
&   \cF_k(   \Psi_{k, \bom(k)} , W_{k, \bpi(k) })  \\
&=   L^{  |\de \Om_{k+1} ^{(k+1)}  | }  \int  \cF_{k+1}\B (  (  \Psi_{k, \bom(k)}  )_{ L^{-1} },    \Psi'_{k+ 1, \de \Om_{k+1}(k+1)}, ( W_{k, \bpi(k) })_{  L^{-1} },
 (W_{k, \Om_{k+1} (k)- \La_{k+1} (k)})_{  L^{-1} } \B) \\ 
& \hs  \hs \hs  D \Psi'_{k+1, \de \Om_{k+1} (k)}  d \mu_{I,\Om_{k+1}(k) - \La_{k+1}(k) } (W_k) \\
 \end{split} 
\ee
Replace $ D \Psi'_{k+1, \de \Om_{k+1} (k)}$ by a Gaussian integral. Then 
 \be \label{alpaca4}
 \begin{split}
&   \cF_k(   \Psi_{k, \bom(k)} , W_{k, \bpi(k) })  \\
&=   L^{  |\de \Om_{k+1} ^{(k+1)}  | }  \int  \cF'_{k+1,L}\B (   \Psi_{k, \bom(k)} ,    \Psi'_{k+ 1, \de \Om_{k+1}(k+1)},  W_{k, \bpi(k)} ,
 W_{k, \Om_{k+1} (k)- \La_{k+1}(k)  }  ) \B) \\ 
&  \exp \B( - \blan \bPsi'_{k+1} ,  \Psi'_{k+1} \bran_ {\de \Om_{k+1}  }\B)  d \mu_{I ,   \de \Om_{k+1} (k+1)}( \Psi'_{k+1}  )             d \mu_{I,\Om_{k+1}(k) - \La_{k+1}(k) } (W_k) \\
 \end{split} 
\ee
where again the prime excludes $\Psi'_{k+1}  $ from scaling.

Now we estimate as before with $(L \bbI_k, 1)   = \bbI_{k+1} $ 
\be
\begin{split} 
\|  \cF_{k} \|_{\bbI_{l}, L\bbI_{k}}
\leq &  \ L^{  |\de \Om_{k+1} ^{(k+1)}  | }  \| \cF'_{k+1,L}      \exp\B(   <\bPsi'_{k+1}, \Psi'_{k+1}   >_{\de \Om _{k+1}}  \B)  \|_{  \bbI_{k},1, L\bbI_{k}, 1}    \\
\leq  & \   L^{  |\de \Om_{k+1} ^{(k+1)}  | }  \| \cF'_{k+1,L }  \|_{  \bbI_{k}, 1,  \bbI_{k+1}}   \exp \B( \| < \bPsi'_{k+1} ,  \Psi'_{k+1} >_ {\de \Om_{k+1} }  \|_1 \B) \\
 \leq  & \  L^{  |\de \Om_{k+1} ^{(k+1)}  | }   \| \cF_{k+1 }  \|_{L  \bbI_{k}, 1,L\bbI_{k+1}}   \exp \B( |  |\de \Om_{k+1} ^{(k+1)}   | \B)   \\
 \leq     &  \   \| \cF_{k+1 }  \|_{  \bbI_{k+1},L \bbI_{k+1} }   \exp \B( C   |\de \Om_{k+1} ^{(k+1)}    | \B) \\
 \end{split}
\ee
 Iterating this inequality  and inserting the bound  (\ref{sometimes}) on $ \| \cF_{K, \bpi} \|_{\bbI_K, L \bbI_K}$ we have
 \be
 \|    \cF_{0}  \|_{\bbI_K}   \leq  \exp \B( C  \sum_{j=1}^{K-1}    | \de \Om^{(j)}_j | \B) \| \cF_{K} \|_{\bbI_K, L\bbI_K }
 \leq   \exp \B(  C  \sum_{j=1 }^{K}        |  \de  \Om^{(j)}_j  |   + \sum_{j=1}^{K}    |\de \La_{j -1}^{(j)}  |\B) 
 \ee
The representation of  $\cJ_{ K, \bpi}$ in terms of  $  \cF_{0, \bpi}  $
is 
\be 
\cJ_{ K, \bpi}   =  \int \cF_{0} (\Psi_0,  L^K \Om_1^c)   
 D \Psi_{0,L^K \Om_1^c}\ 
\ee
This is  also  estimated as a Gaussian integral which gives an additional factor $e^{ | L^K \Om_1^c  | } = e^{ | \Om_1^c | } \equiv   e^{ |\de \Om_0^{(0)}|  }  $.
This completes the proof. 
\bigskip

  We also need a bound on $\sZ'_{k, \bom}(0)$.
  
\begin{lem} \label{apple2}
\be
\frac{\sZ'_{k, \bom}(0) }{  \sZ^f(N,0) }  \leq   e^{(Mr_K)^4 }  \prod_{j=0}^{K-1} \exp \B(  \one  |  \Om^{(j+1),c}_{j+1}| \B)   
\ee
\end{lem}
\bigskip

\pr
We have from (\ref{nuisance}) 
\be 
\sZ'_{K, \bom}(0) =\prod_{j=0}^{K-1}    L^{-8 (s_N - s_{N-j-1} )  }   N_{j+1,L^{K-j} \Om_{j+1}} \de \sZ_{j, L^{K-j} \bom }(0) 
 \ee
 where we recall that  $\de \sZ_{j, \bom }(0) = \de Z_{j, \Om_1, \dots, \Om_{j+1} }(0)$.  Again we define $\bom(j) = L^{K-j} \bom$
 which is a sequence in $\bbT^{-j}_{N-j} $.     
Then taking the expression (\ref{sounder}) for $\sZ^f(N, 0) $ we have
\be 
 \frac{\sZ'_{k, \bom}(0)}{\sZ^f(N, 0)} =   \det (D_K(0) )^{-1} \ \prod_{j=0}^{K-1}
 \frac{ N_{j+1  ,\Om_{j+1}(j) }}{N_{j+1}}  \frac{ \de \sZ_{j,  \bom(j) }(0) }{ \de \sZ_{j}(0) } 
  \ee
By  lemma 10 in \cite{Dim20} and the scale invariance of $ | \Om^{(j)}_{j+1}|$
\be 
\begin{split}
  &  \de \sZ_{j,  \bom(j)}  (0)    
   =      \exp  \B(   \B(    (1 - L^{-3} )     \log  b_j       +  L^{-3}   \log  (     b_j + bL^{-1} )   \B)  | \Om^{(j)}_{j+1}| 
      + \sum_{X \cap \Om_{j+1}(j)  \neq \emptyset } E^d_{j, \bom(j) }(X,0  )  \B)  \\
\end{split}    
\ee
where $| E^d_{j, \bom(j) }(X ,0 ) | \leq CM^3 e^{-\ka d_M(X) } $.    
  We want to compare this with  the global
\be 
    \de \sZ_{j } (0)      
   =     \exp  \B(   \B(    (1 - L^{-3} )     \log  b_j       +  L^{-3}   \log  (     b_j + bL^{-1} )   \B)  | \bbT^{0}_{N-j}|  +
    \sum_{X } E^d_{j}(X,0 )  \B) \\
\ee
But   $E^d_{j, \bom}(X, 0 ) $ is defined   with a random walk expansion in $X$ \cite{Dim20},  and if  
 $X \subset \Om_{j+1}$ this is independent of $\bom$.     So
 $E^d_{j, \bom(j)}(X, 0 ) =E^d_j(X,0  ) $ for  $X \subset \Om_{j+1}(j)$ 
and  we have for the ratio 
\be
\begin{split}
\frac{  \de \sZ_{j,   \bom(j)} (0)      }{\de \sZ_{j } (0)   }
= &   \exp  \B(   \B(    (1 - L^{-3} )     \log  b_j       +  L^{-3}   \log  (     b_j + bL^{-1} )   \B)     | \Om^{(j),c}_{j+1} |  \\
&   +
    \sum_{X \#\Om^c_{j+1}(j) } E^d_{j, \bom(j) }(X,0 )  -\sum_{X \cap \Om^c_{j+1}(j)  \neq \emptyset  }
     E^d_j(X,0  ) \B) \\
    \end{split}
\ee
But the first sum in the exponential is bounded by  
\be
CM^3  \sum_{X \cap \Om^c_{j+1}(j)  \neq \emptyset  }e^{-\ka d_M(X) } \leq CM^3 |  \Om^c_{j+1}(j) |_M  
= C  \Vol ( \Om^c_{j+1}(j) )  =  C |  \Om^{(j),c} _{j+1} | 
\ee
The second sum has the same bound
and so 
\be \label{woe1}
\frac{  \de \sZ_{j,   \bom(j) } (0)      }{\de \sZ_{j } (0)   }
\leq   \exp \B( C   |  \Om^{(j),c}_{j+1}| \B)    
\ee
We also have  the  $\log   N_{j+1,\Om_{j+1}(j) } $ is a constant times   $ | ( \Om_{j+1}(j) )^{(j+1)}| =  |   \Om_{j+1}^{(j+1)}|  $ and so 
\be  \label{woe2}
 \frac{ N_{j+1,\Om_{j+1}(j) }  }{ N_{j+1 } }  =  \exp \B( C   |  \Om^{(j+1),c}_{j+1}| \B)  
\ee
The lemma  follows from  (\ref{woe1}), (\ref{woe2})  and the following result.
\bigskip

\begin{lem} \label{strawberry} 
For   $C_K(0)  =D_K(0) ^{-1} $  satisfies
\be
  |\det C_K(0)|  \leq   \exp(  (Mr_K)  ^4)
\ee
\end{lem}
\bigskip

\pr    $D_K(0) = b_K - b_K^2 Q_k(0) S_K(0) Q^T_K(0)$ is  an operator on $\bbT^0_{N-K}  $   with $L^{N-K} = Mr_K$ sites in each direction.
The inverse  $C_K(0) = D_K(0)^{-1} $ is given by 
\be
C_K(0) = b_K^{-1}  + Q_K(0) ( \fD_0 + \bar m_K )^{-1}Q^T_K(0)
 \ee
This can be verified directly,  or it is  identity in Appendix B in \cite{Dim15b} in the special case $b=0,y=0$.  
The propagator $ ( \fD_0     + \bar  m_K  )^{-1}$ has no projection operator  as did $S_K(0) = ( \fD_0     + \bar  m_K + b_K P_K(0) )^{-1}$.  But now we do not need it 
since we have $\bar m_K =  L^{-2(N-K)}  \bar m $ which is positive and bounded below independent of $N$ by lemma \ref{sync}.
We  note that    since    $  \fD_{0}   =  \ga \cdot \nabla  -  \frac 12   \eta  \De $,  and since $ \ga \cdot \nabla  $ is skew adjoint  we
have  with  $\eta  = L^{-(N-K) }$  and   $L^2$ norms    
\be
 | (f,   ( \fD_{0}     + \bar  m_K ) f) | \geq   | \re (f,   ( \fD_{0}     + \bar  m_K )  f)  | =     \frac 12 \eta  (f, (-  \De + \bar m_K) f)  \geq    \frac 12 \eta \bar m_K  \|f \|^2 
\ee
Hence  $\|  ( \fD_{0}     + \bar  m_K )  f  \|   \geq    \frac 12 \eta \bar m_K  \|f \| $ and so 
\be    
\|  ( \fD_{0}     + \bar  m_K )^{-1}  f  \|   \leq    2 \eta^{-1} \bar m_K^{-1}  \|f \|   = 2 L^{3(N-K )}   \bar m^{-1} \|f\|
\ee
This is the one place where we are assuming  the bare mass $\bar m \neq 0$.  We are not keeping track of how our estimates depend on $\bar m$ and
include the factor $\bar m^{-1}$ in our generic constant $C$.   Since $b_K^{-1} $ and $Q_K(0)$ are bounded we  have
$ \| C_K(0) \| \leq  C  L^{3(N-K)}  = C(Mr_K )^3$.   Hence every matrix element  satisfies $| C_K(0, \sx,\sy)  | \leq   C(Mr_K )^3$. 

We take a crude bound on $\det C_K(0)$ expanding it as a sum of permutations.    There are $((Mr_K)^3) !  \leq \exp( \frac12 (Mr_K)^4) $ permutations 
with the contribution of each permutation  bounded by  $[ C(Mr_K )^3 ]^{(Mr_K)^3} $ which is less than $ \exp( \frac12 (Mr_K)^4)$. 
Altogether we get the stated bound  $\exp(  (Mr_K)^4)$.

 \subsubsection{boson  integral  - large field region}  
 
 The treatment of the boson integral is necessarily quite different from the fermion case since our renormalization group  transformations were not just averaging
 but  contained gauge fixing as well.   In addition this is where we extract the small factors enabling the convergence of the whole expansion
We start with estimates on the large field characteristic functions.   The following result refers to the  characterstic functions as they were when created
after $k$ streps,  and before relabling and scaling in the next $K-k$ step. 

\begin{lem}  \label{presence}  On $\tk$ with axial  $A_k$  on $\tz$   and $A_{k+1} = \cQ A_k$
\label{bowwow}
\be
\begin{split}
 \zeta^0_{k+1}(P_{k+1})     
\leq      &    \exp   \B(    -    p_{k+1}( Mr_{k+1})^{-3}   |P^{(k+1)}_{k+1}|  +  C M^2      p_{k+1}^{-1} \| dA_{k} \|^2_{( \La_k- \Om_{k+1})^{2\nat}}    \B)    \\
  \zeta'_k (Q_{k+1})    
\leq      &    \exp   \B(    -    p^2_{0,k} ( Mr_{k+1})^{-3} |Q^{(k+1)}_{k+1}|   + C M^2    p_{0,k}^{-2} \| dA_{k} \|^2_{( \La_k- \Om_{k+1})^{2\nat}}  \B)    \\
 \zeta_k^{\da} (R_{k+1})
  \leq  &    \exp \B(   -   p^{\frac43} _{0,k}( Mr_{k+1})^{-3}  |R^{(k+1)}_{k+1}|    + Cp_{0,k}^{-\frac43} \| Z_k \|^2_{ (\Om_{k+1}- \La_{k+1})^{\nat} } \B)  \\
  \hat \zeta_k (U_{k+1})
  \leq  &    \exp \B(   -   p_{0,k}( Mr_{k+1})^{-3}  |U ^{(k+1)}_{k+1}|    + C p_{0,k}^{-1}  \| Z_k \|^2_{ \Om_{k+1}- \La_{k+1 } } \B)  \\
 \end{split}   
\ee
\end{lem}  
\bigskip

\pr
We start with  the bound on     $  \zeta^0_{k+1}(P_{k+1} ) $ as defined in (\ref{not}),(\ref{tiger1}).      For  $ \square \subset  P_{k+1}$  the characteristic function
 $   \zeta^0_{k+1}(\sq ) $  enforces   there is at least one bond  in $  \tilde    \sq$ 
where   $| d \cA^0_{k+1,  \bom^+(\sq ) } |  \geq L^{- \frac32}  p_{k+1}  $.  
  We need control over  $d \cA^0_{k+1,  \bom^+(\sq ) }$  and we start with  the closely related $\cA_{k,  \bom(\sq ) }$. 
 The field as defined in (\ref{pool})  is given by
\be
\cA_{k,  \bom(\sq ) } = \cH_{k, \bom(\sq ) } \hat A_{k, \bom} \hs \textrm{ where } \ \ \   \hat A_{k, \bom}   = \cQ_{k, \bom(\sq) } \cQ^{s,T} _k A_k
\ee
By our basic regularity bound,  Theorem  1 in \cite{Dim20},  with weight factors $p_j = L^{\frac12(k-j) } $ we have in $\de \Om_{j'} (\sq) $
\be \label{tinker1} 
| d\cA_{k,  \bom(\sq ) }  | \leq CM  L^{2(k-j')}  \sup_{j}     L^{-2(k-j)}  \| d   \hat A_{k, \bom} \|_ { \infty,   \de  \Om^{(j)} _j (\sq) }
\ee
We are only interested in a bound on $\tilde \sq \subset \Om_k(\sq)$  where the factor $ L^{2(k-j')}  =1$. 
Now $d \hat A_{k, \bom}   = \cQ^{(2)} _{k, \bom(\sq) }d \cQ^{s,T} _k A_k =   \cQ^{(2)} _{k, \bom(\sq) } \cQ^{e,T} _k dA_k $ for edge averaging $ \cQ^{e} _k $.
For an edge plaquette  $p' \in \bbT^{-k}_{N-k}$ the  $(\cQ^{e,T} _k dA_k)  (p')= L^{2k}  dA_k( p) $ where $p$ in $\bbT^0_{N-k} $ is the unique plaquette
containing   $p'$.   Then for $p \in   \de \Om^{j)}_j(\sq)  \subset \bbT^{-j}_{N-k} $  
\be
(d \hat A_{k, \bom} )(p)  =    ( \cQ^{(2)} _j \cQ^{e,T} _k dA_k ) (p) = L^{-5j} \sum_{x \in B_j(y) } \sum_{p' \in p_x}  ( \cQ^{e,T} _k dA_k)(p')   
\ee
where $p_x$ is  $p$ translated so a fixed  corner is  $x$.  There is
only one term in the sum $p' \subset  p_x$ and it is   $ L^{2k}  dA_k( p)$ independent of $x$.      We use the bound $|(\cQ^{e,T} _k dA_k)  (p)| \leq  L^{2k}   \| dA_k \|_{\infty, \Om_1(\sq) }    $ 
and the fact that there are $L^{ 3j}$ terms in the sum over $x$ to give on  $ \de \Om^{j)}_j(\sq) $
\be
|d \hat A_{k, \bom}(\sq)  |  \leq  L^ { 2(k-j)  }\| dA_k \|_{\infty, \Om_1(\sq) } 
\ee
Using this in (\ref{tinker1}) yields on $\tilde \sq$
\be    \label{timely} 
| d\cA_{k,  \bom(\sq ) }  | \leq CM  \| dA_k \|_{\infty, \Om_1(\sq) } 
\ee
Now take this bound for $k+1$  on $\bbT^{-k-1}_{N-k-1} $  and scale up to the bound  for  $LMr_{k+1} $ cubes  $ \sq$ in $\tk$ 
\be \label{timely2} 
|  d\cA^0_{k+1,  \bom^+ (\sq)  } |  \leq CM   \| dA_{k+1}  \|_{\infty,  \Om^+_1(\sq) } 
\ee
But $dA_{k+1} = \cQ^{(2)} dA_k$ and $\Om^+_1(\sq) \subset \sq^{\sim 3} $ so  
$ \| dA_{k+1} \|_{\infty, \Om_1^+(\sq)} \|  \leq \| dA_k \|_{\infty, \sq^{\sim 3} }$. 
Thus we have on $\tilde \sq$
\be \label{sound} 
|  d\cA^0_{k+1,  \bom^+(\sq)   } |  \leq CM   \| dA_k \|_{\infty, \sq^{\sim 3} }\ee

Having established this preliminary bound we  now argue as follows.   
The $   \zeta^0_{k+1}(\sq ) $  enforces that there  must be a bond in $\sq^{\sim 3}$ where
$ | dA_k | \geq   ( CM  ) ^{-1}L^{-\frac32} p_k $ since otherwise  (\ref{sound}) says  $| d \cA^0_{k+1,  \bom^+(\sq ) } |  \leq L^{- \frac32}  p_{k+1}  $
everywhere in $\tilde \sq$ which contradicts the definition of     $ \zeta^0_{k+1}(\sq )$.  Absorb the $L^{-\frac32}$  in the $C^{-1} $ so  $   \zeta^0_{k+1}(\sq ) $  enforces   there is at least one bond  in $  \sq^{\sim 3} $ 
so that  $ | dA_k | \geq   ( CM  ) ^{-1}  p_{k+1} $.   Now we claim that for the unit lattice field $A_k$
  \be  \label{licorice} 
  \zeta^0_{k+1}(\square) \leq  \exp \B(   -      p_{k+1}   +  ( CM)^2  p_{k+1}   ^{-1}   \|   dA_k  \|^2_{ \sq^{\sim 3}}  \B)  
\ee
Indeed if $ \zeta^0_{k+1}(\square)  =0$ the inequality is trivial, while if  $ \zeta^0_{k+1}(\square)  =1$ 
then   $\| d A_k\|^2_{ \sq^{\sim 3} }  \geq  ((CM) ^{-1}   p_{k+1} )^2 $ and the inequality holds.

Now take the product over  $\sq \subset P_{k+1}$ to get a bound on $ \zeta^0_{k+1}(P_{k+1} ) $ .   The volume of $P_{k+1}$ is 
 $| P^{(k) } _{k+1} | = L^3| P^{(k+1) } _{k+1} | $.   Each cube has 
volume  $(LMr_{k+1} )^3$ so the number of cubes in $P_{k+1}$ is  $ (Mr_{k+1} )^{-3} | P^{(k+1) } _{k+1} | $. 
We also use 
\be
 \label{kingkong}
 \sum_{\sq \subset P_{k+1} }  \|  d A_{k} \|_{ \sq^{\sim 3} }^2 =  \sum_{\sq \subset P_{k+1} } \sum_{\sq' \subset   \sq^{\sim 3} }   \|  d A_{k} \|_{ \sq' } ^2 
 = \one   \sum_{\sq' \subset   P^{\sim 3} _{k+ 1} }  \|  d A_{k} \|_{ \sq' } ^2  =  \one   \|  d A_{k} \|^2_{ P^{\sim 3} _{k+1}} 
 \ee
  But $ P_{k+1}  \subset \La_k^{ 5 \nat} $ implies  $ P_{k+1}^{\sim 3}   \subset \La_k^{ 2 \nat} $ (since  $[\cO^{\nat}]^{\sim}  \subset \cO$) 
  and   $ P_{k+1}^{\sim 5} \subset \Om_{k+1} ^c$ implies $  P_{k+1}^{\sim 3} \subset  [  P_{k+1}^{\sim 5}]^{2 \nat}  \subset  [\Om_{k+1} ^c]^{2 \nat} $
  (since $\cO \subset [\tilde \cO ]^{\nat}$).  
  Thus    $P_{k+1} ^{\sim 3} \subset \La_k^{ 2 \nat} \cap     [ \Om_{k+1} ^c ]^{2 \nat} = [ \La_k - \Om_{k+1} ]^{2 \nat}$
   (by  $(\cO_1 \cap \cO_2)^{\nat} =\cO_1^{\nat}  \cap \cO_2^{\nat}$ ), which we use in (\ref{kingkong}). 
 Altogether  the announced bound   on  $ \zeta^0_{k+1}(P_{k+1} ) $  results.

\bigskip 

 Now consider the  $\zeta'_k( Q_{k+1}  )$  bound.  
For $\sq \subset Q_{k+1}$ the   characteristic function $\zeta'_k ( \square )  $  enforces that  there is at least one bond  in $  \tilde    \square$
where  $|A_k  - A^{\min}_{k, \bom(\square)} |  \geq   p_{0,k}^2 $.   We need control of  this field. 
Now    $  A^{\min}_{k, \bom^+(\sq)  } -  A_k$  is  axial and   $ \cQ (A_k-  A^{\min}_{k, \bom^+(\sq)  } )  = A_{k+1} - A_{k+1}  =0$.
It follows from lemma 16 in \cite{Dim20}  that on $\tilde \sq$
\be
| A_k-  A^{\min}_{k, \bom^+(\sq)  }   |   \leq  C  \| d (A_k-  A^{\min}_{k, \bom^+(\sq)  }  )  \|_{\infty, \tilde \sq^+  }  \leq
 C  \| d A_k \|_{\infty, \tilde \sq^+  } +  C  \| d  A^{\min}_{k, \bom^+(\sq) }  \|_{\infty, \tilde \sq^+  }
\ee
where $\tilde \sq^+$ is a slight enlargement of $\tilde \sq^+ $. 
Now    $A^{\min}_{k, \bom^+(\sq)  } = \cQ _k  \cA^{0,\sx} _{k+1, \bom^+(\sq)} $.  On $\tilde \sq ^+  $ we   change to Landau gauge and use the fundamental regularity estimate  (\ref{sound}) 
 to get
\be
| d  A^{\min}_{k, \bom^+(\sq)  }  | =  | \cQ^{(2)} _k d \cA^0_{k+1, \bom^+ (\sq)}  |   \leq  CM \| dA_{k} \|_{\infty,    \sq^{\sim  3 } }
\ee
Therefore on $\tilde \sq$
\be  \label{sound2} 
 |  A_k -  A^{\min}_{k, \bom^+(\sq)  }  |    \leq  CM \| dA_{k } \|_{  \infty,  \sq^{\sim  3 }  } 
\ee
Now $\zeta'_k ( \square )  $  enforces that  there must be a bond  in $\sq^{\sim 3} $ such that $|dA_{k }| \geq (CM)^{-1} p_{0,k}^2$ since otherwise (\ref{sound2}) says  that  $ |  A_k -  A^{\min}_{k, \bom^+(\sq)  }  |  \leq p_{0,k}^2$
everywhere in $\tilde \sq$ which contradicts the definition of   $\zeta_k'( \sq )$.    
Therefore 
\be   \label{quip} 
 | \zeta' _k(\square) |   \leq   \exp \B(   -   p^2_{0,k}     +    (CM )^2 p_{0,k}^{-2}    \| dA_k \|^2_{ \sq^{\sim  3 } }  \B)  
\ee
Take the product over $\sq \subset Q_{k+1} $ to get the result as before.

\bigskip
Now consider the   $ \zeta^{\da} (R_{k+1})$   bound.   For $\sq \subset R_{k+1}$ the  characteristic function  $ \zeta^{\da} (\sq) $  enforces  there is at least one point  in $\tilde   \square$
where   $|C^{\frac12, \loc}_{k, \bom^+ } \tilde Z_k |  \geq   p^{4/3}_{0,k}$.   
However   $|C^{\frac12, \loc}_{k, \bom^+ }  (\Up, \Up') | \leq e^{- \ga d(\Up,\Up') } $ and the operator only connects points in the same $\tilde \sq$.  Therefore 
on $\tilde \sq$ we have  $|C^{\frac12, \loc}_{k, \bom^+ }  \tilde  Z_k| \leq C \| \tilde Z_k \|_{\infty, \tilde \sq}$.
This implies that  there must be a point in $\tilde \sq $ where  $|\tilde Z_k | \geq C^{-1}   p^{4/3}_{0,k}$. Thus  
 \be \label{cursty2}
    \zeta^{\da} (\square)  \leq   \exp \B(   -   p^{\frac43} _{0,k}    +   C^2    p_{0,k} ^{-\frac43}  \| \tilde Z_k \|^2_{ \tilde \sq}   \B)  
\ee
Take the product over $\sq \subset R_{k+1} $.      We have    $\tilde R_{k+1}  \subset \Om_{k+1}^{\nat} $ and $\tilde R_{k+1}  \subset [\La^c_{k+1} ] ^{\nat} $
and therefore  $ \tilde R_{k+1}  \subset [\Om_{k+1}  -\La_{k+1} ] ^{\nat} $ which gives the result.      

\bigskip
Now consider  the $  \hat \zeta (U_{k+1})$ bound.  For $\sq \subset U_{k+1}$ the   characteristic function  $   \hat \zeta (\sq )$
 enforces  there is at least one point  in $  \square$
where   $|\tilde Z_k |  \geq   p_{0,k}$.  
and so 
\be 
\hat     \zeta  (\square)  \leq   \exp \B(   -   p_{0,k}    +    p_{0,k} ^{-1}      \| Z_k \|^2_{  \square}     \B)  
\ee
Again take   the product over $\sq \subset U_{k+1}  $.   Since $U_k \subset \Om_{k+1} - \La_{k+1}$  we have.  the result. This completes the proof. 

\bigskip

\centerline{ ------------------------} 
\bigskip

Now back to the main story.    For all large field regions  we have from (\ref{zL})
   \begin{equation}    
\begin{split}
&    \sZ^{\cL} (N,e)  =      \sum_{\bpi:  \La_K = \emptyset  }   \sZ'_{k, \bom} (0)   \sZ'_{k, \bom}        \\
  &\int   \de (  \cQ^* A_K  )    \de  ( \tau^* A_K)  D   A_K    \   Dm_{K, \bom} ( A )      Dm_{K, \bpi} (Z) \
\B[ \cJ_{K, \bpi}    \  \cC_{K, \bpi} \B]  \exp\B(- \frac12 \|d \cA_{K, \bom} \|^2 \B)      \\
\end{split}
\end{equation}

We bound  the fermion integral  $\cJ_{K, \bpi}$  by  lemma \ref{apple1}.    For $0 \leq j \leq K-1$ we use $
| \de \Om_j^{(j)}| \leq | \Om^{(j),c}_{j+1} | = L^3     | \Om^{(j+1),c}_{j+1} |  \leq    | \La^{(j+1),c}_{j+1} |  
$.   For $j=K$ we use
 $ | \de \Om_K^{(K)}|  =    |   \Om_K^{(K)}|  \leq  |  \bbT^0_{N-K} | = |\La^{(K),c} _K |$
 Hence we have $ | \cJ_{K, \bpi}|  \leq  \exp ( C \sum_{j=1}^K    | \La^{(j),c}_{j} | )  $.     
 Similarly 
 $\sZ'_{k, \bom}(0)    \leq   e^{(Mr_K)^4 }  \sZ^f(N,0)  \prod_{j=0}^{K-1} \exp ( C \sum_{j=1}^K    | \La^{(j),c}_{j} | ) 
$ 
 from  lemma  \ref{apple2}. Thus 
   \begin{equation}     \label{startrep2}
\begin{split}
    \sZ^{\cL} (N,e)   \leq  &   e^{  (Mr_K)^4}  \  \sZ^f(N,0)     \sum_{\bpi:  \La_K = \emptyset  }  \sZ'_{k, \bom}     \exp   \B(  C\sum_{j=1}^K    | \La^{(j),c}_{j}  |  \B)
\\
&   
  \int   \de (  \cQ^* A_K  )    \de  ( \tau^* A_K)   \  DA_K      Dm_{K, \bom} ( A )        Dm_{K, \bpi} ( Z ) \
 \cC_{K, \bpi} \  \exp\B(- \frac12 \|d \cA_{K, \bom} \|^2 \B)   
 \\
 \end{split}
\end{equation}
In $\cC_{K,\bpi} $ we estimate a small field characteristic functions by one.  Taking account the  relabeling and rescaling we have that   $ \cC_{K,\bpi}   $ 
is bounded by   (with  $\cO(j) = L^{K-j} \cO$)  
\be
  \prod_{j=0} ^{K-1}\B[ \zeta^0_{j+1} (  P _{j+1}(j) )\B  ]_{ L^{-(K-j) } } \B[ \zeta'_{j} (  Q _{j+1} (j))\B  ]_{ L^{-(K-j) }}
\B[ \zeta^{\dagger} _{j} (  R _{j+1} (j))\B  ]_{ L^{-(K-j) } }\B[ \hat \zeta _{j} (  U _{j+1}(j) )\B  ]_{ L^{-(K-j) } } 
\ee
Now use the bounds of lemma  \ref{presence}.     In the bounds on   $\zeta^0_{j+1} (  P _{j+1})$  and  $\zeta'_{j+1} (  Q_{j+1})$ 
we use  
\be
\B[  \| dA_{j} \|^2_{( \La_j- \Om_{j +1})^{2\nat}(j) } \B]_{L^{-(K-j) } }= \| dA_{j, L^{(K-j) }} \|^2_{( \La_j- \Om_{j +1})^{2\nat}(j) }  =  \| dA_{j} \|^2_{( \La_j- \Om_{j +1})^{2\nat}  }
 \ee  
 Here $A_j$ is on $\bbT^{-(K-j)}_{N-K} $.
 In the decay factors take the minimum coefficient which is $p_{0,j} \geq 2 p_{0,j+1} $  and shift $j \to j-1$.  Then
  \be  \label{sinister} 
\begin{split}
  \sZ^{\cL} (N,e)  
  \leq   &    e^{  (Mr_K)^4}  \   \sZ^f(N,0)   \sum_{ \bpi_0   :  \La_K  =  \emptyset}   \sZ'_{k, \bom} \exp   \B(    -    \sum_{j=1 }^K  2 p_{0,j}( Mr_{j })^{-3}  ( |P^{(j)}_j|    +  |Q^{(j)}_j|    + |R^{(j)}_j| + |U^{j)}_j |   )   \B) \\
  &
     \exp   \B(  C \sum_{j=1}^K    | \La^{(j),c}_{j} |\B)   \int  DA_K \de (  \cQ^* A_K  )\    \de  ( \tau^* A_K)  \ 
  \cI_{K, \bpi}   (A_K)
     \\
\end{split}
\ee
where 
\be      \label{slim2} 
\begin{split}
&\cI_{K, \bpi}(A_K)
=  \int        \   Dm_{K, \bom} ( A )      \   Dm_{K, \bpi} ( Z )  \exp\B(- \frac12 \|d \cA_{K, \bom} \|^2 \B)  \\
& \exp \B(  \sum_{j=0}^ {K-1}       C M^2   p_{0,j}^{-2} \| dA_{j} \|^2_{( \La_j- \Om_{j +1})^{2\nat}   } 
 + Cp_{0,j}^{-1} \B[ \| \tilde  Z_j    \|^2_{   (\Om_{j+1} - \La_{j+1})^{\nat} (j)     } \B]_{ L^{-(K-j)} }   \B)
  \\
 \end{split} 
\ee

\begin{lem}  \label{singsong}
 \be      \label{slim4} 
\cI_{K, \bpi}(A_K) \leq     \exp   \B(    \sum_{j=1 }^{K}    C   |  \La^{(j),c}_j  |  \B)
    \int    \exp\B(- \frac12 \| d\cA_{K, \bom} \|^2 \B)   \   Dm_{K, \bom} ( A )   
\ee
\end{lem} 
\bigskip

\pr  This factors into an integral over $A$ fields and $Z$ fields.  
For the $Z$ integral we have 
 \be
\begin{split}
&  \int          \   Dm_{K, \bpi} ( Z )  \exp \B(  \sum_{j=0}^ {K-1}     Cp_{0,j}^{-1}  \B[ \| \tilde  Z_j    \|^2_{   (\Om_{j+1} - \La_{j+1})^{\nat} (j)     } \B]_{ L^{-(K-j)} } \B)  \\
=&   \prod_{j=0}^{K-1}  \int  \ \B[  d \mu_{ I,  ( \Om_{j+1} - \La_{j+1})(j)   }(\tilde Z_j ) \B]_{L^{-(K-j) } }    
 \exp \B(  \sum_{j=0}^ {K-1}     Cp_{0,j}^{-1}  \B[ \|\tilde   Z_j    \|^2_{   (\Om_{j+1} - \La_{j+1})^{\nat} (j)     } \B]_{ L^{-(K-j) }   } \B)  \\
=&   \prod_{j=0}^{K-1}  \int  \  d \mu_{ I,  ( \Om_{j+1} - \La_{j+1})(j)   }(\tilde Z_j )    
 \exp \B(  \sum_{j=0}^ {K-1}     Cp_{0,j}^{-1}   \|\tilde   Z_j    \|^2_{   (\Om_{j+1} - \La_{j+1})^{\nat} (j)     }  \B)  \\
\leq  &   \prod_{j=0}^ {K-1}  \exp \B(  Cp_{0,j}^{-1}    |  \Om^{(j)} _{j+1} -  \La^{(j)}_{j+1}  |  \B ) 
\leq    \exp   \B(    \sum_{j=1 }^{K}       |  \La^{(j),c}_j  |  \B)  \\
\end{split} 
\ee
Here $\tilde Z_j$ is on $\bbT^0_{N-j} $ and    $(\Om_{j+1} - \La_{j+1})(j) $ is a subset of $\bbT^{-j}_{N-j} $.   There are
 $| (\Om_{j+1} - \La_{j+1})^{(j) } | $ variables in the integral. 

For the $A$ integral we show 
in appendix \ref{lower}  that 
\be \label{fourthirteen}
\sum_{j=0}^{K-1} \| dA_j \|^2_{( \La_k- \Om_{k+1})^{\nat}} 
\leq    \one \| dA_{K, \bom} \|^2
\ee
by taking   $p_0$ sufficiently large so that    
we can make $  C M^2   p_{0,j}^{-2}$ as small as we like and  the integral is estimated by
 \be \label{samsung} 
 \begin{split}
 & \int    \exp\B(- \frac12 \| \cA_{K, \bom} \|^2 \B)   \exp \B(  \sum_{j=1}^{K}     C M^2   p_{0,j}^{-2} \| dA_j \|^2_{( \La_j - \Om_{j+1})^{\nat} }   \B)
   \   Dm_{K, \bom} ( A )  \\
    &  \leq        \int    \exp\B(- \frac14 \| \cA_{K, \bom} \|^2 \B)    \   Dm_{K, \bom} ( A )              \\
\end{split} 
\ee
Now 
 $
 \|d \cA_{k, \bom} \|^2 )=  <A_{K, \bom}, \De_{K, \bom} A_{K, \bom} >$ so we can change the coefficient $\frac14$  in    (\ref{samsung})  to $\frac12$ 
 by a change of variables   $ A_{K, \bom}  \to \sqrt{2}   A_{K, \bom} $ which means $A_{j, \de \Om_j } \to \sqrt {2} A_{j, \de \Om_j } $ for $j=0 \leq j \leq  K-1$.
If the measure $  Dm_{K, \bom} ( A ) $ had no delta functions  the change of variables  would generate a factor  of $\sqrt 2$
raised to the power  $ \sum_{j=0} ^{K-1} | \de \Om_j^{(j)} |  $.   The delta functions reduce the power but the expression is still a good upper bound, 
as is $  \exp ( \sum_{j=1 }^{K}    C   |  \La^{(j),c}_j  | ) $.     So the $A$ integral is bounded by  
\be
    \exp   \B(      \sum_{j=1 }^{K}  \one      |  \La^{(j),c}_j  |  \B)   \int    \exp\B(- \frac12 \| \cA_{K, \bom} \|^2 \B)    \   Dm_{K, \bom} ( A )  
\ee 
which gives the result.

\begin{lem} 
 \label{omega1}
 For $e$ sufficiently small 
\be  
\sZ^{\cL} (N,e)  \leq   \sZ(N,0)   e_K  
\ee
\end{lem} 
\bigskip

\pr 
Insert the bound on $\cI_{K, \bpi} $ in (\ref{sinister}) and get
\be  \label{sinister2} 
\begin{split}
  \sZ^{\cL} (N,e)  
  \leq  &  \ 
          e^{  (Mr_K)^4}  \   \sZ^f(N,0) 
        \sum_{ \bpi   :  \La_K  =  \emptyset}    \exp   \B(   -    \sum_{j=1 }^K  2 p_{0,j}( Mr_{j })^{-3} ( |P^{(j)}_j|    +  |Q^{(j)}_j|    + |R^{(j)}_j|  + |U^{j)}_j | ) 
       \B) \\
   &
  \exp\B(  C \sum_{j=1  }^{K}     |  \La^{(j),c}_j  | \B)   \B[   \sZ'_{k, \bom}   \int   \de (  \cQ^* A_K  )    \de  ( \tau^* A_K)   DA_K   Dm_{K, \bpi} ( A ) 
   \exp\B(- \frac12 \| \cA_{K, \bom} \|^2 \B)   \     \B] \\
  \end{split}
\ee
But  by (\ref{twice}) the bracketed expression   is identified as the free boson partition function  $  \sZ^b(N,0)   $.
 Also  we   identify  $\sZ(N,0)  =\sZ^b(N,0) \sZ^f(N,0) $ and have
   \be   
\begin{split}
  \sZ^{\cL} (N,e)  
  \leq  &  \ 
        e^{  (Mr_K)^4}  \   \sZ(N,0) \\
 &
       \sum_{ \bpi   :  \La_K  =  \emptyset}    \exp   \B(   -    \sum_{j=1 }^K  2 p_{0,j}( Mr_{j })^{-3}( |P^{(j)}_j|    +  |Q^{(j)}_j|    + |R^{(j)}_j|  + |U^{j)}_j | ) 
        +  C \sum_{j=1  }^{K}    |  ( \La_{j}^{(j),c}  |  \B) \\
    \end{split}
\ee

Without the condition $\La_K \neq \emptyset$
this  is a model independent  sum over  $\{ P_j,Q_j, R_j,U_j\}_{j=1}^K$   first estimated in  \cite{Bal83a} where it is bounded by a constant; see also  \cite{Dim14}.
The key point as before  is that if $p_0$ is sufficiently large then  the constant $ p_{0,j}(Mr_{j })^{-3}$ can be as large power of $(-\log e_K)$  as we like to drive the convergence.    
Since the case  with all the $P_j, Q_j, R_j,U_j   = \emptyset$ is excluded we  can extract  a  tiny factor $\exp ( - 2 p_{0,j} ( Mr_{j })^{-3}) \leq  \exp ( - 2 p_{0,K} (Mr_{K })^{-3})$
from somewhere.   Again assuming  $p_{0,K} $ is sufficiently large,  this is enough to dominate the factor $ \exp(   (Mr_K)^4 ) $ and  any constants and still leave
a factor smaller than  $e_K$.   Then we have the announced 
$ \sZ^{\cL} (N,e)  \leq   \sZ(N,0)    e_K  $. 
\bigskip

\subsubsection{fermion integral  - small field region} 

We  look at the fermion contribution to  $\sZ^{\cS}(N,e)$ defined in (\ref{snuffit}).    With $\La_K  = \bbT^{-K}_{N-K}$ 
and replacing $\fS_K ( \cA_K, \Psi_K, \psi_K (\cA_K) ) $ by $<\bPsi_K,  D_K(\cA_K)  \Psi_K>$ it is    
  \be
\Xi_K(\cA_K)     \equiv   Z'_K(0)  \int   D\Psi_K   \exp  \B(    -  \blan \bPsi_K,  D_K(\cA_K)  \Psi_K  \bran  
  +    E_K(\La_K,  \cA_K,   \psi^\#_K(\cA_k))  \B)   
\ee
We want to compare this with the free fermion partition function from (\ref{sounder}) which is 
\be   \label{fine}
\sZ^f(N,0) =   \sZ'_K(0)  \int   D\Psi_K   \exp  \B(    -  \blan \bPsi_K,  D_K(0)  \Psi_K  \bran     \B)   
\ee
We study this for  $|d\cA_K| \leq p_K $ which (by definition)  is  enforced by $\chi_K $.

\begin{lem} \label{early} For  $|d\cA_K| \leq p_K $ and uniformly in $N$  
\be
 \frac{\Xi_K(\cA_K)  }{ \sZ^f(N,0)}   =  1 + \cO( e_K ^{\frac14 - 8 \ep})
\ee
\end{lem}
\bigskip

\pr  $|d\cA_K| \leq p_{K}$ implies $|dA_k| \leq p_K$ and  hence  $\cA_K \in  e_K^{\frac34 - 4 \ep} \tilde  \cR_k$ as  in (\ref{peachy2}).  So it suffices to study
$\Xi_K(\cA) / \sZ^f(N,0) $  for   $\cA  \in  e_K^{\frac34 - 4 \ep} \tilde  \cR_K$.
We define 
\be
 \begin{split}
 E'_K(\La_K, \cA, \Psi_K) =  & \blan \bPsi_K,  D_K(\cA)  \Psi_K  \bran -\blan \bPsi_K,  D_K(0)  \Psi_K  \bran \\
E''_K(\La_K, \cA, \Psi_K)  =   & E_K ( \La_K, \cA, \cH^{\#}_K (\cA) \Psi_K ) \\ 
\end{split}
\ee
and  $E_K^*(\La_K)  = -E'_K(\La_K) + E_K''(\La_K) $.   Then we have  
 \be
\Xi_K(\cA)   =  Z'_K(0)  \int   D\Psi_K   \exp  \B(    -  \blan \bPsi_K,  D_K(0)  \Psi_K  \bran  +   E^{\star}_K(\La_K)  \B)   
\ee
Comparing this with  (\ref{fine}) we identify    a Gaussian integral with covariance $C_K(0) =   D_K(0) ^{-1}$ introduced in lemma \ref{strawberry}.
Then we have
\be
 \frac{\Xi_K(\cA)  }{ \sZ^f(N,0)}=    \int  e^{   E^{\star}_K (\La_K)} d \mu_{C_K(0)}(\Psi_K) 
\ee

We need estimates on $E'_k(\La_K),  E_K''(\La_K)$ and so $E_K^*(\La_K) $. 
For the first we  use the representation
$
D_K(\cA  )   = b_K - b_K^2Q_K(\cA)S_K(\cA) Q^T_K(-\cA)
$.
Using the decay  bounds for  $S_K(\cA)$ we find for $\cA \in \tilde \cR_K$
\be \label{cicada}
\| \blan \bPsi_K,  D_K(\cA)  \Psi_K  \bran\|_{h_K} \leq C h_K^2 |\La^{(K)} _K|  =  C h_K^2  |\bbT^0_{N-K}  |  
\ee
Indeed this is (\ref{lemon1}), (\ref{lemon2}), but with weight $h_K$,  specialized to $\de \Om_K =  \Om_K =\La_K  = \bbT^{-K} _{N-K} $
and $\de \Om_j = \emptyset$ for $j<K$.   Or see a similar bound in lemma 20 in  \cite{Dim15b}. 
Now we write  
\be \label{plum1}
 E'_K(\La_K, \cA, \Psi_K)  =   \frac{1}{2\pi i } \int_{|t| =e_K^{-\frac34 + 4\ep } } \frac{dt}{t(t-1)} \blan \bPsi_K,  D_K(t \cA)  \Psi_K  \bran  
\ee
Then  (\ref{cicada})   yields the bound
\be \label{plum3}
\|  E'_K(\La_K)   \|_{h_K} \leq  C  e_K^{\frac34  - 4 \ep }h_K^2  |\bbT^0_{N-K}  |  =C  e_K^{\frac14  - 4 \ep }   |\bbT^0_{N-K}  |  
\ee
For the second  (\ref{cinnamon2})  says that there is a constant $C_0$ such that $|\cH^{\#} _K (\cA) f | \leq C_0 \|f\|_{\infty}$. 
Hence  by our basic estimate (\ref{oscar1}),(\ref{oscar2}) 
\be
\|  E''_K(\La_K)   \|_{C_0^{-1} h_K}  \leq \| E_K(\La_K ) \|_{\bh_K}  \leq \one e_K^{\frac14  -7 \ep }   |\bbT^0_{N-K}  |  
\ee
This bound is satisfied by  $E'_K(\La_K) $  and hence $E^*_K(\La_K) $ as well.

Now we write
\be \label{pear1} 
\begin{split}
 \frac{\Xi_K(\cA)  }{ \sZ^f(N,0)}= &1+   \int \B ( \exp  \B(  e^{    E^{\star}_K (\La_K)}   - 1 \B)  d \mu_{C_K(0)}(\Psi_K) \\
= & 1 + \int_0^1 dt  \int    E^{\star}_K (\La_K)  e^{  t  E^{\star}_K (\La_K)} \   d \mu_{C_K(0)}(\Psi_K) \\ 
= & 1 + \int_0^1 dt  \int    E^{\star \star }_K (\La_K)   e^{  t  E^{\star \star }_K (\La_K)} \   d \mu_{I }(\Psi_K) \\ 
\end{split}
\ee
 Here  in the last step we changed to an identity covariance defining
   \be
    E^{\star \star}_K (\La_K), \cA, \bPsi_K, \Psi_K )
    =   E^{\star}_K (\La_K, \cA, \bPsi_K,C_K(0) \Psi_K )
\ee
As noted in lemma \ref{strawberry} $C_K(0)$ has a bounded kernel  $    | C_K  (0,\sx,\sy)  |  \leq    C(Mr_K)^3$
and then  
$|    C_K(0)f| \leq  C (MR_K) ^6  \|f\|_{\infty}.  $.   It follows
that
\be 
\|  E^{\star \star}_K(\La_K)   \|_1 \leq   
\|  E^{\star }_K(\La_K)   \|_{C( Mr_K)^6} \leq   
\|  E^{\star}_K(\La_K)   \|_{C_0^{-1} h_K} \leq 
\one     e_K^{\frac14 - 7 \ep}    |\bbT^0_{N-K}  |  
\ee
This is small since  $   |\bbT^0_{N-K}  |  
 = (Mr_K)^3$ is only logarithmic in $e_K$.
Then   
\be
\sup_{0 \leq t \leq 1} \|   E^{\star \star}_K (\La_K)  \exp \B( t  E^{\star \star}_K (\La_K) \B) \|_1
\leq   \|   E^{\star \star}_K (\La_K) \|_1 \exp\B( \| E^{\star \star}_K (\La_K)\|_1  \B) 
\leq   \one e_K^{\frac14 - 7 \ep}   |\bbT^0_{N-K}  |  
\ee
This implies  $ |\Xi_K(\cA) / \sZ^f(N,0)-1| $ is bounded by   $\one  e_K^{\frac14 - 7 \ep}   |\bbT^0_{N-K}  |    \leq    e_K^{\frac14 - 8 \ep}   $ and hence
the result.

\subsubsection{boson  integral - small field region}

On   $\La_K = \bbT^{-K} _{N-K} $ let 
\be 
Dm_K  =    \exp  \B(   -  \frac12  \| d\cA_K \|^2   \B)\    \de (  \cQ^* A_K  )    \de  ( \tau^* A_K) \   DA_K 
\ee 
Then we have from (\ref{snuffit})
\be  \label{snuffit2}
  \sZ^{\cS}  (N,e)  = \sZ'_K  \  \exp\B( \vep_K^0 | \bbT^0_{N-K} |\B) \    \int   \Xi_K\
   \chi_{K} \    Dm_K     \ee
We want to compare this with
\be
\sZ(N,0) = \sZ^f(N,0)\sZ^b(N,0) 
=  \sZ^f(N,0) \sZ'_K    \int         Dm_K
 \ee

\begin{lem} \label{pittsfield2}   Uniformly in $N$   
\be 
\frac{ \sZ^{\cS}  (N,e) }{\sZ(N,0)}   =  1 + \cO( e_K ^{\frac14 - 8 \ep})
\ee
\end{lem}

\pr   
The ratio is 
 \be \label{silt}
 \begin{split}
 & \frac{ \sZ^{\cS}  (N,e) }{\sZ(N,0)}
=  \exp\B( \vep_K^0 | \bbT^0_{N-K} |\B)  \left[  \frac{ \int    \B(  \Xi_K/   Z^f(N,0) \B)  \chi_K  Dm_K     }{ \int    \chi_K Dm_K} \right]
\left[\frac{ \int   \chi_K  Dm_K      }{ \int   Dm_K} \right] \\
 \end{split}
 \ee
 It suffices to show that each factor is sufficiently close to one.  
 The first factor is say $1 + \cO(e_K^6)$ by (\ref{v2}) and $ |\bbT^0_{N-K}| = (Mr_K)^3 $.
 The second  factor  is  $ 1 + \cO( e_K ^{\frac14 - 8 \ep})$
   by lemma \ref{early}. 
  For the third   factor we define   $\zeta_K = 1- \chi_K$
and write it as 
\be \label{lanky} 
1 -   \frac{ \int   \zeta_K  Dm_K      }{ \int   Dm_K }
\ee
To analyze this we follow lemma \ref{singsong}. 
The characteristic function $ \chi _K$ imposes that  $|d\cA_K| \leq p_K$ , and so $\zeta_K $ imposes that there is a bond
in $\La_K$ such that $|d\cA_K| \geq p_K$.  However   $|d\cA_K|  \leq CM \| dA_K \|_{\infty}$, a special case of (\ref{timely}),  so there must  be a bond where
$|dA_K| \geq (CM)^{-1}  p_K$.  Therefore
\be
 \zeta_K  \leq \exp \B( - p_K  + (CM)^2 p_K^{-1} \| dA_K \| ^2 \B) 
 \ee
Furthermore  $\| dA_K \| ^2 \leq \one \| d\cA_K \| ^2$, a special case of (\ref{sounder3}),   and so for $p_K$ sufficiently large
\be
\int   \zeta_K  \ Dm_K 
\leq  e^{-p_K}   \int  e^{  \frac14 \| d\cA_k \| ^2 }   Dm_K 
\ee
In the last integral   we have       $ \frac14 \| d\cA_k \| ^2  - \frac12 \| d\cA_k \| ^2 = - \frac14 \| d\cA_k \| ^2 $.  We restore the coefficient $-\frac14$ to $-\frac12$
by the change of variables $A_K  \to \sqrt 2 A_K$.     This  introduces a factor of $\sqrt 2$ raised to a power bounded by $|\La_K^{(K)} | = |\bbT^0_{N-K}| = (Mr_K)^3 $.
Thus for $p_K$ sufficiently large 
\be
\int   \zeta_K  \ Dm_K 
\leq  e^{-p_K +  (Mr_K)^3   }   \int   Dm_K   \leq e_K    \int   Dm_K 
\ee
 Thus     (\ref{lanky}) is $1 + \cO(e_K) $ and the result follows.

\subsection{the stability bound}   

Now we can prove the main result
\bigskip

\noindent
\textit{Proof of theorem \ref{theorem2} }: 
   Combining  lemma \ref{omega1}  and  lemma \ref{pittsfield2} we have for $e$ sufficiently small and uniformly in  $N$ 
\be
   \frac{  \sZ(N, e) }{  \sZ(N, 0)}    =   \frac{  \sZ^{\cL}  (N,e)}{   \sZ(N,0)  } +     \frac  { \sZ^{\cS} (N,e) }{ \sZ(N,0) }   = 1 + \cO(e_K^{\frac14- 8 \ep} )  
\ee
This gives the stability bound of theorem \ref{theorem2}  which we recall says 
\be
 \frac12  \leq     \B |  \frac{  \sZ(N, e) }{  \sZ(N, 0)} \B|  \leq  \  \frac32
\ee
It also shows that   $ \sZ(N, e) /  \sZ(N, 0)   \to 1$ as  $e \to 0$.

 \bigskip

\rems
\begin{enumerate}

\item   The results refer to the unit cube   $\bbT^{-N}_0$ with lattice spacing $L^{-N}$.  But  they  could easily be extended  to  a  lattice   $ \bbT^{-N}_{N'}$  with a volume   $L^{3N'}$.    But  controlling the  infinite volume  limit  $N' \to \infty$ is a difficult problem for this massless model.

\item  The restriction to tiny coupling constant $e$ is probably  not essential.  For any $e$ we will still have that $e_0 = L^{-\frac12 N}e$ is tiny  for $N$ sufficiently large.   We can 
still carry out our analysis as long as $e_k = L^{\frac12k}e_0 $ is small.  We would  just have to stop the iteration sooner. 

\item It should be possible to include   source terms in the partition function and thereby generate results for correlation functions.   See for example \cite{BIJ88}. 
In particular one could expect to  show the existence of the continuum limit $N \to \infty $. 

\end{enumerate}
\bigskip

\noindent{\textbf{Acknowledgement:}  I thank John Imbrie and David Brydges for helpful comments }

\newpage

\begin{appendix}

\section{Notation}  \label{notation} 

This is a  guide to the notation used in the text, with references to exact definitions. 
 Everything  refers  to decreasing sequence of small field regions
$ \Om_1 \supset \La_1 \supset \Om_2 \supset \La_2 \supset    \cdots  \supset   \Om_k \supset  \La_k$.

\begin{enumerate}

\item Gauge fields

\noindent
\begin{itemize}

\item   After $k$  steps one has fundamental gauge fields $ (A_0, A_1,  \cdots, A_k) $  with  the final field  $A_k$ on a unit lattice
and  previous fields   $A_j$  scaled down from a unit lattice by $L^{-(k-j)}$.    One is particularly interested in fields  
$ A_{k, \bom}=  ( A_{0, \Om_1^c},A_{1, \de \Om_1},  \cdots, A_{k-1, \de \Om_{k-1}}, A_{k, \Om_k} ) $ localized in $\de \Om_j = \Om_j - \Om_{j+1} $
which play a role in every subsequent step.

\item   $ \cQ_{k, \bom } $  is a multiscale averaging operator defined in (218) in \cite{Dim20}.  The field $\cA_{k, \bom } $ on $\tk$   is the minimizer of $\| d \cA \|^2$ 
subject to $\cQ_{k, \bom}\cA =  A_{k, \bom} $   and a gauge condition \cite{Dim20}.   This could be either Landau gauge or axial gauge with notation $\cA_{k, \bom} $
reserved for the Landau gauge and the axial gauge  denoted $\cA^{\sx} _{k, \bom} $.   They are linear functions $\cA_{k, \bom}  = \cH_{k, \bom} A_{k, \bom} $
of the fundamental fields. 

\item  There are also a single step minimizer  $A^{\min} _{k, \bom^+ } $ on $\tz$  depending on $\bom^+ = ( \bom, \Om_{k+1} )$.  It is    defined as the minimizer of $\| d\cA_{k, \bom} ^{\sx} \|^2 $ in $A_{k, \bom} =A_k$ on $\Om_{k+1}$ 
subject  to $\cQ A_k = A_{k+1} $ and an axial gauge condition,  with $A_{ k, \bom}$ fixed on $\Om^c_{k+1}$.  See (241) in \cite{Dim20}.

\item   If $\bom^+ = ( \bom, \Om_{k+1} )$ and    $ \cA^{(0)}_{k+1, \bom^+ } $  on $\tk$  is the field   $ \cA_{k+1, \bom^+ } $  before scaling to $\bbT^{-k-1}_{N-k-1}$.
We have that     $\cA^{(0)}_{k+1, \bom^+ }$  is gauge equivalent to  $\cH_{k, \bom} A^{\min}_{k, \bom^+ } $.

 \item   Fluctuation fields $Z_k$  on $\Om_{k+1} $ are defined by    $A_{k, \bom} = A^{\min} _{k, \bom^+ }  + Z_k$.   Then 
 $\cA_{k, \bom}$ is gauge equivalent to   $ \cA^{(0)}_{k+1, \bom^+ }  + \cZ_{k, \bom} $ where $\cZ_{k, \bom}  = \cH_{k, \bom} Z_k$. 
 The $Z_k$  satisfy an axial gauge condition  and are parametrized by $Z_k = C \tilde Z_k$, see \cite{Dim15}, \cite{Dim20}.
 
 \item   $ \cA^{(0)}_{k+1, \bom^+  (\sq) }$ and   $A^{\min}_{k, \bom^+  (\sq) } $    are versions of the above operators
approximately localized in $M$-cubes $\sq$.  They are  defined in   (\ref{croc1}),(\ref{croc2}).

\end{itemize}

\item Fermi fields

\begin{itemize} 

\item     After $k$ steps   we have fundamental fermi fields
$(\Psi_0, \Psi_1,  \cdots, \Psi_k ) $  with  the final field  $\Psi_k$ on a unit lattice
and  previous fields   $\Psi_j$  scaled down from a unit lattice by $L^{-(k-j)}$. 
 We are particularly interested in 
 fields  $ \Psi_{k, \bom}=  ( \Psi_{0, \Om_1^c},\Psi_{1, \de \Om_1},  \cdots, \Psi_{k-1, \de \Om_{k-1}}, \Psi_{k, \Om_k} )$.  

\item  $Q_{k, \bom}(\cA)  $ is a multiscale covariant averaging operator defined in (39) in \cite{Dim20}.   The field $\psi_{k, \bom} (\cA) = \psi_{k, \bom} (\cA, \Psi_{k, \bom}, \psi_{\Om_1^c} )   $ is a critical point
of  the action   
\be   \fS_{k, \bom} \B(\cA, \Psi_{k, \bom},  \psi \B) =    \blan  (\bPsi_{ k,  \bom}- Q_{ k,  \bom} (- \cA ) \bpsi),   \bb^{(k)}  (  \Psi_{ k,  \bom}- Q_{ k, \bom} (\cA ) \psi )  \bran_{\Om_1}
  +  \blan   \bpsi,(\fD_{ \cA } + \bar   m_k) \psi  \bran
  \ee 
for some constants  $\bb^{(k)}$, see (51)-(54)  in \cite{Dim20}.   It is linear in  $\Psi_{k, \bom}$.

\item   There are also single step critical points $\Psi^{\crit} _{k, \bom^+ } (\cA) $,   defined as the critical point in $\Psi_{k, \bom}= \Psi_k $  on $\Om_{k+1}$ of
\be     
 bL^{-1}
\blan   \bPsi_{k+1}-Q(-\cA)\bPsi_k, \Psi_{k+1}-Q(\cA)\Psi_k \bran_{\Om_{k+1}}  +     \fS_{k, \bom} \B(\cA, \Psi_{k, \bom},  \psi_{k, \bom} (\cA) \B) 
\ee
with $\Psi_{k, \bom} $ fixed on $\Om_{k+1}^c$.  
See lemma 3  in \cite{Dim20}.    

\item  $\psi^{(0)}_{k+1, \bom^+ }(\cA)  $  on   $\tk$  is the field   $ \psi_{k+1, \bom^+ } (\cA) $  before scaling to $\bbT^{-k-1}_{N-k-1}$.
Then      $\psi^{(0)}_{k+1, \bom^+ }(\cA)  =  \psi _{k, \bom^+ }(\cA, \Psi^{\crit} _{k, \bom^+ }(\cA),\psi_{\Om_1^c}) $, see   lemma 3 in \cite{Dim20} 

\item 
Fluctuation fields $W_k$ are defined on $\Om_{k+1}$ by $\Psi_{k, \bom}   =\Psi^{\crit}_{k, \bom} (\cA)  + W_k$.  Then 
$\psi_{k, \bom }(\cA)  = \psi^{(0)}_{k+1, \bom^+ }(\cA)  + \cW_{k, \bom^+} (\cA)$  where $\cW_{k, \bom^+} (\cA)=  \psi_{k, \bom} (\cA,  W_k,0) $,
see (71)-(73)  in \cite{Dim20}.

\end{itemize}

\item Miscellaneous.

\begin{itemize}

\item   A polymer $X$ is a connected union of $M$-cubes, and $|X|_M$ is the number of cubes.     The quantity $d_M(X) $ is defined by  specifying that $Md_M(X)$ is the length  of a shortest continuum  tree joining the cubes in $X$. 
 
\item If $X$ is specified as a union of $M$ cubes or $Mr_k$ cubes,  then $\tilde X$ is  $X$ enlarged by a layer of cubes of the same type, and $X^{\nat} 
\equiv ((X^c)^{\sim} )^c$  is $X$ shrunk by a layer of cubes. 

\item  $   \cR_{k, \bom}, \tilde   \cR_{k, \bom} $  are analyticity regions for a gauge field defined in (\ref{rk}).

\item  In passing from step $k$ to step $k+1$ four   new characteristic functions are introduced.   They are $\chi_{k+1} $ defined in (\ref{not})  which limits the size of fields 
  $\cA^{(0)} _{k+1, \bom^+(\sq) }$.  This is followed by $\chi'_k$  defined in (\ref{prime})  which limits the size  of  the fluctuation fields  $A_k - A^{\min} _{k, \bom(\sq) } $.
This  is not quite $Z_k$, but this is corrected  by $\chi^{\dagger}_k $  defined in (\ref{sister1})  which gives local bounds on $Z_k$.    The fluctuation integral is changed to unit covariance
which delocalizes the dependence on $Z_k$; the final characteristic function $\hat \chi_k$  defined in (\ref{sister2})  restores  local bounds on $Z_k$. 

\item
Expansions in   $\chi_{k+1},  \chi_k'$ define $\Om_{k+1}$.  Expansions  in    $\chi_k^{\dagger}, \hat \chi_k $  define  $\La_{k+1} $. 

\end{itemize}

\end{enumerate}

\section{Norms}   
\label{A}

The effective actions  in our renormalization group analysis will be expressed in terms of polymer functions which are elements of 
a Grassmann algebra.   We define some norms on such elements.     These can also be combined for mixed versions.

\subsection{single scale}  \label{ss}

 Consider  the unit lattice, say    $\tz$.  
      Fermi fields $ \Psi (\sx)  $ are the generators of a Grassmann algebra   
  indexed by        $\sx =  (x, \beta, \om )   $  with    $x \in  \tz  $,  $1 \leq  \beta \leq 4$, and  $\om =0,1$ and have the form $ \Psi( x, \beta, 0 )  = \Psi_{\beta}(  x) $
  and     $ \Psi( x, \beta, 1 )  = \bPsi_{k,\beta}(  x) $.  We consider  elements of the Grassmann algebra of the form
\be   \label{monty2} 
E(\Psi  )   =     \sum_{n=0}^{\infty}  \frac{1}{n!}  \sum_{\sx_1, \dots, \sx_n  }   E_n(  \sx_1,  \dots,  \sx_n)    
   \Psi( \sx_1 )  \cdots  \Psi( \sx_n) 
 \ee
 This is actually a finite sum since $\tz$ is finite.  
A norm with a parameter  $h>0$    is defined   by   
 \be
 \| E \|_h    =         \sum_{n=0}^{\infty}  \frac{ h^n  }{n!}     \sum_{ \sx_1,  \dots,  \sx_n  } | E_n( \sx_1,  \dots,  \sx_n)  |  
\ee

 \subsection{dressed fields} 
 Now  consider  a   fine   lattice, say   $\tk$.    
  Fermi fields $ \psi_k (\xi) $ are elements   of certain  Grassmann algebras.    
 indexed by    $\xi =  (x, \beta, \om )   $  with    $x \in  \tk  $,  $1 \leq  \beta \leq 4$, and  $\om =0,1$  and have the form
 $\psi (x, \beta, 0) = \psi_{\beta} (x) $ and $\psi (x, \beta, 1) = \bpsi_{\beta} (x) $.
 We  have in mind   smeared functions
 of  the  fundamental fields   like    $\psi  =  \psi_{k, \bom  }(\cA)$.
 We consider elements of the Grassman algebra of the form   
   \be  
 \begin{split}  
 E(  \psi )
   =  &  \sum_{n=0}^{\infty}  \frac{1}{n!}  \int  \  E_{n}(   \xi_1,  \dots,  \xi_n)    \psi( \xi_1 )  \cdots  \psi( \xi_n)  
     d\xi_1 \cdots  d\xi_n        \\
   \end{split}
\ee
Here  with $\eta = L^{-k}$ we define  $  \int d \xi   =  \sum_{x,\beta, \om}  \eta^3 $.   A norm on the kernel is defined by    
\be
\|  E_{n} \|_h =         \sum_{n=0}^{\infty}  \frac{ h^n  }{n!}   \int   | E_{n}(  \xi_1,  \dots,  \xi_n) |     d \xi_1 \cdots d \xi_n
\ee   
     
A   further  variation  allows treatment the   Holder derivative   $\de_{\al} \psi(x,y)$, defined for $|x-y| <1$,     as a separate field.   For $\zeta =(x,y, \beta, \om)$ we define $\de_{\al} \psi (\zeta)$ by 
$\de_{\al} \psi (x,y, \beta, 0 ) = \de_{\al} \psi_{\beta}  (x,y)  $ and $\de_{\al} \psi (x,y, \beta, 1 ) = \de_{\al} \bpsi_{\beta}  (x,y)  $. 
 An integral over $\zeta$ is 
$  \int d \zeta   =  \sum_{|x-y|<1,\beta, \om}  \eta^6 $.
We consider functions of the form 
  \be   \label{monty3} 
 \begin{split}  
 &E(  \psi_k, \de_{\al}  \psi_k   )  =    \sum_{n,m=0}^{\infty}  \frac{1}{n!m!}  \int    E_{nm}( \xi_1,  \dots,  \xi_n, \zeta_1, \dots,  \zeta_m)    \\
    &  \hs \hs   \psi(\xi_1 )  \cdots  \psi( \xi_n) \ \de_{\al}  \psi( \zeta_1 )  \cdots \de_{\al} \psi ( \zeta_m) 
\     d\xi_1 \cdots  d\xi_n    d\zeta_1 \cdots  d\zeta_m  
       \\
   \end{split}
\ee
Norms for the kernels  are defined for a pair of parameters  
  $\bh = (h_1, h_2)$    
by
\be   
   \| E \|_{  \bh }  
      =         \sum_{n,m=0}^{\infty}  \frac{ h_1^n h_2^m }{n!m!}     \int   | E_{nm}(   \xi_1,  \dots,  \xi_n, \zeta_1, \dots,  \zeta_m) |
   d \xi_1 \cdots d \xi_n    d \zeta_1 \cdots d \zeta_m
         \ee

 \subsection{multiscale}  
      Now we  consider a multiscale version .   As in the text   suppose we are given a decreasing sequence of small field  
 regions  $\bom  =   (\Om_1,  \Om_2,  \dots,   \Om_k) $ and  fermi fields    $ \Psi_{k,   \bom}   =  
     (  \Psi_{1, \de  \Om_1},   \dots,   \Psi_{k-1, \de \Om_{k-1}}, \Psi_{k, \de \Om_k})$ with $\de \Om_k = \Om_k$. 
  The   fields   are the generators of a Grassmann algebra indexed by     $\sx =  (x, \beta, \om )   $    as in section \ref{ss}   except that  now    $ \Psi_{j, \de \Om_j} (\sx) $ has  $x \in   \de  \Om_{j}^{(j)} \subset \bbT^{-(k-j)}_{N-k}$.     
 We consider elements of  the form     
\be   \label{monty4} 
E(  \Psi_{k,\bom}  )   =     \sum_{n_1, \cdots, n_k}^{\infty}  \frac{1}{n_1!  \cdots n_k! }  \int   E( \ud{ \sx_1},  \dots,  \ud{\sx_k} )    
   \Psi_{1, \de  \Om_1}(\ud{ \sx_1 ) }  \cdots \Psi_{k, \de \Om_k}( \ud{\sx_n} ) \ d\ud{\sx_1} \cdots,  d \ud{\sx_n }    
 \ee
 where for $\ud{ \sx_j } =( \sx_{ j,1} \cdots  \sx_{j,n_j}  ) $ 
 \be
  \Psi_{j, \de  \Om_j}(\ud{ \sx_j ) }  = \Psi_j ( \sx_{ j,1})  \cdots \Psi_j( \sx_{j,n_j} )  
 \ee
 and where     $\int  d \sx_i    = \sum_{x_i} L^{-3(k-i)} $. 
The norms on the kernels  with a multiweight $\bh = (h_1, \dots, h_k)$ are
 \be \label{monty5}
 \| E \|_ {\bh}   
      =          \sum_{n_1, \cdots, n_k}^{\infty}  \frac{h_1^{n_1} \cdots h_k^{n_k} }{n_1!\cdots n_k!}   \int  |  E( \ud{ \sx_1},  \dots,  \ud{\sx_k} )    |
   \ d\ud{\sx_1} \cdots,  d \ud{\sx_n }   \ee

\section{bound on  $T_{k, \bom}(\cA)$ } \label{B} 

We study the operator  $T_{k, \bom}(\cA)$ which is an approximate left inverse of the minimizer $\cH_{k, \bom}$ and is  defined on functions on $\Om_1 \subset\tk$
 by 
\be 
T_{k, \bom} (\cA)f = (\bb^{(k)})^{-1}Q_{k, \bom}(\cA) \B( \fD_{\cA} + m_k + P_{k, \bom} (\cA) \B)  f
\ee

\begin{lem} For $\cA \in  \tilde \cR_{k, \bom}$ there is a constant $C$ such that
\be
|T_{k, \bom} (\cA)  f | \leq C \| f \|_{\infty}
\ee
\end{lem}

\pr 
We have 
\be
T_{k, \bom} (\cA) =  (\bb^{(k)})^{-1}Q_{k, \bom}(\cA) \fD_{\cA} +    (\bb^{(k)})^{-1}m_k + Q_{k, \bom} (\cA)   
\ee
The last two terms satisfy the bound so it suffices to consider $ (\bb^{(k)})^{-1}Q_{k, \bom}(\cA) \fD_{\cA}$.
On $\Om_k$ this is  $b_k^{-1} Q_k(\cA) \fD_{\cA} $  and in  \cite{Dim15a}, \cite{Dim15b}  we show that
\be
|Q_k(\cA) \fD_{\cA}   f |  
 \leq     \one \B( 1 + e_k \|dA \|_{\infty, \Om_k } \B) \| f \|_{\infty}  
  \leq     \one \B( 1 + e_k^{\frac14 + 2 \ep }\B) \| f \|_{\infty}  
 \ee
where  last step follows  since $|d \cA| \leq   e_k^{- \frac34 + 2 \ep}$ on $ \Om_k$. 

On $\de \Om_j$  the operator $  (\bb^{(k)})^{-1}Q_{k, \bom}(\cA) \fD_{\cA}$  is  
$
b_j^{-1} L^{-(k-j)} Q_j(\cA) \fD_{\cA} 
$
 We  treat this by scaling it to the previous result.
 We have
  \be
  b_j^{-1} L^{-(k-j)} Q_j(\cA) \fD_{\cA}  f=    \B[    b_j^{-1} Q_j(\cA_{L^{k-j}})  \fD_{\cA_{L^{k-j}}}  f_{L^{k-j} } \B]_{L^{-(k-j)}}
\ee
where we used $(  \fD_{\cA} f )_{L^{k-j} } =L^{k-j}   \fD_{  \cA_{L^{k-j}  } }   f_{L^{k-j} }$.   The bracketed expression is now the same as the previous case
but on $\bbT^{-j}_{N-j}$ rather than $\tk$. We have therefore
\be
\begin{split} 
\B| \B[    Q_j(\cA_{L^{k-j}})  \fD_{\cA_{L^{k-j}}}  f_{L^{k-j} } \B]_{L^{-(k-j)}} \B| 
\leq   &       \one \B( 1 + e_j \|d \cA_{L^{k-j}}\|_{\infty, L^{k-j} \de \Om_j } \B) \| f \|_{\infty}  \\
\leq     &   \one \B( 1 + e_j   L^{-\frac32(k-j)} \|d \cA\|_{\infty,  \de \Om_j } \B) \| f \|_{\infty}  \\
\leq     &   \one ( 1 + e_j ^{\frac14 + 2 \ep}  ) \| f \|_{\infty}  \\
\end{split}
\ee
where the last step follows since  $| d \cA |  \leq L^{\frac32(k-j)} e_j^{- \frac34 + 2 \ep} $ on $ \de \Om_j$.

\section{more  bounds}  \label{lower} 

We  give a bound relating $\cA_{k, \bom}$  and the fundamental fields $A_j$.    First a preliminary estimate. 

\begin{lem} \label{flemma}
 Let  $X$ be a   union of unit blocks in $\tk$  with $\tilde X$ 
an enlargement by a layer of unit cubes. 
For a function $F$ on plaquettes in $\tk$ the $L^2$ norms satisfy
\be
\|\cQ_k^{(2)} F \|_X  \leq  \one \| F \|_{ \tilde X }
\ee
\end{lem}
\bigskip

\pr   For   $x \in \tk$  let   $  P_x   =   [ x, x+ e_{\mu}, x+ e_{\mu} + e_{\nu}, x + e_{\nu} ] $.
Then for   $y \in \tz$  we have
\be
       (\cQ_k^{(2)} F )(P_y)   =  \int_{|x-y| \leq \frac12}  L^{-2k}   \sum_{p \in P_x}   F(p)  =  \int_{|x-y| \leq \frac12}     \int  _{p \in P_x}   F(p)  
\ee
By the Schwarz inequality in the $p$ integral
\be    
\B|  \int  _{p \in P_x}   F(p) \B |^2  \leq    \int_{p \in P_x}   |F(p)|^2     
\ee
and    then by the Schwarz inequality in the $x$ integral 
\be
 | (\cQ_k^{(2)} F )(P_y) |^2 \leq  \int_{|x-y| \leq \frac12}    \int_{p \in P_x}   |F(p)|^2   
\ee
Then 
\be 
\begin{split}
& \|\cQ_k^{(2)} F \|^2_X 
  \equiv    \sum_{y:P_y \cap X \neq \emptyset }  | (\cQ_k^{(2)} F )(P_y) |^2 
\leq   \int_{x \in  \tilde  X}  \int_{p \in P_x}   |F(p)|^2   \\
=  & \int_{p \in \tilde X}  \int_{x\in \tilde X:  P_x \ni p}   |F(p)|^2   
\leq     \one \| F \|^2_{\tilde X}
\end{split}
\ee
This completes the proof. 
\bigskip

Now consider  $\cA_{k, \bom}$ which  satisfies $\cQ_{k, \bom}\cA_{k, \bom} = A_{k, \bom}$.  Then 
\be
 dA_{k, \bom}=  d\cQ_{k, \bom} \cA_{k, \bom} =\cQ^{(2)}_{k, \bom} d\cA_{k, \bom} 
\ee
  In $\Om_k$ this says  $dA_{k}=\cQ^{(2)}_{k} d\cA_{k, \bom} $.
It follows by  lemma  \ref{flemma} that 
\be \label{sounder3}
  \| d A_k \|^2_X  \leq \one  \| d \cA_{k, \bom} \|^2_{\tilde X} \hs  \tilde X \subset \Om_{k}
\ee
We  want to drop the restriction to $\Om_k$ here.

\begin{lem}   \label{tinkertoy}
Let  $X = \cup X_j$ where   $  X_j \subset  \de \Om_j$ is a union
of $L^{-(k-j)} $ cubes    whose enlargements at that scale also satisfy  $\tilde  X_j \subset  \de \Om_j$. Then
\be     \sum_{j =0}^k    \|   d A_j  \|^2_{ X_j}  
  \leq \one   \| d \cA_{k, \bom} \|^2
  \ee  
 \end{lem}  
\bigskip

\pr  $A_j$ is a function on $\de \Om^{(j)}_j \subset \bbT^{-(k-j)}_{N-k}$.  Therefore $A_{j,L^{k-j}}$ is a function on a subset of  the unit lattice
 $\bbT^{0}_{N-j}$ and on $L^k \de \Om_j$ we have $  \cQ^{(2)}_j d\cA_{k, \bom, L^{k-j}} = dA_{j, L^{k-j}} $. Then by a bound like  (\ref{sounder3})
\be
 \|dA_j \|^2_{X_j}   = \|dA_{j, L^{k-j}} \|^2_{L^{k-j}X_j} \leq  \one  \| d\cA_{k, \bom, L^{k-j}}  \|^2_{(L^{k-j}X_j)^{\sim}} = \one  \| d\cA_{k, \bom}  \|^2_{\tilde X_j} 
 \ee
Summing over $j$ gives the result.
\bigskip

\end{appendix}

\end{document}